\documentclass[11pt,twoside,british,dutch,english]{article}
\usepackage[T1]{fontenc}
\usepackage[latin9]{inputenc}
\usepackage{float}
\usepackage{amsmath}
\usepackage{amssymb}
\usepackage{cancel}
\usepackage{graphicx}
\usepackage{wasysym}
\usepackage{nomencl}
\pdfoutput=1

\providecommand{\makenomenclature}{\makeglossary}
\makenomenclature

\makeatletter

\providecommand{\tabularnewline}{\\}

\usepackage[margin=1in,headheight=13.6pt,includeheadfoot]{geometry}
\usepackage{fancyhdr,etoolbox}

\renewcommand\subsectionmark[1]{\relax}
\patchcmd{\@part}{\markboth{}{}}{\markboth{\thepart. #1}{}}{\typeout{done patching part!}}{\typeout{oh dear! could not patch part command...}}
\patchcmd{\@spart}{\nobreak}{\nobreak\markboth{\thepart. #1}{}}{\typeout{done patching starred part!}}{\typeout{oh dear! could not patch starred part command...}}

\fancypagestyle{plain}{%
  \fancyhf{}%
}
\fancypagestyle{simple}{%
  \fancyhf{}
  \fancyfoot[C]{\thepage}
}
\@ifundefined{definecolor}
 {\usepackage{color}}{}
\usepackage{pgfplots}
\usepackage[font={small}]{caption}
\makeatother
\usepackage[numbers,sort&compress]{natbib}
\usepackage{babel}
\usepackage[titles]{tocloft}
\usepackage{chngcntr}
\usepackage{tikz}
\usetikzlibrary{trees}
\usetikzlibrary{decorations.pathmorphing}
\usetikzlibrary{decorations.markings}
\usetikzlibrary{shapes.misc}
\numberwithin{equation}{section}
\usepackage[section]{placeins}

\usepackage{hyperref}
\usepackage{tocloft}
{}

\tocloftpagestyle{simple}
\cftpagenumbersoff{part}
\makeatother

\usepackage{babel}
\begin{document}
\selectlanguage{british}%
\begin{titlepage}
\vfil
\begin{figure}[h]
 \centering \includegraphics[scale=0.3]{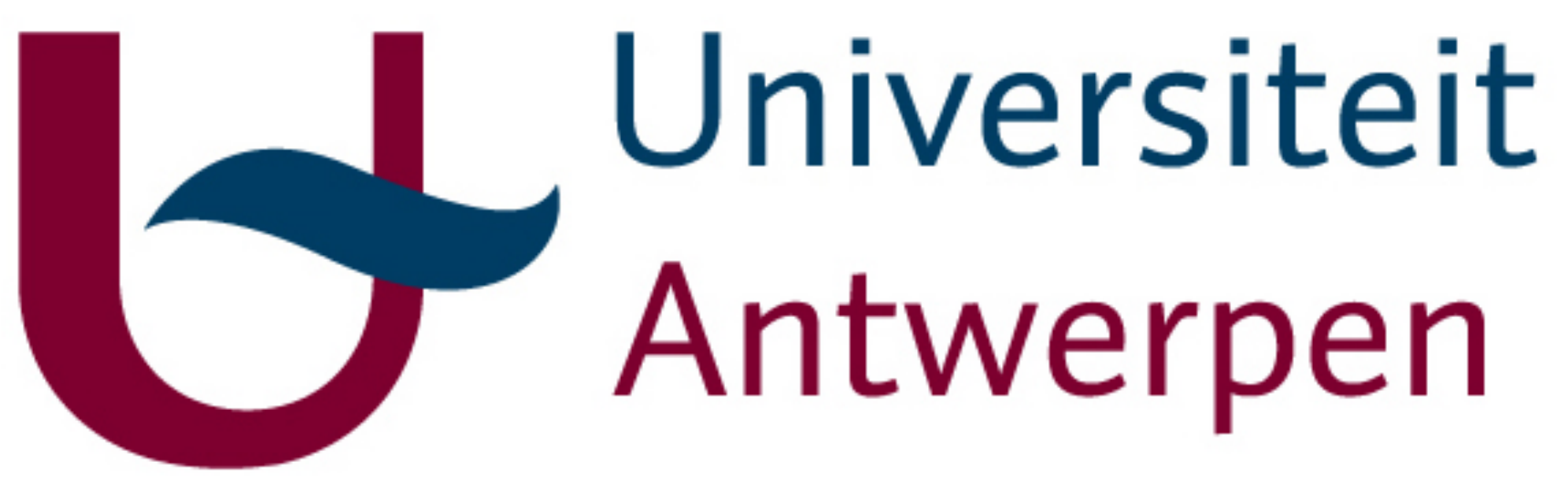} 
\end{figure}

\begin{center}
\begin{Large}Faculteit Wetenschappen \\
Departement Fysica \\
 \end{Large} 
 \par\end{center}
\vfill
\begin{center}
\begin{Huge} Quantum Chromodynamics at small Bjorken-$x$ \\
 \end{Huge}
\par\end{center}

\begin{center}
\begin{LARGE}Kwantumchromodynamica bij kleine Bjorken-$x$\\
 \end{LARGE} 
\par\end{center}

\begin{center}
\begin{Large} Pieter Taels\\
\end{Large}
\par\end{center}

\begin{center}
\vfill{}
\par\end{center}

\begin{flushleft}
\begin{tabular}{l}
\selectlanguage{english}%
\selectlanguage{british}%
\tabularnewline
\selectlanguage{english}%
\selectlanguage{british}%
\tabularnewline
\selectlanguage{english}%
\selectlanguage{british}%

Promotor: Prof. dr. Pierre Van Mechelen\tabularnewline
\selectlanguage{english}%
\selectlanguage{british}%
\tabularnewline
Proefschrift voorgelegd tot het behalen van de \tabularnewline
graad van doctor in de wetenschappen: fysica \tabularnewline
aan de Universiteit Antwerpen\hspace{7.79cm}Antwerpen, juli 2017\tabularnewline
\end{tabular}
\par\end{flushleft}

\end{titlepage} 

\selectlanguage{english}%
\newpage{}\thispagestyle{empty}
\hbox{}

\newpage{}

\thispagestyle{simple}
\pagenumbering{roman}
\cleardoublepage

\newpage{}

\section*{Samenvatting}

\addcontentsline{toc}{section}{Samenvatting}

\selectlanguage{dutch}%
Door de komst van steeds krachtigere deeltjesversnellers, zoals RHIC
en de LHC, wordt het mogelijk om kwantumchromodynamica (QCD) te bestuderen
bij een zeer hoge energie, in botsingen die gevoelig zijn aan de gluondichtheid
van het proton of van de atoomkern. Vaak wordt deze dichtheid zo groot
dat niet-lineaire effecten een rol beginnen te spelen. QCD in dit
regime staat bekend als \textquoteleft small-$x$ QCD', en wordt met
grote nauwkeurigheid beschreven door het zogenaamde Color Glass Condensate
(CGC): een effectieve theorie voor een proton of atoomkern bij hoge
energie. In deze thesis presenteer ik, na een inleiding over small-$x$
QCD en over het CGC, twee projecten waarin het CGC wordt toegepast
op twee verschillende problemen.

In het eerste project wordt de voorwaartse productie van twee zware
quarks in proton-nucleus botsingen bestudeerd. Gewoonlijk wordt zo'n
reactie beschreven binnen een zeker factorisatieschema, waarbij het
\textquoteleft harde deel' van de werkzame doorsnede, d.i. het deel
dat we kunnen berekenen in storingstheorie, wordt afgesplitst van
de niet-perturbative partondichtheden (PDFs), die de structuur van
het hadron in termen van quarks en gluonen beschrijft. Aangezien deze
partondichtheden niet perturbatief zijn, kunnen ze niet analytisch
verkregen worden vanuit de QCD Lagrangiaan, maar moeten ze gemeten
worden in experimenten. Afhankelijk van de precieze configuratie van
de impulsschalen in de reactie, kunnen er twee verschillende factorisatieschemas
op ons probleem worden toegepast: High-Energy Factorization (HEF)
en Transverse Momentum Dependent (TMD) factorisatie. In dit werk wordt
de berekening van de voorwaartse zware-quarkproductie in een ander
framework uitgevoerd, dat van het CGC, en interessant genoeg zitten
de resultaten van zowel het HEF als het TMD schema in deze berekening
vervat. Daarnaast kunnen de \textendash normaal gezien niet-perturbatieve\textendash{}
partondichtheden zowel analytisch als numeriek in het CGC gemodelleerd
worden, waarvan ik de resultaten presenteer.

Het tweede project kadert in de fysica van zware ionen. In een botsing
van twee zware atoomkernen wordt het zogenaamde Quark-Gluon Plasma
(QGP) gevormd: een toestand van QCD waarin quarks en gluonen een fractie
van een seconde vrij zijn, en dit bij een enorm hoge dichtheid en
temperatuur. Energetische \textquoteleft harde' jets, die gevormd
werden tijdens de botsing van de atoomkernen, moeten zich doorheen
het QGP voortbewegen alvorens ze de detector bereiken, en zullen beïnvloed
worden door hun interactie met het medium. Dit fenomeen, of eerder
verzameling van fenomenen, staat bekend als jet quenching, en is een
van de voornaamste hulpmiddelen om het QGP te bestuderen. In dit werk
wordt één bepaald aspect van jet quenching bestudeerd, namelijk de
diffusie in transversale impuls van een energetisch deeltje dat zich
door een nucleair medium voortbeweegt. We gebruiken technieken uit
small-$x$ QCD en het CGC, om de radiatieve correcties aan deze diffusie,
ten gevolge van de straling van zachte gluonen, trachten te hersommeren
in de logaritmische benadering, de zogenaamde leading logarithmic
approximation (LLA). Het resultaat van deze analyse is een niet-lineaire
evolutievergelijking in een medium, die we uiteindelijk enkel kunnen
oplossen in de double leading logarithmic approximation (DLA). We
presenteren een goed onderbouwd theoretisch kader voor dit probleem,
bespreken uitgebreid de verbanden met het CGC, en vergelijken met
de literatuur.

\selectlanguage{english}%
\newpage{}

\thispagestyle{simple}

\section*{Abstract}

\addcontentsline{toc}{section}{Abstract}

With the advent of particle accelerators such as RHIC and the LHC,
which are more and more powerful, it becomes possible to study quantum
chromodynamics (QCD) in very high energy collisions, in which the
gluon content of the proton or nucleus is probed and its density becomes
often large enough for nonlinear effects to play a role. This so-called
small-$x$ regime of QCD is very well described by the Color Glass
Condensate (CGC), an effective theory for a proton or nucleus at high
energies. In this thesis, after an introduction to small-$x$ QCD
and to the CGC, we present two projects in which the CGC is applied
to two different problems.

In the first project, we study forward heavy-quark production in proton-nucleus
collisions. Such a process is usually studied in a certain factorization
scheme, in which the perturbatively calculable \textquoteleft hard
part' of the cross section is separated from the nonperturbative parton
distribution functions (PDFs), which encode the hadron structure in
terms of quarks and gluons. Being nonperturbative, these PDFs cannot
be obtained directly from first principles within QCD, but rather
need to be measured experimentally. Moreover, depending on the precise
configuration of the transverse momentum scales in the problem, two
different factorization schemes are relevant: High-Energy Factorization
(HEF) and Transverse Momentum Dependent (TMD) factorization. Interestingly,
taking an alternative approach and calculating the cross section within
the CGC, we reproduce the results obtained in both the TMD and the
HEF scheme in the appropriate limits. Moreover, since the CGC encompasses
a well-substantiated model for the nucleus, this matching allows us
to make both analytical and numerical predictions for the, in principle,
nonperturbative PDFs.

The second project is devoted to an important problem in heavy-ion
physics. In the collision of heavy nuclei, a state of matter known
as the Quark-Gluon Plasma (QGP) is created, in which quarks and gluons
shortly exist freely in a regime characterized by an extremely high
density and temperature. Hard jets, produced in the scattering of
the two nuclei, have to travel through the QGP before reaching the
detector, and will be attenuated as a result of their interaction
with this medium. This phenomenon, or rather set of phenomena, is
known as jet quenching and is one of the principal probes to investigate
the properties of the QGP. To study jet quenching, we focus on the
transverse momentum broadening of a hard particle traveling through
a nuclear medium, and we employ small-$x$ techniques to attempt to
resum the single leading logarithmic corrections, due to soft gluon
radiation, to this broadening. Although, ultimately, we can only solve
the resulting in-medium evolution equation in the double leading logarithmic
approximation, we do present a concise framework for the problem,
and draw a detailed comparison with the CGC and with the existing
literature.

\newpage{}
\thispagestyle{simple}
\makeatletter
\renewcommand{\l@section}{\@dottedtocline{1}{1.5em}{2.6em}} 
\renewcommand{\l@subsection}{\@dottedtocline{2}{4.0em}{3.6em}} 
\renewcommand{\l@subsubsection}{\@dottedtocline{3}{7.4em}{4.5em}} 
\makeatother

\tableofcontents{}

\thispagestyle{simple}

\newpage{}

\thispagestyle{simple}

\section*{Introduction}

\addcontentsline{toc}{section}{Introduction}

The main topic of this thesis is the study of the structure of hadrons,
such as the proton or the neutron. Hadrons are composite particles
built from quarks and gluons: the fundamental elementary particles
that participate in the strong interaction, which, in the Standard
Model, is described by the theory of quantum chromodynamics (QCD).

Quantum chromodynamics is a non-Abelian gauge theory, characterized
by two main features. First, the strong coupling constant $\alpha_{s}$
is not actually a constant, but rather a function of the momentum
scale of the process at hand. For large exchanges of momentum, $\alpha_{s}$
is small and one can apply perturbation theory, while in the opposite
regime of small exchanged momenta, $\alpha_{s}$ is large and therefore
perturbation theory fails. Unfortunately, most information on the
hadron structure is \textquoteleft hidden' in this nonperturbative
domain of QCD.

A second feature of QCD is confinement: the property that the degrees
of freedom of the theory, i.e. quarks and gluons or, collectively,
\textquoteleft partons', cannot be observed separately, but are always
grouped into hadrons. 

The way to deal with these particular intricacies of QCD is through
factorization, which is a consistent framework to separate the perturbative
from the nonperturbative part of a cross section. A well-known example
is deep-inelastic scattering (DIS), for example in electron-proton
collisions. The electron interacts with the proton through the emission
of an energetic and highly virtual photon, which doesn't scatter off
the proton as a whole, but rather kicks one of the proton's constituents
out. The cross section can then be written as the product of the partonic
part, which involves the electron-quark or electron-gluon interaction,
on the one hand, and the structure function $F_{2}$ on the other
hand. The virtuality of the photon provides the aforementioned large
momentum scale, at which the coupling constant is small enough for
the partonic part to be calculable in perturbation theory, up to a
certain order in $\alpha_{s}$. Therefore, the nonperturbative information
on the process is now only contained in the structure function $F_{2}$,
which can be written as a function of the so-called parton distribution
functions (PDFs). Roughly speaking, these PDFs describe the probability
to find a quark or gluon with a fraction $x$ of the energy of the
proton.

Information on the structure of the proton, and of hadrons in general,
in terms of the QCD degrees of freedom, is thus generally encoded
inside nonperturbative parton distribution functions. These PDFs are
beyond the reach of perturbation theory, but rather need to be extracted
from the experiment. They are, however, scale-dependent, and their
evolution of one transverse momentum scale to the other can be calculated
with great precision in perturbative QCD. The importance of the ensuing
\textquoteleft evolution equations' cannot be overstated, since they
allow to plug in the PDFs, measured in a certain experiment and at
a certain scale, into any other factorizable cross section.

In this thesis, we will concentrate in particular on hadron collisions
at very high energies, in which those constituents are probed that
carry a very small energy fraction (Bjorken-$x$) of their parent.
This regime of QCD is mainly dominated by gluons, because the probability
of a gluon emission scales with the inverse of the energy, and hence
favors arbitrarily low gluon energies. An important consequence is
that, at large enough collider energies, a high-density regime of
QCD is reached. Moreover, at a certain scale, the gluon density becomes
so high that nonlinear effects start \newpage{}

\thispagestyle{simple}to play a role. In particular, the fast growth
of the gluon density towards small energy fractions $x$ is damped
due to the rising importance of gluon recombinations, resulting in
the so-called \textquoteleft saturation' of the gluon density function. 

Luckily for us, a high-density regime implies large momenta, which
in turn corresponds to a small coupling constant. Hence, the applicability
of perturbative QCD is extended to the description of the hadron structure
in or near the saturation regime. The corresponding effective theory
is known as the Color Glass Condensate (CGC), and combines the McLerran-Venugopalan
(MV) model: a semi-classical description of a large nucleus or highly
energetic proton, with a field theoretical nonlinear evolution equation
called BK-JIMWLK, after its authors.

The first project we present in this work, in collaboration with Dr.
Cyrille Marquet and Dr. Claude Roiesnel, is related to the forward
production of two heavy quarks in proton-nucleus collisions. If the
quarks have large transverse momenta and are almost back-to-back in
the transverse plane, this process is not characterized by one, but
by two momentum scales: the large transverse momentum of a single
outgoing quark, and the small imbalance between both momenta. Such
a process, with two ordered scales, can be described within \textquoteleft transverse-momentum
dependent' (TMD) factorization, in which one works with so-called
TMD PDFs, or simply TMDs, which extend the regular PDFs by including
the information on the transverse momentum of the parton inside the
proton or, in our case, nucleus. TMD factorization is a framework
which contains many intricacies, in particular, the TMDs become dependent
on the process under consideration. As a result, not one but six different
gluon TMD distribution functions play a role in the cross section
for forward heavy diquark production. Interestingly, three of them
only couple through the mass of the heavy quarks, and correspond to
linearly polarized gluons within the unpolarized nucleus. 

In the small-$x$ limit, which \textendash as we will show\textendash{}
applies when the diquark production is forward, we bring in information
from the point of view of the Color Glass Condensate. Not only can
we calculate the cross section within the CGC, we can also model the
gluon TMDs which, again, are in general not perturbatively calculable.
Doing so, we obtain analytical expressions for the gluon TMDs in the
MV model, as well as numerical results for the evolution of the TMDs
with a lattice QCD implementation of JIMWLK.

It was my task to extend the earlier work of Marquet and collaborators
(Ref. \protect\cite{Cyrille}) on forward dijet production, by including the
heavy quark mass. I derived analytically the CGC cross section, Eq.
(\ref{eq:finalCGCcrosssection}), as well as the expressions for the
gluon TMDs in the MV model at finite-$N_{c}$, four of which are new
(Eqs. (\ref{eq:F1finiteNc-1}), (\ref{eq:H1finiteNc-1}), (\ref{eq:F2finiteNc-1})
and (\ref{eq:H2finiteNc-1})). Furthermore, I implemented them in
the numerical code, written by Claude Roiesnel, and analyzed the results
of the simulations (Figs. \ref{fig:AllgTMDs0}, \ref{fig:AllgTMDs1000}).

The second project, in collaboration with Dr. Edmond Iancu, is another
application of small-$x$ physics, this time to a problem within heavy-ion
physics. A central concept in heavy-ion physics is \textquoteleft jet
quenching': the attenuation of energetic partons when traveling through
a nuclear medium, such as the quark-gluon plasma which is created
in high-energy collisions of two nuclei. The way these partons are
influenced by their interaction with the medium, and the resulting
signatures in the detector, are an important probe to the properties
of the plasma itself. We focus on one aspect of jet quenching, namely
\textquoteleft transverse momentum broadening': the average transverse
momentum obtained by the energetic parton as a result of the scatterings
off the medium constituents. \newpage \thispagestyle{simple}Generalizing the JIMWLK equation to be
applicable to an extended medium, large logarithmic corrections due
to gluon radiation can be resummed and absorbed
into a renormalization of the transverse momentum broadening, or rather,
the \textquoteleft jet quenching parameter' $\hat{q}$ associated
with it. It was the aim of the project to extend the currently available
calculations (Refs. \protect\cite{Al,Blaizot2014RG,Iancu}) to a greater precision,
i.e. resumming all the single leading logarithmic corrections rather
than the double large logarithms. We were, however, unable to do this,
and at present it is not even clear whether such a thing is feasible
and/or meaningful. I therefore limit myself to a careful explanation
of \textendash my understanding of\textendash{} the approach of Iancu
to the problem (Ref. \protect\cite{Iancu}) supplemented with an outline of,
and comparison with, the work in Ref. \protect\cite{Al}. Even without reaching
our original goal, the project calls for an in-depth analysis of gluon
radiation, as well as its interaction with a nuclear medium, and as
such it may deepen the understanding of the matters in the preceding
parts of the thesis.

Before presenting both projects in Parts \ref{part:projectCyrille}
and \ref{part:edmond}, I introduce the basics of small-$x$ physics
through the example of DIS, in Part \ref{part:DIS}, followed by an
introduction on the Color Glass Condensate in Part \ref{part:CGC}.
For these first parts, I closely follow the very educational review
on the CGC in Ref. \protect\cite{EdmondRev}, supplemented with the book by
Kovchegov and Levin, Ref. \protect\cite{KovchegovLevin}, the review by Gelis
\protect\cite{Gelis2013}, and of course what I learned during the many discussions
with Dr. Iancu and Dr. Marquet. Further references are specified in
the main text. 

\section*{Acknowledgements}

\addcontentsline{toc}{section}{Acknowledgements}

First, I am incredibly indebted to Prof. Pierre Van Mechelen, for
accepting me in the Particle Physics Group in Antwerp and allowing
me to make a PhD. He always states that, since he is primarily an
experimentalist, he was merely my official advisor and the actual
guidance was done by others, but that is way too modest. I learned
a tremendous amount from him, not only from being his assistant for
the courses on special relativity and particle physics, but most importantly
from the countless discussions we had on both the theoretical and
experimental aspects of QCD. Second, I am grateful to Dr. Igor Cherednikov,
who warmly welcomed me in his small theory group during the first
years of my PhD, and patiently introduced me to QCD. I also thank
my fellow students in this group, Dr. Frederik Van der Veken and Dr.
Tom Mertens, for the pleasant atmosphere, and for their help and support. 

Furthermore, this thesis would never have been written without the
help of Dr. Edmond Iancu and Dr. Cyrille Marquet. Each of them welcomed
me for half a year in their lab in Paris, taught me about small-$x$
physics and the CGC, involved me in their research, answered hundreds
of questions and read and corrected my thesis meticulously. Indeed,
the bulk of the guidance was done by them, for which I am incredibly
grateful.

For one year, I had the pleasure to share an office in Antwerp with
Dr. Cristian Pisano. I am not only thankful for his friendship, but
also for introducing me to the physics of TMDs and for involving me
in his research. Cristian was, and still is, another victim of my
emails loaded with questions, which he always answered patiently and
in great detail, through which I learned a lot. In addition, I am
very grateful to him for inviting me to Pavia, for his help preparing
my talk in Trento, and for correcting parts of my thesis. I am also
thankful for the pleasant collaboration with Prof. Umberto D'Alesio
and Prof. Francesco Murgia, who invited me and Cristian to Cagliari,
which resulted in a nice project together. 

I gratefully acknowledge the help of Prof. Al Mueller, explaining
me his work on the radiative corrections to transverse momentum broadening,
and of Dr. Stéphane Peigné, who explained me with admirable patience
the BDMPS formalism. Furthermore, I am very grateful to Dr. Krzysztof
Kutak for kindly welcoming me for an internship in his group in Krakow,
for involving me in his work, and for the many stimulating discussions.
Thanks as well to my fellow PhD student Merijn van de Klundert, for
reading parts of my thesis and for the many enlightening conversations
that ensued, and to Dr. Hans Van Haevermaet for answering many questions
on QCD. Thanks to Prof. Nick Van Remortel for the numerous interesting
discussions, and in particular for the very nice trip we made to India,
and thanks to Thomas Epelbaum and his wife, Véronique Mariotti, for
their friendship and for inviting me in Montréal.

Thanks to the members of the jury that I didn't mention yet: Prof.
Francesco Hautmann, Prof. Etienne Goovaerts and the chairman: Prof.
Michiel Wouters, for carefully reading the manuscript and giving numerous
helpful comments and corrections.

A very special thanks goes to Sarah Van Mierlo, the secretary of our
group. For every question that wasn't related to physics, every practical
or even emotional problem, every remark or joke, I could go to her.
Thanks, Sarah, for all your help and all your kindness, and for making
our group such a pleasant place.

Thanks as well to Hilde Evans, the secretary of the physics department,
for kindly taking care of all the practical and administrative issues
which writing a thesis brings.

Thanks to all my friends in physics, making a day spent at the university
a nice day: thanks Ben, Christophe, Mario, Matthias, Maxim, Maya,
Merijn, Nikolas, Roeland, Ruben and Sam \textendash I am sure we stay
in touch!

During the largest part of my PhD, I lived in Roeach, one of the houses
of L'Arche Antwerp: a community in which people with and without an
intellectual disability live together\footnote{See http://www.larche.org (international) or http://www.arkantwerpen.be
(Antwerp).}. For more than four years, Roeach has been my house, and for much
longer L'Arche has been a home. The list of people to acknowledge
is endless, so let me just thank my dear housemates and friends of
Roeach: Tony, Korneel, Carina, Sonja, Sabien and Greta.

Finally, thanks to my close group of friends from Antwerp (and Leuven
and The Netherlands, nowadays), thanks to my loving parents and my
dear two sisters, and of course thanks to my beloved Aurélie, whom
I love even more than physics!
\begin{flushright}
June 26th, 2017 
\par\end{flushright}

\thispagestyle{simple}

\newpage{}

\thispagestyle{simple}
\pagenumbering{arabic}

\part{\label{part:DIS}Deep-inelastic scattering at small Bjorken-$x$}

\section{Deep-inelastic scattering}

We start this work with a quick review of deep-inelastic scattering
(DIS) (Refs. \protect\cite{HalzenMartin,Mueller01,Peskin,EdmondRev}), in
which a nucleon or nucleus is probed by a highly energetic lepton.
Not only did DIS play a crucial role in the early development of QCD
in the seventies: much later, in the early nineties, the HERA measurements
of DIS at high energies sparked a large interest in the phenomenology
of small-$x$ physics. In addition, DIS is a particular suited case
to introduce the concept of evolution, which will play a central role
in this thesis.

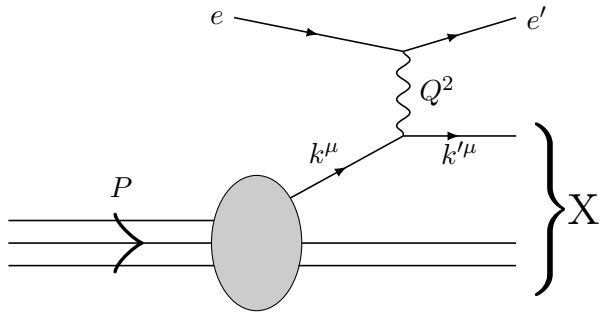
\begin{figure}[t]
\begin{centering}
\begin{tikzpicture}[scale=1.5] 

\tikzset{photon/.style={decorate,decoration={snake}},
		electron/.style={ postaction={decorate},decoration={markings,mark=at position .5 with {\arrow[draw]{latex}}}},      	gluon/.style={decorate,decoration={coil,amplitude=4pt, segment length=5pt}}}

\draw[semithick] (0,0) --++ (2.05,0);
\draw[semithick,postaction={decorate},decoration={markings,mark=at position .6 with {\arrow[very thin,scale=8]{to}}}] (0,-.2) --++ (2,0);
\draw[semithick] (0,-.4) --++ (2.05,0);
\node at (1,.3) {$P$};
\filldraw [fill=black!20] (2.2,-.2) ellipse (.4 and .6);
\node [rotate=90,scale=.8] at (2.2,-.1) {$$};
\draw[semithick] (2.6,-.2) -- (4.5,-.2);
\draw[semithick] (2.57,-.4) -- (4.5,-.4);

\draw[semithick,photon] (3.5,1.5) --++ (0,-.75);
\node at (3.8,1.15) {$Q^2$};
\draw[electron,semithick] (2.5,.2) -- (3.5,.75);
\node at (2.8,.6) {$k^\mu$};
\draw[electron,semithick] (3.5,.75) --++(1,0);
\node at (4,.6) {$k'^\mu$};

\node [scale=2,ultra thin] at (4.8,.1) {$\Bigg\}$};
\node [scale=1.5,ultra thin] at (5.1,.1) {X};

\draw[electron,semithick]  (2,1.8)  node[left] {$e$}-- (3.5,1.5);
\draw[electron,semithick] (3.5,1.5) --++(1,.3) node[right]{$e'$};

\end{tikzpicture} 
\par\end{centering}
\caption{\label{fig:DIS}Inclusive deep-inelastic scattering (DIS) }
\end{figure}

For definiteness, let us look at the case of deep-inelastic scattering
of an electron off a proton, as is illustrated in Fig. \ref{fig:DIS}.
The electron interacts with the proton through the exchange of a highly
virtual photon, which kicks out one of the proton's constituents,
thus breaking the proton and probing its structure. In pre-QCD language,
these constituents are collectively called \textquoteleft partons',
but we know them of course as quarks and gluons. 

In the cross section, one can separate the lepton current from the
hadronic part of the interaction, and therefore we can just consider
the virtual photon-proton part of the scattering process.
Furthermore, we work in a so-called infinite momentum frame (IMF),
in which the proton has a very large momentum $P$ along the $z$-axis:
\begin{equation}
\begin{aligned}P^{\mu} & =\left(\sqrt{P^{2}+M^{2}},\mathbf{0}_{\perp},P\right)\simeq\left(P,\mathbf{0}_{\perp},P\right),\end{aligned}
\end{equation}
and we choose the frame such that the virtual photon has momentum
\begin{equation}
\begin{aligned}q^{\mu} & =\left(q_{0},\mathbf{q}_{\perp},0\right).\end{aligned}
\end{equation}
The kinematics of the photon-proton scattering is fully determined
by two parameters: the virtuality of the photon $Q^{2}=-q^{2}$, and
the dimensionless Bjorken-$x$ variable, defined as:
\begin{equation}
\begin{aligned}x & \equiv\frac{Q^{2}}{2P\cdot q}=\frac{Q^{2}}{s+Q^{2}-M^{2}}\simeq\frac{Q^{2}}{s+Q^{2}}\simeq\frac{Q^{2}}{s},\end{aligned}
\label{eq:Bjorkenx}
\end{equation}
where $s=\left(P+q\right)^{2}$ is the center-of-mass energy, and
where the last equality holds in the high-energy regime or \textquoteleft Regge
limit' $s\gg Q^{2}$. Since in the lab frame $Q^{2}=2EE'\left(1-\cos\theta\right)$,
with $E$ and $E'$ the energy of the electron before and after the
scattering with an angle $\theta$, $Q^{2}$ can be measured experimentally,
and so can $x$, via Eq. (\ref{eq:Bjorkenx}).

In Eq. (\ref{eq:Bjorkenx}), we assumed that the virtuality of the
photon is much larger than the proton mass: $Q^{2}\gg M^{2}$. In
this case, the scattering will be inelastic and the proton is broken
up: the photon interacts not with the proton as a whole, but with
one single parton which it kicks out. In comparison with the large
virtuality of the photon, the virtualities $k^{2}$ and $k'^{2}$
of the parton can be neglected, and the parton is assumed to be on
shell before and after the interaction. Furthermore, throughout this
analysis (and, in fact, the largest part of this thesis) the partons
are taken to be massless. Assuming that $k^{\mu}\simeq\xi P^{\mu}$,
i.e. introducing the fraction $\xi$ of the momentum of the proton
that is carried by the parton and neglecting the transverse momentum
of the latter, we obtain the following requirement:
\begin{equation}
\begin{aligned}k'^{2}=\left(\xi P^{\mu}+q^{\mu}\right)^{2} & \simeq0,\end{aligned}
\end{equation}
from which we easily find that:
\begin{equation}
\xi=x.\label{eq:Bjorkenxisxi}
\end{equation}
The Bjorken-$x$ variable (\ref{eq:Bjorkenx}), a priori a kinematic
parameter which can be controlled in the experiment, therefore turns
out to describe the momentum fraction of the proton that is carried
by the parton that participates in the scattering. 

The time scale over which the photon-parton interaction takes place
is estimated as the inverse of the energy of the virtual photon:
\begin{equation}
\begin{aligned}\Delta t_{\mathrm{coll}} & \sim\frac{1}{q^{0}}=\frac{2xP}{Q^{2}}.\end{aligned}
\label{eq:DIScolltime}
\end{equation}
Each of the partons, however, have their own characteristic lifetime.
In order to compute it, we have to reintroduce the transverse momentum
of the parton, which we previously neglected, and which is naturally
estimated as being of the same order as $\Lambda_{\mathrm{QCD}}$,
since this is the only scale in the proton at this moment: 
\begin{equation}
k^{\mu}\simeq\left(\omega\equiv xP,\mathbf{k}_{\perp},xP\right)\quad\mathrm{with}\quad k_{\perp}\sim\Lambda_{\mathrm{QCD}}.\label{eq:partonk}
\end{equation}
The parton lifetime is now, by virtue of Heisenberg's principle, given
by the inverse of its virtual mass $m^{*}=\sqrt{|k_{\mu}k^{\mu}|}=k_{\perp}$,
times the Lorentz contraction factor:

\begin{equation}
\Delta t_{\mathrm{fluc}}\simeq\gamma\frac{2}{m^{*}}=\frac{\omega}{m^{*}}\frac{2}{m^{*}}\simeq\frac{2\omega}{k_{\perp}^{2}}.\label{eq:DISfluctime}
\end{equation}
Since a parton in the proton can only absorb a virtual photon with
a shorter lifetime than its own, we require that:
\begin{equation}
\begin{aligned}\Delta t_{\mathrm{coll}} & \lesssim\Delta t_{\mathrm{fluc}}.\end{aligned}
\end{equation}
In other words: the parton dynamics is \textquoteleft frozen' with
respect to the interaction with the photon. Comparing expression (\ref{eq:DIScolltime})
with Eqs. (\ref{eq:DISfluctime}) and (\ref{eq:partonk}), we see
that the above requirement is tantamount to the statement that the
photon interacts with partons with a transverse momentum smaller than
its virtuality:
\begin{equation}
\begin{aligned}k_{\perp}^{2} & \lesssim Q^{2}.\end{aligned}
\label{eq:DISmomentumrequirement}
\end{equation}
From Heisenberg's uncertainty relation, these partons are localized
over an area $r_{\perp}^{2}\sim1/Q^{2}$ in the transverse plane,
hence $Q^{2}$ can be interpreted as the resolution with which the
proton's constituents are probed.

\begin{figure}[t]
\begin{centering}
\includegraphics[scale=0.4]{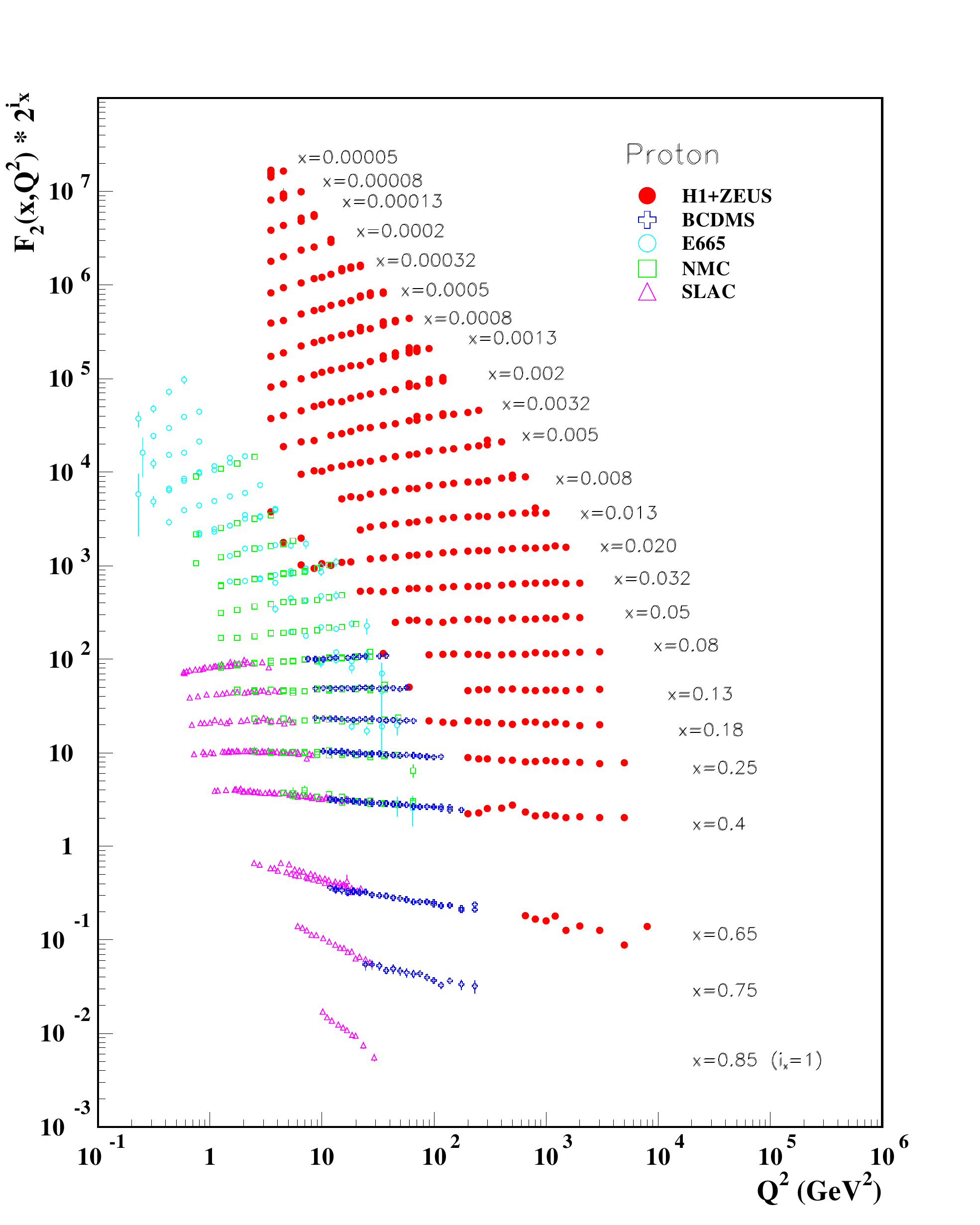}
\par\end{centering}
\caption{\label{fig:HERA F2}The proton structure function $F_{2}$, for clarity
multiplied by $2^{i_{x}}$, with $i_{x}$ the number of the bin in
Bjorken-$x$ from $i_{x}=1$ (for $x=0.85$) to $i_{x}=24$ (for $x=5\cdot10^{-5}$)
\protect\cite{PDG}}
\end{figure}

In the high-energy regime $s\gg Q^{2}$, the virtual photon-proton
cross section can be written as follows (see Refs. \protect\cite{HalzenMartin,Nachtmann2008}):
\begin{equation}
\begin{aligned}\sigma_{\gamma^{*}p} & =\frac{4\pi^{2}\alpha_{\mathrm{em}}}{Q^{2}}F_{2}\left(x,Q^{2}\right).\end{aligned}
\label{eq:photonprotoncrosssection}
\end{equation}
In the above formula, $F_{2}\left(x,Q^{2}\right)$ is the proton structure
function, which parametrizes our ignorance about the precise details
of its interior. The combined data on the $F_{2}$ structure function
is shown in Fig. \ref{fig:HERA F2}. It is extracted from measurements
at the H1 and Zeus $e^{-}p$ and $e^{+}p$ deep-inelastic scattering
experiments at HERA, as well as some fixed target electron and muon
scattering experiments at SLAC, BCDMS \& EMC (CERN) and E665 (Fermilab).
At fairly large values of $x$, of the order $x\sim0.1$, the structure
function is approximately constant with varying $Q^{2}$. This phenomenon
was predicted by Bjorken (Ref. \protect\cite{Bjorken1969}) and is therefore
known as Bjorken scaling. Shortly after, Feynman established the so-called
parton model (Ref. \protect\cite{Feynman1971}), in which he interpreted this
scaling as a consequence of the fact that the proton is composed of
asymptotically free partons, which are point like (hence the scattering
is independent of the resolution $1/Q^{2}$) and which carry an electrical
charge. The structure function can then be written as the following
function of the quark and antiquark parton distribution functions
(PDFs) $q_{f}\left(x,Q^{2}\right)$ and $\bar{q}_{f}\left(x,Q^{2}\right)$:
\begin{equation}
\begin{aligned}F_{2}\left(x,Q^{2}\right) & =\sum_{f}e_{q_{f}}^{2}\left(xq_{f}\left(x,Q^{2}\right)+x\bar{q}_{f}\left(x,Q^{2}\right)\right),\end{aligned}
\label{eq:partonmodel}
\end{equation}
where $e_{q_{f}}^{2}$ is the fractional electromagnetic charge carried
by the quark with flavor $f$. It is however clear from the data that
Bjorken scaling is violated, which we also suspect from our heuristic
argument that increasing $Q^{2}$ is tantamount to increasing the
phase space for partons to participate in the scattering, see Eq.
(\ref{eq:DISmomentumrequirement}). This argument is in fact deeply
rooted in QCD: to make this more clear, and to elucidate on the connection
with scaling violation, it is worth to do a small calculation.

\section{The soft Bremsstrahlung law\label{subsec:The-soft-Bremsstrahlung}}

Let us compute (following Ref. \protect\cite{Salgadolectures}) the probability
in QCD for a quark to radiate a gluon in the high-energy limit. This
probability can be easily extracted from the cross section of the
following process, depicted in Fig. \ref{fig:Bremms},
\begin{figure}[t]
\begin{centering}
\begin{tikzpicture}[scale=2] 

\tikzset{electron/.style={ postaction={decorate},decoration={markings,mark=at position .5 with {\arrow[]{latex}}}},      	gluon/.style={decorate,decoration={coil,amplitude=4pt, segment length=5pt}}}
\draw (0,0) circle(.3);
\node at (0,0) {$\mathcal{M}_{h}$};
\draw[semithick,electron] (.3,0)  -- ++ (1,0);
\draw[semithick,gluon] (1,0) ..controls (1.5,-.3) and (1.5,-.3) .. (2,-.4) node [right]{$a,\lambda$};
\draw[semithick,electron] (1.3,0) --++ (.7,0) node [right]{$j$};
\node at (.7,.2) {$p+k$};
\node at (1.5,.2) {$p$};
\node at (1.5,-.5) {$k$};

\end{tikzpicture} 
\par\end{centering}
\caption{\label{fig:Bremms}A quark, created in some hard process $\mathcal{M}_{h}$,
loses its virtuality by emitting a gluon. From this process, the Bremsstrahlung
law Eq. (\ref{eq:Bremsstrahlung law}) can be extracted.}
\end{figure}
in which a quark, which was created in some hard process $\mathcal{M}_{h}$,
emits a gluon. The amplitude corresponding to this process is:
\begin{equation}
\begin{aligned}\mathcal{M} & =\bar{u}(p)ig_{s}t^{a}\cancel{\epsilon}^{\lambda}(k)\frac{i\left(\cancel{p}+\cancel{k}\right)}{\left(p+k\right)^{2}+i\epsilon}\mathcal{M}_{h},\end{aligned}
\label{eq:FeynEikParton}
\end{equation}
where $\bar{u}\left(p\right)$ is the spinor of the outgoing quark,
$t^{a}$ is the generator of the $SU\left(3\right)$ group in the
fundamental representation, $g_{s}$ is the strong coupling constant,
$\mathcal{M}_{h}$ is the matrix element of the hard process, $\epsilon_{\mu}^{\lambda}\left(k\right)$
for $\lambda=1,2$ are the polarization vectors of the gluon, and
where the Feynman slash denotes multiplication with the gamma matrices:
$\cancel{p}=p_{\mu}\gamma^{\mu}$. Since we are interested in the
high-energy limit, in which the quark is\emph{ }traveling at almost
the speed of light, the process can be best described using light-cone
(LC)\emph{ }coordinates\emph{,} defined as follows:
\begin{equation}
\begin{aligned}\begin{aligned}x^{+}\end{aligned}
 & \equiv\frac{x^{0}+x^{3}}{\sqrt{2}},\quad\begin{aligned}x^{-}\end{aligned}
\equiv\frac{x^{0}-x^{3}}{\sqrt{2}},\quad\mathbf{x}_{\perp}\equiv\left(x^{1},x^{2}\right),\end{aligned}
\end{equation}
where the scalar product of two vectors reads:
\begin{equation}
x\cdot y=x^{+}y^{-}+x^{-}y^{+}-\mathbf{x}_{\perp}\cdot\mathbf{y}_{\perp}.\label{eq:LCscalarproduct}
\end{equation}
In these coordinates, the momentum of a highly energetic particle
which is traveling in the positive $z$-direction (a \textquoteleft right-mover')
has a large \textquoteleft $+$' component $q^{+}=\left(E+q_{z}\right)/\sqrt{2}\simeq\sqrt{2}q_{z}$,
while its \textquoteleft $-$' component $q^{-}=\left(E-q_{z}\right)/\sqrt{2}\simeq q_{\perp}^{2}/2q^{+}$
(the last equality holds only when the particle is on shell) is very
small:
\begin{equation}
q^{\mu}\simeq\left(q^{+},q^{-},\mathbf{q}_{\perp}\right)\quad\mathrm{with}\quad q^{+}>q_{\perp}>q^{-}.
\end{equation}
For a right-mover, $x^{+}\simeq\sqrt{2}x^{0}=\sqrt{2}t$ plays the
role of light-cone time, while the particle is sharply located around
the longitudinal coordinate $x^{-}\ll1$. Likewise, the $q^{-}$ momentum,
conjugate to $x^{+}$, can be regarded as the light-cone energy with
$x^{+}\sim1/q^{-}$, while $q^{+}$ is the light-cone momentum and
is conjugate to $x^{-}$, i.e. $x^{-}\sim1/q^{+}$. For a \textquoteleft left-mover',
which travels in the negative $z$-direction with a large \textquoteleft $-$'
momentum component, these roles are reversed. 

The momenta of the quark (neglecting its mass and transverse momentum,
since it is on shell this implies $p^{-}=p_{\perp}^{2}/2p^{+}=0$)
and the gluon in Eq. (\ref{eq:FeynEikParton}) are then given by:
\begin{equation}
\begin{aligned}p^{\mu} & =\left(p^{+},0,\mathbf{0}_{\perp}\right),\qquad k^{\mu}=\left(k^{+},k^{-},\mathbf{k}_{\perp}\right).\end{aligned}
\end{equation}
The two polarization vectors $\epsilon_{\mu}^{\lambda}(k)$ are defined
to be transverse with respect to the gluon momentum $k^{\mu}$: 
\begin{equation}
\begin{aligned}\epsilon_{\mu}^{\lambda}\left(k\right) & =\left(0,\frac{\mathbf{k}_{\perp}\cdot\boldsymbol{\epsilon}_{\perp}^{\lambda}}{k^{+}},\mathbf{\boldsymbol{\epsilon}}_{\perp}^{\lambda}\right)\qquad\lambda=0,1,\\
\sum_{\lambda}\epsilon_{i}^{\lambda\dagger}\epsilon_{j}^{\lambda} & =\delta^{ij},\quad\mathbf{\boldsymbol{\epsilon}}_{\perp}^{\lambda\dagger}\cdot\mathbf{\boldsymbol{\epsilon}}_{\perp}^{\lambda'}=\delta^{\lambda\lambda'},\quad k^{\mu}\epsilon_{\mu}^{\lambda}\left(k\right)=0.
\end{aligned}
\label{eq:polarization vectors}
\end{equation}
In order to evaluate the amplitude Eq. (\ref{eq:FeynEikParton}),
let us first set $\cancel{k}\simeq0$ in the numerator of the propagator,
since the momentum $p^{\mu}$ of the highly energetic quark dominates.
Next, the gamma matrices can be interchanged according to $\left\{ \gamma^{\mu},\gamma^{\nu}\right\} =2g^{\mu\nu}$,
after which we apply the Dirac equation $\bar{u}(k)\cancel{k}=\bar{u}(k)m\simeq0$:
\begin{equation}
\begin{aligned}\mathcal{M} & \simeq-2g_{s}t^{a}\bar{u}(p)p{}^{\mu}\epsilon_{\mu}^{\lambda}(k)\frac{1}{\left(p+k\right)^{2}+i\epsilon}\mathcal{M}_{h},\\
 & =-2g_{s}t^{a}\bar{u}(p)p{}^{+}\frac{\mathbf{k}_{\perp}\cdot\boldsymbol{\epsilon}_{\perp}^{\lambda}}{k^{+}}\frac{1}{\left(p+k\right)^{2}+i\epsilon}\mathcal{M}_{h}.
\end{aligned}
\end{equation}
Furthermore:
\begin{equation}
\begin{aligned}\mathcal{M} & =-2g_{s}t^{a}\bar{u}(p)p^{+}\frac{\mathbf{k}_{\perp}\cdot\boldsymbol{\epsilon}_{\perp}^{\lambda}}{k^{+}}\frac{1}{2p\cdot k}\mathcal{M}_{h},\\
 & =-2g_{s}t^{a}\bar{u}(p)\frac{\mathbf{k}_{\perp}\cdot\boldsymbol{\epsilon}_{\perp}^{\lambda}}{2k^{-}k^{+}}\mathcal{M}_{h},\\
 & =-2g_{s}t^{a}\frac{\mathbf{k}_{\perp}\cdot\boldsymbol{\epsilon}_{\perp}^{\lambda}}{k_{\perp}^{2}}\bar{u}(p)\mathcal{M}_{h},
\end{aligned}
\label{eq:amplitudeeikonalgluon}
\end{equation}
where, in the last equality, we assumed that the gluon is on shell
hence $k^{-}=k_{\perp}^{2}/2k^{+}$. From this amplitude, the cross
section is obtained by taking the absolute value squared, averaging
over the colors of the quark, summing over the gluon polarization,
integrating over the momentum of the gluon, and imposing the mass-shell
condition:
\begin{equation}
\begin{aligned}\sigma & =\int\frac{\mathrm{d}k^{+}\mathrm{d}k^{-}\mathrm{d}^{2}\mathbf{k}_{\perp}}{\left(2\pi\right)^{4}}2\pi\delta\left(k^{2}\right)\left|\mathcal{M}\right|^{2},\\
 & =\int\frac{\mathrm{d}k^{+}\mathrm{d}^{2}\mathbf{k}_{\perp}}{\left(2\pi\right)^{3}2k^{+}}4g_{s}^{2}C_{F}\frac{1}{k_{\perp}^{2}}\left|\bar{u}(p)\mathcal{M}_{h}\right|^{2}.
\end{aligned}
\label{eq:sigmabremss}
\end{equation}
In the above formula, $C_{F}=\mathrm{Tr}\left(t^{a}t^{a}\right)/N_{c}=\left(N_{c}^{2}-1\right)/2N_{c}$
is the Casimir operator for the fundamental representation. Forgetting
about the generic hard part, we can extract from the cross section
above the differential probability for a quark to radiate a soft gluon,
which we will call the \textquoteleft soft\emph{ }Bremsstrahlung\emph{
}law':
\begin{equation}
\begin{aligned}\frac{\mathrm{d}P_{\mathrm{Brems}}}{\mathrm{d}x\mathrm{d}^{2}\mathbf{k}_{\perp}} & =\frac{\alpha_{s}C_{F}}{\pi}\frac{1}{x}\frac{1}{\pi k_{\perp}^{2}},\end{aligned}
\label{eq:Bremsstrahlung law}
\end{equation}
with $\alpha_{s}\equiv g_{s}^{2}/4\pi$ and where $x\equiv k^{+}/p^{+}$.
When the emitter is a gluon, rather than a quark, the result is exactly
the same, apart from the Casimir operator $C_{F}$ which should be
replaced with the Casimir operator for the adjoint representation:
$C_{F}\to N_{c}$. Although the above formula is a simple result,
from a straightforward calculation, its importance cannot be overstated.
First, it should be noted that the Bremsstrahlung law is very general,
and holds beyond the process in Fig. \ref{fig:Bremms}. In particular,
the outgoing particles do not necessarily have to be on shell, but
can be further involved in other processes, or emit radiation themselves.
Second, Eq. (\ref{eq:Bremsstrahlung law}) is clearly singular both
in the limit $x\to0$, where the energy of the emitted gluon goes
to zero, and in the limit $k_{\perp}^{2}\to0$, in which the gluon's
transverse momentum goes to zero. These singularities, called the
soft and the collinear divergency, respectively, can be dealt with
by including the appropriate infrared cutoffs. However, when integrating
over Eq. (\ref{eq:Bremsstrahlung law}), which happens for example
in the cross section Eq. (\ref{eq:sigmabremss}), and when there is
a large interval of energy or transverse momentum available, the Bremsstrahlung
law gives rise to large logarithms: 
\begin{equation}
\begin{aligned}P_{\mathrm{Brems}} & =\frac{\alpha_{s}C_{F}}{\pi}\int_{\mu^{2}}^{k_{\perp}^{2}}\frac{\mathrm{d}k_{\perp}^{'2}}{k_{\perp}^{'2}}\int_{x}^{1}\frac{\mathrm{d}x'}{x'},\\
 & =\frac{\alpha_{s}C_{F}}{\pi}\ln\frac{k_{\perp}^{2}}{\mu^{2}}\ln\frac{1}{x},
\end{aligned}
\end{equation}
where $\mu$ is an infrared cutoff for the transverse momentum, and
where the upper bound for the energy fraction of the emitted gluon
is simply $x_{\mathrm{max}}=1$. If either one of the above logarithms
is large enough to compensate for the small coupling constant:
\begin{equation}
\begin{aligned}\alpha_{s}\ln\frac{k_{\perp}^{2}}{\mu^{2}} & \sim1\quad\mathrm{or}\quad\alpha_{s}\ln\frac{1}{x}\sim1,\end{aligned}
\end{equation}
then the probability of a single gluon emission becomes of order one
or larger, and hence perturbation theory loses its meaning. At this
point, it is not correct anymore to regard the probability for two
or more emissions as a correction to a Feynman diagram with one single
emission. Rather, multiple emissions take place which, to render the
theory consistent again, have to be resummed to all orders in $\alpha_{s}$.
This can be done by dividing the problematic part of the phase space
in small intervals, such that $\alpha_{s}\ln\left(x_{i+1}/x_{i}\right)\apprle1$
and $\alpha_{s}\ln\left(k_{\perp i+1}^{2}/k_{\perp i}^{2}\right)\apprle1$:
\begin{equation}
x'\in\left[1,...,x_{i},x_{i+1},...,x_{\mathrm{min}}\right]\quad\mathrm{and}\quad k_{\perp}^{2}\in\left[k_{\perp\mathrm{min}}^{2},...,k_{\perp i}^{2},k_{\perp i+1}^{2},...,k_{\perp\mathrm{max}}^{2}\right].
\end{equation}
In addition, we impose that consecutive emissions have to be strongly
ordered: each emitted gluon has a much larger transverse momentum
and much smaller energy than the previous one:
\begin{equation}
k_{\perp i-1}^{2}\ll k_{\perp i}^{2}\ll k_{\perp i+1}^{2}\quad\mathrm{and}\quad x_{i-1}\gg x_{i}\gg x_{i+1}.
\end{equation}
The result of this condition is that only those emissions are taken
into account that are enhanced by a large logarithm both in the transverse
momentum and in the energy, and which are therefore the dominant contributions.
For example, the probability for two successive strongly-ordered emissions
is:
\begin{equation}
\begin{aligned}P_{\mathrm{Brems}}^{\left(2\right)} & =\left(\frac{\alpha_{s}C_{F}}{\pi}\right)^{2}\int_{k_{\perp\mathrm{min}}^{2}}^{k_{\perp\mathrm{max}}^{2}}\frac{\mathrm{d}k_{\perp}^{2}}{k_{\perp}^{2}}\int_{x_{\mathrm{min}}}^{1}\frac{\mathrm{d}x}{x}\int_{k_{\perp\mathrm{min}}^{2}}^{k_{\perp}^{2}}\frac{\mathrm{d}k_{\perp}^{'2}}{k_{\perp}^{'2}}\int_{x}^{1}\frac{\mathrm{d}x'}{x'},\\
 & =\frac{1}{2!2!}\left(\frac{\alpha_{s}C_{F}}{\pi}\right)^{2}\ln^{2}\frac{k_{\perp\mathrm{max}}^{2}}{k_{\perp\mathrm{min}}^{2}}\ln^{2}\frac{1}{x_{\mathrm{min}}},
\end{aligned}
\end{equation}
and in general, for $n$ such emissions: 
\begin{equation}
\begin{aligned}P_{\mathrm{Brems}}^{\left(n\right)} & =\frac{1}{n!n!}\left(\frac{\alpha_{s}C_{F}}{\pi}\ln\frac{k_{\perp\mathrm{max}}^{2}}{k_{\perp\mathrm{min}}^{2}}\ln\frac{1}{x_{\mathrm{min}}}\right)^{n}.\end{aligned}
\end{equation}
Summing all these probabilities, we obtain a modified Bessel function
of the first kind (see Refs. \protect\cite{Jung,KovchegovLevin}):
\begin{equation}
\begin{aligned}P_{\mathrm{Brems}} & =\sum_{n=0}^{\infty}P_{\mathrm{Brems}}^{\left(n\right)},\\
 & =I_{0}\left(2\sqrt{\frac{\alpha_{s}C_{F}}{\pi}\ln\frac{k_{\perp\mathrm{max}}^{2}}{k_{\perp\mathrm{min}}^{2}}\ln\frac{1}{x_{\mathrm{min}}}}\right).
\end{aligned}
\label{eq:BremsDLA}
\end{equation}
In this expression, soft and collinear gluon emissions are resummed
to all orders in $\alpha_{s}$ in what is known as the double leading
logarithmic approximation (DLA or DLLA). For the sake of this discussion,
the upper and lower bounds on the phase space were left generic: we
will see soon that in practice the appropriate limits are imposed
by the physical problem under consideration. 

However, computing both the transverse momentum and the energy merely
to logarithmic accuracy is a very crude approximation. Often, either
integrating over the energies, or integrating over the transverse
momenta leads to large logarithms, but not at the same time. In such
a case, it is more accurate to work in the so-called leading logarithmic
approximation (LLA) in which strong ordering is only imposed on the
problematic part of phase space, resumming the ensuing large logarithms,
while treating the other part exactly. For instance, when computing
radiative corrections to DIS at moderate values of $x$, there is
only a large phase space for the transverse momenta available. Hence,
the transverse momenta of emitted partons are required to be strongly
ordered, such that logarithms $\alpha_{s}\ln\left(k_{\perp i}^{2}/k_{\perp i-1}^{2}\right)$
can be resummed, but the energy dependence of the emissions is treated
exactly, in particular beyond the approximation $\mathrm{d}P_{\mathrm{Brems}}\propto\alpha_{s}\mathrm{d}x/x$
which only holds for gluons in the limit $x\ll1$. This is the main
idea behind the DGLAP evolution equation, which we will study in the
following section, while the BFKL equation, which is the subject of
Sec. \ref{sec:BKBFKL}, is applicable in the opposite case of a large
phase space for the energy. The DLA approximation to gluon radiation
can thus be seen as the common limit of the DGLAP and BFKL evolution
equations.

\section{\label{subsec:DGLAP}DGLAP}

We are now ready to compute the various leading-order QCD corrections
to the parton model of DIS, Eqs. (\ref{eq:photonprotoncrosssection})
and (\ref{eq:partonmodel}), in which, before scattering with the
virtual photon, the parton coming from the proton radiates. As announced
earlier, the transverse momenta of the splittings will be treated
in the leading logarithmic approximation, while we compute the energy-dependence
exactly.
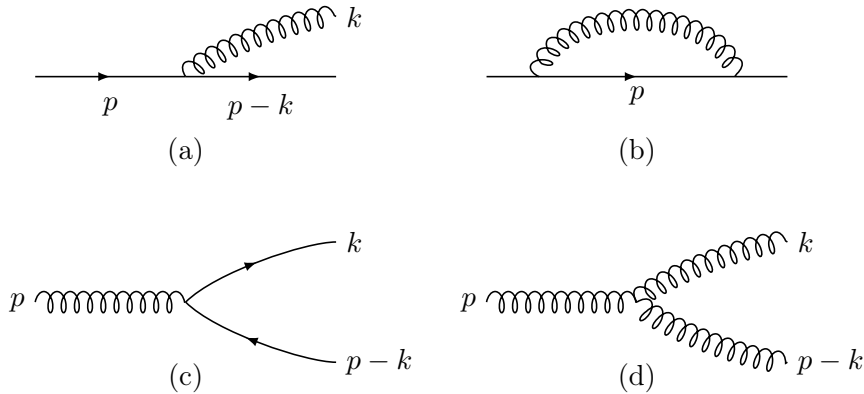
\begin{figure}[t]
\centering{}\begin{tikzpicture}[scale=2] 
\tikzset{electron/.style={ postaction={decorate},decoration={markings,mark=at position .5 with {\arrow[]{latex}}}},	positron/.style={ postaction={decorate},decoration={markings,mark=at position .5 with {\arrow[]{latex reversed}}}},      	gluon/.style={decorate,decoration={coil,amplitude=4pt, segment length=5pt}}}
\draw[semithick,electron] (0,0)  -- ++ (1,0);
\draw[semithick,gluon] (1,0).. controls (1.2,.2) and (1.8,.4)  .. (2,.4) node [right]{$k$};
\draw[semithick,electron] (1,0) --++ (1,0) node [right]{$$};
\node at (.5,-.2) {$p$};
\node at (1.5,-.2) {$p-k$};
\node at (1,-.5) {(a)};

\draw[semithick,electron] (3,0)  --++(1,0) node[below]{$p$} --++ (1,0);
\draw[semithick,gluon] (3.35,0) .. controls (3.45,.5) and (4.55,.5) .. (4.65,0);
\node at (4,-.5) {(b)};

\draw[semithick,gluon] (0,-1.5) node [left]{$p$} --(1,-1.5);
\draw[semithick,electron] (1,-1.5).. controls (1.2,-1.3) and (1.8,-1.1)  .. (2,-1.1) node [right]{$k$};
\draw[semithick,positron] (1,-1.5).. controls (1.2,-1.7) and (1.8,-1.9)  .. (2,-1.9) node [right]{$p-k$};
\node at (1,-2) {(c)};

\draw[semithick,gluon] (3,-1.5) node [left]{$p$} --++(1,0);
\draw[semithick,gluon] (4,-1.5).. controls (4.2,-1.3) and (4.8,-1.1)  .. (5,-1.1) node [right]{$k$};
\draw[semithick,gluon] (4,-1.5).. controls (4.2,-1.7) and (4.8,-1.9)  .. (5,-1.9) node [right]{$p-k$};
\node at (4,-2) {(d)};
\end{tikzpicture} \caption{\label{fig:splittings}The four DGLAP splitting functions.}
\end{figure}
First, consider the process in which a quark emits a gluon (Fig. \ref{fig:splittings},
a). Unlike the similar calculation that we performed in the previous
section, this time we work beyond the high-energy limit, hence the
computation becomes a bit more complicated since we explicitly have
to keep track of the different helicity configurations. However, instead
of performing the computation from scratch (see for instance Ref.
\protect\cite{Peskin,KovchegovLevin}), let us take a shortcut. As it happens,
we have some results from light-cone perturbation theory (LCPT) at
our disposal, which we will use later in this work in order to compute
dijet production in proton-nucleus collisions. Without going into
the details of LCPT, for which we refer to Ref. \protect\cite{Brodsky1998},
let us simply remark that it is a reformulation of perturbation theory,
similar to time-ordered or old-fashioned perturbation theory (see
for instance Ref. \protect\cite{Sterman1993}), which is completely equivalent
to the more standard covariant approach, but has some useful advantages.
In particular, in LCPT the notion of Fock space is reintroduced, hence
one can count particle states. The probability for a certain splitting
process to occur is then encoded in the corresponding light-cone wave
function. In section \ref{subsec:q->qg} of the appendix, the wave
function for the $q\to gq$ process is calculated, which turns out
to be:
\begin{equation}
\begin{aligned}\psi_{\alpha\beta}^{\lambda}\left(p,k\right) & =\frac{1}{\sqrt{k^{+}}}\frac{1}{\mathbf{k}_{\perp}^{2}+z^{2}m^{2}}\begin{cases}
\sqrt{2}\mathbf{k}_{\perp}\cdot\boldsymbol{\epsilon}_{\perp}^{1}\left[\delta_{\alpha-}\delta_{\beta-}+\left(1-z\right)\delta_{\alpha+}\delta_{\beta+}\right]+mz^{2}\delta_{\alpha+}\delta_{\beta-} & \lambda=1\\
\sqrt{2}\mathbf{k}_{\perp}\cdot\boldsymbol{\epsilon}_{\perp}^{2}\left[\delta_{\alpha+}\delta_{\beta+}+\left(1-z\right)\delta_{\alpha-}\delta_{\beta-}\right]-mz^{2}\delta_{\alpha-}\delta_{\beta+} & \lambda=2
\end{cases}.\end{aligned}
\label{eq:Psidressedquark}
\end{equation}
In the above expression, $p^{\mu}$ is the four-momentum of the incoming
quark, $k^{\mu}$ is the momentum of the outgoing gluon and $z\equiv k^{+}/p^{+}$
is the longitudinal momentum fraction carried by the radiated gluon.
The helicities of the incoming and outgoing quark are denoted by $\alpha$
and $\beta$, respectively, while the gluon has a polarization $\lambda$
and color $c$. The full dynamics of the emission process is present
in Eq. (\ref{eq:Psidressedquark}), and we can extract it (see for
instance \protect\cite{thesisStephane}) by taking the absolute value squared,
Eq. (\ref{eq:wavegqqsquared}), from which we obtain the differential
probability to find a gluon inside a dressed quark:
\begin{equation}
\begin{aligned}\frac{\mathrm{d}P_{gq}}{\mathrm{d}k^{+}\mathrm{d}^{2}\mathbf{k}_{\perp}} & =\frac{g_{s}^{2}}{\left(2\pi\right)^{3}}\frac{1}{N_{c}}\mathrm{Tr}\left(t^{c}t^{c}\right)\left|\psi^{q\rightarrow qg}\left(p,k\right)\right|^{2}.\end{aligned}
\label{eq:gluoninquarkdefinition}
\end{equation}
In the massless limit, this simplifies to (see Eq. (\ref{eq:gqqsquared})):
\begin{equation}
\begin{aligned}\frac{\mathrm{d}P_{gq}}{\mathrm{d}z\mathrm{d}^{2}\mathbf{k}_{\perp}} & =\frac{\alpha_{s}}{2\pi}\frac{1}{\pi k_{\perp}^{2}}\mathcal{P}_{gq}\left(z\right),\end{aligned}
\label{eq:dPqg}
\end{equation}
where we expressed the probability in terms of $z$ instead of $k^{+}$,
and where we introduced the Altarelli-Parisi splitting function:
\begin{equation}
\mathcal{P}_{gq}\left(z\right)\equiv C_{F}\frac{1+\left(1-z\right)^{2}}{z}.
\end{equation}
Clearly, in the limit in which the gluon is soft with respect to its
parent: $z\rightarrow0$, expression (\ref{eq:dPqg}) reduces to the
soft Bremsstrahlung law, Eq. (\ref{eq:Bremsstrahlung law}). 

Next, we want to compute the probability to find a quark, with a momentum
fraction $z$, within a dressed quark (which is, in leading order,
a quark-gluon pair). Naively, one would expect the probability to
be simply:
\begin{equation}
\begin{aligned}\frac{\mathrm{d}P_{qq}}{\mathrm{d}z\mathrm{d}^{2}\mathbf{k}_{\perp}} & =\frac{\alpha_{s}}{2\pi}\frac{1}{\pi k_{\perp}^{2}}\mathcal{P}_{qq}\left(z\right),\\
\mathcal{P}_{qq}\left(z\right) & \equiv\mathcal{P}_{gq}\left(1-z\right)=C_{F}\frac{1+z^{2}}{1-z}.
\end{aligned}
\label{eq:pqq}
\end{equation}
This answer is however only partially true: in contrast to the previous
case, in which diagram (a) of Fig. \ref{fig:splittings} is the only
leading-order contribution to the probability to find a gluon in a
dressed quark, we now have to include the probability that the quark
does not to radiate at all, or that it emits and re-absorbs a virtual
gluon (diagram (b) in Fig. \ref{fig:splittings}). One can, however,
avoid calculating these contributions explicitly, by looking at the
integrated probability:
\begin{equation}
\begin{aligned}\mathcal{F}_{qq}\left(z,Q^{2}\right) & =\int_{\mu^{2}}^{Q^{2}}\mathrm{d}^{2}k_{\perp}\frac{\mathrm{d}P_{qq}}{\mathrm{d}z\mathrm{d}^{2}\mathbf{k}_{\perp}},\end{aligned}
\label{eq:Pqqintegrated}
\end{equation}
where $\mu$ is some infrared cutoff. Indeed, demanding the total
probability to find a quark inside a dressed quark to be equal to
one:
\begin{align}
\int_{0}^{1}\mathrm{\mathrm{d}}z\mathcal{F}_{qq}\left(z,Q^{2}\right) & =1,
\end{align}
Eq. (\ref{eq:Pqqintegrated}) can be written as follows:
\begin{equation}
\begin{aligned}\mathcal{F}_{qq}\left(z,Q^{2}\right) & =\delta\left(1-z\right)+\frac{\alpha_{s}}{2\pi}\ln\frac{Q^{2}}{\mu^{2}}\left(\mathcal{P}_{qq}\left(z\right)-\mathrm{virtual}\right),\end{aligned}
\end{equation}
and hence the real and virtual leading-order contributions are required
to cancel:
\begin{align}
\int_{0}^{1}\mathrm{\mathrm{d}}z\left(C_{F}\frac{1+z^{2}}{1-z}-\mathrm{virtual}\right) & =0.\label{eq:plusprescriptioncancel}
\end{align}
Obviously, the first term of this expression, corresponding to the
real emission of a gluon, is divergent in the soft limit $z\rightarrow1$.
However, this singularity will be cancelled by the virtual correction,
as is guaranteed by the Kinoshita-Lee-Nauenberg theorem. Knowing this,
we can force the integral over the first term to converge by introducing
the so-called plus notation:
\begin{align}
\int_{0}^{1}\mathrm{\mathrm{d}}z\frac{f\left(z\right)}{\left(1-z\right)_{+}} & \equiv\int_{0}^{1}\mathrm{\mathrm{d}}z\frac{f\left(z\right)-f\left(1\right)}{1-z},
\end{align}
yielding:
\begin{equation}
\int_{0}^{1}\mathrm{\mathrm{d}}z\frac{1+z^{2}}{\left(1-z\right)_{+}}=\int_{0}^{1}\mathrm{\mathrm{d}}z\frac{z^{2}-1}{1-z}=-\frac{3}{2}.
\end{equation}
The result of this procedure is that we extracted the regular part
of the integration over the emission term in Eq. (\ref{eq:plusprescriptioncancel}).
We are therefore left with the singular piece (or rather, the piece
that will yield a singularity after integration), which, from Eq.
(\ref{eq:plusprescriptioncancel}), together with the virtual piece
should give: 
\begin{equation}
\int_{0}^{1}\mathrm{\mathrm{d}}z\left(\mathrm{singular}+\mathrm{virtual}\right)=-\frac{3}{2}C_{F}.
\end{equation}
Since the cancellation of the problematic terms takes place when $z=1,$
we infer:
\begin{equation}
\begin{aligned}\mathrm{singular}+\mathrm{virtual} & =-\frac{3}{2}C_{F}\delta\left(1-z\right),\end{aligned}
\end{equation}
and finally obtain:
\begin{equation}
\begin{aligned}\frac{\mathrm{d}P_{gq}}{\mathrm{d}z\mathrm{d}^{2}\mathbf{k}_{\perp}} & =\frac{\alpha_{s}}{2\pi}\frac{1}{\pi k_{\perp}^{2}}\tilde{\mathcal{P}}_{qq}\left(z\right),\\
\mathcal{\tilde{P}}_{qq}\left(z\right) & \equiv C_{F}\frac{1+z^{2}}{\left(1-z\right)_{+}}+\frac{3}{2}C_{F}\delta\left(1-z\right),
\end{aligned}
\end{equation}
where $\mathcal{\tilde{P}}_{qq}\left(z\right)$ is the regularized
quark-quark splitting function (as opposed to the unregularized splitting
function Eq. (\ref{eq:pqq}) which contains only the real emission).

The next step in our analysis is to include the probability that a
gluon splits into a quark-antiquark pair, for which there is at leading
order again only one diagram, see (c) in Fig. \ref{fig:splittings}.
In analogy with the Eq. (\ref{eq:gluoninquarkdefinition}) for the
$q\to qg$ case, we can write the probability distribution:
\begin{equation}
\begin{aligned}\frac{\mathrm{d}P_{qg}}{\mathrm{d}k^{+}\mathrm{d}^{2}\mathbf{k}_{\perp}} & =\frac{g_{s}^{2}}{\left(2\pi\right)^{3}}n_{f}\frac{1}{N_{c}^{2}-1}\mathrm{Tr}\left(t^{c}t^{c}\right)\left|\psi^{g\rightarrow q\bar{q}}\left(p,k\right)\right|^{2},\end{aligned}
\end{equation}
where $n_{f}$ is the number of quark flavors. With the help of the
result for the squared $g\rightarrow q\bar{q}$ wave function, Eq.
(\ref{eq:gqqsquared}) from appendix \ref{subsec:gqqwave}, we obtain:
\begin{equation}
\begin{aligned}\frac{\mathrm{d}P_{qg}}{\mathrm{d}z\mathrm{d}^{2}\mathbf{k}_{\perp}} & =\frac{\alpha_{s}}{2\pi}\frac{1}{\pi k_{\perp}^{2}}\mathcal{P}_{qg}\left(z\right),\end{aligned}
\end{equation}
with the splitting function:
\begin{equation}
\begin{aligned}\mathcal{P}_{qg}\left(z\right) & =n_{f}\frac{z^{2}+\left(1-z\right)^{2}}{2}.\end{aligned}
\end{equation}
Finally, let us take the splitting of a gluon into two other gluons
into account, which is depicted in Fig. \ref{fig:splittings}, d.
This process corresponds to the following probability:
\begin{equation}
\begin{aligned}\frac{\mathrm{d}P_{gg}}{\mathrm{d}z\mathrm{d}^{2}\mathbf{k}_{\perp}} & =\frac{\alpha_{s}}{2\pi}\frac{1}{\pi k_{\perp}^{2}}\mathcal{\tilde{P}}_{gg}\left(z\right),\end{aligned}
\end{equation}
with the regularized splitting function given by (see, for instance
\protect\cite{KovchegovLevin,thesisStephane}):
\begin{equation}
\begin{aligned}\tilde{\mathcal{P}}_{gg}\left(z\right) & \equiv2N_{c}\left[\frac{1-z}{z}+\frac{z}{\left(1-z\right)_{+}}+z\left(1-z\right)+\left(\frac{11}{12}-\frac{n_{f}}{18}\right)\delta\left(1-z\right)\right].\end{aligned}
\end{equation}

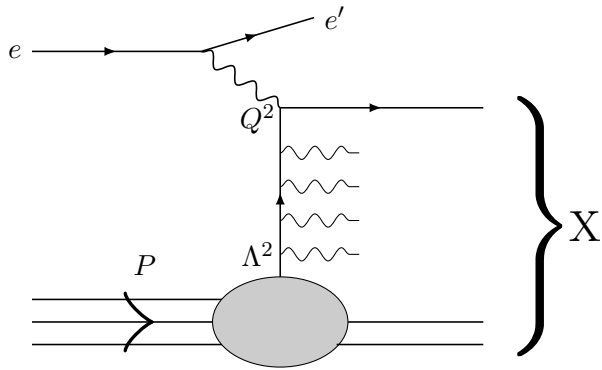
\begin{figure}[t]
\begin{centering}
\begin{tikzpicture}[scale=1.5] 

\tikzset{photon/.style={decorate,decoration={snake}},
		electron/.style={ postaction={decorate},decoration={markings,mark=at position .5 with {\arrow[draw]{latex}}}},      	gluon/.style={decorate,decoration={coil,amplitude=4pt, segment length=5pt}}}

\draw[semithick] (0,0) --++ (1.85,0);
\draw[semithick,postaction={decorate},decoration={markings,mark=at position .6 with {\arrow[very thin,scale=8]{to}}}] (0,-.2) --++ (1.8,0);
\draw[semithick] (0,-.4) --++ (1.85,0);
\node at (1,.3) {$P$};
\filldraw [fill=black!20] (2.2,-.2) ellipse (.6 and .4);
\node [rotate=90,scale=.8] at (2.2,-.1) {$$};
\draw[semithick] (2.8,-.2) -- (4,-.2);
\draw[semithick] (2.7,-.4) -- (4,-.4);

\draw[electron,semithick] (2.2,.2) --++(0,1.5);
\node at (2,.4) {$\Lambda^2$};
\draw[photon] (2.2,.4) --++ (.7,0);
\draw[photon] (2.2,.7) --++ (.7,0);
\draw[photon] (2.2,1.0) --++ (.7,0);
\draw[photon] (2.2,1.3) --++ (.7,0);
\node at (2,1.6) {$Q^2$};

\draw[semithick,photon] (1.5,2.2) -- (2.2,1.7);
\draw[electron,semithick] (0,2.2) node[left] {$e$} -- (1.5,2.2);
\draw[electron,semithick] (1.5,2.2) --++(1,.3) node[right]{$e'$};

\draw[electron,semithick] (2.2,1.7) --(4,1.7);

\node [scale=3,ultra thin] at (4.5,.65) {$\Bigg\}$};
\node [scale=1.5,ultra thin] at (4.9,.65) {X};

\end{tikzpicture} 
\par\end{centering}
\caption{\label{fig:DISQCD}Inclusive deep-inelastic scattering in the QCD-improved
parton model. The multiple collinear emissions, resummed by the DGLAP
equations and depicted by the wiggly lines, can involve either gluons
or (anti)quarks.}
\end{figure}

\begin{figure}[t]
\begin{centering}
\begin{tikzpicture}[scale=2] 

\tikzset{photon/.style={decorate,decoration={snake}},electron/.style={semithick, postaction={decorate},decoration={markings,mark=at position .5 with {\arrow[]{latex}}}},positron/.style={semithick, postaction={decorate},decoration={markings,mark=at position .5 with {\arrow[]{latex reversed}}}},      	gluon/.style={semithick, decorate,decoration={coil,amplitude=4pt, segment length=5pt}}}
\draw[electron] (0,0)  --++(.5,0);
\draw[electron] (.5,0)  --++(1,0);
\draw[gluon] (1.5,0)  --++(.5,0);
\draw[gluon] (2,0)  --++(.5,0);
\draw[electron] (2.5,0)  --++(1,0);
\draw[electron] (3.5,0)  --++(.5,0);
\draw[gluon] (.5,0) --++ (0,-.5) node [below]{$\Lambda^2$};
\node at (1,-.3) {$...$};
\draw[electron] (1.5,0) --++ (0,-.5) node [below]{$k^2_{\perp i-1}$};
\draw[gluon] (2,0) --++ (0,-.5) node [below]{$k^2_{\perp i}$};
\draw[positron] (2.5,0) --++ (0,-.5) node [below]{$k^2_{\perp i+1}$};
\node at (3,-.3) {$...$};
\draw[gluon] (3.5,0) --++ (0,-.5) node [below]{$Q^2$};
\end{tikzpicture} 
\par\end{centering}
\caption{\label{fig:DGLAPladder}A DGLAP ladder in detail, depicting quark-
and gluon emissions with increasing transverse momenta. }
\end{figure}
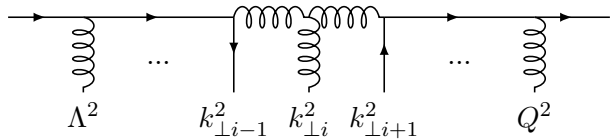

We now have all the ingredients to resum, in the leading logarithmic
approximation, the leading-order radiative contributions to DIS and
absorb them into the parton distribution functions. Starting from,
for instance, the gluon PDF $x\mathcal{G}\left(x,Q^{2}\right)$, we
can calculate the same distribution at a slightly higher resolution
$Q^{2}+\Delta Q^{2}$, by incorporating the probabilities that the
gluon we observe was emitted earlier by another quark or gluon. In
formulas:
\begin{equation}
\begin{aligned}x\mathcal{G}\left(x,Q^{2}+\Delta Q^{2}\right) & =x\mathcal{G}\left(x,Q^{2}\right)+\int_{0}^{1}\mathrm{d}z\int_{0}^{1}\mathrm{d}z'\delta\left(\frac{x}{z}-z'\right)\\
 & \left[\sum_{f}\frac{\mathrm{d}P_{gq}}{\mathrm{d}z}\left(z'q_{f}\left(z',Q^{2}\right)+z'\bar{q}_{f}\left(z',Q^{2}\right)\right)+\frac{\mathrm{d}P_{gg}}{\mathrm{d}z}z'\mathcal{G}\left(z',Q^{2}\right)\right],
\end{aligned}
\end{equation}
or, writing the differential:
\begin{equation}
\begin{aligned}x\mathcal{G}\left(x,Q^{2}+\Delta Q^{2}\right) & =x\mathcal{G}\left(x,Q^{2}\right)+\frac{\alpha_{s}}{2\pi}\frac{\Delta Q^{2}}{Q^{2}}\int_{0}^{1}\mathrm{d}z\int_{0}^{1}\mathrm{d}z'\delta\left(\frac{x}{z}-z'\right)\\
 & \left[\sum_{f}\tilde{\mathcal{P}}_{gq}\left(z\right)\left(z'q_{f}\left(z',Q^{2}\right)+z'\bar{q}_{f}\left(z',Q^{2}\right)\right)+\tilde{\mathcal{P}}_{gg}z'\mathcal{G}\left(z',Q^{2}\right)\right],
\end{aligned}
\label{eq:DGLAPttstp}
\end{equation}
where we used that:
\begin{equation}
\begin{aligned}\frac{\mathrm{d}P_{pp}}{\mathrm{d}z} & =\frac{\alpha_{s}}{2\pi}\frac{\mathrm{d}Q^{2}}{Q^{2}}\mathcal{\tilde{P}}_{pp}\left(z\right).\end{aligned}
\end{equation}
By looking at small steps in the virtuality, we effectively cut the
problematic transverse-momentum phase space into pieces, as we explained
in the previous chapter. From Eq. (\ref{eq:DGLAPttstp}), we obtain
an integro-differential equation in the logarithm of the virtuality:
\begin{equation}
\begin{aligned} & \frac{\mathrm{d}}{\mathrm{d}\ln Q^{2}}x\mathcal{G}\left(x,Q^{2}\right)\\
 & =\frac{\alpha_{s}}{2\pi}\int_{x}^{1}\mathrm{d}z\left[\sum_{f}\tilde{\mathcal{P}}_{gq}\left(z\right)\left(\frac{x}{z}q_{f}\left(\frac{x}{z},Q^{2}\right)+\frac{x}{z}\bar{q}_{f}\left(\frac{x}{z},Q^{2}\right)\right)+\tilde{\mathcal{P}}_{gg}\left(z\right)\frac{x}{z}\mathcal{G}\left(\frac{x}{z},Q^{2}\right)\right].
\end{aligned}
\label{eq:DGLAPxG}
\end{equation}
Analogously, one finds for the quark and antiquark distributions:
\begin{equation}
\begin{aligned}\frac{\mathrm{d}}{\mathrm{d}\ln Q^{2}}xq_{f}\left(x,Q^{2}\right) & =\frac{\alpha_{s}}{2\pi}\int_{x}^{1}\mathrm{d}z\left[\tilde{\mathcal{P}}_{qq}\left(z\right)\frac{x}{z}q_{f}\left(\frac{x}{z},Q^{2}\right)+\tilde{\mathcal{P}}_{qg}\left(z\right)\frac{x}{z}\mathcal{G}\left(\frac{x}{z},Q^{2}\right)\right],\\
\frac{\mathrm{d}}{\mathrm{d}\ln Q^{2}}x\bar{q}_{f}\left(x,Q^{2}\right) & =\frac{\alpha_{s}}{2\pi}\int_{x}^{1}\mathrm{d}z\left[\tilde{\mathcal{P}}_{qq}\left(z\right)\frac{x}{z}\bar{q}_{f}\left(\frac{x}{z},Q^{2}\right)+\tilde{\mathcal{P}}_{qg}\left(z\right)\frac{x}{z}\mathcal{G}\left(\frac{x}{z},Q^{2}\right)\right].
\end{aligned}
\end{equation}
These are the famous Dokshitzer-Gribov-Lipatov-Altarelli-Parisi or
DGLAP equations (Refs. \protect\cite{DGLAP1,DGLAP2,DGLAP3}). They are differential
equations which describe to leading-logarithmic accuracy the change
in the parton distribution functions when changing $Q^{2}$, and are
an important example of what one calls evolution equations in quantum
field theory. Solving them amounts to the resummation of all the leading-order
collinear QCD corrections to DIS into so-called parton ladders, see
Figs. \ref{fig:DISQCD} and \ref{fig:DGLAPladder}. Equivalently,
the DGLAP equations can be regarded as renormalization-group equations
(see e.g. Ref. \protect\cite{HalzenMartin,Jung}) which renormalize the parton
densities with respect to the scale $Q^{2}$. 

The DGLAP equation allows us to finally explain the phenomenon of
the scaling violation of the proton structure function $F_{2}$ in
Fig. \ref{fig:HERA F2}. Indeed, before scattering with the photon,
the parton emits quarks and gluons through DGLAP radiation (see Figs.
\ref{fig:DISQCD} and \ref{fig:DGLAPladder}). The virtuality $Q^{2}$
of the photon is the natural upper bound for the transverse momentum
of these emitted partons, hence increasing $Q^{2}$ amounts to increasing
the available transverse phase space for the emissions. The more emissions,
the more the parton loses energy, hence when $Q^{2}$ increases one
observes a shift of the parton distributions, and thus of $F_{2}$,
towards smaller values of $x$. Furthermore, according to Eq. (\ref{eq:Bremsstrahlung law}),
when $x$ becomes very small, gluons are emitted in abundance. Even
though DIS is only sensitive to gluons at next-to-leading order, the
growth in the gluon distribution is large enough to magnify the scaling
violation dramatically towards smaller values of $x$.

To establish the connection with the previous section, let us revert
to the double leading logarithmic approximation (DLA), in which we
only keep those contributions that are enhanced by an additional large
logarithm in the energy. Since
\begin{equation}
\ln\frac{p^{+}}{k^{+}}=\ln\frac{1}{x},
\end{equation}
this approximation is valid in the small-$x$ limit of DGLAP. In this
regime, as we mentioned already before, the dynamics is dominated
by gluons, hence we will solve DGLAP for the gluon distribution, Eq.
(\ref{eq:DGLAPxG}), which for $x\ll1$ reduces to:
\begin{equation}
\begin{aligned}\frac{\mathrm{d}}{\mathrm{d}\ln Q^{2}}x\mathcal{G}\left(x,Q^{2}\right) & =\frac{\alpha_{s}N_{c}}{\pi}\int_{x}^{1}\frac{\mathrm{d}z}{z}\frac{x}{z}\mathcal{G}\left(\frac{x}{z},Q^{2}\right),\\
 & =\frac{\alpha_{s}N_{c}}{\pi}\int_{x}^{1}\frac{\mathrm{d}z}{z}z\mathcal{G}\left(z,Q^{2}\right).
\end{aligned}
\end{equation}
Rewriting this equation in integral form, we obtain:
\begin{equation}
\begin{aligned}x\mathcal{G}\left(x,Q^{2}\right) & =x\mathcal{G}\left(x,Q_{0}^{2}\right)+\bar{\alpha}\int_{Q_{0}^{2}}^{Q^{2}}\frac{\mathrm{d}k_{\perp}^{2}}{k_{\perp}^{2}}\int_{x}^{1}\frac{\mathrm{d}z}{z}z\mathcal{G}\left(z,k_{\perp}^{2}\right),\end{aligned}
\end{equation}
where we introduced the notation $\bar{\alpha}\equiv\alpha_{s}N_{c}/\pi$.
The second term in the r.h.s. of the above equation is of course nothing
else than the convolution of the gluon density with the Bremsstrahlung
law, which we resummed in the previous section. Indeed, solving the
above equation iteratively, assuming the initial value to be scale
independent:
\begin{equation}
\begin{aligned}x\mathcal{G}^{\left(0\right)}\left(x,Q^{2}\right) & =x\mathcal{G}^{\left(0\right)},\end{aligned}
\end{equation}
we obtain, in accordance with Eq. (\ref{eq:BremsDLA}):
\begin{equation}
\begin{aligned}x\mathcal{G}\left(x,Q^{2}\right) & =x\mathcal{G}^{\left(0\right)}I_{0}\left(2\sqrt{\bar{\alpha}\ln\frac{Q^{2}}{Q_{0}^{2}}\ln\frac{1}{x}}\right).\end{aligned}
\label{eq:gluon asymptotic}
\end{equation}
It should be noted, however, that to obtain the above result we neglected
the running of the coupling. This is by no means justified, since
the coupling is also enhanced by a large logarithm in the virtuality:
\begin{equation}
\alpha_{s}\left(Q^{2}\right)\overset{\mathrm{1-loop}}{=}\frac{1}{b_{0}\ln\left(Q^{2}/\Lambda_{\mathrm{QCD}}^{2}\right)},\quad b_{0}=\frac{11N_{c}-2n_{f}}{12\pi}.
\end{equation}
A more careful analysis in Mellin space (see e.g. Refs. \protect\cite{qcdandcollider,KovchegovLevin})
shows that including this correction, Eq. (\ref{eq:gluon asymptotic})
becomes:
\begin{equation}
\begin{aligned}x\mathcal{G}\left(x,Q^{2}\right) & \sim x\mathcal{G}^{\left(0\right)}\exp\sqrt{\frac{4N_{c}}{\pi b_{0}}\ln\frac{\ln\left(Q^{2}/\Lambda_{\mathrm{QCD}}^{2}\right)}{\ln\left(Q_{0}^{2}/\Lambda_{\mathrm{QCD}}^{2}\right)}\ln\frac{1}{x}}.\end{aligned}
\label{eq:xGDLLA}
\end{equation}
As expected, the DLA gluon distribution above grows with decreasing
$x$ and increasing $Q^{2}$. The growth with $x$ is faster than
$\ln1/x$ but slower than any power of $1/x$. We will see shortly
that, in the small-$x$ regime, by virtue of the BFKL equation, which
resums the logarithms $\ln1/x$, the gluon distribution turns out
to grow even faster, with a power-like behavior in $1/x$:
\begin{equation}
x\mathcal{G}\left(x,Q^{2}\right)\sim\frac{1}{x^{2.77\times\bar{\alpha}}}.
\end{equation}

\begin{figure}[t]
\begin{centering}
\includegraphics[scale=0.4]{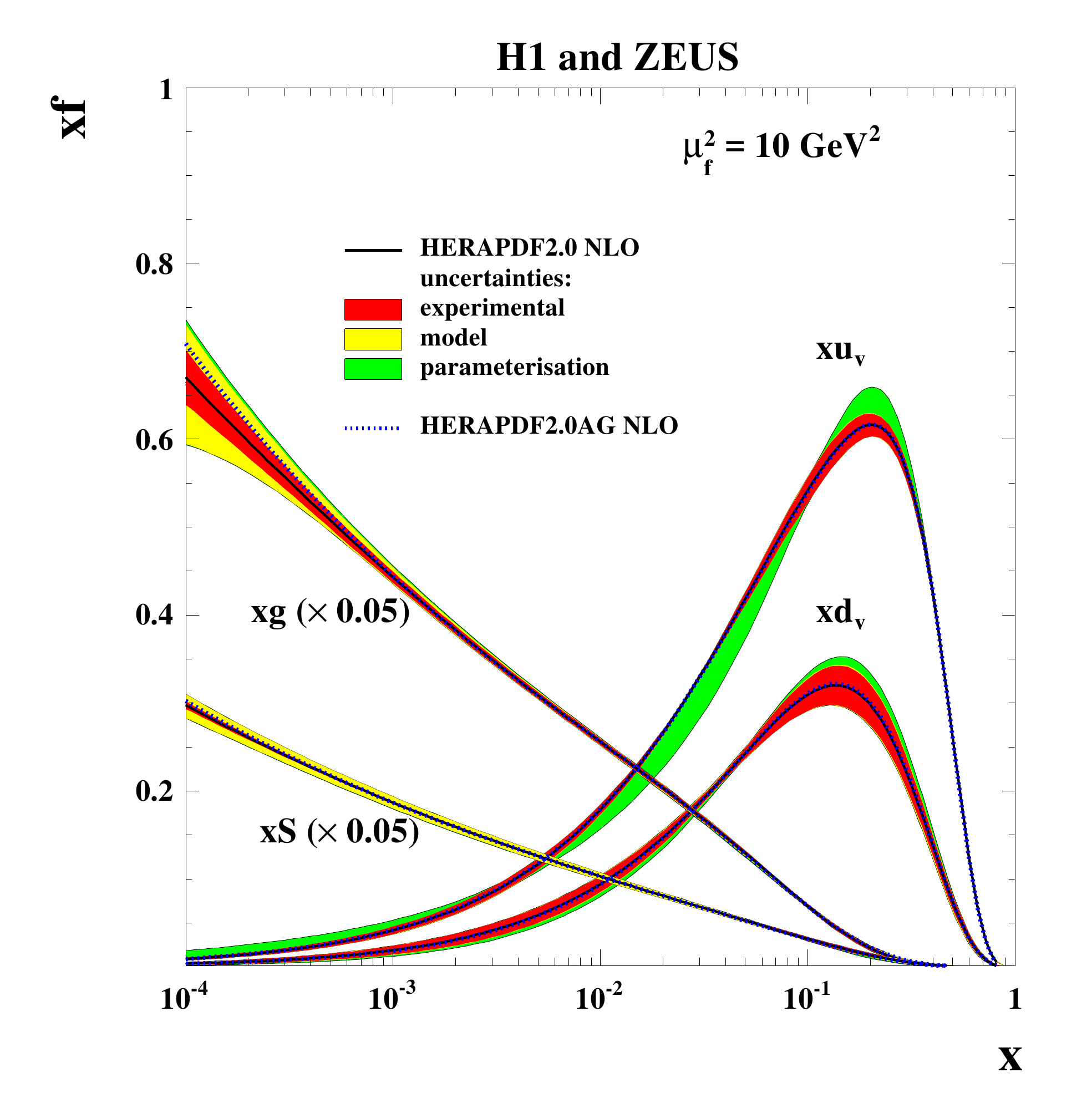}
\par\end{centering}
\caption{\label{fig:HERA pdfs}The parton densities at $Q^{2}=10\,\mathrm{GeV}^{2}$,
extracted from H1 and Zeus data with NLO DGLAP fits \protect\cite{H1Zeus}.
The valence quark densities are defined as $xu_{v}=xu-x\bar{u}$ and
$xd_{v}=xd-x\bar{d}$. For clarity, the sea quark distribution $xS=xu+x\bar{u}+xd+x\bar{d}+xs+x\bar{s}$
and the gluon distribution $xg=x\mathcal{G}$ have been multiplied
with a factor $0.05$.}
\end{figure}

With the help of sum rules (see e.g. Ref. \protect\cite{qcdandcollider})
and DGLAP fits at next-to-leading order, the parton distributions
have been extracted with great accuracy from DIS experiments, see
Fig. \ref{fig:HERA pdfs}. Clearly, this approach works very well,
except at small values of $x$, where, as expected, the gluon distribution
dominates and large energy logarithms become important.

\section{\label{sec:The-dipole-picture-and-saturation}DIS at small-$x$:
the dipole picture}

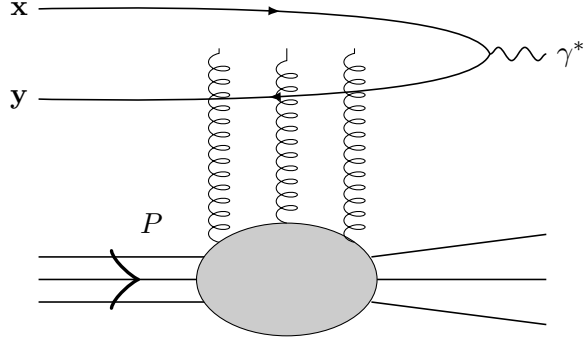
\begin{figure}[t]
\begin{centering}
\begin{tikzpicture}[scale=1.5] 

\tikzset{photon/.style={semithick,decorate,decoration={snake}},
		electron/.style={semithick,postaction={decorate},decoration={markings,mark=at position .5 with {\arrow[draw]{latex}}}}, 	positron/.style={semithick,postaction={decorate},decoration={markings,mark=at position .5 with {\arrow[draw]{latex reversed}}}},     	gluon/.style={decorate,decoration={coil,amplitude=4pt, segment length=5pt}}}

\draw[semithick] (0,0) --++ (2.05,0);
\draw[semithick,postaction={decorate},decoration={markings,mark=at position .6 with {\arrow[very thin,scale=8]{to}}}] (0,-.2) --++ (1.5,0);
\draw[semithick] (0,-.4) --++ (2.05,0);
\node at (1,.3) {$P$};
\filldraw [fill=black!20] (2.2,-.2) ellipse (.8 and .5);
\node [rotate=90,scale=.8] at (2.2,-.1) {$$};
\draw[semithick] (3,-.2) -- (4.5,-.2);
\draw[semithick] (2.95,-.4) -- (4.5,-.6);
\draw[semithick] (2.95,0) -- (4.5,.2);

\draw[gluon] (1.6,0.13) -- (1.6,1.85);
\draw[gluon] (2.2,.3) -- (2.2,1.85);
\draw[gluon] (2.8,0.13) -- (2.8,1.85);

\draw[photon,semithick] (4.5,1.8) node[right] {$\gamma^*$}--++(-.5,0);
\draw[positron] (4,1.8).. controls (3.8,2.3) and (0,2.2)  .. (0,2.2) node [left]{$\mathbf{x}$};
\draw[electron] (4,1.8).. controls (3.8,1.3) and (0,1.4)  .. (0,1.4) node [left]{$\mathbf{y}$};

\end{tikzpicture} 
\par\end{centering}
\caption{\label{fig:DISDP}Inclusive deep-inelastic scattering in the dipole
picture}
\end{figure}
To describe DIS in the high-energy limit, which, according to Eq.
(\ref{eq:Bjorkenx}), corresponds to small values of Bjorken-$x$,
it is very useful to perform a boost from the infinite momentum frame
to the so-called dipole frame, in which the photon-quark vertex is
pulled out of the hadron (see Refs. \protect\cite{Mueller99,Mueller01,BaronePredazzi,EdmondRev,Nachtmann2008,KovchegovLevin}).
DIS in this \textquoteleft dipole picture' is then viewed as a two-step
process, which is illustrated in Fig. \ref{fig:DISDP}: before the
interaction, the photon splits up in a color-singlet quark-antiquark
pair, called a dipole, which subsequently scatters off the gluon content
of the hadron. We will show that in the small-$x$ limit, the lifetime
of this dipole is much longer than the interaction time with the proton,
and the scattering can be treated as eikonal such that the transverse
coordinates of the quark and antiquark do not change.

Let us substantiate these claims with some small calculations. First,
we write for the momenta $P$ and $q$ of the proton and the photon,
respectively:
\begin{equation}
\begin{aligned}P^{\mu} & \simeq\left(P,\mathbf{0}_{\perp},P\right),\quad q^{\mu}=\left(\sqrt{q^{2}-Q^{2}},\mathbf{0}_{\perp},-q\right).\end{aligned}
\end{equation}
Requiring that the photon is highly energetic: $q\gg Q$, the above
momenta can be written in light-cone coordinates as follows:
\begin{equation}
\begin{aligned}P^{\mu} & \simeq\left(P^{+},0,\mathbf{0}_{\perp}\right),\quad q^{\mu}=\left(-\frac{Q^{2}}{2q^{-}},q^{-},\mathbf{0}_{\perp}\right),\end{aligned}
\end{equation}
where $q^{-}\simeq q/\sqrt{2}$ and $P^{+}=\sqrt{2}P$. Since the
photon is a left-mover, its lifetime can be estimated as:
\begin{equation}
\Delta x_{\mathrm{dip}}^{-}\sim\frac{1}{\left|q^{+}\right|}=\frac{2q^{-}}{Q^{2}}.
\end{equation}
The longitudinal extent of the proton is:
\begin{equation}
\Delta x_{P}^{-}\sim\frac{2R}{\gamma}=2R\frac{m}{P}\sim\frac{1}{P},
\end{equation}
hence, from Eq. (\ref{eq:Bjorkenx}), requiring that the lifetime
of the dipole is much larger than the proton size is tantamount to
the requirement that $x$ is small:
\begin{equation}
\Delta x_{\mathrm{dip}}^{-}\gg\Delta x_{P}^{-}\quad\leftrightarrow\quad x\ll1.
\end{equation}
Note that, since in the dipole frame the proton (or nucleus) is squeezed
together due to Lorentz dilation, one often speaks of a \textquoteleft shockwave'
in the literature.

One of the big advantages of the dipole frame is that the interaction
of the dipole with the proton diagonalizes, in this sense that the
interaction becomes eikonal: the transverse positions $\mathbf{x}$
and $\mathbf{y}$ of the quark and antiquark do not change as a result
of the scattering. That this is the case is a simple corollary of
the above estimates: if the transverse dipole size is approximately
equal to $r_{\perp}^{2}\sim1/Q^{2}$, and changes by an amount $\Delta r_{\perp}=\Delta x_{P}^{-}k_{\perp}/q^{0}$
over the course of the interaction, we have that:
\begin{equation}
\frac{\Delta r_{\perp}}{r_{\perp}}\simeq\Delta x_{P}^{-}\frac{k_{\perp}^{2}}{q^{0}}\sim x\ll1.
\end{equation}
One can then show that the photoabsorption cross section can be factorized
as follows (see for instance Ref. \protect\cite{Nikolaev1991,Mueller99NPB}):
\begin{equation}
\begin{aligned}\sigma_{\gamma^{*}p}\left(Y,Q^{2}\right) & =\frac{\alpha_{em}e_{q}^{2}}{8\pi^{4}}p^{+}\int_{0}^{1}\mathrm{d}z\int\mathrm{d}^{2}\mathbf{r}\left|\psi^{\gamma\to q\bar{q}}\left(p^{+},z,\mathbf{r};Q^{2}\right)\right|^{2}\sigma_{\mathrm{dip}}\left(Y,\mathbf{r}\right),\end{aligned}
\label{eq:DISdipolecrosssection}
\end{equation}
where the wave function $\psi^{\gamma\to q\bar{q}}$ for the photon
with longitudinal momentum $p^{+}$\footnote{The factor $p^{+}$ cancels with the $1/p^{+}$ dependence of the
wave function squared.} and virtuality $Q^{2}$ to split in a quark-antiquark pair is convolved
with the dipole cross section $\sigma_{\mathrm{dip}}\left(Y,\mathbf{r}\right)$,
which encodes the eikonal interaction of the dipole with the proton.
The $\gamma\to q\bar{q}$ wave function can be calculated in light-cone
perturbation theory, and is in fact equal to the $g\to q\bar{q}$
wave function (see Sec. \ref{subsec:gqqwave}), up to the coupling
constant and a color factor. Note the appearance of the \textquoteleft rapidity'
$Y\equiv\ln1/x$ in Eq. (\ref{eq:DISdipolecrosssection}), which,
as it turns out, is a natural variable to describe small-$x$ processes.
It can be viewed as the difference in rapidity between the virtual
photon and the proton:
\begin{equation}
\begin{aligned}\Delta y & \equiv\frac{1}{2}\ln\frac{q^{-}}{q^{+}}-\frac{1}{2}\ln\frac{P^{-}}{P^{+}},\\
 & =\frac{1}{2}\ln\frac{4\left(q^{-}\right)^{2}\left(P^{+}\right)^{2}}{Q^{2}M^{2}},\\
 & =\ln\frac{1}{x}+\ln\frac{Q}{M}\sim Y\:\mathrm{if}\:x\ll1.
\end{aligned}
\end{equation}

\begin{figure}[t]
\centering{}\includegraphics[scale=0.3]{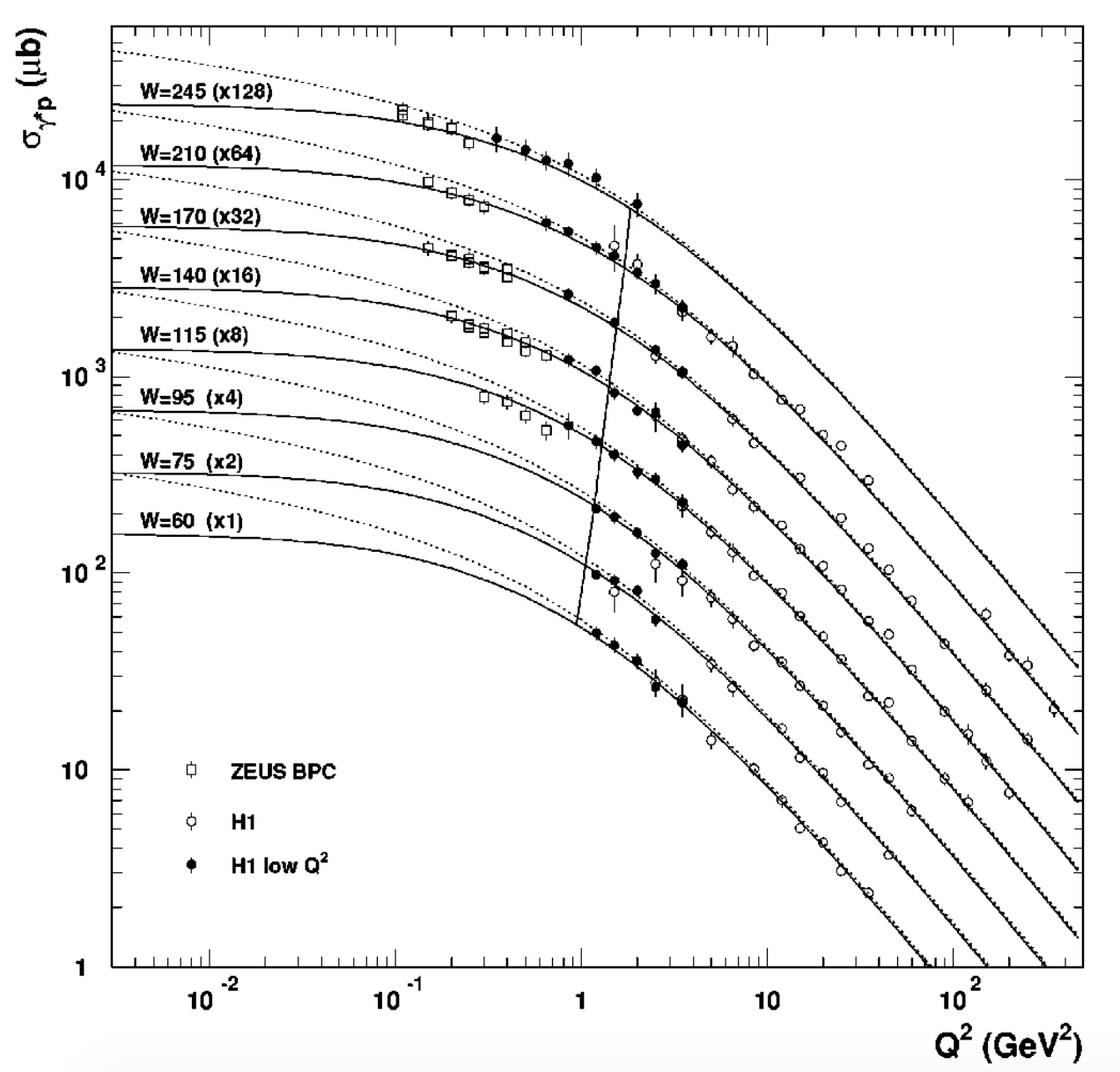}\includegraphics[scale=0.33]{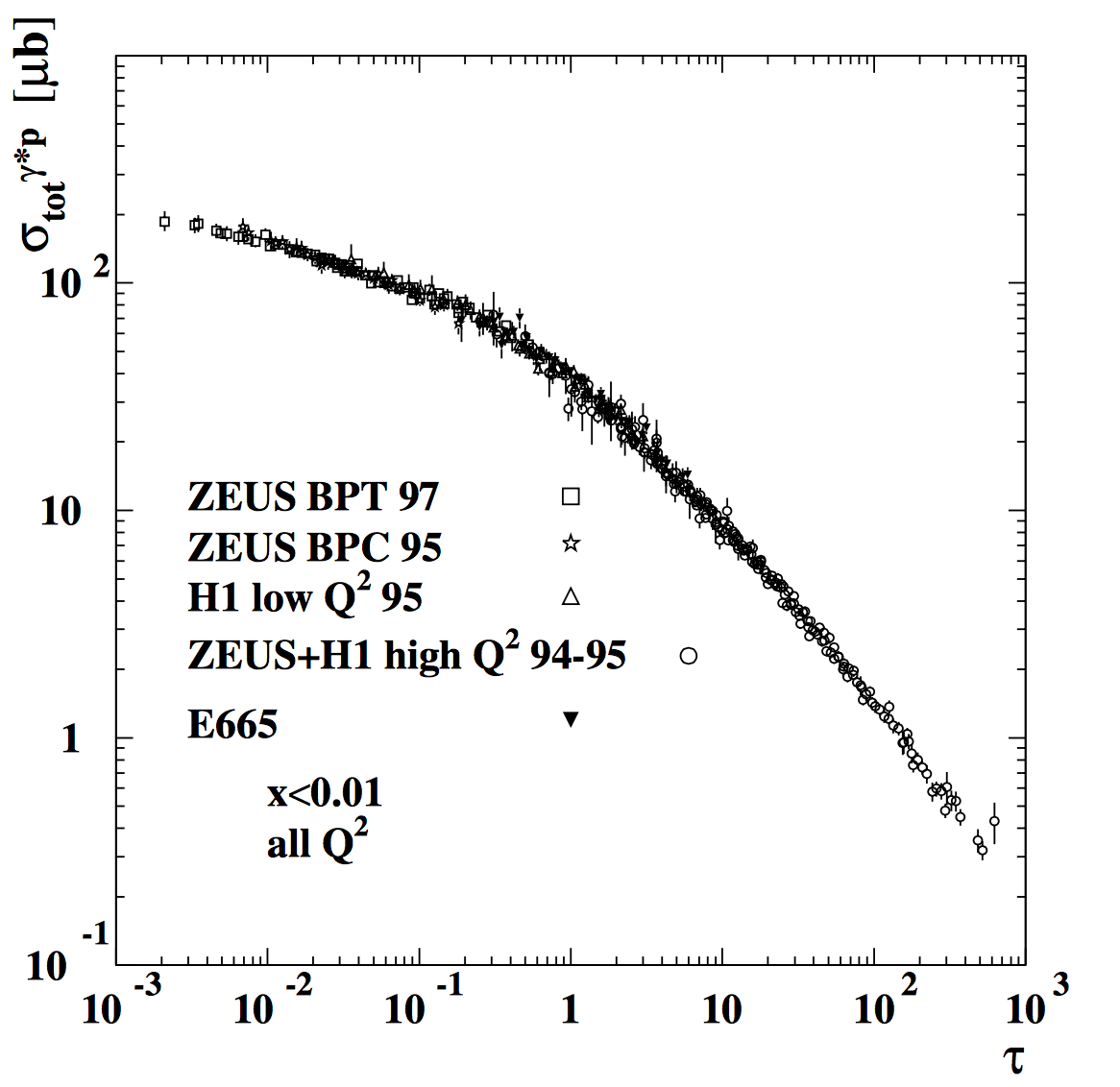}\caption{\label{fig:GBW}First hints of saturation physics in early HERA data.
Left: the Golec-Biernat and Wüsthoff fit \protect\cite{GBW1998} to Zeus and
H1 data on the photoabsorption cross section. $W=\sqrt{s}$, the center-of-mass
energy of the photon-proton scattering (in $\mathrm{GeV}$). The dotted
lines are the fits in which the three lightest quarks are massless,
and the full lines the ones in which they have a common mass of $m=140\,\mathrm{MeV}.$
Right: the dipole cross section $\sigma_{\gamma*p}$ in the small-$x$
($x<10^{-2}$) region, as a function of the scaling variable $\tau=Q^{2}R_{0}^{2}\left(x\right)$
(from Ref. \protect\cite{Stasto2001}).}
\end{figure}

Obviously, we cannot expect to be able to calculate the dipole cross
section $\sigma_{\mathrm{dip}}\left(Y,\mathbf{r}\right)$ in Eq. (\ref{eq:DISdipolecrosssection})
exactly, since it contains the information on the structure of the
proton, which is nonperturbative and hence beyond the reach of perturbative
QCD. The expression for $\sigma_{\mathrm{dip}}\left(Y,\mathbf{r}\right)$
is thus dependent on the way one models the small-$x$ gluon content
of the proton. A very well-known phenomenological parametrization
is the Golec-Biernat and Wüsthoff (GBW) model (Ref. \protect\cite{GBW1998}),
in which the dipole cross section is written as follows:
\begin{equation}
\begin{aligned}\sigma_{\mathrm{dip}}^{\mathrm{GBW}}\left(x,\mathbf{r}\right) & =\sigma_{0}\left(1-e^{-r^{2}/4R_{0}^{2}\left(x\right)}\right),\\
R_{0}^{2}\left(x\right) & =\left(\frac{x}{x_{0}}\right)^{\lambda}\mathrm{GeV}^{-2}.
\end{aligned}
\label{eq:GBW}
\end{equation}
In the above formulas, $R_{0}$ is the so-called correlation length,
and its inverse is known as the saturation scale $Q_{s}=1/R_{0}$.
The limit in which the dipole size $r$ is very small corresponds
to color transparency, in which the legs of the dipole close, upon
which the dipole becomes color neutral. In that case, the dipole cannot
interact and the cross section vanishes. In the opposite limit, for
a very large dipole size $r$, the dipole is completely absorbed by
the target, and the cross section saturates to the maximal possible
geometric cross section $\sigma_{0}\simeq2\pi R^{2}$ known as the
black disk limit. Golec-Biernat and Wüsthoff fitted their model to
the HERA data (see Fig. \ref{fig:GBW}), and obtained the following
parameters (for the light quark flavors $u$, $d$ and $s$):
\begin{equation}
\sigma_{0}=23.03\,\mathrm{mb},\quad\lambda=0.288,\quad x_{0}=3.04\cdot10^{-4}.
\end{equation}
\begin{figure}[t]
\begin{centering}
\includegraphics[scale=0.3]{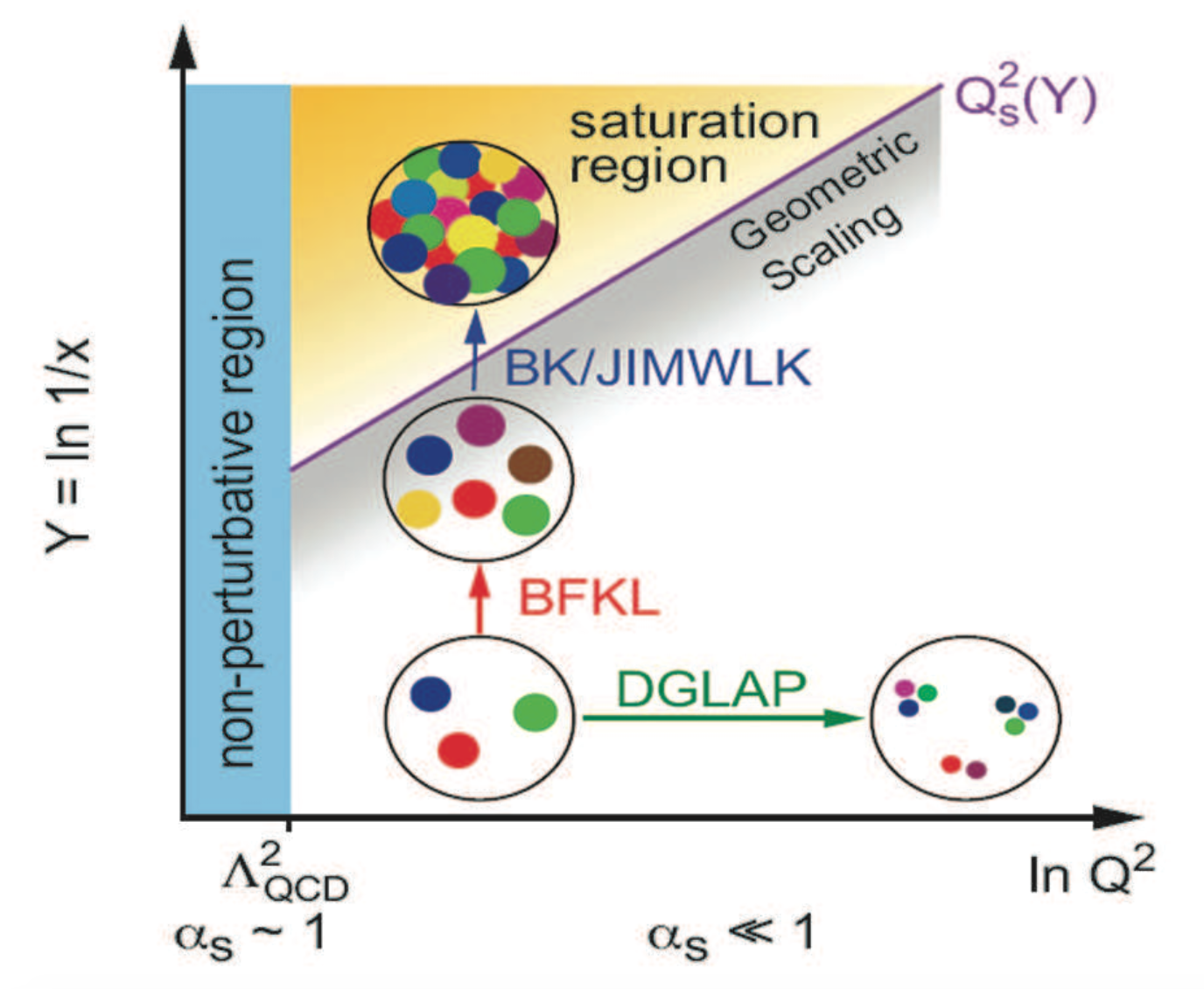}
\par\end{centering}
\caption{\label{fig:QCDevolution}The phase space for DIS, taken from Ref.
\protect\cite{Prokudin2012}. The transverse area of the partons are depicted
by the colored dots. As we will explain later, BK and JIMWLK are nonlinear
evolution equations in $1/x$, which are reduced to BFKL in the dilute
limit. Geometric scaling is an effect of the saturation boundary on
the evolution in the dilute regime. }
\end{figure}

In order to properly interpret Eq. (\ref{eq:GBW}), we should elaborate
on the saturation scale $Q_{s}\left(x\right)=1/R_{0}\left(x\right)$,
which plays a central role in this thesis. As we argued in the previous
section, small-$x$ physics is dominated by gluons, whose distribution
grows very fast towards smaller values of $x$: $x\mathcal{G}\left(x,Q^{2}\right)\sim\left(1/x\right)^{2.77\bar{\alpha}}$.
However, we saw as well that each gluon has a transverse extent of
the order $\sim1/Q^{2}$. Introducing the ratio of the total area
$x\mathcal{G}\left(x,Q^{2}\right)/Q^{2}$ occupied by all the gluons,
and the transverse hadron size $\pi R^{2}$: 
\begin{equation}
\begin{aligned}\varphi\left(x,Q^{2}\right) & \simeq\frac{1}{N_{c}^{2}-1}\frac{1}{\pi R^{2}}\frac{x\mathcal{G}\left(x,Q^{2}\right)}{Q^{2}},\end{aligned}
\label{eq:phioverlap}
\end{equation}
the gluons will overlap when the above so-called gluon overlapping
factor $\varphi\left(x,Q^{2}\right)$ is of order one. The system
is then in a high density regime, in which the gluon occupation factors
become so large that one expects the system to reach an equilibrium
state. Indeed, although suppressed by powers of $\alpha_{s}$, the
relative importance of nonlinear gluon merging effects $gg\to g$
increases due to the high density. These recombinations will act against
the very fast radiative growth of the gluon distribution, until both
effects compensate for each other and the gluon density saturates.
There are two ways to reach this highly dense so-called \textquoteleft saturation'
regime of QCD (see Fig. \ref{fig:QCDevolution}): first, fixing $Q^{2}$
and increasing the energy, thus decreasing $x$, the gluon distribution
$x\mathcal{G}\left(x,Q^{2}\right)$ rapidly grows, while each gluon
has more or less the same transverse extent $\sim1/Q^{2}$, and while
the hadron area $\pi R^{2}$ grows only logarithmically (see the following
section, Eq. (\ref{eq:Rlns}), for a heuristic derivation). At a certain
point the gluon overlapping factor becomes of order one, and saturation
kicks in. Second, one could keep $x$ fixed and decrease $Q^{2}$.
Indeed, even though the gluon distribution decreases towards lower
values of $Q^{2}$ since there is less phase space for radiation available,
$\varphi\left(x,Q^{2}\right)$ keeps growing because the transverse
area $1/Q^{2}$ of the gluons increases a lot faster, until the high
density regime is reached. The transverse momentum scale that marks
the onset of the saturation regime is precisely the saturation scale
$Q_{s}$:
\begin{equation}
\varphi\left(x,Q^{2}\right)\gtrsim\frac{1}{\alpha_{s}N_{c}}\quad\leftrightarrow\quad Q^{2}\lesssim Q_{s}^{2}.
\end{equation}
The opposite regime, where $Q^{2}>Q_{s}^{2}$, is known as the dilute
regime in which the gluon distribution keeps growing explosively.
An approximate expression for $Q_{s}$ is found by estimating the
rate $\Gamma\left(x,Q^{2}\right)$ of gluon recombinations:
\begin{equation}
\Gamma\left(x,Q^{2}\right)\sim\frac{\alpha_{s}N_{c}}{N_{c}^{2}-1}\frac{1}{\pi R^{2}}\frac{x\mathcal{G}\left(x,Q^{2}\right)}{Q^{2}},\label{eq:recombinationrate}
\end{equation}
which is of order one when $\varphi\left(x,Q^{2}\right)\sim1/\alpha_{s}N_{c}$.
Requiring $\Gamma\left(x,Q^{2}=Q_{s}^{2}\right)\sim1$ we then find:
\begin{equation}
Q_{s}^{2}\left(x\right)\sim\frac{\alpha_{s}N_{c}}{N_{c}^{2}-1}\frac{1}{\pi R^{2}}x\mathcal{G}\left(x,Q_{s}^{2}\right).
\end{equation}
From the above expression, we expect a power-like growth for the saturation
scale, similar to the growth of the gluon density in the dilute regime:
\begin{equation}
Q_{s}^{2}\left(x\right)\sim\left(\frac{1}{x}\right)^{\bar{\alpha}c_{s}},
\end{equation}
which is encoded as well in the GBW model, Eq. (\ref{eq:GBW}), and
which defines a straight \textquoteleft saturation line' $\ln Q_{s}^{2}\left(x\right)/Q_{0}^{2}$
in the kinematical ($\ln Q^{2}$,$Y$) plane for DIS, as is shown
in Fig. \ref{fig:QCDevolution} (see also Refs. \protect\cite{Gribov1983,Iancu2002,MuellerTriantafyllopoulos2002}).

The Golec-Biernat and Wüsthoff fit, as well as later more advanced
phenomenological models (see e.g. Refs. \protect\cite{Bartels2002,Kowalski2003,Iancu2004,Rezaeian2013})
are of great historical importance, since their quality made a strong
case for saturation physics. However, at present, small-$x$ evolution
equations such as BFKL and BK (the subjects of the next section) are
known to sufficient theoretical accuracy (including the running coupling
\protect\cite{Albacete2007,Kovchegov2007,Kovchegov2007bis,Balitsky2007},
collinear improvements \protect\cite{Iancu2016} or even at full NLO accuracy
\protect\cite{Balitsky2008,Balitsky2013,Kovner2014}) to make such model-dependent
fits obsolete. Instead, small-$x$ data can be fitted the same way
as in the case of DGLAP: one only models the initial condition (e.g.
the MV model, see Part. \ref{part:CGC}) and the impact-parameter
dependence (Ref. \protect\cite{Rezaeian2013}), after which the evolution
is completely fixed by the theory (see Refs. \protect\cite{Albacete2009,Albacete2011,KutakSapeta,Kuokkanen2012,Iancu2015}).
Despite these theoretical successes, it is however too optimistic
to state that saturation has really been observed at HERA: the data
points at very small-$x$ lie dangerously close to the nonperturbative
region. 

There is, however, another striking phenomenon in the DIS data from
HERA: geometric scaling (see Ref. \protect\cite{Stasto2001}). This is the
property that, at small-$x$, the $\gamma^{*}p$ cross section only
depends on $x$ and $Q^{2}\sim1/r^{2}$ via the dimensionless combination
$r^{2}/R_{0}^{2}\left(x\right)=Q_{s}^{2}\left(x\right)/Q^{2}$. This
scaling is explicitly encoded in the GBW-parametrization Eq. (\ref{eq:GBW}),
and indeed seems to hold, at least approximately, for $x<10^{-2}$
and virtualities $Q^{2}\leq450\,\mathrm{GeV}^{2}$ (see Fig. \ref{fig:GBW}).
In the saturation regime, scaling is a natural consequence of the
fact that $Q_{s}$ is the only relevant scale in this region of phase
space. Moreover, theoretical studies (see Ref. \protect\cite{Mueller2002,Iancu2002,Triantafyllopoulos2003})
have found that geometric scaling is also consistent with saturation
physics in the dilute regime $Q_{s}^{2}\ll Q^{2}\ll Q_{s}^{4}\left(x\right)/\Lambda_{\mathrm{QCD}}^{2}$,
where the linear evolution is influenced by the boundary condition
of saturation. 

In Part \ref{part:CGC}, we will describe in detail the physics of
the hadron structure in the saturation regime. First, however, we
should study the small-$x$ evolution equations from the point of
view of the projectile. Indeed, the latter will also be sensitive
to saturation, via the unitarization of the amplitude for multiple
scattering.

\section{\label{sec:BKBFKL}BFKL and BK evolution equations}

\begin{figure}[t]
\begin{centering}
\begin{tikzpicture}[scale=1.3] 

\tikzset{photon/.style={semithick,decorate,decoration={snake}},
		electron/.style={semithick,postaction={decorate},decoration={markings,mark=at position .515 with {\arrow[draw]{latex}}}}, 	positron/.style={semithick,postaction={decorate},decoration={markings,mark=at position .505 with {\arrow[draw]{latex reversed}}}},     	gluon/.style={decorate,decoration={coil,amplitude=4pt, segment length=5pt}}}

\fill[black!30!white, rounded corners] (1.8,.9) rectangle(2.2,2.7);
\draw[photon,semithick] (4.5,1.8) node[right] {$$}--++(-.5,0);
\draw[positron,semithick] (4,1.8).. controls (3.8,2.3) and (0,2.2)  .. (0,1.8) node [left]{$$};
\draw[electron,semithick] (4,1.8).. controls (3.8,1.3) and (0,1.4)  .. (0,1.8) node [left]{$$};
\draw[photon,semithick] (0,1.8) node[right] {$$}--++(-.5,0);
\draw[gluon] (.5,2.02) .. controls (.7,1.7) and (1.3, 1.8) .. (1.5, 2.13);
\node at (0,2.3) {(a)};

\fill[black!30!white, rounded corners] (7.3,.9) rectangle(7.7,2.7);
\draw[photon,semithick] (10,1.8) node[right] {$$}--++(-.5,0);
\draw[positron,semithick] (9.5,1.8).. controls (9.3,2.3) and (5.5,2.2)  .. (5.5,1.8) node [left]{$$};
\draw[electron,semithick] (9.5,1.8).. controls (9.3,1.3) and (5.5,1.4)  .. (5.5,1.8) node [left]{$$};
\draw[photon,semithick] (5.5,1.8) node[right] {$$}--++(-.5,0);
\draw[gluon] (6.5,2.1) -- (8.5,1.5);
\node at (5.5,2.3) {(b)};

\fill[black!30!white, rounded corners] (1.8,-1.5) rectangle(2.2,.3);
\draw[photon,semithick] (4.5,-0.6) node[right] {$$}--++(-.5,0);
\draw[positron,semithick] (4,-.6).. controls (3.8,-.1) and (0,-.2)  .. (0,-.6) node [left]{$$};
\draw[electron,semithick] (4,-.6).. controls (3.8,-1.1) and (0,-1)  .. (0,-.6) node [left]{$$};
\draw[photon,semithick] (0,-.6) node[right] {$$}--++(-.5,0);
\draw[gluon] (1.2,-.285) --++ (0,-.625);
\node at (0,-.1) {(c)};

\fill[black!30!white, rounded corners] (7.3,-1.5) rectangle(7.7,.3);
\draw[photon,semithick] (10,-0.6) node[right] {$$}--++(-.5,0);
\draw[positron,semithick] (9.5,-.6).. controls (9.3,-.1) and (5.5,-.2)  .. (5.5,-.6) node [left]{$$};
\draw[electron,semithick] (9.5,-.6).. controls (9.3,-1.1) and (5.5,-1)  .. (5.5,-.6) node [left]{$$};
\draw[photon,semithick] (5.5,-.6) node[right] {$$}--++(-.5,0);
\draw[gluon] (6.5,-.31) .. controls (6.7,-.8) and (8.2,-.8) .. (8.5,-.3);
\node at (5.5,-.1) {(d)};
\end{tikzpicture} 
\par\end{centering}
\caption{\label{fig:BK}Contributions to the BK equation.}
\end{figure}
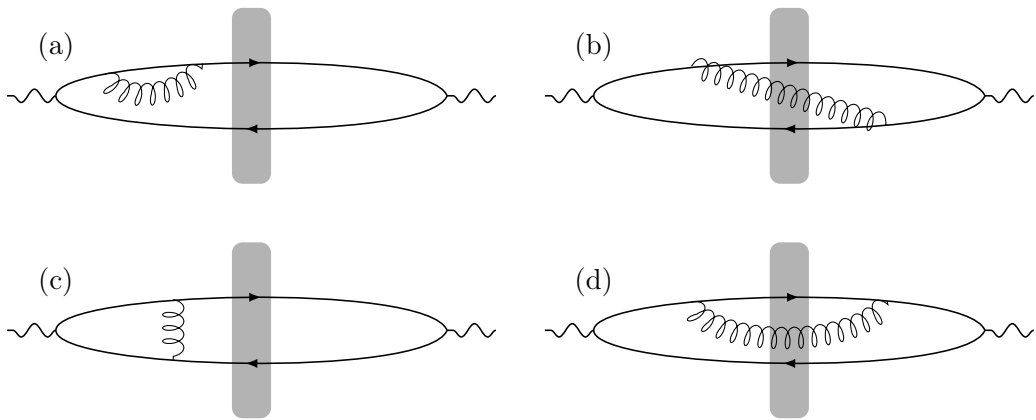

We have seen that, when calculating radiative QCD corrections to a
certain process, in our case DIS, large logarithms in transverse momentum
threaten to jeopardize perturbation theory, and hence have to be resummed
in the DGLAP evolution equations. However, as is clear from the Bremsstrahlung
law Eq. (\ref{eq:Bremsstrahlung law}), in the small-$x$ regime the
logarithms in the energy might become a lot more problematic than
the ones in transverse momentum:
\begin{equation}
\ln\frac{1}{x}\gg\ln\frac{Q^{2}}{Q_{0}^{2}}.
\end{equation}
In this case, the situation is opposite to the one we encountered
in DGLAP: now, the large logarithms $\alpha_{s}\ln1/x$ have to be
resummed, while the $k_{\perp}$-dependence can be computed exactly.
The result of such a procedure is the famous Balitsky-Fadin-Kuraev-Lipatov
or BFKL equation (Refs. \protect\cite{BFKL1,BFKL2,BFKL3,BFKL4}), which holds
at asymptotic small-$x$, and hence involves only gluons. Being a
LLA equation, the successive gluon emissions are strongly ordered
in energy:
\begin{equation}
x_{0}\gg...\gg x_{i}\gg x_{i+1}\gg...\gg x_{n}.
\end{equation}
Building gluon ladders from the proton to the partonic scattering
process, like we did for DGLAP (cf. Figs. \ref{fig:DGLAPladder} and
\ref{fig:DISQCD}), turns out to be significantly more involved in
the case of BFKL, and involves reshuffling perturbation theory by
trading the three-gluon vertices for the so-called Lipatov vertices,
which effectively sum different radiative contributions. However,
nothing prevents us to go to a Lorentz frame in which the major part
of the energy is contained in the dipole, rather than in the proton.
In such a frame, one can construct an evolution equation for the dipole
(we follow Ref. \protect\cite{Gelis2013}), which turns out to be much simpler
than constructing BFKL from the point of view of the hadron.

One of the starting points of the calculation is the observation that
the multiple scatterings of the dipole off the Lorentz-contracted
proton or \textquoteleft shockwave' are eikonal, as we already argued
in the previous chapter. This implies that they can be resummed (see
e.g. \protect\cite{Peskin}) into Wilson lines which are directed along each
leg of the dipole, and which are defined as:
\begin{equation}
U=\mathcal{P}e^{ig_{s}\int\mathrm{d}z_{\mu}A_{a}^{\mu}\left(z\right)t^{a}}.
\end{equation}
In the above formula, $\mathcal{P}$ is the path ordering operator,
which places the color matrices $t^{a}$ from left to right in the
order of their appearance along the integration path. If the dipole
is a left-mover, as in Fig. \ref{fig:DISDP}, the Wilson lines are
directed along the $x^{-}$-axis, and hence the quark and antiquark
couple only to the $A^{+}$ fields of the proton:
\begin{equation}
U\left(\mathbf{x}\right)=\mathcal{P}e^{ig_{s}\int\mathrm{d}z^{-}A_{a}^{+}\left(z^{-},\mathbf{x}\right)t^{a}}.
\end{equation}
In function of these Wilson lines, the interaction of the dipole with
the shockwave can be written as:
\begin{equation}
\begin{aligned}D_{Y}\left(\mathbf{x}-\mathbf{y}\right) & \equiv\langle S\left(\mathbf{x},\mathbf{y}\right)\rangle_{Y},\qquad S\left(\mathbf{x},\mathbf{y}\right)\equiv\frac{1}{N_{c}}\mathrm{Tr}\left(U\left(\mathbf{x}\right)U^{\dagger}\left(\mathbf{y}\right)\right),\end{aligned}
\label{eq:dipole}
\end{equation}
where $D_{Y}\left(\mathbf{x}-\mathbf{y}\right)$ is the $S$-matrix
of dipole-shockwave scattering, and where the information on the structure
of the proton is now contained in the averages $\left\langle ...\right\rangle _{Y}$
over the gluon fields inside the Wilson lines. We will often call
$D_{Y}\left(\mathbf{x}-\mathbf{y}\right)$ simply a \textquoteleft dipole'.
In the above formula, $U\left(\mathbf{x}\right)$ corresponds to the
quark leg and $U^{\dagger}\left(\mathbf{y}\right)$ to the antiquark
leg, and we take a trace because the dipole is in a singlet configuration.
In addition, we assume translation symmetry in the transverse plane.
As an example, the dipole cross section from Eq. (\ref{eq:DISdipolecrosssection})
can be written with the help of the optical theorem as:
\begin{equation}
\begin{aligned}\sigma_{\mathrm{dip}}\left(Y,\mathbf{r}\right) & =2S_{\perp}\langle T\left(\mathbf{r},\mathbf{0}\right)\rangle_{Y},\qquad T\left(\mathbf{x},\mathbf{y}\right)\equiv1-S\left(\mathbf{x},\mathbf{y}\right),\end{aligned}
\label{eq:dipolecrossection}
\end{equation}
where $S_{\perp}=\pi R^{2}$ is the transverse area of the proton,
and where $T\left(\mathbf{x},\mathbf{y}\right)$ is the forward scattering
amplitude\footnote{$T$ is defined as the imaginary part of the usual scattering amplitude
$\mathcal{A}$, which is in turn defined through the $S$-matrix as
follows: $S=1+i\mathcal{A}$. }. 

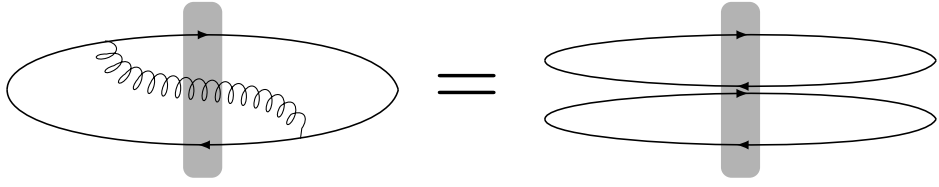
\begin{figure}[t]
\begin{centering}
\begin{tikzpicture}[scale=1.3] 

\tikzset{photon/.style={semithick,decorate,decoration={snake}},
		electron/.style={semithick,postaction={decorate},decoration={markings,mark=at position .51 with {\arrow[draw]{latex}}}}, 	positron/.style={semithick,postaction={decorate},decoration={markings,mark=at position .51 with {\arrow[draw]{latex reversed}}}},     	gluon/.style={decorate,decoration={coil,amplitude=4pt, segment length=5pt}}}

\fill[black!30!white, rounded corners] (7.3,.9) rectangle(7.7,2.7);
\draw[positron,semithick] (9.5,2.1).. controls (9.3,2.5) and (5.5,2.4)  .. (5.5,2.1) node [left]{$$};
\draw[electron,semithick] (9.5,2.1).. controls (9.3,1.7) and (5.5,1.8)  .. (5.5,2.1) node [left]{$$};
\draw[positron,semithick] (9.5,1.5).. controls (9.3,1.9) and (5.5,1.8)  .. (5.5,1.5) node [left]{$$};
\draw[electron,semithick] (9.5,1.5).. controls (9.3,1.1) and (5.5,1.2)  .. (5.5,1.5) node [left]{$$};

\node[scale=3] at (4.7,1.8) {$=$};

\fill[black!30!white, rounded corners] (1.8,.9) rectangle(2.2,2.7);

\draw[positron,semithick] (4,1.8).. controls (3.8,2.6) and (0,2.5)  .. (0,1.8) node [left]{$$};
\draw[electron,semithick] (4,1.8).. controls (3.8,1.) and (0,1.1)  .. (0,1.8) node [left]{$$};

\draw[gluon] (1,2.3) .. controls (1.1,1.5) and (2.9,2.1) .. (3,1.3);

\end{tikzpicture} 
\par\end{centering}
\caption{\label{fig:BKevolution}In the large-$N_{c}$ limit, a dipole emitting
a real gluon can be treated as a system of two dipoles.}
\end{figure}

Let us now compute the leading-order corrections to the dipole, Eq.
(\ref{eq:dipole}), when going to smaller values of $x$, or equivalently,
when increasing the rapidity $Y$. We already calculated the emission
of an eikonal gluon from a quark, and found that it gives an effective
vertex (see Eq. (\ref{eq:amplitudeeikonalgluon})): 
\begin{equation}
-2g_{s}t^{a}\frac{\mathbf{k}_{\perp}\cdot\boldsymbol{\epsilon}_{\perp}^{\lambda}}{k_{\perp}^{2}}.
\end{equation}
Since such an eikonal emission preserves the transverse position of
the quark or antiquark leg, it is more appropriate to work in the
mixed Fourier space, in which we only transform the transverse momenta
to coordinate space. For the above vertex, this yields:
\begin{equation}
-2g_{s}t^{a}\int\frac{\mathrm{d}^{2}\mathbf{k}_{\perp}}{\left(2\pi\right)^{2}}e^{i\mathbf{k}_{\perp}\left(\mathbf{x}-\mathbf{z}\right)}\frac{\mathbf{k}_{\perp}\cdot\boldsymbol{\epsilon}_{\perp}^{\lambda}}{k_{\perp}^{2}}=-i\frac{g_{s}}{\pi}t^{a}\frac{\left(\mathbf{x}-\mathbf{z}\right)\cdot\boldsymbol{\epsilon}_{\perp}^{\lambda}}{\left(\mathbf{x}-\mathbf{z}\right)^{2}},
\end{equation}
where we made use of formula (\ref{eq:WWfieldrelation}). Using $\sum_{\lambda}\epsilon_{i}^{\lambda\dagger}\epsilon_{j}^{\lambda}=\delta_{ij}$
(from Eq. (\ref{eq:polarization vectors})), we obtain the following
result for the self-energy corrections from Fig. \ref{fig:BK} (a)
(compare with Eq. (\ref{eq:sigmabremss})):
\begin{equation}
\begin{aligned}\left(\mathrm{a}\right) & =-2\alpha_{s}\Delta Y\int\frac{\mathrm{d}^{2}\mathbf{z}}{\left(2\pi\right)^{2}}\frac{1}{\left(\mathbf{x}-\mathbf{z}\right)^{2}}\mathrm{Tr}\langle t^{a}t^{a}U\left(\mathbf{x}\right)U^{\dagger}\left(\mathbf{y}\right)\rangle_{Y}.\end{aligned}
\end{equation}
The other virtual correction, see Fig. \ref{fig:BK} (c), in which
the legs of the dipole exchange a gluon before or after the interaction
with the target, yields:
\begin{equation}
\begin{aligned}\left(\mathrm{c}\right) & =4\alpha_{s}\Delta Y\int\frac{\mathrm{d}^{2}\mathbf{z}}{\left(2\pi\right)^{2}}\frac{\left(\mathbf{x}-\mathbf{z}\right)\cdot\left(\mathbf{y}-\mathbf{z}\right)}{\left(\mathbf{x}-\mathbf{z}\right)^{2}\left(\mathbf{y}-\mathbf{z}\right)^{2}}\mathrm{Tr}\langle U\left(\mathbf{x}\right)t^{a}t^{a}U^{\dagger}\left(\mathbf{y}\right)\rangle_{Y}.\end{aligned}
\end{equation}
Combining the four possible self-energy corrections (a), and the two
diagrams like (c), we obtain:
\begin{equation}
\begin{aligned}\mathrm{virtual}\;\mathrm{corrections} & =-4\alpha_{s}\Delta Y\mathrm{Tr}\langle t^{a}t^{a}U\left(\mathbf{x}\right)U^{\dagger}\left(\mathbf{y}\right)\rangle_{Y}\\
 & \times\int\frac{\mathrm{d}^{2}\mathbf{z}}{\left(2\pi\right)^{2}}\left(\frac{1}{\left(\mathbf{x}-\mathbf{z}\right)^{2}}+\frac{1}{\left(\mathbf{y}-\mathbf{z}\right)^{2}}-2\frac{\left(\mathbf{x}-\mathbf{z}\right)\cdot\left(\mathbf{y}-\mathbf{z}\right)}{\left(\mathbf{x}-\mathbf{z}\right)^{2}\left(\mathbf{y}-\mathbf{z}\right)^{2}}\right),\\
 & =-4\alpha_{s}C_{F}N_{c}\Delta Y\,D_{Y}\left(\mathbf{x}-\mathbf{y}\right)\\
 & \times\int\frac{\mathrm{d}^{2}\mathbf{z}}{\left(2\pi\right)^{2}}\frac{\left(\mathbf{x}-\mathbf{y}\right)^{2}}{\left(\mathbf{x}-\mathbf{z}\right)^{2}\left(\mathbf{y}-\mathbf{z}\right)^{2}},
\end{aligned}
\label{eq:BKvirtual}
\end{equation}
where we made use of definition of the Casimir operator (\ref{eq:Casimirfund}).
Likewise, the real correction in which the emitted gluon crosses the
shockwave (Fig. \ref{fig:BK}, b) yields:
\begin{equation}
\begin{aligned}\left(\mathrm{b}\right) & =-4\alpha_{s}\Delta Y\int\frac{\mathrm{d}^{2}\mathbf{z}}{\left(2\pi\right)^{2}}\frac{\left(\mathbf{x}-\mathbf{z}\right)\cdot\left(\mathbf{y}-\mathbf{z}\right)}{\left(\mathbf{x}-\mathbf{z}\right)^{2}\left(\mathbf{y}-\mathbf{z}\right)^{2}}\mathrm{Tr}\langle t^{a}U\left(\mathbf{x}\right)t^{b}U^{\dagger}\left(\mathbf{y}\right)W_{ab}\left(\mathbf{z}\right)\rangle_{Y},\end{aligned}
\end{equation}
where $W\left(\mathbf{z}\right)$ is a Wilson line in the adjoint
representation:
\begin{equation}
W\left(\mathbf{z}\right)=\mathcal{P}e^{ig_{s}\int\mathrm{d}z^{-}A_{a}^{+}\left(z^{-},\mathbf{z}\right)T^{a}},
\end{equation}
encoding the multiple eikonal scatterings of the soft gluon on the
target. Using Eq. (\ref{eq:U2W}), the expression for (b) can be rewritten
as:
\begin{equation}
\begin{aligned}\left(\mathrm{b}\right) & =-4\alpha_{s}\Delta Y\int\frac{\mathrm{d}^{2}\mathbf{z}}{\left(2\pi\right)^{2}}\frac{\left(\mathbf{x}-\mathbf{z}\right)\cdot\left(\mathbf{y}-\mathbf{z}\right)}{\left(\mathbf{x}-\mathbf{z}\right)^{2}\left(\mathbf{y}-\mathbf{z}\right)^{2}}\mathrm{Tr}\langle t^{a}U\left(\mathbf{x}\right)U^{\dagger}\left(\mathbf{z}\right)t^{a}U\left(\mathbf{z}\right)U^{\dagger}\left(\mathbf{y}\right)\rangle_{Y},\\
 & =-4\alpha_{s}\Delta Y\int\frac{\mathrm{d}^{2}\mathbf{z}}{\left(2\pi\right)^{2}}\frac{\left(\mathbf{x}-\mathbf{z}\right)\cdot\left(\mathbf{y}-\mathbf{z}\right)}{\left(\mathbf{x}-\mathbf{z}\right)^{2}\left(\mathbf{y}-\mathbf{z}\right)^{2}}\\
 & \times\left(\frac{1}{2}\langle\mathrm{Tr}\left(U\left(\mathbf{x}\right)U^{\dagger}\left(\mathbf{z}\right)\right)\mathrm{Tr}\left(U\left(\mathbf{z}\right)U^{\dagger}\left(\mathbf{y}\right)\right)\rangle_{Y}-\frac{1}{2N_{c}}\mathrm{Tr}\langle U\left(\mathbf{x}\right)U^{\dagger}\left(\mathbf{y}\right)\rangle_{Y}\right),
\end{aligned}
\end{equation}
where we again made use of Eq. (\ref{eq:Fierz}). Finally, the real
correction in Fig. \ref{fig:BK} (d) gives:
\begin{equation}
\begin{aligned}\left(\mathrm{d}\right) & =2\alpha_{s}\Delta Y\int\frac{\mathrm{d}^{2}\mathbf{z}}{\left(2\pi\right)^{2}}\frac{1}{\left(\mathbf{x}-\mathbf{z}\right)^{2}}\\
 & \times\left(\frac{1}{2}\langle\mathrm{Tr}\left(U\left(\mathbf{x}\right)U^{\dagger}\left(\mathbf{z}\right)\right)\mathrm{Tr}\left(U\left(\mathbf{z}\right)U^{\dagger}\left(\mathbf{y}\right)\right)\rangle_{Y}-\frac{1}{2N_{c}}\mathrm{Tr}\langle U\left(\mathbf{x}\right)U^{\dagger}\left(\mathbf{y}\right)\rangle_{Y}\right).
\end{aligned}
\end{equation}
Summing the real emissions, we obtain:

\begin{equation}
\begin{aligned}\mathrm{real}\;\mathrm{emissions} & =2\alpha_{s}\Delta Y\int\frac{\mathrm{d}^{2}\mathbf{z}}{\left(2\pi\right)^{2}}\frac{\left(\mathbf{x}-\mathbf{y}\right)^{2}}{\left(\mathbf{x}-\mathbf{z}\right)^{2}\left(\mathbf{y}-\mathbf{z}\right)^{2}}\\
 & \times\left(\frac{1}{N_{c}^{2}}\langle\mathrm{Tr}\left(U\left(\mathbf{x}\right)U^{\dagger}\left(\mathbf{z}\right)\right)\mathrm{Tr}\left(U\left(\mathbf{z}\right)U^{\dagger}\left(\mathbf{y}\right)\right)\rangle_{Y}-D_{Y}\left(\mathbf{x}-\mathbf{y}\right)\right),
\end{aligned}
\label{eq:BKreal}
\end{equation}
and the sum of the real and virtual contributions is:
\begin{equation}
-2\alpha_{s}N_{c}^{2}\Delta Y\int\frac{\mathrm{d}^{2}\mathbf{z}}{\left(2\pi\right)^{2}}\mathcal{M}_{\mathbf{xyz}}\left(D_{Y}\left(\mathbf{x}-\mathbf{y}\right)-\frac{1}{N_{c}^{2}}\langle\mathrm{Tr}\left(U\left(\mathbf{x}\right)U^{\dagger}\left(\mathbf{z}\right)\right)\mathrm{Tr}\left(U\left(\mathbf{z}\right)U^{\dagger}\left(\mathbf{y}\right)\right)\rangle_{Y}\right),\label{eq:BKrhs}
\end{equation}
where $\mathcal{M}_{\mathbf{xyz}}$ is the so-called dipole kernel:
\begin{equation}
\mathcal{M}_{\mathbf{xyz}}\equiv\frac{\left(\mathbf{x}-\mathbf{y}\right)^{2}}{\left(\mathbf{x}-\mathbf{z}\right)^{2}\left(\mathbf{y}-\mathbf{z}\right)^{2}}.\label{eq:dipolekernel}
\end{equation}
Eq. (\ref{eq:BKrhs}) is the sum of the leading-order radiative corrections
to the multiple eikonal interactions of a dipole with the shockwave
when increasing the phase space in rapidity. Hence, we obtain the
following evolution equation:
\begin{equation}
\begin{aligned}\frac{\partial}{\partial Y}D_{Y}\left(\mathbf{x}-\mathbf{y}\right) & =-\frac{\bar{\alpha}}{2\pi}\int_{\mathbf{z}}\mathcal{M}_{\mathbf{xyz}}\left(D_{Y}\left(\mathbf{x}-\mathbf{y}\right)-\frac{1}{N_{c}^{2}}\langle\mathrm{Tr}\left(U\left(\mathbf{x}\right)U^{\dagger}\left(\mathbf{z}\right)\right)\mathrm{Tr}\left(U\left(\mathbf{z}\right)U^{\dagger}\left(\mathbf{y}\right)\right)\rangle_{Y}\right),\end{aligned}
\label{eq:Balitksy}
\end{equation}
which a non-closed equation, the first in the so-called Balitsky hierarchy
(see Ref. \protect\cite{Balitksy1996}). It should be noted that the dipole
kernel is singular for \textbf{$\mathbf{z}=\mathbf{x}$ }and $\mathbf{z}=\mathbf{y}$,
causing possible ultraviolet divergences when integrating over $\mathbf{z}$.
However, since $U\left(\mathbf{x}\right)U^{\dagger}\left(\mathbf{x}\right)=1$,
in these cases both terms in Eq. (\ref{eq:Balitksy}) cancel exactly.
It was therefore crucial to correctly account for the real and virtual
contributions in order to make the equation regular. Equation (\ref{eq:Balitksy})
can be closed by going to the large-$N_{c}$ limit, in which the medium
average of two dipoles factorizes :
\begin{equation}
\frac{1}{N_{c}^{2}}\langle\mathrm{Tr}\left(U\left(\mathbf{x}\right)U^{\dagger}\left(\mathbf{z}\right)\right)\mathrm{Tr}\left(U\left(\mathbf{z}\right)U^{\dagger}\left(\mathbf{y}\right)\right)\rangle_{Y}=\frac{1}{N_{c}}\langle\mathrm{Tr}\left(U\left(\mathbf{x}\right)U^{\dagger}\left(\mathbf{z}\right)\right)\rangle_{Y}\frac{1}{N_{c}}\langle\mathrm{Tr}\left(U\left(\mathbf{z}\right)U^{\dagger}\left(\mathbf{y}\right)\right)\rangle_{Y},\label{eq:largeNcfactorization}
\end{equation}
sacrificing\footnote{Interestingly, it turns out, however, that the differences between
the evolution of a dipole with the Balitsky-JIMWLK equation and with
BK are remarkably small, of the order of $\sim0.1\%$, see Ref. \protect\cite{Kovchegov2009}.} some of the correlations between two dipoles which are built up during
the evolution. The result is the Balitsky-Kovchegov (BK) equation
(Refs. \protect\cite{Kovchegov1999,Kovchegov2000}):
\begin{equation}
\begin{aligned}\frac{\partial}{\partial Y}D_{Y}\left(\mathbf{x}-\mathbf{y}\right) & =-\frac{\bar{\alpha}}{2\pi}\int_{\mathbf{z}}\mathcal{M}_{\mathbf{xyz}}\Bigl(D_{Y}\left(\mathbf{x}-\mathbf{y}\right)-D_{Y}\left(\mathbf{x}-\mathbf{z}\right)D_{Y}\left(\mathbf{y}-\mathbf{z}\right)\Bigr).\end{aligned}
\label{eq:BKweakfield}
\end{equation}
The interpretation of the BK equation is now very straightforward:
opening the phase space, by increasing the rapidity, allows for the
emission of a real gluon by one of the legs of the dipole. In the
large-$N_{c}$ limit, this gluon can be seen as a quark-antiquark
pair, and effectively a system $D_{Y}\left(\mathbf{x}-\mathbf{z}\right)D_{Y}\left(\mathbf{y}-\mathbf{z}\right)$
of two \textquoteleft child' dipoles scatters off the shockwave (see
Fig. \ref{fig:BKevolution}). The virtual term, in which the gluon
is emitted and reabsorbed before or after the scattering, contributes
with a term $D_{Y}\left(\mathbf{x}-\mathbf{y}\right)$ in the r.h.s.
of Eq. (\ref{eq:BKweakfield}), conserving probability and ensuring
the cancellation of the ultraviolet divergences.

In the dilute limit $\left(\mathbf{x}-\mathbf{y}\right)^{2}\ll1/Q_{s}^{2}$,
the probability for the two child dipoles to simultaneously interact
with the shockwave can be neglected. The BK equation Eq. (\ref{eq:BKweakfield})
can simply be linearized by rewriting it in function of the scattering
amplitude $T\left(\mathbf{x},\mathbf{y}\right)$, defined earlier
in Eq. (\ref{eq:dipolecrossection}):
\begin{equation}
\begin{aligned}\frac{\partial}{\partial Y}\left(1-\left\langle T\left(\mathbf{x},\mathbf{y}\right)\right\rangle _{\tau}\right) & =-\frac{\bar{\alpha}}{2\pi}\int_{\mathbf{z}}\mathcal{M}_{\mathbf{xyz}}\Bigl(1-\left\langle T\left(\mathbf{x},\mathbf{y}\right)\right\rangle _{Y}-\left(1-\left\langle T\left(\mathbf{x},\mathbf{z}\right)\right\rangle _{Y}\right)\left(1-\left\langle T\left(\mathbf{y},\mathbf{z}\right)\right\rangle _{Y}\right)\Bigr),\end{aligned}
\end{equation}
or:
\begin{equation}
\begin{aligned}\frac{\partial}{\partial Y}\left\langle T\left(\mathbf{x},\mathbf{y}\right)\right\rangle _{\tau} & =\frac{\bar{\alpha}}{2\pi}\int_{\mathbf{z}}\mathcal{M}_{\mathbf{xyz}}\left(\left\langle T\left(\mathbf{x},\mathbf{z}\right)\right\rangle _{Y}+\left\langle T\left(\mathbf{y},\mathbf{z}\right)\right\rangle _{Y}-\left\langle T\left(\mathbf{x},\mathbf{y}\right)\right\rangle _{Y}-\left\langle T\left(\mathbf{x},\mathbf{z}\right)\right\rangle _{Y}\left\langle T\left(\mathbf{y},\mathbf{z}\right)\right\rangle _{Y}\right).\end{aligned}
\label{eq:BK4T}
\end{equation}
To linearize the above formulation of the BK equation, one simply
neglects the quadratic term:
\begin{equation}
\begin{aligned}\frac{\partial}{\partial Y}\left\langle T\left(\mathbf{x},\mathbf{y}\right)\right\rangle _{Y} & =\frac{\bar{\alpha}}{2\pi}\int_{\mathbf{z}}\mathcal{M}_{\mathbf{xyz}}\left(\left\langle T\left(\mathbf{x},\mathbf{z}\right)\right\rangle _{Y}+\left\langle T\left(\mathbf{y},\mathbf{z}\right)\right\rangle _{Y}-\left\langle T\left(\mathbf{x},\mathbf{y}\right)\right\rangle _{Y}\right).\end{aligned}
\label{eq:BFKL}
\end{equation}
This is the BFKL equation in coordinate space, describing the linear
small-$x$ evolution of the projectile (the dipole-hadron scattering
amplitude). It was first derived by Mueller (Refs. \protect\cite{Mueller1994,Mueller1995})
for the scattering of two quarkonia.

From Eq. (\ref{eq:BFKL}), one can obtain the more traditional formulation
of BFKL as an evolution equation for the gluon distribution in the
proton. More specifically, since BFKL treats the $k_{\perp}$-dependence
of the splittings exactly, it pertains to the unintegrated gluon distribution
$f\left(x,k_{\perp}\right)$ defined as:
\begin{equation}
f\left(x,k_{\perp}\right)=\pi k_{\perp}^{2}\frac{\mathrm{d}N}{\mathrm{d}Y\mathrm{d}^{2}\mathbf{k}_{\perp}},\label{eq:unintgluondistrspectrum}
\end{equation}
which is related to the integrated gluon distribution as follows:
\begin{equation}
\begin{aligned}x\mathcal{G}\left(x,Q^{2}\right) & \equiv\int^{Q^{2}}\frac{\mathrm{d}^{2}\mathbf{k}_{\perp}}{\pi k_{\perp}^{2}}f\left(x,k_{\perp}\right)\quad\mathrm{or}\quad f\left(x,k_{\perp}\right)\equiv\frac{\partial x\mathcal{G}\left(x,k_{\perp}^{2}\right)}{\partial\ln k_{\perp}^{2}}.\end{aligned}
\label{eq:unintgluondist}
\end{equation}
In the dilute limit, $f\left(x,k_{\perp}\right)$ is linearly related
to the dipole cross section, as we will prove in the following section:
\begin{equation}
\begin{aligned}\sigma_{\mathrm{dip}}\left(x,\mathbf{r}\right) & \simeq\frac{4\pi\alpha_{s}}{N_{c}}\int\frac{\mathrm{d}^{2}\mathbf{k}_{\perp}}{\mathbf{k}_{\perp}^{4}}f\left(x,k_{\perp}\right)\left(1-e^{i\mathbf{k}_{\perp}\mathbf{r}}\right).\end{aligned}
\label{eq:sigmadipdilute}
\end{equation}
With the help of the above formula, the BFKL equation (\ref{eq:BFKL})
for the dipole scattering amplitude can be written as an evolution
equation for $f\left(x,k_{\perp}\right)$ (this is done by going to
Mellin space, see e.g. Ref. \protect\cite{kutakstasto2005} and references
therein):
\begin{equation}
\begin{aligned}\frac{\mathrm{d}f\left(Y,k_{\perp}\right)}{\mathrm{d}Y} & =\bar{\alpha}\int\frac{\mathrm{d}^{2}\mathbf{p}_{\perp}}{\pi}\frac{k_{\perp}^{2}}{p_{\perp}^{2}\left(\mathbf{k}_{\perp}-\mathbf{p}_{\perp}\right)^{2}}\left(f\left(Y,p_{\perp}\right)-\frac{1}{2}f\left(Y,k_{\perp}\right)\right).\end{aligned}
\label{eq:BFKLforf}
\end{equation}
As one can easily check, the above equation is convergent in the infrared
due to the fact that the singularity for $p_{\perp}=0$ exactly cancels
with the one for $\mathbf{p}_{\perp}=\mathbf{k}_{\perp}$. Furthermore,
Eq. (\ref{eq:BFKLforf}) can be solved analytically, with the initial
condition inspired by the Bremsstrahlung law:
\begin{equation}
f_{0}\left(Y,k_{\perp}\right)\simeq\frac{\alpha_{s}C_{F}}{\pi},
\end{equation}
yielding (see e.g. Refs. \protect\cite{Forshaw1997,KovchegovLevin}):
\begin{equation}
f\left(Y,k_{\perp}\right)\simeq\sqrt{\frac{k_{\perp}^{2}}{\Lambda^{2}}}\frac{e^{\bar{\alpha}\omega_{0}Y}}{\sqrt{2\pi\beta_{0}\bar{\alpha}Y}}\exp\left(-\frac{\ln^{2}\left(k_{\perp}^{2}/\Lambda^{2}\right)}{2\beta_{0}\bar{\alpha}Y}\right),\label{eq:BFKLsolution}
\end{equation}
where $\omega_{0}=4\ln2\simeq2.77$ and $\beta_{0}\simeq33.7$. The
most important feature of the above result is that $f\left(Y,k_{\perp}\right)$
exhibits a very fast asymptotic growth with the rapidity:
\begin{equation}
\begin{aligned}f\left(Y,k_{\perp}\right) & \sim e^{\bar{\alpha}\omega_{0}Y},\\
 & =\frac{1}{x^{\bar{\alpha}\omega_{0}}}\sim s^{\alpha_{I\!\!P}-1},
\end{aligned}
\label{eq:BFKLpomeron}
\end{equation}
where $\alpha_{I\!\!P}=1+\bar{\alpha}4\ln2$ is known as the BFKL
intercept. In the literature, this solution is often called the hard
pomeron, since it is reminiscent of the soft pomeron trajectory in
Regge theory, which corresponds to a $t$-channel exchange with the
quantum numbers of the vacuum and exhibits a similar, but much slower,
power-like growth of the cross section with $s$ ($\alpha_{I\!\!P}=1.081$
in the famous Donnachie-Landshoff fit, Ref. \protect\cite{Donnachie1992}).
\begin{figure}[t]
\begin{centering}
\begin{tikzpicture}[scale=1.3] 

\tikzset{photon/.style={semithick,decorate,decoration={snake}},
		electron/.style={semithick,postaction={decorate},decoration={markings,mark=at position .5 with {\arrow[draw]{latex}}}}, 	positron/.style={semithick,postaction={decorate},decoration={markings,mark=at position .5 with {\arrow[draw]{latex reversed}}}},     	gluon/.style={decorate,decoration={coil,amplitude=4pt, segment length=5pt}}}

\draw[->] (0,.3) --++ (4,0)node[right] {$x^+$} ;
\draw[semithick,gluon] (0,0) -- (4,0) node[right] {$$};
\draw[semithick,gluon] (.5,0).. controls (.5,-.7) and (3,-.5)  .. (4,-.5) node[right] {$$};
\draw[semithick,gluon] (1,-.5).. controls (1,-1.2) and (3,-1)  .. (4,-1) node[right] {$$};
\draw[semithick,gluon] (1.5,-1).. controls (1.5,-1.7) and (3,-1.5)  .. (4,-1.5) node[right] {$$};
\draw[semithick,gluon] (2,-1.5).. controls (2,-2.2) and (3,-2)  .. (4,-2) node[right] {$$};

\end{tikzpicture} 
\par\end{centering}
\caption{\label{fig:BFKLgrowth}\textquoteleft Explosive' growth of the gluon
distribution in BFKL.}
\end{figure}
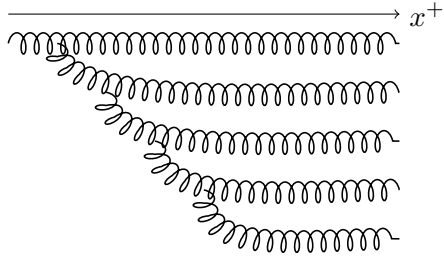
Note that we could have expected this power-like growth on some simple
general grounds. Indeed, since the lifetime of the gluons is given
by $\Delta x^{+}\sim2k^{+}/k_{\perp}^{2}$, the strong ordering of
the energies implies that:
\begin{equation}
x_{0}\gg...\gg x_{i}\gg x_{i+1}\gg...\gg x_{n}\quad\leftrightarrow\quad\Delta x_{0}^{+}\gg...\gg\Delta x_{i}^{+}\gg\Delta x_{i+1}^{+}\gg...\gg\Delta x_{n}^{+},
\end{equation}
hence the \textquoteleft parent' gluons always live to see the subsequent
emissions, as is illustrated in Fig. \ref{fig:BFKLgrowth}. The gluons
in a BFKL cascade are thus coherent with each other, and the number
of gluons per unit of rapidity is, from the resummation of the Bremsstrahlung
law:
\begin{equation}
\begin{aligned}\frac{\mathrm{d}N}{\mathrm{d}Y} & \sim\sum_{n}\bar{\alpha}^{n}\int_{x_{1}}^{1}\frac{\mathrm{d}x_{0}}{x_{0}}\int_{x_{2}}^{x_{0}}\frac{\mathrm{d}x_{1}}{x_{1}}...\int_{x_{n}}^{x_{n-2}}\frac{\mathrm{d}x_{n-1}}{x_{n-1}},\\
 & =\sum_{n}\frac{1}{n!}\left(\bar{\alpha}Y\right)^{n}=e^{\omega\bar{\alpha}Y},
\end{aligned}
\end{equation}
where $\omega$ is some constant which cannot be determined from this
heuristic derivation. The BFKL equation (\ref{eq:BFKLforf}) and its
solution (\ref{eq:BFKLpomeron}) raises two important problems. The
first problem is that, through the relation (\ref{eq:sigmadipdilute}),
the power-like growth of the gluon distribution transfers to the dipole-hadron
cross section:
\begin{equation}
\begin{aligned}\sigma_{\mathrm{dip}}\left(x\ll1,\mathbf{r}\right)=2\pi R^{2}\langle T\left(\mathbf{r},\mathbf{0}\right)\rangle_{Y\gg1} & \sim s^{\alpha_{I\!\!P}-1}.\end{aligned}
\label{eq:sigmadipasymptotic}
\end{equation}
Moreover, from the comparison of the black disc limit: 
\begin{equation}
\begin{aligned}\sigma_{\mathrm{tot}} & \leq2\pi R^{2},\end{aligned}
\end{equation}
with the well-known Froissart bound for total cross sections (Refs.
\protect\cite{Froissart1961,Martin1969}):
\begin{equation}
\begin{aligned}\sigma_{\mathrm{tot}} & \leq\sigma_{0}\ln^{2}s,\end{aligned}
\end{equation}
one can deduce that the hadron radius grows as a logarithm of the
center-of-mass energy:
\begin{equation}
R\sim\ln s.\label{eq:Rlns}
\end{equation}
Therefore, we see from Eq. (\ref{eq:sigmadipasymptotic}) that the
dipole scattering amplitude grows like a power of $s$ as well:
\begin{equation}
\begin{aligned}\langle T\left(\mathbf{r},\mathbf{0}\right)\rangle_{Y} & \sim s^{\alpha_{I\!\!P}-1}\simeq e^{\left(\alpha_{I\!\!P}-1\right)Y},\end{aligned}
\end{equation}
eventually violating the unitarity bound $\langle T\left(\mathbf{r},\mathbf{0}\right)\rangle_{Y}\leq1$.
This problem is not present in the BK equation (\ref{eq:BK4T}), where
$T=1$ is a fixed point, which turns out to be stable (see Ref. \protect\cite{Gelis2013}).
The nonlinear term in the BK equation therefore provides an unitarization
mechanism.

A second problem of BFKL is known as infrared diffusion, and can be
explained by comparing with DGLAP. In the latter case, one evolves
towards large values of $Q^{2}$, and since the coupling constant
decreases with increasing $Q^{2}$, this implies that DGLAP evolves
into more and more perturbative terrain. For BFKL, however, this is
not the case: there is no mechanism to prevent deviations into the
nonperturbative domain $k_{\perp}^{2}\lesssim\Lambda_{\mathrm{QCD}}^{2}$.
Worse, the more one evolves in rapidity, the more the gluon distribution
will receive contributions from the nonperturbative sector (see Refs.
\protect\cite{Bartels1993,Bartels1996}, and the discussions in Refs. \protect\cite{EdmondRev,KovchegovLevin}).
Once again, saturation physics provides a solution: the dynamically
generated saturation scale $Q_{s}$, which is hard and thus perturbative,
can be regarded as the inverse of the distance $1/Q_{s}$ over which
the gluons arrange themselves to achieve color neutrality (Refs. \protect\cite{Mueller2002,Iancu2002b}).
The gluon spectrum decreases for transverse momenta below $Q_{s}$,
and is therefore not sensitive anymore to nonperturbative physics. 

For consistency, we should remark that if one approximates the $k_{\perp}$-dependence
in the BFKL equation (\ref{eq:BFKLforf}) logarithmically, and rewrites
it as an equation for $x\mathcal{G}\left(x,Q^{2}\right)$, its solution
coincides with the (fixed-coupling) result Eq. (\ref{eq:gluon asymptotic})
of DGLAP for the gluon distribution in the DLA. Hence, as we expect
from our discussion in Sec. \ref{subsec:The-soft-Bremsstrahlung},
in which we argued that BFKL and DGLAP are in a sense dual, in the
double logarithmic approximation the two evolution equations are in
fact the same. In particular, the DLA solution Eq. (\ref{eq:xGDLLA})
also violates the unitarity bound at large energies.

\section{\label{sec:GluonTMDs}Transverse momentum dependent gluon distributions}

Before we proceed, a small digression is in place. Indeed, in the
preceding section, we introduced the unintegrated gluon distribution
$f\left(x,k_{\perp}\right)$, Eq. (\ref{eq:unintgluondistrspectrum}).
Up to a normalization, $f\left(x,k_{\perp}\right)$ is the same as
the so-called Weizsäcker-Williams gluon distribution $\mathcal{F}_{gg}^{\left(3\right)}\left(x,k_{\perp}\right)$
(Refs. \protect\cite{Catani1991,Collins1991,Kovchegov1998,McLerran1999}):
\begin{equation}
f\left(x,k_{\perp}\right)=\pi k_{\perp}^{2}\mathcal{F}_{gg}^{\left(3\right)}\left(x,k_{\perp}\right),\label{eq:unintegratedWW}
\end{equation}
which is defined as the number of gluons per unit of rapidity and
per unit of area in transverse momentum space:
\begin{equation}
\mathcal{F}_{gg}^{\left(3\right)}\left(x,k_{\perp}\right)\equiv\frac{\mathrm{d}N}{\mathrm{d}\ln\left(1/x\right)\mathrm{d}^{2}\mathbf{k}_{\perp}}.
\end{equation}
This definition becomes explicit in the light-cone gauge $A^{+}=0$,
in which $\mathcal{F}_{gg}^{\left(3\right)}\left(x,k_{\perp}\right)$
can be written as the number operator in Fock space (Refs. \protect\cite{Brodsky1998,Kovchegov1998,thesisStephane,EdmondRev}):
\begin{equation}
\begin{aligned}\mathcal{F}_{gg}^{\left(3\right)}\left(x,k_{\perp}\right) & =k^{+}\sum_{\lambda}\Bigl\langle a_{a}^{\lambda\dagger}\left(x^{+},\vec{k}\right)a_{a}^{\lambda}\left(x^{+},\vec{k}\right)\Bigr\rangle_{x},\end{aligned}
\label{eq:WWcounting}
\end{equation}
where we introduced the notation $\vec{k}=\left(k^{+},\mathbf{k}_{\perp}\right)$,
and where $a_{a}^{\lambda\dagger}$ and $a_{a}^{\lambda}$ are the
light-cone creation and annihilation operators, with $\lambda$ the
polarization and $a$ the color of the gluon field. They obey the
equal-LC-time commutation relation:
\begin{equation}
\left[a_{a}^{\lambda}\left(x^{+},\vec{k}\right),a_{b}^{\lambda'}\left(x^{+},\vec{p}\right)\right]=2k^{+}\delta_{ab}\delta_{\lambda\lambda'}\left(2\pi\right)^{3}\delta^{\left(3\right)}\left(\vec{k}-\vec{p}\right).
\end{equation}
The gluon field can be written as the following decomposition in function
of $a_{a}^{\lambda\dagger}$ and $a_{a}^{\lambda}$ 
\begin{equation}
A_{c}^{i}\left(x^{+},\vec{x}\right)=\sum_{\lambda}\int\frac{\mathrm{d}^{3}k}{\left(2\pi\right)^{3}2k^{+}}\theta\left(k^{+}\right)\left(a_{c}^{\lambda}\left(x^{+},\vec{k}\right)\epsilon_{\lambda}^{i}\left(\vec{k}\right)e^{i\vec{k}\cdot\vec{x}}+a_{c}^{\lambda\dagger}\left(x^{+},\vec{k}\right)\epsilon_{\lambda}^{i*}\left(\vec{k}\right)e^{-i\vec{k}\cdot\vec{x}}\right),\label{eq:gluondecomposition}
\end{equation}
with $\vec{x}=\left(x^{-},\mathbf{x}\right)$ and $\vec{k}\cdot\vec{x}=k^{+}x^{-}-\mathbf{k}_{\perp}\cdot\mathbf{x}$.
Still in the LC gauge, the field strength has the following form:
\begin{equation}
F_{a}^{i+}\left(\vec{k}\right)=ik^{+}A_{a}^{i}\left(\vec{k}\right),
\end{equation}
hence combining Eqs. (\ref{eq:gluondecomposition}) and (\ref{eq:WWcounting}),
we find that $\mathcal{F}_{gg}^{\left(3\right)}\left(x,k_{\perp}\right)$
can be written as a correlator of field strengths:
\begin{equation}
\begin{aligned}\mathcal{F}_{gg}^{\left(3\right)}\left(x,k_{\perp}\right) & =\frac{\Bigl\langle F_{a}^{i+}\left(\vec{k}\right)F_{a}^{i+}\left(-\vec{k}\right)\Bigr\rangle_{x}}{4\pi^{3}},\end{aligned}
\end{equation}
or, in Fourier space:
\begin{equation}
\begin{aligned}\mathcal{F}_{gg}^{\left(3\right)}\left(x,k_{\perp}\right) & =4\int\frac{\mathrm{d}^{3}v\mathrm{d}^{3}w}{\left(2\pi\right)^{3}}e^{i\vec{k}\cdot\left(\vec{v}-\vec{w}\right)}\mathrm{Tr}\Bigl\langle F^{i+}\left(\vec{v}\right)F^{i+}\left(\vec{w}\right)\Bigr\rangle_{x},\end{aligned}
\label{eq:WWFFFourier}
\end{equation}
where we used that $\mathrm{Tr}\left(t^{a}t^{b}\right)=\left(1/2\right)\delta^{ab}$.
In the small-$x$ limit, the phase $\exp\left(ik^{+}\left(v^{-}-w^{-}\right)\right)=\exp\left(ixp^{+}\left(v^{-}-w^{-}\right)\right)\simeq1$
can be neglected, and the rapidity-dependence is only left inside
the medium averages. Finally, the light-cone Fock states are related
to the hadronic states $|A\rangle$ as follows (see Ref. \protect\cite{fabio,Cyrille}):
\begin{equation}
\langle\mathcal{O}\rangle_{x}=\frac{\langle A|\mathcal{O}|A\rangle}{\langle A|A\rangle},\label{eq:Fock2hadronic}
\end{equation}
where the latter are normalized as: $\langle A|A\rangle=\left(2\pi\right)^{3}2p_{A}^{+}\delta^{\left(3\right)}\left(\vec{0}\right)$,
with $p_{A}$ the momentum of the hadron. Therefore, we can write
\begin{equation}
\begin{aligned}\mathcal{F}_{gg}^{\left(3\right)}\left(x,k_{\perp}\right) & =2\int\frac{\mathrm{d}^{3}v\mathrm{d}^{3}w}{\left(2\pi\right)^{3}p_{A}^{+}}e^{i\vec{k}\cdot\left(\vec{v}-\vec{w}\right)}\mathrm{Tr}\Bigl\langle A\Bigr|F^{i+}\left(\vec{v}\right)F^{i+}\left(\vec{w}\right)\Bigl|A\Bigr\rangle,\end{aligned}
\end{equation}
and finally, using that $k^{+}=xp_{A}^{+}$ and making use of translational
invariance, we obtain the following so-called operator definition
of $\mathcal{F}_{gg}^{\left(3\right)}\left(x,k_{\perp}\right)$:
\begin{equation}
\begin{aligned}\mathcal{F}_{gg}^{\left(3\right)}\left(x,k_{\perp}\right) & \equiv2\int\frac{\mathrm{d}^{3}\xi}{\left(2\pi\right)^{3}p_{A}^{+}}e^{ixp_{A}^{+}\xi^{-}}e^{-i\mathbf{k}_{\perp}\boldsymbol{\xi}}\mathrm{Tr}\Bigl\langle A\Bigr|F^{i+}\left(\vec{\xi}\right)F^{i+}\left(\vec{0}\right)\Bigl|A\Bigr\rangle.\end{aligned}
\label{eq:WWoperator}
\end{equation}
This expression, however, only holds in the light-cone gauge. To render
it gauge invariant, we can insert Wilson lines that connect the points
$\vec{0}$ and $\vec{\xi}$ (see Refs. \protect\cite{Peskin,collins,Cherednikov2014}).
In contrast to the integrated gluon distribution, whose gauge-invariant
operator definition reads:
\begin{equation}
\begin{aligned}x\mathcal{G}\left(x,Q^{2}\right) & \equiv\int\frac{\mathrm{d}\xi^{-}}{2\pi p_{A}^{+}}e^{ixp_{A}^{+}\xi^{-}}\mathrm{Tr}\Bigl\langle A\Bigr|F^{i+}\left(\xi^{-}\right)U\left(\xi^{-},0^{-};\mathbf{0}\right)F^{i+}\left(0^{-}\right)U\left(0^{-},\xi^{-};\mathbf{0}\right)\Bigl|A\Bigr\rangle,\end{aligned}
\label{eq:xGoperator}
\end{equation}
with
\begin{equation}
U\left(\xi^{-},0^{-};\mathbf{0}\right)\equiv\mathcal{P}e^{ig_{s}\int_{\xi^{-}}^{0^{-}}\mathrm{d}z^{-}A_{a}^{+}\left(z^{-},\mathbf{0}\right)t^{a}},
\end{equation}
in the case of Eq. (\ref{eq:WWoperator}), $\vec{0}$ and $\vec{\xi}$
are points in three-dimensional coordinate space, and hence there
are different possible paths to connect them. In the case of the Weizsäcker-Williams
distribution, one chooses (Ref. \protect\cite{Boer2003,fabio}):
\begin{equation}
\begin{aligned}\mathcal{F}_{gg}^{\left(3\right)}\left(x,k_{\perp}\right) & \equiv2\int\frac{\mathrm{d}^{3}\xi}{\left(2\pi\right)^{3}p_{A}^{+}}e^{ixp_{A}^{+}\xi^{-}}e^{-i\mathbf{k}_{\perp}\boldsymbol{\xi}}\mathrm{Tr}\Bigl\langle A\Bigr|F^{i+}\left(\vec{\xi}\right)U^{\left[+\right]\dagger}F^{i+}\left(\vec{0}\right)U^{\left[+\right]}\Bigl|A\Bigr\rangle,\end{aligned}
\label{eq:WeizsackerWilliamsdef}
\end{equation}
where $U^{\left[+\right]}$, is a so-called staple gauge link (see
Fig. \ref{fig:staple}):
\begin{equation}
\begin{aligned}U^{\left[+\right]} & \equiv U\left(0^{-},+\infty;\boldsymbol{0}\right)U\left(+\infty,\xi^{-};\boldsymbol{\xi}\right),\\
U^{\left[+\right]\dagger} & \equiv U\left(\xi^{-},+\infty;\boldsymbol{\xi}\right)U\left(+\infty,0^{-};\boldsymbol{0}\right).
\end{aligned}
\end{equation}
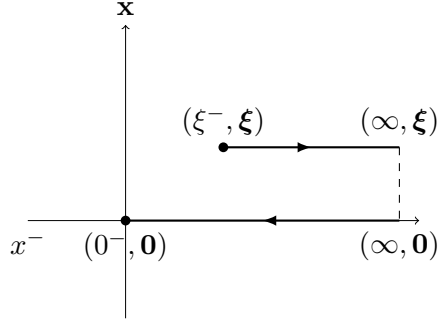
\begin{figure}[t]
\begin{centering}
\begin{tikzpicture}[scale=1.3] 

\tikzset{photon/.style={thick,decorate,decoration={snake}},
		electron/.style={thick,postaction={decorate},decoration={markings,mark=at position .5 with {\arrow[draw]{latex}}}}, 	positron/.style={semithick,postaction={decorate},decoration={markings,mark=at position .5 with {\arrow[draw]{latex reversed}}}},     	gluon/.style={decorate,decoration={coil,amplitude=4pt, segment length=5pt}}}

\draw[very thin,->] (0,-1) --++ (0,3)node[above] {$\bf{x}$};
\draw[very thin,->] (-1,0) node[below] {$x^-$}--++ (4,0) ;
\draw[electron] (1,.75) node [above]{$(\xi^-,\boldsymbol{\xi})$} --++(1.8,0) node [above]{$(\infty,\boldsymbol{\xi})$};
\draw[dashed] (2.8,.75)  --++(0,-.75);
\draw[electron] (2.8,0) node [below]{$(\infty,\mathbf{0})$}-- (0,0) node [below]{$(0^-,\mathbf{0})$};
\fill[black] (1,.75) circle(.05);
\fill[black] (0,0) circle(.05);

\end{tikzpicture} 
\par\end{centering}
\caption{\label{fig:staple}\textquoteleft Staple' gauge link used in the gauge-invariant
definition of the Weizsäcker-Williams gluon TMD.}

\end{figure}
Since these Wilson lines run along the $x^{-}$-axis and hence only
resum $A_{a}^{+}$ gluon fields, the gauge-invariant definition (\ref{eq:WeizsackerWilliamsdef})
is reduced to Eq. (\ref{eq:WWoperator}) when choosing the LC gauge. 

As we already said, there are different possible Wilson line configurations
to render the operator definition (\ref{eq:WWoperator}) gauge invariant.
Therefore, a multitude of so-called transverse momentum dependent
(TMD) PDFs, TMDs in short, exists, of which the Weizsäcker-Williams
distribution $\mathcal{F}_{gg}^{\left(3\right)}\left(x,k_{\perp}\right)$
is one example (and, incidentally, the only TMD for which the \textquoteleft gluon
counting' interpretation Eq. (\ref{eq:WWcounting}) holds). In general,
it is the physical process under consideration which determines which
Wilson-line path to chose, and hence which TMD PDF plays a role. In
Part \ref{part:projectCyrille} of this thesis we will compute the
cross section for forward dijet production in proton-nucleus collisions,
and analyze in the small-$x$ limit the various gluon TMDs that play
a role. Just like the nonlinear BK equation is reduced to BFKL in
the dilute regime, in the limit of large transverse momenta, the Wilson-line
structure of the gluon TMDs simplifies, and the latter either disappear
or become equal to the Weizsäcker-Williams/unintegrated gluon distribution.
That this reduction does indeed take place, can be understood from
expression (\ref{eq:WeizsackerWilliamsdef}): for large momenta $k_{\perp}$,
the transverse coordinate $\boldsymbol{\xi}$ becomes small, and the
two Wilson lines in Fig. \ref{fig:staple} overlap, cancel (since
$U\left(a,b,\mathbf{x}\right)U^{\dagger}\left(a,b,\mathbf{x}\right)=1$),
upon which the whole path is reduced to the simple one-dimensional
path $U\left(\xi^{-},0^{-};\mathbf{0}\right)$ of Eq. (\ref{eq:xGoperator}).

We should stress that in principle the TMDs still depend on the hard
scale $Q^{2}$. However, in the small-$x$ limit this dependence is
trivial, which is why we do not include it in the notation. Approaches
exist (see e.g. Refs. \protect\cite{Hautmann2014,Hautmann2014bis,TMDlib})
which do take this hard-scale dependence of TMDs into account in the
evolution, allowing one to interpolate between the DGLAP and the BFKL
regime. These approaches are usually based on the well-known CCFM
(Catani-Ciafaloni-Fiorani-Marchesini, Refs. \protect\cite{CCFM1,CCFM2,CCFM3})
evolution equation, in which angular ordering is imposed on the parton
splittings. 

Finally, note that we have all the ingredients to prove Eq. (\ref{eq:sigmadipdilute}),
i.e. the fact that in the dilute limit the dipole amplitude is linearly
related to the unintegrated gluon distribution. Indeed, in the dilute
regime, the multiple scatterings of the dipole off the target can
be approximated by a single scattering, involving a two-gluon exchange
in the forward amplitude $\langle T\left(\mathbf{x},\mathbf{y}\right)\rangle_{Y}$.
This is implemented by expanding the Wilson lines, which resum the
gluon fields of the target, to second order:
\begin{equation}
\begin{aligned}U\left(\mathbf{x}\right) & \simeq1+ig_{s}\int\mathrm{d}z^{-}A_{a}^{+}\left(z^{-},\mathbf{x}\right)t^{a}+\frac{\left(ig_{s}\right)^{2}}{2!}\mathcal{P}\int\mathrm{d}z_{1}^{-}\mathrm{d}z_{2}^{-}A_{a}^{+}\left(z_{1}^{-},\mathbf{x}\right)A_{b}^{+}\left(z_{2}^{-},\mathbf{x}\right)t^{a}t^{b}.\end{aligned}
\label{eq:WLexpansion}
\end{equation}
To second-order accuracy, the path ordering doesn't play a role, since
$\mathrm{Tr}\left(t^{a}t^{b}\right)=\mathrm{Tr}\left(t^{b}t^{a}\right)=\left(1/2\right)\delta^{ab}$.
Consequently, in the dilute regime, governed by the BFKL equation,
a strict separation holds between the longitudinal and the transverse
dynamics. We can therefore introduce the short-hand notation:
\begin{equation}
\alpha_{\mathbf{x}}^{a}\equiv\int\mathrm{d}z^{-}A_{a}^{+}\left(z^{-},\mathbf{x}\right),\label{eq:nolongstructure-2}
\end{equation}
and the dipole scattering amplitude from definition (\ref{eq:dipolecrossection})
can be written as follows:
\begin{equation}
\langle T\left(\mathbf{x},\mathbf{y}\right)\rangle_{Y}=1-\frac{1}{N_{c}}\mathrm{Tr}\langle U\left(\mathbf{x}\right)U^{\dagger}\left(\mathbf{y}\right)\rangle_{Y}\simeq\frac{g_{s}^{2}}{4N_{c}}\Bigl\langle\left(\alpha_{\mathbf{x}}^{a}-\alpha_{\mathbf{y}}^{a}\right)^{2}\Bigr\rangle_{Y}.\label{eq:Txy-1}
\end{equation}
On the other hand, in a covariant gauge, the field strength and the
gluon fields obey the relation:
\begin{equation}
F_{a}^{i+}\left(\vec{x}\right)=\partial^{i}A_{a}^{+}\left(\vec{x}\right).
\end{equation}
With the help of this property, we find from Eqs. (\ref{eq:unintegratedWW})
and (\ref{eq:WWFFFourier}) (using partial integration as well as
the Casimir property Eq. (\ref{eq:tracetatb}), and setting $\exp\left(-ik^{+}\cdot\left(v^{-}-w^{-}\right)\right)=\exp\left(-ixp^{+}\cdot\left(v^{-}-w^{-}\right)\right)=1$):
\begin{equation}
\begin{aligned}f\left(x,k_{\perp}\right) & =2\pi\mathbf{k}_{\perp}^{4}\int\frac{\mathrm{d}^{3}v\mathrm{d}^{3}w}{\left(2\pi\right)^{3}}e^{i\mathbf{k}\cdot\left(\mathbf{v}-\mathbf{w}\right)}\Bigl\langle A_{a}\left(v^{-},\mathbf{v}\right)A_{a}\left(w^{-},\mathbf{w}\right)\Bigr\rangle_{Y}.\end{aligned}
\end{equation}
Plugging this expression into the r.h.s. of Eq. (\ref{eq:sigmadipdilute}),
we recover expression (\ref{eq:sigmadipdilute}):
\begin{equation}
\begin{aligned} & \frac{4\pi\alpha_{s}}{N_{c}}\int\frac{\mathrm{d}^{2}\mathbf{k}_{\perp}}{k_{\perp}^{4}}f\left(x,k_{\perp}\right)\left(1-e^{i\mathbf{k}_{\perp}\mathbf{r}}\right)\\
 & =\frac{8\pi^{2}\alpha_{s}}{N_{c}}\int\frac{\mathrm{d}^{3}v\mathrm{d}^{3}w}{2\pi}\left(\delta^{\left(2\right)}\left(\mathbf{v}-\mathbf{w}\right)-\delta^{\left(2\right)}\left(\mathbf{r}+\mathbf{v}-\mathbf{w}\right)\right)\Bigl\langle A_{a}\left(v^{-},\mathbf{v}\right)A_{a}\left(w^{-},\mathbf{w}\right)\Bigr\rangle_{Y},\\
 & =\frac{2\pi\alpha_{s}}{N_{c}}\int\mathrm{d}^{2}\mathbf{v}\Bigl\langle\left(\alpha_{\mathbf{r}}^{a}-\alpha_{\mathbf{0}}^{a}\right)^{2}\Bigr\rangle_{Y}=\frac{2\pi\alpha_{s}}{N_{c}}S_{\perp}\Bigl\langle\left(\alpha_{\mathbf{r}}^{a}-\alpha_{\mathbf{0}}^{a}\right)^{2}\Bigr\rangle_{Y}\\
 & =2S_{\perp}\langle T\left(\mathbf{r}\right)\rangle_{Y}=\sigma_{\mathrm{dip}}\left(x,\mathbf{r}\right),
\end{aligned}
\end{equation}
making use of the translational invariance of the medium averages
in the transverse plane, and restricting the transverse coordinates
to the area $S_{\perp}$ of the target:
\begin{equation}
S_{\perp}=\int\mathrm{d}^{2}\mathbf{v}.
\end{equation}
Combining Eq. (\ref{eq:sigmadipdilute}) with the BFKL solution for
$f\left(x,k_{\perp}\right)$, Eq. (\ref{eq:BFKLsolution}), and with
expression (\ref{eq:DISdipolecrosssection}) for the photon-proton
cross section, we obtain the DIS cross section as found in the BFKL
approach (see Ref. \protect\cite{Forshaw1997}).

Incidentally, from Eq. (\ref{eq:sigmadipdilute}) it is also apparent
that the dipole-hadron cross section can be seen as a direct probe
of the gluon distribution. Indeed, in the limit of small dipole sizes,
which, since $r_{\perp}^{2}\sim1/Q^{2}$, is again consistent with
the dilute approximation, the exponential can be expanded, yielding:
\begin{equation}
\sigma_{\mathrm{dip}}\left(x,\mathbf{r}\right)\simeq r^{2}\frac{\alpha_{s}\pi^{2}}{N_{c}}x\mathcal{G}\left(x,1/r^{2}\right).\label{eq:sigmaxG}
\end{equation}
This implies that, in the dilute regime, the dipole scattering amplitude
scales approximately quadratically with the size $\mathbf{r}$ of
the dipole:
\begin{equation}
\left\langle T\left(\mathbf{r}\right)\right\rangle _{x}\propto r^{2}.
\end{equation}
One can show (see Ref. \protect\cite{EdmondRev}, and references therein)
that the high-energy evolution towards the saturation regime (but
still in the dilute region $r^{2}\ll1/Q_{s}^{2}$, hence inside the
window for geometric scaling) introduces a nonperturbative anomalous
dimension $\gamma_{s}\simeq0.63$, which changes this scaling to:
\begin{equation}
\left\langle T\left(\mathbf{r}\right)\right\rangle _{x}\propto r^{2\gamma_{s}}.
\end{equation}

Finally, from Eq. (\ref{eq:Txy-1}) we see that there is another way
to view the unitarity of the cross section. Indeed, written as a correlator
of Wilson lines, the forward scattering amplitude is manifestly unitary,
which is broken by expanding the Wilson lines in the single-scattering
approximation. It is therefore multiple scattering that renders the
cross section unitary. Multiple scattering becomes important when
the exponent of the Wilson lines is of order one, which is equivalent
to the statement that the gluon fields are strong:
\begin{equation}
g_{s}\int\mathrm{d}x^{-}A^{+}\sim1,
\end{equation}
and that the gluon distribution in the target saturates. This equivalence
is highly nontrivial, but arises naturally in the Color Glass Condensate,
which is the subject of the following part of the thesis.\newpage{}

\thispagestyle{simple}

\part{\label{part:CGC}Color Glass Condensate}

\section{Introduction}

In the previous part, we went through the basics of QCD at small-$x$,
and gave a physical motivation for the saturation of the gluon distribution
in the high density regime. We saw that there are not only phenomenological
hints for saturation, but also that saturation physics solves in a
natural way the problems of unitarity breaking and infrared diffusion
in the small-$x$ limit of QCD.

In this part, we discuss the Color Glass Condensate, which is an effective
theory for the proton or nucleus in the saturation regime. Its initial
condition is the McLerran-Venugopalan (MV) model, a classical theory
for the partons in a large nucleus. It is based on the fact that,
in the infinite-momentum frame, the valence quarks, which carry a
longitudinal momentum fraction $x_{\mathrm{val}}\sim1$ of their parent
nucleon, have a considerably longer lifetime than the so-called wee
gluons which carry a momentum fraction $10^{-2}\lesssim x_{\mathrm{wee}}\lesssim10^{-1}$
(the lower limit guarantees that soft gluon radiation is small enough
to be negligible). The resolution $Q^{2}$, with which the nucleus
is probed, has to be high enough to penetrate the nucleons, but small
enough to see a dense collection of color charges, leading to the
requirement: $\Lambda_{\mathrm{QCD}}^{2}\lesssim Q^{2}\lesssim\Lambda_{\mathrm{QCD}}^{2}A^{1/3}$,
where $A$ is the atomic number of the nucleus. In this region of
phase space, the valence quarks are practically frozen with respect
to the dynamics of the wee gluons, and act as recoilless color charges
$\rho_{a}$ which generate the gluons through the Yang-Mills equations.
Since the scattering of a projectile off the nucleus is described
by Wilson lines, which resum the multiple interactions with the soft
gluons, scattering in this sense boils down to probing the correlations
of the color charges. The latter are distributed according to a Gaussian
distribution or weight function $\mathcal{W}_{A}\left[\rho\right]$,
over which all observables have to be averaged. 

It is to this kinematical separation between frozen valence quarks
on the one hand, and short-living wee gluons on the other hand, that
the Color Glass Condensate owes its name, since dynamics that takes
place over two very different time scales is the hallmark of \textquoteleft glassy'
behavior. In addition, in the saturation regime the gluons behave
effectively like a Bose condensate. The typical transverse momentum
$k_{\perp}^{2}$ of the gluons is of the order of the saturation scale
$Q_{s}^{2}$ which, as we will see, scales linearly with $\mu_{A}$:
the density of the color charge squared per color and per unit area.
Hence, parametrically we have that $k_{\perp}^{2}\sim Q_{s}^{2}\sim\mu_{A}$,
implying that in the high-density regime, the typical momenta of the
hadron constituents are high enough to describe the saturated hadron
system within perturbation theory.

To logarithmic accuracy, the MV model can be extended towards smaller
values of Bjorken-$x$, by pushing the scale $\Lambda^{+}$, which
separates valence and wee partons, down towards a smaller scale $\Lambda'^{+}$.
The so-called semi-fast classical gluons, which have energies in the
strip $\Lambda^{+}\gg k^{+}\gg\Lambda'^{+}$, are then promoted again
to quantum gluons, and subsequently integrated out and absorbed into
the new sources which account for the hard partons ($k^{+}\gg\Lambda'^{+}$,
so they include the valence partons). After this procedure, which
corresponds to one step in the rapidity evolution, one has again a
classical theory in which the static color sources generate soft ($\Lambda'^{+}\gg k^{+}$)
classical gluon fields by virtue of the Yang-Mills equations. The
kinematical \textquoteleft glassy' structure of the MV model is hence
preserved throughout the evolution. Finally, the nonlinear evolution
equation for the weight function $\mathcal{W}_{\Lambda'}\left[\rho\right]$
of the color charges, which one obtains through the renormalization
group procedure we just described, is the well-known JIMWLK equation.

\section{\label{sec:McLerran-Venugopalan-model}McLerran-Venugopalan model}

As already said above, the McLerran-Venugopalan model (Refs. \protect\cite{MV1,MV2,MV3})
is a classical effective theory for a large nucleus, and is the initial
condition of the Color Glass Condensate. Its starting point is the
observation that, if one considers a nucleus in the infinite momentum
frame, its parton content is kinematically separated in rapidity.
Indeed, in light-cone coordinates, taking the nucleus to be a right-mover,
the longitudinal extent of a parton with momentum $p^{+}$ is given
by:
\begin{equation}
\Delta x^{-}\sim\frac{1}{p^{+}}=\frac{1}{xP^{+}},\label{eq:Dx-parton}
\end{equation}
where $x$ is the fraction the parton carries of the momentum $P^{+}$
of the parent nucleon. The nucleus itself, with a radius $R_{A}\simeq R_{0}A^{1/3}\simeq\Lambda_{\mathrm{QCD}}^{-1}A^{1/3}$,
is squeezed together, due to Lorentz contraction, to a thickness
\begin{equation}
\Delta x_{A}^{-}\sim\frac{2R_{A}}{\gamma}=\frac{2R_{A}}{P^{+}}m_{N},
\end{equation}
where we used that $\gamma=P_{A}^{+}/M_{A}\simeq P^{+}/m_{N}$, with
$P_{A}$ and $M_{A}$ the momentum and mass of the nucleus, and $m_{N}$
the mass of an individual nucleon. Comparing with Eq. (\ref{eq:Dx-parton}),
we can label partons according to their energy fraction $x$ and therefore
their longitudinal extent. Indeed, valence partons, which carry a
large fraction of the nucleus' momentum, are localized on the $x^{-}$-axis,
and appear as a thin sheet in contrast to the \textquoteleft wee'
partons with $x\ll1$, which are delocalized, well beyond the longitudinal
radius of the nucleus. In formulas:
\begin{equation}
\Delta x_{\mathrm{val}}^{-}\ll\Delta x_{A}^{-}\ll\Delta x_{\mathrm{wee}}^{-},
\end{equation}
where:
\begin{align}
x_{\mathrm{val}}\sim1,\quad x_{\mathrm{wee}}\ll & \frac{1}{R_{A}m_{N}}\sim A^{-1/3}.\label{eq:MVxupperlimit}
\end{align}
In addition, for the classical description to remain valid, radiative
effects should be negligible. From the Bremsstrahlung law, the probability
for partons with $x'$ between $x$ and $1$ to radiate is:
\begin{equation}
\int_{x}^{1}\mathrm{d}P_{\mathrm{Bremsstr}}\simeq\frac{\alpha_{s}N_{c}}{\pi}\ln\frac{1}{x},
\end{equation}
hence requiring that this probability is much smaller than one amounts
to the lower limit:
\begin{equation}
x_{\mathrm{wee}}\gg e^{-1/\bar{\alpha}}.\label{eq:MVxlowerlimit}
\end{equation}
As an example, for lead, with $A=206$, and a coupling $\bar{\alpha}=0.3$,
we find the following, rather small, phase space for the wee partons:
$0.04\ll x_{\mathrm{wee}}\ll0.2$. 

Of course, the opposite effect is also true: the partons have a lifetime
of the order
\begin{equation}
\Delta x^{+}\sim\frac{1}{p^{-}}=\frac{2xP^{+}}{p_{\perp}^{2}},
\end{equation}
hence the valence partons live considerably longer than the wee partons:
\begin{equation}
\Delta x_{\mathrm{wee}}^{+}\ll\Delta x_{\mathrm{val}}^{+}.
\end{equation}
Thus, with respect to the characteristic time scale of the wee partons,
the valence partons appear to be \textquoteleft frozen' in time. In
addition, since their momenta are much larger, they can radiate gluons
without experiencing the effects of this radiation, hence without
recoil. For these two reasons, the valence partons can be treated
to a good approximation as static currents of color charge:
\begin{equation}
\begin{aligned}J_{a}^{\mu} & =\delta^{\mu+}\rho_{a}\left(\vec{x}\right),\end{aligned}
\end{equation}
where $\vec{x}=\left(x^{-},\mathbf{x}\right)$, and where the valence
parton color charge density $\rho_{a}$ is random, static (it has
no $x^{+}$ LC-time dependence), and sharply localized around $x^{-}\sim1/p^{+}$.
The wee partons are generated by these currents by virtue of the classical
Yang-Mills equation:
\begin{equation}
\begin{aligned}\left[D_{\nu},F^{\nu\mu}\right]\left(\vec{x}\right) & =\delta^{\mu+}\rho_{a}\left(\vec{x}\right)T^{a},\\
\left(\partial_{\nu}F_{a}^{\nu\mu}+g_{s}f^{abc}A_{\nu}^{b}F_{c}^{\nu\mu}\right)\left(\vec{x}\right) & =\delta^{\mu+}\rho_{a}\left(\vec{x}\right).
\end{aligned}
\label{eq:Yang-Mills}
\end{equation}

with the covariant derivative (in the adjoint representation):
\begin{equation}
\begin{aligned}D_{\mu} & \equiv\partial_{\mu}-ig_{s}A_{\mu}^{a}T^{a},\end{aligned}
\end{equation}
and the field strength:
\begin{equation}
\begin{aligned}F^{\mu\nu} & \equiv\partial^{\mu}A^{\nu}-\partial^{\nu}A^{\mu}-ig_{s}\left[A^{\mu},A^{\nu}\right],\\
 & =\left(\partial^{\mu}A_{a}^{\nu}-\partial^{\nu}A_{a}^{\mu}+g_{s}f^{abc}A_{b}^{\mu}A_{c}^{\nu}\right)T^{a}.
\end{aligned}
\end{equation}
To solve the Yang-Mills equation, Eq. (\ref{eq:Yang-Mills}), we follow
the method described in Ref. \protect\cite{JIMWLK4}. First, since the source
$\rho\left(x\right)$ is static, we expect our solution to be static
as well, hence we require $\partial_{+}A^{\mu}=0$. In addition, we
make the ansatz that $A^{-}=0$, which ensures that the current is
covariantly conserved: $\left[D^{-},J^{+}\right]=0$. With these requirements,
the only non-zero components of the field strength are $F^{i+}$ and
$F^{ij}$. However, since we want a solution that vanishes in the
absence of a source $\rho\left(x\right)$, it follows from the $\mu=i$
component of the Yang-Mills equation:
\begin{equation}
\begin{aligned}\left[D_{\nu},F^{\nu i}\right]\left(x\right) & =\left[D_{j},F^{ji}\right]\left(x\right)=0,\end{aligned}
\end{equation}
that $F^{ij}=0$. This implies that the gauge fields form a pure gauge
in the two transverse dimensions, i.e.:
\begin{equation}
\begin{aligned}A^{i} & =\frac{i}{g_{s}}\mathcal{S}\partial^{i}\mathcal{S}^{\dagger},\end{aligned}
\label{eq:puregauge}
\end{equation}
with $\mathcal{S}=\mathcal{S}\left(x^{-},\mathbf{x}\right)\in SU\left(N_{c}\right)$. 

By fixing the gauge, the remaining degrees of freedom will be further
reduced to one, as we will now demonstrate. First, let us choose a
gauge in which $\tilde{A}^{i}=0,$ gauge transforming Eq. (\ref{eq:puregauge})
as follows:
\begin{equation}
\begin{aligned}\tilde{A}^{i} & =\mathcal{S}^{\dagger}\left(A^{i}+\frac{i}{g}\partial^{i}\right)\mathcal{S},\\
 & =\frac{i}{g_{s}}\mathcal{S}^{\dagger}\mathcal{S}\left(\partial^{i}\mathcal{S}^{\dagger}\right)\mathcal{S}+\frac{i}{g_{s}}\mathcal{S}^{\dagger}\partial^{i}\mathcal{S},\\
 & =0,
\end{aligned}
\end{equation}
where we used the fact that $\left(\partial^{i}\mathcal{S}^{\dagger}\right)\mathcal{S}=-\mathcal{S}^{\dagger}\partial^{i}\mathcal{S}$.
This gauge choice is a covariant gauge since clearly $\partial_{\mu}\tilde{A}^{\mu}=0$.
The only field left is now $\alpha_{a}\left(\vec{x}\right)\equiv\tilde{A}_{a}^{+}\left(\vec{x}\right)$,
for which the Yang-Mills equation (\ref{eq:Yang-Mills}) yields:
\begin{equation}
\begin{aligned}\partial_{\mu}\partial^{\mu}\alpha_{a}\left(\vec{x}\right)=-\nabla_{\perp}^{2}\alpha_{a}\left(\vec{x}\right) & =\rho_{a}\left(\vec{x}\right).\end{aligned}
\label{eq:YMCOV}
\end{equation}
Its solution is easily found by going to Fourier space, and yields:
\begin{equation}
\begin{aligned}\alpha_{a}\left(x^{-},\mathbf{x}\right) & =\int\mathrm{d}^{2}\mathbf{y}\int\frac{\mathrm{d}^{2}\mathbf{k}_{\perp}}{\left(2\pi\right)^{2}}\frac{e^{i\mathbf{k}_{\perp}\cdot\left(\mathbf{x}-\mathbf{y}\right)}}{k_{\perp}^{2}}\rho_{a}\left(x^{-},\mathbf{y}\right),\\
 & =\frac{1}{4\pi}\int\mathrm{d}^{2}\mathbf{y}\ln\frac{1}{\left(\mathbf{x}-\mathbf{y}\right)^{2}\Lambda^{2}}\rho_{a}\left(x^{-},\mathbf{y}\right),
\end{aligned}
\label{eq:A+}
\end{equation}
where we made use of Eq. (\ref{eq:fourier2log}), and where $\Lambda^{2}$
is some infrared cutoff, such as $\Lambda_{\mathrm{QCD}}$. The associated
field strength is:
\begin{equation}
\begin{aligned}F_{\mathrm{COV}}^{i+}\left(x^{-},\mathbf{x}\right) & =\partial^{i}\alpha\left(x^{-},\mathbf{x}\right),\\
 & =\int\mathrm{d}^{2}\mathbf{y}\int\frac{\mathrm{d}^{2}\mathbf{k}_{\perp}}{\left(2\pi\right)^{2}}\frac{ik_{\perp}^{i}}{k_{\perp}^{2}}e^{i\mathbf{k}_{\perp}\cdot\left(\mathbf{x}-\mathbf{y}\right)}\rho\left(x^{-},\mathbf{y}\right),\\
 & =-\frac{1}{2\pi}\int\mathrm{d}^{2}\mathbf{y}\frac{\left(x-y\right)^{i}}{\left(\mathbf{x}-\mathbf{y}\right)^{2}}\rho\left(x^{-},\mathbf{y}\right).
\end{aligned}
\end{equation}

However, the MV model is most naturally formulated in the light-cone
gauge $A^{+}=0$. From the solution of the Yang-Mills equation in
covariant gauge, one can obtain the classical gluon field $\mathcal{A}^{\mu}$
in LC gauge with the help of another gauge transformation $\mathcal{V}^{\dagger}\left(\vec{x}\right)$:
\begin{equation}
\begin{aligned}\mathcal{A}^{\mu} & =\mathcal{V}^{\dagger}\left(\tilde{A}^{\mu}+\frac{i}{g_{s}}\partial^{\mu}\right)\mathcal{V}.\end{aligned}
\label{eq:ACOVtoLC}
\end{equation}
For the case $\mu=+$, we have:
\begin{equation}
\mathcal{V}^{\dagger}\left(\alpha+\frac{i}{g_{s}}\partial^{+}\right)\mathcal{V}=0,
\end{equation}
which is solved by the Wilson line in the adjoint representation:
\begin{equation}
\begin{aligned}\mathcal{V}\left(x^{-},\mathbf{x}\right) & \equiv\mathcal{P}e^{ig_{s}\int_{-\infty}^{x^{-}}\mathrm{d}z^{-}\alpha_{a}\left(z^{-},\mathbf{x}\right)T^{a}}=W\left(x^{-},\mathbf{x}\right).\end{aligned}
\label{eq:COVtoLC}
\end{equation}
Hence, from Eq. (\ref{eq:ACOVtoLC}) we obtain the gauge field $\mathcal{A}^{i}$
in the LC gauge, as a pure gauge field with the gauge rotation (\ref{eq:COVtoLC}):
\begin{equation}
\begin{aligned}\mathcal{A}^{i}\left(x^{-},\mathbf{x}\right) & =\frac{i}{g_{s}}W^{\dagger}\left(x^{-},\mathbf{x}\right)\partial_{-}W\left(x^{-},\mathbf{x}\right),\end{aligned}
\end{equation}
and the associated field strength:
\begin{align}
F_{\mathrm{LC}}^{i+}\left(\vec{x}\right) & =W^{\dagger}\left(\vec{x}\right)F_{\mathrm{COV}}^{i+}\left(\vec{x}\right)W\left(\vec{x}\right)=W_{ba}\left(\vec{x}\right)\partial^{i}\alpha_{b}\left(\vec{x}\right)t^{a},\label{eq:fieldstrenghtLCMV}
\end{align}
where we used the property (\ref{eq:adjoint2adjoint}). In the literature,
$\mathcal{A}^{i}$ is often called the non-Abelian Weizsäcker-Williams
field (see e.g. Ref. \protect\cite{Kovchegov1996}, or Ref. \protect\cite{Wessels}
for an instructive comparison with the Abelian case).

Whether they are formulated in the covariant gauge ($\mathcal{A}^{+}$)
or in the LC gauge ($\mathcal{A}^{i}$), the classical wee gluon fields
in the MV model are functions of the color sources $\rho_{a}\left(x\right)$
which generate them. In order to proceed, we need to study these color
sources and their distribution closer. To do so, let us consider deep-inelastic
scattering, in which a quark-antiquark dipole probes the nucleus,
with a transverse resolution $1/Q^{2}$. The number of sources seen
by the projectile is then given by $\Delta N=n/Q^{2}$, with $n\equiv N_{c}A/\pi R_{A}^{2}$
the transverse density of valence quarks. Again using that $R_{A}\simeq\Lambda_{\mathrm{QCD}}^{-1}A^{1/3}$,
we find:
\begin{equation}
\Delta N=\frac{\Lambda_{\mathrm{QCD}}^{2}}{Q^{2}}\frac{N_{c}A^{1/3}}{\pi},
\end{equation}
and therefore, if the virtuality of the photon is small enough: $Q^{2}\ll\Lambda_{\mathrm{QCD}}^{2}A^{1/3}$
(but still large enough, such that $Q^{2}\gg\Lambda_{\mathrm{QCD}}^{2}$),
a large number $\Delta N$ of color sources is probed. Since these
sources are associated with valence quarks, which belong to different
nucleons in the large nucleus, it is safe to assume that they are
randomly distributed in the transverse plane, and hence the average
of the total color charge $Q_{a}$ seen by the probe is zero:
\begin{equation}
\langle\mathcal{Q}_{a}\rangle_{A}=0,\label{eq:Qa1pnt}
\end{equation}
while the two-point function of the color charges is given by:
\begin{equation}
\langle\mathcal{Q}_{a}\mathcal{Q}_{a}\rangle_{A}=g_{s}^{2}C_{F}\Delta N,\label{eq:Qa2pnt}
\end{equation}
and all the higher-point functions are assumed to vanish. Furthermore,
note that, in order for boundary effects to be negligible, it is assumed
that the transverse extent of the nucleus is infinite. 

Since Eqs. (\ref{eq:Qa1pnt}) and (\ref{eq:Qa2pnt}) are expected
to hold irrespectively of the precise value of the resolution $1/Q^{2}$,
the color charges $\mathcal{Q}_{a}$ are naturally associated with
a continuous distribution of color charge with density $\rho_{a}$:
\begin{equation}
\mathcal{Q}_{a}\equiv\int_{1/Q^{2}}\mathrm{d}^{2}\mathbf{x}\int\mathrm{d}x^{-}\rho_{a}\left(x^{-},\mathbf{x}\right).
\end{equation}
Using $\int_{1/Q^{2}}\mathrm{d}^{2}\mathbf{x}=1/Q^{2}$, as well as
the definition of $C_{F}\equiv\left(N_{c}^{2}-1\right)/2N_{c}$, correlators
(\ref{eq:Qa1pnt}) and (\ref{eq:Qa2pnt}) can be written in function
of the sources, yielding:
\begin{equation}
\begin{aligned}\langle\rho_{a}\left(x^{-},\mathbf{x}\right)\rangle_{A} & =0,\\
\langle\rho_{a}\left(x^{-},\mathbf{x}\right)\rho_{b}\left(y^{-},\mathbf{y}\right)\rangle_{A} & =g_{s}^{2}\delta_{ab}\delta\left(x^{-}-y^{-}\right)\delta^{\left(2\right)}\left(\mathbf{x}-\mathbf{y}\right)\lambda_{A}\left(x^{-}\right).
\end{aligned}
\label{eq:rhocorrelators}
\end{equation}
where
\begin{equation}
\mu_{A}\equiv g_{s}^{2}\int\mathrm{d}x^{-}\lambda_{A}\left(x^{-}\right)=\frac{g_{s}^{2}A}{2\pi R_{A}^{2}}\label{eq:muA}
\end{equation}
is the density of color charge squared of the valence quarks, per
unit area and per color. Finally, the correlators can be generated
from the following weight function, which is a Gaussian distribution
of the color charges:
\begin{equation}
\mathcal{W}_{A}\left[\rho\right]=\mathcal{N}\exp\left(-\frac{1}{2}\int\mathrm{d}^{3}x\frac{\rho_{a}\left(\vec{x}\right)\rho_{a}\left(\vec{x}\right)}{\lambda_{A}\left(x^{-}\right)}\right),
\end{equation}
and which can be used to write the averages over observables explicitly:
\begin{equation}
\langle\mathcal{O}\rangle_{A}=\frac{\int\mathcal{D}\left[\rho\right]\mathcal{W}_{A}\left[\rho\right]\mathcal{O}}{\int\mathcal{D}\left[\rho\right]\mathcal{W}_{A}\left[\rho\right]}.\label{eq:MVcorrelator}
\end{equation}
Note that these averages are purely classical, similar to a Boltzmann
average in statistical physics.

The correlators Eq. (\ref{eq:rhocorrelators}) can be written in function
of the gluon fields $\alpha_{a}\left(\mathbf{x}\right)$ in the covariant
gauge, using Eqs. (\ref{eq:YMCOV}) and (\ref{eq:A+}):
\begin{equation}
\begin{aligned}\langle\alpha_{a}\left(x^{-},\mathbf{x}\right)\alpha_{b}\left(y^{-},\mathbf{y}\right)\rangle_{A} & =\frac{1}{g_{s}^{2}}\delta_{ab}\delta\left(x^{-}-y^{-}\right)\lambda_{A}\left(x^{-}\right)L\left(\mathbf{x}-\mathbf{y}\right),\end{aligned}
\label{eq:MV2point}
\end{equation}
where:
\begin{equation}
\begin{aligned}L\left(\mathbf{x}-\mathbf{y}\right) & =g_{s}^{4}\int\frac{\mathrm{d}^{2}\mathbf{k}_{\perp}}{\left(2\pi\right)^{2}}\frac{e^{i\mathbf{k}_{\perp}\left(\mathbf{x}-\mathbf{y}\right)}}{k_{\perp}^{4}}.\end{aligned}
\label{eq:L}
\end{equation}

It is a very instructive exercise to evaluate the dipole in the MV
model:
\begin{equation}
\begin{aligned}D\left(\mathbf{x}-\mathbf{y}\right) & \equiv\frac{1}{N_{c}}\mathrm{Tr}\left\langle U\left(\mathbf{x}\right)U^{\dagger}\left(\mathbf{y}\right)\right\rangle _{A},\end{aligned}
\label{eq:DipoleDef}
\end{equation}
which we already encountered earlier in the discussion on deep-inelastic
scattering, and which will continue to play an important role in what
follows. Note that we do not write an index $Y$, since the MV model
comes with a fixed validity range for $x$ (see Eqs. (\ref{eq:MVxupperlimit})
and (\ref{eq:MVxlowerlimit})). Let us introduce the following definitions:
\begin{equation}
\begin{aligned}D\left(L,\mathbf{x}-\mathbf{y}\right) & \equiv\frac{1}{N_{c}}\mathrm{Tr}\left\langle U\left(L,\mathbf{x}\right)U^{\dagger}\left(L,\mathbf{y}\right)\right\rangle _{A},\\
U\left(L,\mathbf{x}\right) & \equiv\mathcal{P}e^{ig_{s}\int_{-\infty}^{L}\mathrm{d}z^{-}\alpha_{a}\left(z^{-},\mathbf{x}\right)t^{a}},
\end{aligned}
\end{equation}
where the dipole and the Wilson lines have support $x^{-}\in]-\infty,L]$.
Increasing $L$ with an infinitesimal distance $\epsilon$, one obtains:
\begin{equation}
\begin{aligned} & D\left(L+\epsilon,\mathbf{x}-\mathbf{y}\right)=\frac{1}{N_{c}}\mathrm{Tr}\Bigl\langle U\left(L+\epsilon,\mathbf{x}\right)U^{\dagger}\left(L+\epsilon,\mathbf{y}\right)\Bigr\rangle_{A},\\
 & \simeq\frac{1}{N_{c}}\mathrm{Tr}\Bigl\langle U\left(L,\mathbf{x}\right)\left(1+ig_{s}\epsilon\alpha_{a}\left(L+\epsilon,\mathbf{x}\right)t^{a}+\frac{1}{2!}\left(ig_{s}\right)^{2}\epsilon^{2}\left(\alpha_{a}\left(L+\epsilon,\mathbf{x}\right)t^{a}\right)^{2}\right)\\
 & \times\left(1-ig_{s}\epsilon\alpha_{a}\left(L+\epsilon,\mathbf{y}\right)t^{a}+\frac{1}{2!}\left(-ig_{s}\right)^{2}\epsilon^{2}\left(\alpha_{a}\left(L+\epsilon,\mathbf{y}\right)t^{a}\right)^{2}\right)U^{\dagger}\left(L,\mathbf{y}\right)\Bigr\rangle_{A},\\
 & =D\left(L,\mathbf{x}-\mathbf{y}\right)+D\left(L,\mathbf{x}-\mathbf{y}\right)\\
 & \times\frac{-g_{s}^{2}}{2N_{c}}\epsilon^{2}\mathrm{Tr}\Bigl\langle\left(\alpha_{a}\left(L,\mathbf{y}\right)t^{a}\right)^{2}+\left(\alpha_{a}\left(L,\mathbf{x}\right)t^{a}\right)^{2}-2\alpha_{a}\left(L,\mathbf{x}\right)t^{a}\alpha_{b}\left(L,\mathbf{y}\right)t^{b}\Bigr\rangle_{A},
\end{aligned}
\label{eq:dipolecalcu}
\end{equation}
where, in the last line, we factorized $D\left(L,\mathbf{x}-\mathbf{y}\right)$
from the parts that are evaluated at $x^{-}=L+\epsilon$, since, from
Eq. (\ref{eq:MV2point}), the correlator is local in $x^{-}$. In
addition, we used that $\alpha_{a}\left(L,\mathbf{x}\right)\simeq\alpha_{a}\left(L+\epsilon,\mathbf{x}\right)$,
and we kept only the quadratic terms in the gluon fields since, from
Eq. (\ref{eq:rhocorrelators}), the one-point function disappears.
Furthermore, plugging in the discrete version of Eq. (\ref{eq:MV2point}),
in which an extra factor $1/\epsilon$ appears (since the delta function
becomes upon discretization $\delta\left(x^{-}-y^{-}\right)=\delta\left(\left(n-m\right)\epsilon\right)=\delta_{nm}/\epsilon$),
we obtain:
\begin{equation}
\begin{aligned}D\left(L+\epsilon,\mathbf{x}-\mathbf{y}\right)-D\left(L,\mathbf{x}-\mathbf{y}\right) & =D\left(L,\mathbf{x}-\mathbf{y}\right)\times\frac{-g_{s}^{2}C_{F}}{2\mu_{A}}\epsilon\lambda_{A}\left(L\right)\Gamma\left(\mathbf{x}-\mathbf{y}\right),\end{aligned}
\label{eq:dipoleMVttstp}
\end{equation}
where we introduced the dimensionless quantity:
\begin{equation}
\begin{aligned}\Gamma\left(\mathbf{x}-\mathbf{y}\right) & \equiv\frac{\mu_{A}}{g_{s}^{2}}\left(L_{\mathbf{xx}}+L_{\mathbf{yy}}-2L_{\mathbf{xy}}\right)=2\frac{\mu_{A}}{g_{s}^{2}}\left(L\left(\mathbf{0}\right)-L\left(\mathbf{x}-\mathbf{y}\right)\right),\\
 & =2\mu_{A}g_{s}^{2}\int\frac{\mathrm{d}^{2}\mathbf{k}_{\perp}}{\left(2\pi\right)^{2}}\frac{1}{k_{\perp}^{4}}\left(1-e^{i\mathbf{k}_{\perp}\left(\mathbf{x}-\mathbf{y}\right)}\right).
\end{aligned}
\label{eq:Gamma}
\end{equation}
Eq. (\ref{eq:dipoleMVttstp}) can be cast in the form of a differential
equation:
\begin{equation}
\begin{aligned}\frac{\partial}{\partial x^{-}}D\left(x^{-},\mathbf{x}-\mathbf{y}\right) & =-\frac{g_{s}^{2}}{\mu_{A}}\frac{C_{F}}{2}\lambda_{A}\left(x^{-}\right)\Gamma\left(\mathbf{x}-\mathbf{y}\right)D\left(x^{-},\mathbf{x}-\mathbf{y}\right),\end{aligned}
\label{eq:derivD}
\end{equation}
with the solution:
\begin{equation}
\begin{aligned}D\left(x^{-},\mathbf{x}-\mathbf{y}\right) & =\exp\left(-\frac{g_{s}^{2}}{\mu_{A}}\frac{C_{F}}{2}\int_{-\infty}^{x^{-}}\mathrm{d}z^{-}\lambda_{A}\left(z^{-}\right)\Gamma\left(\mathbf{x}-\mathbf{y}\right)\right),\end{aligned}
\label{eq:dipoleMVx-dependent}
\end{equation}
or, taking $x^{-}\to\infty$:
\begin{equation}
\begin{aligned}D\left(\mathbf{r}\right) & =e^{-\frac{C_{F}}{2}\Gamma\left(\mathbf{r}\right)}\simeq e^{-\frac{r^{2}}{4}Q_{s}^{2}\left(r^{2}\right)},\end{aligned}
\label{eq:DipoleMVGamma}
\end{equation}
The integral in the expression for $\Gamma\left(\mathbf{x}\right)$
is dominated by small momenta. It can therefore be evaluated, to logarithmic
accuracy, by expanding the numerator:
\begin{equation}
\begin{aligned}\Gamma\left(\mathbf{r}\right) & \simeq2\mu_{A}g_{s}^{2}\int_{\Lambda^{2}}^{1/r^{2}}\frac{\mathrm{d}^{2}\mathbf{k}_{\perp}}{\left(2\pi\right)^{2}}\frac{1}{k_{\perp}^{4}}\frac{k_{\perp}^{2}r^{2}\cos^{2}\theta}{2!},\\
 & =\frac{r^{2}}{2}\alpha_{s}\mu_{A}\ln\frac{1}{r^{2}\Lambda^{2}}=\frac{r^{2}}{2}\frac{1}{C_{F}}Q_{s}^{2}\left(r^{2}\right)=\frac{r^{2}}{2}\frac{1}{N_{c}}Q_{sg}^{2}\left(r^{2}\right),
\end{aligned}
\label{eq:GammaQ}
\end{equation}
and we recover indeed a large logarithm, assuming that $r\ll1/\Lambda$,
where $\Lambda$ is an infrared cutoff $\Lambda$ such as $\Lambda_{\mathrm{QCD}}$.
In the last line of the above expression, we introduced the transverse
momentum scales $Q_{s}$ and $Q_{sg}$, which are the saturation scales
experienced by a quark or a gluon, respectively. In the MV model,
they are defined as follows:
\begin{equation}
\begin{aligned}Q_{s}^{2}\left(r^{2}\right) & \equiv\alpha_{s}C_{F}\mu_{A}\ln\frac{1}{r^{2}\Lambda^{2}},\\
Q_{sg}^{2}\left(r^{2}\right) & \equiv\alpha_{s}N_{c}\mu_{A}\ln\frac{1}{r^{2}\Lambda^{2}},
\end{aligned}
\label{eq:SaturationScale}
\end{equation}
where $\Lambda$ is the infrared cutoff, most commonly taken to be
$\Lambda_{\mathrm{QCD}}$. From the definition of $\mu_{A}$, Eq.
(\ref{eq:muA}), it is clear that the saturation scale rises with
the number of nucleons as $Q_{s}^{2}\sim A^{1/3}$. We will show in
a few moments how the scales $Q_{s}$ and $Q_{sg}$ are related to
the phenomenon of saturation. First, let us remark that the above
results allow us to write the dipole in the MV model as follows:
\begin{equation}
\begin{aligned}D\left(\mathbf{r}\right) & \simeq e^{-\frac{r^{2}}{4}Q_{s}^{2}\left(r^{2}\right)}.\end{aligned}
\label{eq:dipoleMVQs}
\end{equation}
Moreover, it is easy to see that, for a dipole in the adjoint representation:
\begin{equation}
\begin{aligned}D_{A}\left(\mathbf{x}-\mathbf{y}\right) & \equiv\frac{1}{N_{c}^{2}-1}\mathrm{Tr}\left\langle W\left(\mathbf{x}\right)W^{\dagger}\left(\mathbf{y}\right)\right\rangle _{A},\end{aligned}
\label{eq:DA}
\end{equation}
the calculation leading to Eq. (\ref{eq:dipoleMVx-dependent}) can
be repeated and one obtains exactly the same result, except from the
color factor $C_{F}\to N_{c}$:
\begin{equation}
\begin{aligned}D_{A}\left(\mathbf{r}\right) & =e^{-\frac{N_{c}}{2}\Gamma\left(\mathbf{r}\right)}\simeq e^{-\frac{r^{2}}{4}Q_{sg}^{2}\left(r^{2}\right)}.\end{aligned}
\label{eq:AdjointDipoleMV}
\end{equation}

We would now like to evaluate the Weizsäcker-Williams gluon distribution
$\mathcal{F}_{gg}^{\left(3\right)}\left(x,q_{\perp}\right)$, introduced
in Sec. \ref{sec:GluonTMDs}, in the MV model. To do so, we insert
the light-cone field strengths, Eq. (\ref{eq:fieldstrenghtLCMV}),
into expression (\ref{eq:WWFFFourier}), which yields:
\begin{equation}
\begin{aligned}\mathcal{F}_{gg}^{\left(3\right)}\left(x,q_{\perp}\right) & =4\int\frac{\mathrm{d}^{3}v\mathrm{d}^{3}w}{\left(2\pi\right)^{3}}e^{-i\mathbf{q}_{\perp}\left(\mathbf{v}-\mathbf{w}\right)}\mathrm{Tr}\Bigl\langle\left(W_{ba}\partial^{i}\alpha_{b}t^{a}\right)\left(\vec{v}\right)\left(W_{cd}\partial^{i}\alpha_{c}t^{d}\right)\left(\vec{w}\right)\Bigr\rangle_{A}.\end{aligned}
\end{equation}
Due to the locality of the correlators in $x^{-}$, see Eq. (\ref{eq:MV2point}),
the correlator can be factorized as follows:
\begin{equation}
\begin{aligned} & \mathrm{Tr}\Bigl\langle\left(W_{ba}\partial^{i}\alpha_{b}t^{a}\right)\left(\vec{v}\right)\left(W_{cd}\partial^{i}\alpha_{c}t^{d}\right)\left(\vec{w}\right)\Bigr\rangle_{A}\\
 & =\frac{1}{2}\delta^{ad}\Bigl\langle W_{ba}\left(\vec{v}\right)W_{cd}\left(\vec{w}\right)\Bigr\rangle_{A}\times\Bigl\langle\partial^{i}\left(\alpha_{b}\left(\vec{v}\right)\right)\partial^{i}\alpha_{c}\left(\vec{w}\right)\Bigr\rangle_{A},\\
 & =-\frac{1}{2g_{s}^{2}}\delta\left(v^{-}-w^{-}\right)\lambda_{A}\left(v^{-}-w^{-}\right)\mathrm{Tr}\Bigl\langle W\left(\vec{v}\right)W^{\dagger}\left(\vec{w}\right)\Bigr\rangle_{A}\partial_{\perp}^{2}L\left(\mathbf{v}-\mathbf{w}\right),
\end{aligned}
\label{eq:WWMVtss}
\end{equation}
where we used the fact that $W_{ab}\left(\vec{x}\right)$ is real.
From the definition of $L\left(\mathbf{x}\right)$, see Eq. (\ref{eq:L}),
we have that:
\begin{equation}
\begin{aligned}\partial_{\perp}^{2}L\left(\mathbf{v}-\mathbf{w}\right) & =-\frac{g_{s}^{4}}{4\pi}\ln\frac{1}{\left(\mathbf{v}-\mathbf{w}\right)^{2}\Lambda_{\mathrm{QCD}}^{2}}.\end{aligned}
\end{equation}
Using Eqs. (\ref{eq:doublederivquad}) (with $C_{F}\to N_{c}$, since
we now work in the adjoint representation), (\ref{eq:GammaQ}) and
(\ref{eq:SaturationScale}), the intermediate result Eq. (\ref{eq:WWMVtss})
can be written as:
\begin{equation}
\begin{aligned} & \mathrm{Tr}\Bigl\langle\left(W_{ba}\partial^{i}\alpha_{b}t^{a}\right)\left(\vec{v}\right)\left(W_{cd}\partial^{i}\alpha_{c}t^{d}\right)\left(\vec{w}\right)\Bigr\rangle_{A}\\
 & =\frac{4C_{F}}{g_{s}^{2}}\delta\left(v^{-}-w^{-}\right)\frac{1}{\left(\mathbf{v}-\mathbf{w}\right)^{2}}\frac{\partial}{\partial v^{-}}\left(1-D_{A}\left(v^{-},\mathbf{v}-\mathbf{w}\right)\right),
\end{aligned}
\end{equation}
from which we finally obtain:
\begin{equation}
\begin{aligned}\mathcal{F}_{gg}^{\left(3\right)}\left(x,q_{\perp}\right) & =\frac{2S_{\perp}C_{F}}{\alpha_{s}\pi^{2}}\int\frac{\mathrm{d}^{2}\mathbf{r}}{\left(2\pi\right)^{2}}\frac{e^{-i\mathbf{q}_{\perp}\mathbf{r}}}{r^{2}}\left(1-D_{A}\left(r^{2}\right)\right),\\
 & =\frac{2S_{\perp}C_{F}}{\alpha_{s}\pi^{2}}\int\frac{\mathrm{d}^{2}\mathbf{r}}{\left(2\pi\right)^{2}}\biggl(\frac{e^{-i\mathbf{q}_{\perp}\mathbf{r}}}{r^{2}}1-e^{-\frac{r^{2}Q_{sg}^{2}\left(r^{2}\right)}{4}}\biggr),
\end{aligned}
\label{eq:WWMV}
\end{equation}
where, again, we made use of the fact that the integrals over transverse
coordinates are restricted to the transverse surface $S_{\perp}$
of the nucleus:
\begin{equation}
S_{\perp}\equiv\int\mathrm{d}^{2}\mathbf{x}.
\end{equation}
Although the result (\ref{eq:WWMV}) for the Weizsäcker-Williams gluon
distribution is purely classical, and holds only in the MV model,
we can use it to illustrate many of the properties of the gluon density
in the small-$x$ regime, in particular the meaning of the saturation
scale $Q_{s}$ and $Q_{sg}$. For instance, in the dilute limit, defined
as $q_{\perp}^{2}\gg Q_{sg}^{2}$, the rapidly oscillating exponent
will kill most of the integral, except at very small values of $r$:
$r^{2}\ll1/q_{\perp}^{2}$. Therefore, the exponential can be expanded
in $Q_{sg}^{2}/q_{\perp}^{2}$, yielding to next-to-leading order:
\begin{equation}
\begin{aligned}\mathcal{F}_{gg}^{\left(3\right)}\left(x,q_{\perp}\right) & \simeq\frac{2S_{\perp}C_{F}}{\alpha_{s}\pi^{2}}\int\frac{\mathrm{d}^{2}\mathbf{r}}{\left(2\pi\right)^{2}}\frac{e^{-i\mathbf{q}_{\perp}\mathbf{r}}}{r^{2}}\left(\frac{r^{2}Q_{sg}^{2}}{4}-\frac{1}{2}\frac{r^{4}Q_{sg}^{4}}{16}+\mathcal{O}\left(r^{6}Q_{sg}^{6}\right)\right),\\
 & \simeq\frac{S_{\perp}C_{F}}{2\alpha_{s}\pi^{3}}\frac{Q_{sg}^{2}}{q_{\perp}^{2}}\left(1+\frac{Q_{sg}^{2}}{q_{\perp}^{2}}\left[\ln\frac{q_{\perp}^{2}}{\Lambda_{\mathrm{QCD}}^{2}}+2\gamma_{E}-2\right]\right),
\end{aligned}
\label{eq:WWlargekT}
\end{equation}
where we used the so-called harmonic approximation:
\begin{equation}
\begin{aligned}D_{A}\left(\mathbf{r}\right) & \simeq e^{-\frac{r^{2}}{4}Q_{sg}^{2}},\end{aligned}
\end{equation}
in which the logarithmic dependence of the saturation scale on the
dipole size $r^{2}$ is neglected. The result (\ref{eq:WWlargekT})
becomes more clear once we insert the expression for the saturation
scale, Eq. (\ref{eq:SaturationScale}) (again, without the logarithm),
and keep only the leading term:
\begin{equation}
\begin{aligned}\mathcal{F}_{gg}^{\left(3\right)}\left(x,q_{\perp}\right) & \simeq AN_{c}\times\frac{\alpha_{s}C_{F}}{\pi^{2}q_{\perp}^{2}}=AN_{c}\times x\frac{\mathrm{d}P_{\mathrm{Brems}}}{\mathrm{d}x\mathrm{d}^{2}q_{\perp}}.\end{aligned}
\label{eq:WWdilute}
\end{equation}
In the dilute limit, we thus recover a simple BFKL-like picture of
the gluon density, in which each of the $AN_{c}$ valence quarks generate
the Bremsstrahlung spectrum. It is easy to check that the above expression
is equivalent to the dilute approximation for the dipole-hadron amplitude
$\sigma_{\mathrm{dip}}\sim r^{2}$, Eq. (\ref{eq:sigmaxG}).

In the opposite limit of small transverse momenta: $q_{\perp}^{2}\ll Q_{sg}^{2}\left(\mathbf{r}^{2}\right)$,
the integral is dominated by large values of $r$: $r^{2}\gg1/q_{\perp}^{2}$,
and the exponential can simply be neglected, yielding:
\begin{equation}
\begin{aligned}\mathcal{F}_{gg}^{\left(3\right)}\left(x,q_{\perp}\right) & \simeq\frac{2S_{\perp}C_{F}}{\alpha_{s}\pi^{2}}\int\frac{\mathrm{d}^{2}\mathbf{r}}{\left(2\pi\right)^{2}}\frac{e^{-i\mathbf{q}_{\perp}\mathbf{r}}}{\mathbf{r}^{2}}=\frac{S_{\perp}C_{F}}{\alpha_{s}\pi^{3}}\int_{2/Q_{sg}}\mathrm{d}r\frac{J_{0}\left(rq_{\perp}\right)}{r},\\
 & \simeq\frac{S_{\perp}C_{F}}{2\alpha_{s}\pi^{3}}\ln\frac{Q_{sg}^{2}}{q_{\perp}^{2}}.
\end{aligned}
\label{eq:WWsaturation}
\end{equation}
Comparing Eqs. (\ref{eq:WWlargekT}) and (\ref{eq:WWsaturation}),
which hold in the dilute and the saturation regime, respectively,
we see that in the dilute regime $q_{\perp}^{2}\gg Q_{sg}^{2}$ the
gluon spectrum rapidly grows towards smaller transverse momenta, like
$Q_{sg}^{2}/q_{\perp}^{2}$. However, around transverse momenta of
the order of the saturation scale $Q_{sg}^{2}$, the behavior of the
spectrum changes, and the growth is damped to a much slower logarithmic
growth $\sim\ln Q_{sg}^{2}/q_{\perp}^{2}$. The gluon spectrum thus
saturates due to the rising importance of nonlinear effects. In addition,
Eq. (\ref{eq:WWsaturation}) manifests an example of geometric scaling,
which is the property that in the saturation regime, $x$ and $q_{\perp}^{2}$
only appear in the combination $Q_{s}^{2}\left(x\right)/q_{\perp}^{2}$:
a consequence of the fact that $Q_{s}^{2}$ is the only meaningful
scale in this region of phase space. This is in contrast to the dilute
case Eq. (\ref{eq:WWlargekT}), in which the soft scale $\Lambda_{\mathrm{QCD}}$
appears, acting as an infrared cutoff for the gluon radiation. 

\section{\label{sec:JIMWLK}Evolution of the CGC}

In the previous section, we described the classical McLerran-Venugopalan
model for a large nucleus, and argued that it is valid for wee gluons
with Bjorken-$x$ in the range $x_{\mathrm{wee}}\sim10^{-2}-10^{-1}$.
These kinematical limits were obtained by requiring, on the one hand,
that $x_{\mathrm{wee}}\ll x_{\mathrm{val}}\sim1$, such that the valence
quarks appear to be frozen with respect to the time scale $x^{+}\simeq2x_{\mathrm{wee}}P^{+}/p_{\perp}^{2}$
of the wee partons. On the other hand, we wanted the radiative effects
to be small enough for the wee gluons to be adequately described by
classical fields, hence:
\begin{equation}
\int_{x_{0}}^{x_{\mathrm{val}}}\mathrm{d}P_{\mathrm{Brems}}=\frac{\alpha_{s}N_{c}}{\pi}\ln\frac{x_{\mathrm{val}}}{x_{0}}\ll1,
\end{equation}
resulting in a lower limit $x_{0}\simeq e^{-1/\alpha_{s}}$ for the
wee partons. From the above formula it is clear that, if we want to
apply our theory to smaller values of $x$, lowering the limit $x_{0}$,
the upper limit of the phase space for the wee partons has to go down
as well. If we denote this limit with $x'$, we have that:
\begin{equation}
x_{\mathrm{val}}\sim1,\quad x_{0}\ll x_{\mathrm{wee}}\ll x_{\mathrm{val}}\overset{x_{0}\,\mathrm{smaller}}{\longrightarrow}x'\ll x_{\mathrm{val}}\ll1,\quad x_{0}\ll x_{\mathrm{wee}}\ll x',
\end{equation}
hence what we mean with valence or wee partons changes depending on
the choice of $x'$: with wee partons we now refer to partons that
live in a softer region of phase space, down to $x_{0}$, while our
definition of valence partons is expanded to include lower-energy
partons as well. For our classical description, in which valence partons
are treated as sources which generate the wee gluons, to hold, these
\textquoteleft new' valence partons have to be promoted to color sources
as well. This can be done integrating the degrees of freedom in the
strip $x'\ll x_{\mathrm{val}}$ out, and including them in the renormalization
of the distribution of the sources $\mathcal{W}_{x'}\left[\rho\right]$,
which becomes now dependent on the scale $x'$.

In order to describe this renormalization process more precisely,
let us revisit the mechanism we mentioned above, but this time we
will be a bit more general. If $P^{+}$ is the total longitudinal
momentum of the hadron we want to describe, one can distinguish between
the \textquoteleft hard' partons with momenta $k^{+}>\Lambda^{+}$,
and the \textquoteleft soft' partons with momenta $k^{+}<\Lambda^{+}$,
where we introduced a certain momentum scale $\Lambda^{+}\equiv xP^{+}$.
The hard and the soft partons (which are in this context gluons: quarks
only play a role of significance as the sources in the MV model) are
the generalizations of what we called the valence and the wee partons
in the MV model. For each choice of the scale $\Lambda^{+}$, we have
an effective classical theory similar to the MV model, in which the
hard gluons are described by static color charges, which generate
the soft gluons through the Yang-Mills equations. Given this classical
theory at a certain scale $\Lambda^{+}$, we can evolve to a smaller
scale $\Lambda'^{+}$ by integrating the so-called \textquoteleft semi-fast'
modes in the strip $\Lambda^{+}\gg k^{+}\gg\Lambda'^{+}$ out and
including them in the new sources $\rho'=\rho+\delta\rho$. The result
of this procedure is a renormalization group equation for the distribution
function $\mathcal{W}_{x}\left[\rho\right]$. 

Since the new color sources are generated through radiation, the evolution
of $\mathcal{W}_{x}\left[\rho\right]$ is quantum mechanical. As a
consequence, whereas the two-point function for the classical theory
at a scale $\Lambda^{+}$ is given by a classical, statistical average:
\begin{equation}
\begin{aligned}\langle\rho\left(\vec{x}\right)\rho\left(\vec{y}\right)\rangle_{\Lambda^{+}} & =\int\mathcal{D}\left[\rho\right]\mathcal{W}_{\Lambda^{+}}\left[\rho\right]\rho\left(\vec{x}\right)\rho\left(\vec{y}\right),\end{aligned}
\label{eq:classaverage}
\end{equation}
just like in the MV model, this is not the case anymore if we want
to evaluate the correlator at a lower scale $\Lambda'^{+}$. Indeed,
in order to integrate them out, the semi-fast gluons in the region
$\Lambda^{+}\gg k^{+}\gg\Lambda'^{+}$ have to be made quantum mechanical
again, and the two-point function of the new sources $\rho'=\rho+\delta\rho$
becomes:
\begin{equation}
\langle\rho'\left(\vec{x}\right)\rho'\left(\vec{y}\right)\rangle_{\Lambda'^{+}}=\int\mathcal{D}\left[\rho\right]\mathcal{W}_{\Lambda^{+}}\left[\rho\right]\frac{\int_{\Lambda'^{+}}^{\Lambda^{+}}\mathcal{D}\left[a\right]\rho'\left(\vec{x}\right)\rho'\left(\vec{y}\right)e^{iS\left[A,\rho\right]}}{\int_{\Lambda'^{+}}^{\Lambda^{+}}\mathcal{D}\left[a\right]e^{iS\left[A,\rho\right]}}.\label{eq:class+qmaverage}
\end{equation}
In the above formula, we used the following expansion of the gluon
field:
\begin{equation}
A^{\mu}=\mathcal{A}^{\mu}\left[\rho\right]+a^{\mu}+\delta A^{\mu},\label{eq:gluonfieldexpansion}
\end{equation}
where $\mathcal{A}^{\mu}\left[\rho\right]$ is the classical background
field generated through the Yang-Mills equations by the sources $\rho$,
and $a^{\mu}$ is the field corresponding to the quantum fluctuations
of the semi-fast modes. The soft ($k^{+}\ll\Lambda'^{+}$) fields,
which we denote by $\delta A^{\mu}$, do not appear explicitly in
the correlator (\ref{eq:class+qmaverage}), but are generated by the
sources, as is the case in the MV model. Note that we work in the
light-cone gauge, hence it is always understood that $A^{+}=0$. The
action $S\left[A,\rho\right]$ encodes the dynamics of the gluon fields,
in particular of the semi-fast modes $a^{\mu}$, in the presence of
the sources, and quantum correlations between these gluon fields (and
by extension the \textquoteleft new' sources $\rho'=\rho+\delta\rho$
in which the semi-fast modes will be absorbed) are calculated with
the help of a path integral. In addition, everything still needs to
be statistically averaged over the \textquoteleft old' sources $\rho$
at the scale $\Lambda^{+}$. Note that, when $\Lambda'^{+}\sim\Lambda^{+}$,
the quantum effects are negligible, and the path integrals can be
evaluated in the saddle point approximation: $S\left[A,\rho\right]\to S_{\mathrm{classical}}\left[\mathcal{A},\rho\right]$.
The correlator of Eq. (\ref{eq:class+qmaverage}), which contains
both a quantum mechanical path integration and a classical statistical
average, is in this approximation reduced to the purely classical
correlator Eq. (\ref{eq:classaverage}). 

Once the quantum mechanical evolution to the smaller scale $\Lambda'^{+}$
is carried out, we again have a classical theory in which the sources,
distributed according to the renormalized weight function $\mathcal{W}_{\Lambda'^{+}}\left[\rho\right]$
and corresponding to hard gluons with $k^{+}\gg\Lambda'^{+}$, generate
classical soft gluon fields with $k^{+}\ll\Lambda'^{+}$. The relevant
two-point function is then:
\begin{equation}
\begin{aligned}\langle\rho'\left(\vec{x}\right)\rho'\left(\vec{y}\right)\rangle_{\Lambda'^{+}} & =\int\mathcal{D}\left[\rho'\right]\mathcal{W}_{\Lambda'^{+}}\left[\rho'\right]\rho'\left(\vec{x}\right)\rho'\left(\vec{y}\right).\end{aligned}
\label{eq:classaverageEVO}
\end{equation}
Matching this correlator with Eq. (\ref{eq:class+qmaverage}), we
will be able to deduce the evolution equation for the distribution
function $\mathcal{W}_{\Lambda^{+}}\left[\rho\right]$.

To proceed, let us have a closer look at the action $S\left[A,\rho\right]$.
It is given by:
\begin{equation}
\begin{aligned}S\left[A,\rho\right] & =-\frac{1}{4}\int\mathrm{d}^{4}x\,F_{\mu\nu}^{a}F_{a}^{\mu\nu}+\frac{i}{g_{s}N_{c}}\int\mathrm{d}^{3}x\,\mathrm{Tr}\left(\rho\left(\vec{x}\right)W_{T}\left(\vec{y}\right)\right),\end{aligned}
\label{eq:SArho}
\end{equation}
where $W_{T}\left(\vec{x}\right)$ is a temporal Wilson line, running
along the time direction $x^{+}$:
\begin{equation}
\begin{aligned}W_{T}\left(\vec{x}\right) & =\mathrm{T}\exp\left(ig_{s}\int\mathrm{d}x^{+}A^{-}\left(x^{+},\vec{x}\right)\right),\end{aligned}
\end{equation}
with $\mathrm{T}$ the time-ordering operator. Applying the Euler-Lagrange
equations $\partial_{\mu}\delta S/\delta\partial_{\mu}A^{\nu}=\delta S/\delta A^{\nu}$,
one recovers the Yang-Mills equations: 
\begin{equation}
\begin{aligned}\left[D_{\nu},F^{\nu\mu}\right]\left(x\right) & =J^{\mu},\end{aligned}
\end{equation}
with the current:
\begin{equation}
\begin{aligned}J^{\mu} & \equiv\delta^{\mu+}W_{T}\left(x^{+},\vec{x}\right)\rho\left(\vec{x}\right)W_{T}^{\dagger}\left(x^{+},\vec{x}\right),\\
W_{T}\left(x^{+},\vec{x}\right) & \equiv\mathrm{T}\exp\left(ig_{s}\int^{x^{+}}\mathrm{d}z^{+}A^{-}\left(z^{+},\vec{x}\right)\right).
\end{aligned}
\end{equation}
We mentioned already that, in the classical case, one can always find
a solution with $A^{-}=0$, for which one recovers the current in
Eq. (\ref{eq:Yang-Mills}). In contrast to the MV model, however,
the action $S\left[A,\rho\right]$ pertains to quantum gluons, which
is why we need the gauge-invariant expression Eq. (\ref{eq:SArho}).

We can expand the action to second order in the weak field $a^{\mu}$,
around the background $\mathcal{A}^{\mu}$, as follows:
\begin{equation}
\begin{aligned}S\left[A,\rho\right] & \simeq S\left[\mathcal{A},\rho\right]+\int\mathrm{d}^{4}x\left.\frac{\delta S}{\delta A_{a}^{\mu}}\right|_{\mathcal{A}}a_{a}^{\mu}\left(x\right)+\frac{1}{2}\int\mathrm{d}^{4}x\mathrm{d}^{4}y\left.\frac{\delta^{2}S}{\delta A_{a}^{\mu}\delta A_{b}^{\nu}}\right|_{\mathcal{A}}a_{a}^{\mu}\left(x\right)a_{b}^{\nu}\left(y\right).\end{aligned}
\end{equation}
From this expression, one obtains the following propagator for the
semi-fast gluon fields in the presence of the background field $\mathcal{A}^{\mu}$
and the sources $\rho$:
\begin{equation}
\langle a_{a}^{\mu}\left(x\right)a_{b}^{\nu}\left(y\right)\rangle_{\rho}=\frac{1}{2}\left.\frac{\delta^{2}S}{\delta A_{a}^{\mu}\delta A_{b}^{\nu}}\right|_{\mathcal{A}}^{-1}.\label{eq:semifastpropagator}
\end{equation}
Finally, we need to calculate the new contributions $\delta\rho$
to the sources, due to the semi-fast gluons $a^{\mu}$. These sources
couple linearly, as $\delta\rho_{a}\delta A_{a}^{-}$, to the soft
gluon fields $\delta A_{a}^{-}$, hence, from the expansion (\ref{eq:gluonfieldexpansion}),
one finds:
\begin{equation}
\delta\rho_{a}\left(x\right)=\left.\frac{\delta S}{\delta A_{a}^{-}\left(x\right)}\right|_{\mathcal{A}+a}.\label{eq:effectiveCGCrhoa}
\end{equation}
We thus have an effective quantum field theory, describing the dynamics
of semi-fast gluon fields $a^{\mu}$ in the presence of a background
$\mathcal{A}^{\mu}$. Explicit expressions for the propagator (\ref{eq:semifastpropagator})
and the effective sources (\ref{eq:effectiveCGCrhoa}) can be found
in Ref. \protect\cite{JIMWLK1}, for example.

\begin{figure}[t]
\begin{centering}
\begin{tikzpicture}[scale=1.7] 

\tikzset{photon/.style={decorate,decoration={snake}},
		electron/.style={ postaction={decorate},decoration={markings,mark=at position .5 with {\arrow[draw]{>}}}},      	gluon/.style={decorate,decoration={coil,amplitude=4pt, segment length=5pt}}}

\draw[thick] (0,0) --++ (0,.8) node [at end, cross out, draw, solid,  inner sep=2.5 pt]{} node[above]{$\mathcal{F}^{+i}(x)$};
\draw[thick] (2,0) --++ (0,.8) node [at end, cross out, draw, solid, inner sep=2.5 pt]{} node[above]{$\mathcal{F}^{+j}(y)$};
\draw[gluon] (0,0)--++(2,0);
\node at (1,.3) {$\langle a^i(x) a^j(y)\rangle $};
\draw[dashed] (0,0) --++ (-.3,-.5) node[below]{$\delta A^-(x)$};
\draw[dashed] (2,0) --++ (.3,-.5) node[below]{$\delta A^-(y)$};
\node at (1,-.8) {(a)};

\draw[thick] (4,.7) --++ (1.5,0) node[right]{$\mathcal{F}^{+i}(z)$};
\draw[gluon] (4.3,.7) node[above]{$x^+$}.. controls (4.5,.2) and (5,.2)  .. (5.2,.7) node[above]{$y^+$};
\draw[dashed] (4.75,.2) --++ (0,-.45) node[below]{$\delta A^-(z)$};
\node at (4.75,-.8) {(b)};
\end{tikzpicture} 
\par\end{centering}
\caption{\label{fig:JIMWLKdiagrams}Contributions to the charge-charge correlator
$\chi_{bc}^{ij}$, (a), and to the induced color charge density $\sigma_{a}$,
(b).}
\end{figure}
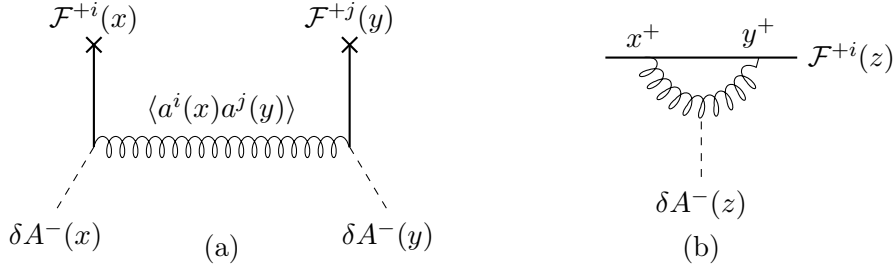

Let us simply state the results of the analysis of the effective theory
(see e.g. Ref. \protect\cite{JIMWLK1}). It turns out that, to logarithmic
accuracy, only the one-point function and the two-point function play
a role of importance:
\begin{equation}
\langle\delta\rho_{a}\left(\vec{x}\right)\rangle_{\rho}\propto\sigma_{a}\left(\vec{x}\right),\quad\langle\delta\rho_{a}\left(\vec{x}\right)\delta\rho_{b}\left(\vec{y}\right)\rangle_{\rho}\propto\chi_{ab}\left(\vec{x},\vec{y}\right).
\end{equation}
The two-point function $\langle\delta\rho_{a}\left(\vec{x}\right)\delta\rho_{b}\left(\vec{y}\right)\rangle_{\rho}$
is known as the induced charge-charge correlator, and describes the
\textquoteleft real' emission of semi-fast gluons by the color charges
(see Fig. \ref{fig:JIMWLKdiagrams}, a). The one-point function $\langle\delta\rho_{a}\left(\vec{x}\right)\rangle_{\rho}$
is the color charge density induced by the semi-fast fields, and provides
virtual corrections to the gluon cascades in the semi-fast region.
These virtual corrections can be both one-loop corrections to the
emission vertices (see Fig. \ref{fig:JIMWLKdiagrams}, b) and self-energy
corrections. Both $\sigma_{a}$ and $\chi_{ab}$ and nonlinear in
$\rho$, and thus contain diagrams such as gluon mergings, as we expect
in the saturation regime. Furthermore, they are $x^{+}$-independent.
We will give their explicit expressions later.

First, we should scrutinize the longitudinal structure of our theory,
which is an important aspect which we previously glossed over. Whereas
in the MV model the valence partons were localized in a very sharp
peak around $x^{-}=0$, the longitudinal extent of the sources becomes
larger for each step in the evolution. Indeed, each strip of semi-fast
gluons with $\Lambda'^{+}=b\Lambda^{+}\ll k^{+}\ll\Lambda^{+}$, included
in the renormalized sources $\rho'=\rho+\delta\rho$, is delocalized
over a region $1/\Lambda^{+}\apprle x^{-}\apprle1/\Lambda'^{+}$,
which becomes bigger as the momentum scales become smaller. When integrating
these semi-fast gluons out, one has to keep track of their multiple
scatterings with the background field $\mathcal{A}^{\mu}$, which
are encoded in Wilson lines along the $x^{-}$-axis. Due to the color
structure of QCD, these scatterings are non-commutative, and hence
the longitudinal structure of $\rho$ cannot simply be neglected.
In particular, when calculating the induced color charge density for
a given configuration of the background sources $\rho$, the result
turns out to be:
\begin{align}
\langle\delta\rho_{a}\left(x^{-},\mathbf{x}\right)\rangle_{\rho} & =F_{\Lambda}\left(x^{-}\right)\sigma_{a}\left(\mathbf{x}\right),
\end{align}
with
\begin{equation}
\begin{aligned}F_{\Lambda}\left(x^{-}\right) & =\theta\left(x^{-}\right)\frac{e^{-i\Lambda'^{+}x^{-}}-e^{-i\Lambda^{+}x^{-}}}{x^{-}},\quad\int\mathrm{d}x^{-}F_{\Lambda}\left(x^{-}\right)=\ln\frac{\Lambda^{+}}{\Lambda'^{+}}=\ln\frac{1}{b}\equiv\mathrm{d}Y.\end{aligned}
\end{equation}
The above equations illustrate how, integrating the semi-fast modes
out in layers of $\Lambda^{+}\gg k^{+}\gg b\Lambda{}^{+}$, layers
of an extent $\mathrm{d}Y\equiv\ln1/b$ along the $x^{-}$-axis are
added. The natural way to describe the evolution is therefore in discrete
rapidity steps $\mathrm{d}\tau$ of the order $1/\alpha_{s}$, since
due to the Bremsstrahlung law:
\begin{equation}
\int_{b\Lambda'^{+}}^{\Lambda^{+}}\mathrm{d}P_{\mathrm{Brems}}=\alpha_{s}\int_{b\Lambda'^{+}}^{\Lambda^{+}}\frac{\mathrm{d}k^{+}}{k^{+}}=\alpha_{s}\mathrm{d}Y,
\end{equation}
resummation becomes necessary when $P\sim1$. Defining the evolution
\textquoteleft time' $Y\equiv\ln P^{+}/\Lambda^{+}$, we can write
\begin{equation}
\delta\rho_{Y}^{a}\left(\mathbf{x}\right)\mathrm{d}Y\equiv\int\mathrm{d}x^{-}\delta\rho^{a}\left(x^{-},\mathbf{x}\right),
\end{equation}
with $\delta\rho_{Y}^{a}\left(\mathbf{x}\right)$ the induced color
charge by integrating out the interval $\Lambda^{+}\gg k^{+}\gg b\Lambda{}^{+}$
or equivalently the rapidity bin $Y\ll\ln1/x\ll Y+\mathrm{d}Y$. To
the desired logarithmic accuracy, the longitudinal structure of each
layer $\mathrm{d}Y$ is thus not important, only the relative ordering
of the different rapidity layers. We then have that:
\begin{equation}
\langle\delta\rho_{Y}^{a}\left(\mathbf{x}\right)\rangle_{\rho}\equiv\sigma_{Y}^{a}\left(\mathbf{x}\right).\label{eq:sigma}
\end{equation}
Similarly, for the induced charge-charge correlator:
\begin{align}
\langle\delta\rho_{Y}^{a}\left(\mathbf{x}\right)\delta\rho_{Y}^{b}\left(\mathbf{y}\right)\rangle_{\rho} & =\frac{1}{\mathrm{d}Y}\chi_{Y}^{ab}\left(\mathbf{x},\mathbf{y}\right),\quad\chi_{Y}^{ab}\left(\mathbf{x},\mathbf{y}\right)\mathrm{d}Y\equiv\int\mathrm{d}x^{-}\mathrm{d}y^{-}\langle\delta\rho_{Y}^{a}\left(x^{-},\mathbf{x}\right)\delta\rho_{Y}^{b}\left(y^{-},\mathbf{y}\right)\rangle.\label{eq:chi}
\end{align}

We are now finally ready to perform the matching procedure between
Eqs. (\ref{eq:class+qmaverage}) and (\ref{eq:classaverageEVO}) that
we alluded to earlier. Indeed, given a source $\rho_{Y}^{a}\left(\mathbf{x}\right)$,
the induced source in the next rapidity bin $Y+\mathrm{d}Y$ is a
variable distributed according to a Gaussian distribution $\mathcal{W}_{\mathrm{d}Y}\left[\delta\rho|\rho\right]$.
Since we integrated the semi-fast modes out and incorporated them
in the new distribution function, the correlator Eq. (\ref{eq:class+qmaverage})
becomes:
\begin{equation}
\langle\rho\left(\mathbf{x}\right)\rho\left(\mathbf{y}\right)\rangle_{Y+\mathrm{d}Y}=\int\mathcal{D}\left[\rho\right]\mathcal{W}_{Y}\left[\rho\right]\int\mathcal{D}\left[\delta\rho\right]\mathcal{W}_{\mathrm{d}Y}\left[\delta\rho|\rho\right]\left(\rho+\delta\rho\right)\left(\mathbf{x}\right)\left(\rho+\delta\rho\right)\left(\mathbf{y}\right),\label{eq:rhorho1}
\end{equation}
which has to be equal to the two-point function of the classical theory
at rapidity $Y+\mathrm{d}Y$ (cf.. Eq. (\ref{eq:classaverageEVO})):
\begin{equation}
\langle\rho\left(\mathbf{x}\right)\rho\left(\mathbf{y}\right)\rangle_{Y+\mathrm{d}Y}=\int\mathcal{D}\left[\rho'\right]\mathcal{W}_{Y+\mathrm{d}Y}\left[\rho'\right]\rho'\left(\mathbf{x}\right)\rho'\left(\mathbf{y}\right).\label{eq:rhorho2}
\end{equation}
One important subtlety is the fact that the support on the $x^{-}$-axis
for the sources in Eqs. (\ref{eq:rhorho1}) and (\ref{eq:rhorho2})
is not the same: in the first correlator, $\rho$ is defined for $\ln\left(1/x\right)<Y$,
while the renormalized $\rho'$ in the second correlator involves
the additional bin $\mathrm{d}Y$ and hence has support $\ln\left(1/x\right)<Y+\mathrm{d}Y$.
In what follows, it is therefore understood that $\rho$ is defined
for arbitrary values of $y$, but that instead the weight functions
$\mathcal{W}_{Y}\left[\rho\right]$ take care of the correct support:
\begin{equation}
\mathcal{W}_{Y}\left[\rho\right]=\mathcal{W}_{Y}\left[\rho\left(\ln\frac{1}{x}<Y\right)\right]\delta\left[\rho\left(\ln\frac{1}{x}>Y\right)\right].\label{eq:weightsupport}
\end{equation}
With this redefinition, we can make a change of variables $\rho\to\rho-\delta\rho$
in Eq. (\ref{eq:rhorho1}):
\begin{equation}
\langle\rho\left(\mathbf{x}\right)\rho\left(\mathbf{y}\right)\rangle_{Y+\mathrm{d}Y}=\int\mathcal{D}\left[\rho\right]\int\mathcal{D}\left[\delta\rho\right]\mathcal{W}_{Y}\left[\rho-\delta\rho\right]\mathcal{W}_{\mathrm{d}Y}\left[\delta\rho|\rho-\delta\rho\right]\rho\left(\mathbf{x}\right)\rho\left(\mathbf{y}\right),
\end{equation}
and comparing with Eq. (\ref{eq:rhorho2}), we find that:
\begin{equation}
\mathcal{W}_{Y+\mathrm{d}Y}\left[\rho\right]=\int\mathcal{D}\left[\delta\rho\right]\mathcal{W}_{Y}\left[\rho-\delta\rho\right]\mathcal{W}_{\mathrm{d}Y}\left[\delta\rho|\rho-\delta\rho\right].
\end{equation}
Expanding the integrand:
\begin{equation}
\begin{aligned}\mathcal{W}_{Y}\left[\rho-\delta\rho\right]\mathcal{W}_{\mathrm{d}Y}\left[\delta\rho|\rho-\delta\rho\right] & \simeq\biggl(1-\mathrm{d}Y\int_{\mathbf{x}}\delta\rho_{Y}^{a}\left(\mathbf{x}\right)\frac{\delta}{\delta\rho_{Y}^{a}\left(\mathbf{x}\right)}\\
 & +\frac{\left(\mathrm{d}Y\right)^{2}}{2}\int_{\mathbf{x}\mathbf{y}}\delta\rho_{Y}^{a}\left(\mathbf{x}\right)\delta\rho_{Y}^{b}\left(\mathbf{y}\right)\frac{\delta^{2}}{\delta\rho_{Y}^{a}\left(\mathbf{x}\right)\delta\rho_{Y}^{b}\left(\mathbf{y}\right)}\biggr)\mathcal{W}_{Y}\left[\rho\right]\mathcal{W}_{\mathrm{d}Y}\left[\delta\rho|\rho\right],
\end{aligned}
\end{equation}
we have:
\begin{equation}
\begin{aligned}\mathcal{W}_{Y+\mathrm{d}Y}\left[\rho\right]-\mathcal{W}_{Y}\left[\rho\right] & =-\mathrm{d}Y\int_{\mathbf{x}}\frac{\delta}{\delta\rho_{Y}^{a}\left(\mathbf{x}\right)}\mathcal{W}_{Y}\left[\rho\right]\sigma_{Y}^{a}\left(\mathbf{x}\right)+\frac{\mathrm{d}Y}{2}\int_{\mathbf{x}\mathbf{y}}\frac{\delta^{2}}{\delta\rho_{Y}^{a}\left(\mathbf{x}\right)\delta\rho_{Y}^{b}\left(\mathbf{y}\right)}\mathcal{W}_{Y}\left[\rho\right]\chi_{Y}^{ab}\left(\mathbf{x},\mathbf{y}\right),\end{aligned}
\end{equation}
where we used that:
\begin{equation}
\begin{aligned}\int\mathcal{D}\left[\delta\rho\right]\mathcal{W}_{\mathrm{d}Y}\left[\delta\rho|\rho\right] & =1,\\
\int\mathcal{D}\left[\delta\rho\right]\delta\rho_{Y}^{a}\left(\mathbf{x}\right)\mathcal{W}_{\mathrm{d}Y}\left[\delta\rho|\rho\right] & =\sigma_{Y}^{a}\left(\mathbf{x}\right),\\
\int\mathcal{D}\left[\delta\rho\right]\delta\rho_{Y}^{a}\left(\mathbf{x}\right)\delta\rho_{Y}^{b}\left(\mathbf{y}\right)\mathcal{W}_{\mathrm{d}Y}\left[\delta\rho|\rho\right] & =\frac{1}{\mathrm{d}Y}\chi_{Y}^{ab}\left(\mathbf{x},\mathbf{y}\right).
\end{aligned}
\end{equation}
Hence, we finally obtain a renormalization group equation for $\mathcal{W}_{Y}\left[\rho\right]$:
\begin{equation}
\begin{aligned}\frac{\partial\mathcal{W}_{Y}\left[\rho\right]}{\partial Y} & =\frac{1}{2}\int_{\mathbf{x}\mathbf{y}}\frac{\delta^{2}}{\delta\rho_{Y}^{a}\left(\mathbf{x}\right)\delta\rho_{Y}^{b}\left(\mathbf{y}\right)}\mathcal{W}_{Y}\left[\rho\right]\chi_{Y}^{ab}\left(\mathbf{x},\mathbf{y}\right)-\int_{\mathbf{x}}\frac{\delta}{\delta\rho_{Y}^{a}\left(\mathbf{x}\right)}\mathcal{W}_{Y}\left[\rho\right]\sigma_{Y}^{a}\left(\mathbf{x}\right).\end{aligned}
\end{equation}
It turns out to be simpler to express the above equation in terms
of the solutions $\alpha_{a}\equiv\tilde{A}_{a}^{+}$ of the Yang-Mills
equations in the covariant gauge, Eq. (\ref{eq:YMCOV}) (where $\Phi,\sigma$
and $\chi$ are now understood to be functionals of $\alpha$):
\begin{equation}
\begin{aligned}\frac{\partial\mathcal{W}_{Y}\left[\alpha\right]}{\partial Y} & =\frac{1}{2}\int_{\mathbf{x}\mathbf{y}}\frac{\delta^{2}}{\delta\alpha_{Y}^{a}\left(\mathbf{x}\right)\delta\alpha_{Y}^{b}\left(\mathbf{y}\right)}\mathcal{W}_{Y}\left[\alpha\right]\chi^{ab}\left(\mathbf{x},\mathbf{y}\right)-\int_{\mathbf{x}}\frac{\delta}{\delta\alpha_{Y}^{a}\left(\mathbf{x}\right)}\mathcal{W}_{Y}\left[\alpha\right]\sigma^{a}\left(\mathbf{x}\right).\end{aligned}
\label{eq:JIMWLK}
\end{equation}
The induced color charge $\sigma^{a}\left(\mathbf{x}\right)$ and
charge-charge correlator $\chi^{ab}\left(\mathbf{x},\mathbf{y}\right)$
(we can omit the index $Y$, since the information on the longitudinal
support is carried by the charge distribution $\mathcal{W}_{Y}\left[\alpha\right]$)
are, after explicit evaluation of all the diagrams in the effective
theory, found to be (Refs. \protect\cite{JIMWLK4,JIMWLK5}):
\begin{equation}
\begin{aligned}\sigma^{a}\left(\mathbf{x}\right) & =i\frac{g_{s}}{2\pi}\int\frac{\mathrm{d}^{2}\mathbf{z}}{\left(2\pi\right)^{2}}\frac{1}{\left(\mathbf{x}-\mathbf{z}\right)^{2}}\mathrm{Tr}\left(T^{a}W^{\dagger}\left(\mathbf{x}\right)W\left(\mathbf{z}\right)\right),\\
\chi^{ab}\left(\mathbf{x},\mathbf{y}\right) & =\frac{1}{\pi}\int\frac{\mathrm{d}^{2}\mathbf{z}}{\left(2\pi\right)^{2}}\mathcal{K}_{\mathbf{xyz}}\left(1+W^{\dagger}\left(\mathbf{x}\right)W\left(\mathbf{y}\right)-W^{\dagger}\left(\mathbf{x}\right)W\left(\mathbf{z}\right)-W^{\dagger}\left(\mathbf{z}\right)W\left(\mathbf{y}\right)\right)^{ab},
\end{aligned}
\label{eq:sigmaandchi}
\end{equation}
with the transverse kernel:
\begin{equation}
\mathcal{K}_{\mathbf{xyz}}\equiv\frac{\left(x-z\right)^{i}\left(y-z\right)^{i}}{\left(\mathbf{x}-\mathbf{z}\right)^{2}\left(\mathbf{y}-\mathbf{z}\right)^{2}},\label{eq:transversekernelK}
\end{equation}
and the Wilson lines $W\left(\mathbf{x}\right)$:
\begin{equation}
\begin{aligned}W\left(\mathbf{x}\right) & =\mathcal{P}e^{ig_{s}\int\mathrm{d}z^{-}\alpha_{a}\left(z^{-},\mathbf{x}\right)T^{a}}.\end{aligned}
\end{equation}
Note that, due to the fact that the longitudinal support of the sources,
and therefore of the fields $\alpha_{y}$, is encoded in the weight
function Eq. (\ref{eq:weightsupport}), the integration path of the
Wilson line is effectively cut of at rapidity $\ln\left(1/x\right)=Y$:
\begin{equation}
\begin{aligned}W\left(\mathbf{x}\right) & =\mathcal{P}e^{ig_{s}\int^{Y}\mathrm{d}x\,\alpha_{x}^{a}\left(\mathbf{x}\right)T^{a}}.\end{aligned}
\end{equation}
The renormalization group equation (\ref{eq:JIMWLK}) is known as
the JIMWLK evolution equation, named after its authors Jalilian-Marian,
Iancu, McLerran, Weigert, Leonidov and Kovner (see Refs. \protect\cite{JIMWLK1,JIMWLK2,JIMWLK3,JIMWLK4,JIMWLK5,JIMWLK6,JIMWLK7,JIMWLK8}). 

\section{Properties of the JIMWLK equation}

The properties presented in this section are not commonly found in
the literature, I mainly relied on Refs. \protect\cite{Hatta2005,EdmondRev}.

The JIMWLK equation, Eq. (\ref{eq:JIMWLK}), can be written in a more
convenient Hamiltonian form, using the following relation between
$\chi^{ab}$ and $\sigma^{a}$:
\begin{equation}
\begin{aligned}\frac{1}{2}\int_{\mathbf{y}}\frac{\delta\chi^{ab}\left(\mathbf{x},\mathbf{y}\right)}{\delta\alpha_{Y}^{b}\left(\mathbf{y}\right)} & =\sigma^{a}\left(\mathbf{x}\right).\end{aligned}
\label{eq:chi2sigma}
\end{equation}
To prove the above formula, note that we have: 
\begin{equation}
\begin{aligned}\frac{\delta W^{\dagger}\left(\mathbf{x}\right)}{\delta\alpha_{Y}^{a}\left(\mathbf{y}\right)} & =\frac{\delta}{\delta\alpha_{Y}^{a}\left(\mathbf{y}\right)}\mathcal{\bar{P}}e^{-ig_{s}\int^{Y}\mathrm{d}x\,\alpha_{x}^{a}\left(\mathbf{x}\right)T^{a}}=-ig_{s}\delta_{\mathbf{x}\mathbf{y}}^{\left(2\right)}T^{a}W^{\dagger}\left(\mathbf{x}\right),\\
\frac{\delta W\left(\mathbf{x}\right)}{\delta\alpha_{Y}^{a}\left(\mathbf{y}\right)} & =ig_{s}\delta_{\mathbf{x}\mathbf{y}}^{\left(2\right)}W\left(\mathbf{x}\right)T^{a}.
\end{aligned}
\end{equation}
Using these properties, we find that
\begin{equation}
\begin{aligned}\frac{\delta W_{ac}^{\dagger}\left(\mathbf{x}\right)W_{cb}\left(\mathbf{y}\right)}{\delta\alpha_{Y}^{b}\left(\mathbf{y}\right)} & =-ig_{s}\delta_{\mathbf{x}\mathbf{y}}^{\left(2\right)}T_{ad}^{b}W_{dc}^{\dagger}\left(\mathbf{x}\right)W_{cb}\left(\mathbf{y}\right)+W_{ac}^{\dagger}\left(\mathbf{x}\right)ig_{s}\delta_{\mathbf{y}\mathbf{y}}^{\left(2\right)}W_{cd}\left(\mathbf{x}\right)T_{db}^{a},\\
 & =-ig_{s}\delta_{\mathbf{x}\mathbf{y}}^{\left(2\right)}T_{ad}^{b}\delta_{db}+ig_{s}\delta_{\mathbf{y}\mathbf{y}}^{\left(2\right)}\delta_{ad}T_{db}^{a}=0,
\end{aligned}
\end{equation}
where we made use of the property that $W_{ac}^{\dagger}\left(\mathbf{x}\right)W_{cb}\left(\mathbf{x}\right)=\delta_{ab}$,
and where both terms vanish due to the antisymmetry of the color matrices.
Similarly, one obtains:
\begin{equation}
\begin{aligned}\frac{\delta W_{ac}^{\dagger}\left(\mathbf{x}\right)W_{cb}\left(\mathbf{z}\right)}{\delta\alpha_{Y}^{b}\left(\mathbf{y}\right)} & =-ig_{s}\delta_{\mathbf{x}\mathbf{y}}^{\left(2\right)}T_{ad}^{b}W_{dc}^{\dagger}\left(\mathbf{x}\right)W_{cb}\left(\mathbf{z}\right)+W_{ac}^{\dagger}\left(\mathbf{x}\right)ig_{s}\delta_{\mathbf{y}\mathbf{z}}^{\left(2\right)}W_{cd}\left(\mathbf{z}\right)T_{db}^{b},\\
 & =ig_{s}\delta_{\mathbf{x}\mathbf{y}}^{\left(2\right)}T_{bd}^{a}W_{dc}^{\dagger}\left(\mathbf{x}\right)W_{cb}\left(\mathbf{z}\right)=ig_{s}\delta_{\mathbf{x}\mathbf{y}}^{\left(2\right)}\mathrm{Tr}\left(T^{a}W^{\dagger}\left(\mathbf{x}\right)W\left(\mathbf{z}\right)\right).
\end{aligned}
\end{equation}
Therefore, we find:
\begin{equation}
\begin{aligned}\frac{1}{2}\int_{\mathbf{y}}\frac{\delta\chi^{ab}\left(\mathbf{x},\mathbf{y}\right)}{\delta\alpha_{Y}^{b}\left(\mathbf{y}\right)} & =\frac{ig_{s}}{2\pi}\int\mathrm{d}^{2}\mathbf{y}\int\frac{\mathrm{d}^{2}\mathbf{z}}{\left(2\pi\right)^{2}}\mathcal{K}_{\mathbf{xyz}}\delta_{\mathbf{x}\mathbf{y}}^{\left(2\right)}\mathrm{Tr}\left(T^{a}W^{\dagger}\left(\mathbf{x}\right)W\left(\mathbf{z}\right)\right)=\sigma^{a}\left(\mathbf{x}\right),\end{aligned}
\end{equation}
proving Eq. (\ref{eq:chi2sigma}). Plugging this relation into the
JIMWLK equation, Eq.(\ref{eq:JIMWLK}), yields:
\begin{equation}
\begin{aligned}\frac{\partial\mathcal{W}_{Y}\left[\alpha\right]}{\partial Y} & =-H_{\mathrm{JIMWLK}}\mathcal{W}_{Y}\left[\alpha\right],\\
H_{\mathrm{JIMWLK}} & \equiv-\frac{1}{2}\int_{\mathbf{x}\mathbf{y}}\frac{\delta}{\delta\alpha_{Y}^{a}\left(\mathbf{x}\right)}\chi^{ab}\left(\mathbf{x},\mathbf{y}\right)\frac{\delta}{\delta\alpha_{Y}^{b}\left(\mathbf{y}\right)},
\end{aligned}
\label{eq:JIMWLKH}
\end{equation}
where we defined the JIMWLK Hamiltonian $H_{\mathrm{JIMWLK}}$.

We can apply the JIMWLK equation Eq. (\ref{eq:JIMWLKH}) for the weight
function $\Phi_{Y}\left[\alpha\right]$, to obtain an evolution equation
for a generic observable $\mathcal{O}$. Indeed, at every value of
rapidity $Y$, the CGC provides a classical theory, in which the expectation
value for a certain observable can be obtained as follows:
\begin{equation}
\langle\mathcal{O}\rangle_{Y}=\int\mathcal{D}\left[\alpha\right]\mathcal{W}_{Y}\left[\alpha\right]\mathcal{O}\left[\alpha\right].
\end{equation}
In the above formula, the rapidity dependence is completely encoded
in the weight function, and can therefore easily be turned into an
evolution equation for $\mathcal{O}$ by taking the derivative with
respect to $Y$:
\begin{equation}
\begin{aligned}\frac{\partial}{\partial Y}\langle\mathcal{O}\rangle_{Y} & =\int\mathcal{D}\left[\alpha\right]\frac{\partial\mathcal{W}_{Y}\left[\alpha\right]}{\partial Y}\mathcal{O}\left[\alpha\right],\\
 & =\int\mathcal{D}\left[\alpha\right]\frac{1}{2}\int_{\mathbf{x}\mathbf{y}}\frac{\delta}{\delta\alpha_{Y}^{a}\left(\mathbf{x}\right)}\left(\chi^{ab}\left(\mathbf{x},\mathbf{y}\right)\frac{\delta}{\delta\alpha_{Y}^{b}\left(\mathbf{y}\right)}\mathcal{W}_{Y}\left[\alpha\right]\right)\mathcal{O}\left[\alpha\right],\\
 & =-\int\mathcal{D}\left[\alpha\right]\frac{1}{2}\int_{\mathbf{x}\mathbf{y}}\chi^{ab}\left(\mathbf{x},\mathbf{y}\right)\left(\frac{\delta}{\delta\alpha_{Y}^{b}\left(\mathbf{y}\right)}\mathcal{W}_{Y}\left[\alpha\right]\right)\frac{\delta}{\delta\alpha_{Y}^{a}\left(\mathbf{x}\right)}\mathcal{O}\left[\alpha\right],\\
 & =\int\mathcal{D}\left[\alpha\right]\frac{1}{2}\int_{\mathbf{x}\mathbf{y}}\mathcal{W}_{Y}\left[\alpha\right]\frac{\delta}{\delta\alpha_{Y}^{b}\left(\mathbf{y}\right)}\chi^{ab}\left(\mathbf{x},\mathbf{y}\right)\frac{\delta}{\delta\alpha_{Y}^{a}\left(\mathbf{x}\right)}\mathcal{O}\left[\alpha\right],\\
 & =-\langle H_{\mathrm{JIMWLK}}\mathcal{O}\rangle_{Y}.
\end{aligned}
\label{eq:JIMWLKO}
\end{equation}
In the last equality, we made use of the fact that:
\begin{equation}
\chi^{ab}\left(\mathbf{x},\mathbf{y}\right)=\chi^{ba}\left(\mathbf{y},\mathbf{x}\right),\label{eq:chisymm}
\end{equation}
which follows immediately from expression (\ref{eq:sigmaandchi})
for $\chi^{ab}$:
\begin{equation}
\begin{aligned}\chi^{ab}\left(\mathbf{x},\mathbf{y}\right) & =\frac{1}{\pi}\int\frac{\mathrm{d}^{2}\mathbf{z}}{\left(2\pi\right)^{2}}\mathcal{K}_{\mathbf{xyz}}\left(1+W^{\dagger}\left(\mathbf{y}\right)W\left(\mathbf{x}\right)-W^{\dagger}\left(\mathbf{z}\right)W\left(\mathbf{x}\right)-W^{\dagger}\left(\mathbf{y}\right)W\left(\mathbf{z}\right)\right)^{ab},\\
 & =\frac{1}{\pi}\int\frac{\mathrm{d}^{2}\mathbf{z}}{\left(2\pi\right)^{2}}\mathcal{K}_{\mathbf{xyz}}\left(\left(1+W^{\dagger}\left(\mathbf{y}\right)W\left(\mathbf{x}\right)-W^{\dagger}\left(\mathbf{z}\right)W\left(\mathbf{x}\right)-W^{\dagger}\left(\mathbf{y}\right)W\left(\mathbf{z}\right)\right)^{\dagger}\right)^{ba},\\
 & =\frac{1}{\pi}\int\frac{\mathrm{d}^{2}\mathbf{z}}{\left(2\pi\right)^{2}}\mathcal{K}_{\mathbf{xyz}}\left(1+W^{\dagger}\left(\mathbf{x}\right)W\left(\mathbf{y}\right)-W^{\dagger}\left(\mathbf{x}\right)W\left(\mathbf{z}\right)-W^{\dagger}\left(\mathbf{z}\right)W\left(\mathbf{y}\right)\right)^{ba},\\
 & =\chi^{ba}\left(\mathbf{y},\mathbf{x}\right),
\end{aligned}
\end{equation}
where, in the second equality, we made use of the fact that Wilson
lines in the adjoint representation are real. The property (\ref{eq:chisymm})
is even more apparent if one casts $\chi^{ab}\left(\mathbf{x},\mathbf{y}\right)$
in the following form:
\begin{equation}
\begin{aligned}\chi^{ab}\left(\mathbf{x},\mathbf{y}\right) & =\frac{1}{\pi}\int\frac{\mathrm{d}^{2}\mathbf{z}}{\left(2\pi\right)^{2}}\mathcal{K}_{\mathbf{xyz}}\left(1-W^{\dagger}\left(\mathbf{x}\right)W\left(\mathbf{z}\right)\right)^{ac}\left(1-W^{\dagger}\left(\mathbf{z}\right)W\left(\mathbf{y}\right)\right)^{cb}.\end{aligned}
\label{eq:chifactorized}
\end{equation}

Before moving to applications of the JIMWLK equation, we should scrutinize
the kernel $\mathcal{K}_{\mathbf{xyz}}$, Eq. (\ref{eq:transversekernelK}),
which has two poles at $\mathbf{z}=\mathbf{x}$ and \textbf{$\mathbf{z}=\mathbf{y}$}
which could give rise to short-distance singularities when integrating
over $\mathbf{z}$. However, as one can see from the above expression
for $\chi_{\mathbf{x}\mathbf{y}}^{ab}$, the residues of these poles
are zero. The JIMWLK Hamiltonian is thus free of ultraviolet divergences,
but it is not immediately clear that this is also the case for soft
divergences. Indeed, in the limit $z\to\infty$ the kernel behaves
as $\mathcal{K}_{\mathbf{xyz}}\to1/z^{2}$, which after integration
could lead to a logarithmic singularity. It turns out that these soft
divergences do in fact cancel, but only in the case of gauge-invariant
observables $\mathcal{O}$, as we will now demonstrate. First, let
us illustrate this cancellation in an example, considering the dilute
limit of the JIMWLK equation, which will turn out to be the BFKL equation,
which is explicitly infrared safe. To compute $\chi^{ab}\left(\mathbf{x},\mathbf{y}\right)$
in the dilute regime, starting from Eq. (\ref{eq:chifactorized}),
it is sufficient to expand the Wilson lines to leading order (this
is a simplification due to the factorized form of Eq. (\ref{eq:chifactorized}).
If we were to start from Eq. (\ref{eq:sigmaandchi}), we should expand
quadratically like usual, see e.g. the discussion around Eq. (\ref{eq:WLexpansion}))
:
\begin{equation}
\begin{aligned}W\left(\mathbf{x}\right) & \simeq1+ig_{s}\alpha_{\mathbf{x}}^{a}T^{a},\end{aligned}
\end{equation}
where we introduced the short-hand notation:
\begin{equation}
\alpha_{\mathbf{x}}^{a}\equiv\int\mathrm{d}z^{-}\alpha_{a}\left(z^{-},\mathbf{x}\right).\label{eq:nolongstructure}
\end{equation}
The induced charge-charge correlator then becomes:
\begin{equation}
\begin{aligned}\chi^{ab}\left(\mathbf{x},\mathbf{y}\right) & \simeq-\frac{g_{s}^{2}}{\pi}\int\frac{\mathrm{d}^{2}\mathbf{z}}{\left(2\pi\right)^{2}}\mathcal{K}_{\mathbf{xyz}}\left(\alpha_{\mathbf{x}}^{d}-\alpha_{\mathbf{z}}^{d}\right)\left(\alpha_{\mathbf{z}}^{e}-\alpha_{\mathbf{y}}^{e}\right)T_{ac}^{d}T_{cb}^{e},\end{aligned}
\end{equation}
from which we find the following Hamiltonian
\begin{equation}
\begin{aligned}H_{BFKL} & \equiv-\frac{\alpha_{s}}{2\pi^{2}}\int_{\mathbf{x}\mathbf{y}\mathbf{z}}\mathcal{K}_{\mathbf{xyz}}T_{ac}^{d}T_{cb}^{e}\left(\alpha_{\mathbf{x}}^{d}-\alpha_{\mathbf{z}}^{d}\right)\frac{\delta}{\delta\alpha_{\mathbf{x}}^{a}}\left(\alpha_{\mathbf{z}}^{e}-\alpha_{\mathbf{y}}^{e}\right)\frac{\delta}{\delta\alpha_{\mathbf{y}}^{b}}.\end{aligned}
\label{eq:JIMWLK2BFKL}
\end{equation}
It is very important to note that, writing (\ref{eq:nolongstructure})
and making the changes $\delta\alpha_{Y}^{a}\left(\mathbf{x}\right)\to\delta\alpha_{\mathbf{x}}^{a}$,
we threw away the longitudinal structure about the theory, which we
carefully constructed in the previous paragraph. This is tantamount
to the statement that in the dilute regime, also called the BFKL approximation
or the regime of $k_{\perp}$-factorization, multiple scattering can
be neglected and a complete factorization holds between the longitudinal
and the transverse dynamics. To show that Eq. (\ref{eq:JIMWLK2BFKL})
is indeed the BFKL Hamiltonian, let us apply it to a gauge invariant
quantity, say the elastic scattering amplitude of a color dipole off
the CGC, Eq. (\ref{eq:dipolecrossection}): 
\begin{equation}
\sigma_{\mathrm{dip}}\left(x,\mathbf{r}\right)=2\pi R_{A}^{2}T\left(r\right),\label{eq:dipolecrosssection}
\end{equation}
with:
\begin{align}
\langle T\left(\mathbf{x},\mathbf{y}\right)\rangle_{Y} & =1-D\left(\mathbf{x},\mathbf{y}\right)=1-\frac{1}{N_{c}}\mathrm{Tr}\langle U\left(\mathbf{x}\right)U^{\dagger}\left(\mathbf{y}\right)\rangle_{Y}.
\end{align}
In the weak-field approximation, the scattering amplitude is given
by:
\begin{equation}
\begin{aligned}\langle T\left(\mathbf{x},\mathbf{y}\right)\rangle_{Y} & \simeq\frac{g_{s}^{2}}{4N_{c}}\langle\left(\alpha_{\mathbf{x}}^{a}-\alpha_{\mathbf{y}}^{a}\right)^{2}\rangle_{Y},\\
 & =\frac{g_{s}^{2}}{4N_{c}}\langle\alpha_{\mathbf{x}}^{a}\alpha_{\mathbf{x}}^{a}+\alpha_{\mathbf{y}}^{a}\alpha_{\mathbf{y}}^{a}-2\alpha_{\mathbf{x}}^{a}\alpha_{\mathbf{y}}^{a}\rangle_{Y},
\end{aligned}
\label{eq:Txy}
\end{equation}
Applying the BFKL Hamiltonian, Eq. (\ref{eq:JIMWLK2BFKL}), on the
substructure $\langle\alpha_{\mathbf{x}}^{a}\alpha_{\mathbf{y}}^{a}\rangle_{Y}$
yields (we frequently made use of Eq. (\ref{eq:Casimiradj})):
\begin{equation}
\begin{aligned}\frac{\partial}{\partial Y}\langle\alpha_{\mathbf{x}}^{a}\alpha_{\mathbf{y}}^{a}\rangle_{Y} & =-\langle H_{BFKL}\alpha_{\mathbf{x}}^{a}\alpha_{\mathbf{y}}^{a}\rangle_{Y}\\
 & =-\frac{\alpha_{s}}{2\pi^{2}}\int_{\mathbf{v}\mathbf{w}\mathbf{z}}\mathcal{K}_{\mathbf{vwz}}T_{a'c}^{d}T_{cb}^{e}\Bigl\langle\left(\alpha_{\mathbf{v}}^{d}-\alpha_{\mathbf{z}}^{d}\right)\frac{\delta}{\delta\alpha_{\mathbf{v}}^{a'}}\left(\alpha_{\mathbf{z}}^{e}-\alpha_{\mathbf{w}}^{e}\right)\frac{\delta}{\delta\alpha_{\mathbf{w}}^{b}}\alpha_{\mathbf{x}}^{a}\alpha_{\mathbf{y}}^{a}\Bigr\rangle_{Y}\\
 & =-\frac{\alpha_{s}N_{c}}{2\pi^{2}}\int_{\mathbf{z}}\Bigl\langle\mathcal{K}_{\mathbf{xxz}}\left(\alpha_{\mathbf{x}}^{a}-\alpha_{\mathbf{z}}^{a}\right)\alpha_{\mathbf{y}}^{a}+\mathcal{K}_{\mathbf{yyz}}\left(\alpha_{\mathbf{y}}^{a}-\alpha_{\mathbf{z}}^{a}\right)\alpha_{\mathbf{x}}^{a}\\
 & +2\mathcal{K}_{\mathbf{xyz}}\left(\alpha_{\mathbf{z}}^{a}-\alpha_{\mathbf{x}}^{a}\right)\left(\alpha_{\mathbf{y}}^{a}-\alpha_{\mathbf{z}}^{a}\right)\Bigr\rangle_{Y}\\
 & =-\frac{\alpha_{s}N_{c}}{2\pi^{2}}\int_{\mathbf{z}}\Bigl\langle\mathcal{M}_{\mathbf{xyz}}\alpha_{\mathbf{x}}^{a}\alpha_{\mathbf{y}}^{a}-\mathcal{K}_{\mathbf{xxz}}\alpha_{\mathbf{z}}^{a}\alpha_{\mathbf{y}}^{a}\\
 & -\mathcal{K}_{\mathbf{yyz}}\alpha_{\mathbf{x}}^{a}\alpha_{\mathbf{z}}^{a}+2\mathcal{K}_{\mathbf{xyz}}\left(\alpha_{\mathbf{y}}^{a}\alpha_{\mathbf{z}}^{a}-\alpha_{\mathbf{z}}^{a}\alpha_{\mathbf{z}}^{a}+\alpha_{\mathbf{z}}^{a}\alpha_{\mathbf{x}}^{a}\right)\Bigr\rangle_{Y}
\end{aligned}
\end{equation}
In the first term of the last line in the above expression, the kernels
combined in the familiar dipole kernel:
\begin{equation}
\begin{aligned}\mathcal{M}_{\mathbf{xyz}} & \equiv\frac{\left(\mathbf{x}-\mathbf{y}\right)^{2}}{\left(\mathbf{x}-\mathbf{z}\right)^{2}\left(\mathbf{y}-\mathbf{z}\right)^{2}},\\
 & =\mathcal{K}_{\mathbf{xxz}}+\mathcal{K}_{\mathbf{yyz}}-2\mathcal{K}_{\mathbf{xyz}}.
\end{aligned}
\label{eq:M2K}
\end{equation}
In the soft limit $z\to\infty$, this kernel behaves as $\mathcal{M}_{\mathbf{xyz}}\to1/z^{4}$
and therefore goes to zero fast enough for the integral over $z$
to converge. The other terms in Eq. (\ref{eq:Txy}) are still problematic,
though. As we predicted, however, they cancel when looking at the
evolution of the full gauge invariant object $T\left(\mathbf{x},\mathbf{y}\right)$
\begin{equation}
\begin{aligned}\frac{\partial}{\partial\tau}\Bigl\langle\left(\alpha_{\mathbf{x}}^{a}-\alpha_{\mathbf{y}}^{a}\right)^{2}\Bigr\rangle_{Y} & =-\frac{\alpha_{s}N_{c}}{2\pi^{2}}\int_{\mathbf{z}}\Bigl\langle2\mathcal{M}_{\mathbf{xyz}}\left(\alpha_{\mathbf{y}}^{a}\alpha_{\mathbf{z}}^{a}-\alpha_{\mathbf{z}}^{a}\alpha_{\mathbf{z}}^{a}+\alpha_{\mathbf{z}}^{a}\alpha_{\mathbf{x}}^{a}-\alpha_{\mathbf{x}}^{a}\alpha_{\mathbf{y}}^{a}\right)\Bigr\rangle_{Y},\\
 & =\frac{\alpha_{s}N_{c}}{2\pi^{2}}\int_{\mathbf{z}}\Bigl\langle\mathcal{M}_{\mathbf{xyz}}\left(-\left(\alpha_{\mathbf{x}}^{a}-\alpha_{\mathbf{y}}^{a}\right)^{2}+\left(\alpha_{\mathbf{x}}^{a}-\alpha_{\mathbf{z}}^{a}\right)^{2}+\left(\alpha_{\mathbf{y}}^{a}-\alpha_{\mathbf{z}}^{a}\right)^{2}\right)\Bigr\rangle_{Y},
\end{aligned}
\end{equation}
yielding the familiar BFKL evolution equation for the dipole cross
section (\ref{eq:dipolecrosssection}) which is explicitly infrared
finite:
\begin{equation}
\begin{aligned}\frac{\partial}{\partial Y}T_{Y}\left(\mathbf{x}-\mathbf{y}\right) & =\frac{\bar{\alpha}}{2\pi}\int_{\mathbf{z}}\mathcal{M}_{\mathbf{xyz}}\left(T_{Y}\left(\mathbf{x}-\mathbf{z}\right)+T_{Y}\left(\mathbf{y}-\mathbf{z}\right)-T_{Y}\left(\mathbf{x}-\mathbf{y}\right)\right).\end{aligned}
\end{equation}

With the BFKL equation as an example, we showed that gauge invariance
is essential if we want the evolution equation to be free of infrared
divergences. 

To extend this claim beyond the dilute limit, we should specify a
bit what we mean with a gauge invariant operator in the context of
the CGC. Since the CGC is formulated in the light-cone gauge, with
the help of the classical field solutions in the covariant gauge,
an operator is gauge invariant if it can be written as a function
of Wilson loops in the covariant gauge, i.e.:
\begin{equation}
\begin{aligned}\mathcal{O}\left[\alpha\right] & =\mathrm{Tr}\left(U_{\mathbf{x}}^{\dagger}U_{\mathbf{y}}W_{\mathbf{v}}^{\dagger}W_{\mathbf{w}}...\right),\\
U\left(\mathbf{x}\right) & \equiv\mathcal{P}e^{ig\int\mathrm{d}z^{-}\alpha_{a}\left(z^{-},\mathbf{x}\right)t^{a}},\\
W\left(\mathbf{x}\right) & \equiv\mathcal{P}e^{ig\int\mathrm{d}z^{-}\alpha_{a}\left(z^{-},\mathbf{x}\right)T^{a}},
\end{aligned}
\end{equation}
where the classical gauge fields thus have the structure: $A_{a}^{\mu}\left(x\right)=\delta^{\mu+}\alpha_{a}\left(x\right)$.
To quantify this requirement on a generic operator $\mathcal{O}\left[\alpha\right]$,
one can make use of the observation that this specific structure of
the gauge fields is preserved under residual gauge transformations
$\Omega\left(x^{-}\right)=\exp\left(ig_{s}\omega^{a}\left(x^{-}\right)t^{a}\right)$:
\begin{equation}
\alpha_{a}\left(x^{-},\mathbf{x}\right)\to\Omega\left(x^{-}\right)\left(\alpha_{a}\left(x^{-},\mathbf{x}\right)+\frac{i}{g_{s}}\partial^{+}\right)\Omega^{\dagger}\left(x^{-}\right).
\end{equation}
An operator $\mathcal{O}\left[\alpha\right]$ is therefore gauge invariant
if it does not change under the above gauge transformation, which
for a Wilson line amounts to:
\begin{equation}
U^{\dagger}\left(\mathbf{x}\right)\to\Omega\left(x^{-}=+\infty\right)U^{\dagger}\left(\mathbf{x}\right)\Omega^{\dagger}\left(x^{-}=-\infty\right),
\end{equation}
or
\begin{equation}
U^{\dagger}\left(\mathbf{x}\right)\to\Omega_{L}U^{\dagger}\left(\mathbf{x}\right)\Omega_{R}^{\dagger}.
\end{equation}
It is easy to show that these independent and global (since they take
place beyond the longitudinal support of the nucleus) color rotations,
are generated by the following operators:
\begin{equation}
\begin{aligned}\mathcal{G}_{L} & \equiv\int\mathrm{d}^{2}\mathbf{x}\frac{\delta}{\delta\alpha_{Y}^{a}\left(\mathbf{x}\right)},\quad\mathcal{G}_{R}\equiv-\int\mathrm{d}^{2}\mathbf{x}W_{\mathbf{x}}^{ab}\frac{\delta}{\delta\alpha_{Y}^{b}\left(\mathbf{x}\right)}.\end{aligned}
\end{equation}
As an example, let us show that an adjoint dipole is gauge invariant,
according to the above criteria. Indeed, we have:
\begin{equation}
\begin{aligned}\mathcal{G}_{L}D_{A}\left(\mathbf{x}-\mathbf{y}\right) & =\frac{1}{N_{c}^{2}-1}\int\mathrm{d}^{2}\mathbf{z}\frac{\delta}{\delta\alpha_{Y}^{a}\left(\mathbf{z}\right)}\mathrm{Tr}\Bigl\langle W\left(\mathbf{x}\right)W^{\dagger}\left(\mathbf{y}\right)\Bigr\rangle_{Y},\\
 & =\frac{1}{N_{c}^{2}-1}\int\mathrm{d}^{2}\mathbf{z}\mathrm{Tr}\Bigl\langle ig_{s}\delta_{\mathbf{x}\mathbf{z}}W\left(\mathbf{x}\right)T^{a}W^{\dagger}\left(\mathbf{y}\right)-ig_{s}\delta_{\mathbf{y}\mathbf{z}}W\left(\mathbf{x}\right)T^{a}W^{\dagger}\left(\mathbf{y}\right)\Bigr\rangle_{Y},\\
 & =0,
\end{aligned}
\end{equation}
and
\begin{equation}
\begin{aligned} & \mathcal{G}_{R}D_{A}\left(\mathbf{x}-\mathbf{y}\right)=\frac{-1}{N_{c}^{2}-1}\int\mathrm{d}^{2}\mathbf{z}W_{\mathbf{z}}^{ab}\frac{\delta}{\delta\alpha_{Y}^{b}\left(\mathbf{z}\right)}\mathrm{Tr}\Bigl\langle W\left(\mathbf{x}\right)W^{\dagger}\left(\mathbf{y}\right)\Bigr\rangle_{Y},\\
 & =\frac{-1}{N_{c}^{2}-1}\int\mathrm{d}^{2}\mathbf{z}\mathrm{Tr}\Bigl\langle ig_{s}\delta_{\mathbf{x}\mathbf{z}}W\left(\mathbf{x}\right)W_{\mathbf{z}}^{ab}T^{b}W^{\dagger}\left(\mathbf{y}\right)-ig_{s}\delta_{\mathbf{y}\mathbf{z}}W\left(\mathbf{x}\right)W_{\mathbf{z}}^{ab}T^{b}W^{\dagger}\left(\mathbf{y}\right)\Bigr\rangle_{Y},\\
 & =\frac{-ig_{s}}{N_{c}^{2}-1}\int\mathrm{d}^{2}\mathbf{z}\mathrm{Tr}\Bigl\langle\delta_{\mathbf{x}\mathbf{z}}W\left(\mathbf{x}\right)W^{\dagger}\left(\mathbf{z}\right)T^{a}W\left(\mathbf{z}\right)W^{\dagger}\left(\mathbf{y}\right)\\
 & -\delta_{\mathbf{y}\mathbf{z}}W\left(\mathbf{x}\right)W^{\dagger}\left(\mathbf{z}\right)T^{a}W\left(\mathbf{z}\right)W^{\dagger}\left(\mathbf{y}\right)\Bigr\rangle_{Y},\\
 & =\frac{-ig_{s}}{N_{c}^{2}-1}\mathrm{Tr}\Bigl\langle T^{a}W\left(\mathbf{x}\right)W^{\dagger}\left(\mathbf{y}\right)-W\left(\mathbf{x}\right)W^{\dagger}\left(\mathbf{y}\right)T^{a}\Bigr\rangle_{Y}=0.
\end{aligned}
\end{equation}
We can use this to rewrite the JIMWLK equation in a simpler dipole
form, which is explicitly infrared safe. To do this, observe that
the relation between the transverse kernel and the dipole kernel,
Eq. (\ref{eq:M2K}), can be inverted:
\begin{equation}
\mathcal{K}_{\mathbf{xyz}}=\frac{1}{2}\left(\frac{1}{\left(\mathbf{x}-\mathbf{z}\right)^{2}}+\frac{1}{\left(\mathbf{y}-\mathbf{z}\right)^{2}}-\frac{\left(\mathbf{x}-\mathbf{y}\right)^{2}}{\left(\mathbf{x}-\mathbf{z}\right)^{2}\left(\mathbf{y}-\mathbf{z}\right)^{2}}\right).
\end{equation}
Furthermore, one can show (see Ref. \protect\cite{Hatta2005}) that for gauge
invariant observables, the contributions of the first two terms in
the above equation, which are potentially divergent, disappear. One
thus obtains:
\begin{equation}
\begin{aligned}H_{\mathrm{JIMWLK}} & \equiv\frac{1}{16\pi^{3}}\int_{\mathbf{x}\mathbf{y}\mathbf{z}}\mathcal{M}_{\mathbf{xyz}}\left(1+W_{\mathbf{x}}^{\dagger}W_{\mathbf{y}}-W_{\mathbf{x}}^{\dagger}W_{\mathbf{z}}-W_{\mathbf{z}}^{\dagger}W_{\mathbf{y}}\right)^{ab}\frac{\delta}{\delta\alpha_{Y}^{a}\left(\mathbf{x}\right)}\frac{\delta}{\delta\alpha_{Y}^{b}\left(\mathbf{y}\right)}.\end{aligned}
\label{eq:JIMWLKDP}
\end{equation}
Note that in the above dipole form, one can freely change around the
functional derivative with respect to $\alpha_{Y}^{a}\left(\mathbf{x}\right)$,
since:
\begin{equation}
\begin{aligned} & \frac{\delta}{\delta\alpha_{Y}^{a}\left(\mathbf{x}\right)}\left(1+W_{\mathbf{x}}^{\dagger}W_{\mathbf{y}}-W_{\mathbf{x}}^{\dagger}W_{\mathbf{z}}-W_{\mathbf{z}}^{\dagger}W_{\mathbf{y}}\right)^{ab}\\
 & =ig_{s}\left(\delta_{\mathbf{xy}}W_{\mathbf{x}}^{\dagger}W_{\mathbf{y}}T^{a}-\delta_{\mathbf{zx}}W_{\mathbf{x}}^{\dagger}W_{\mathbf{z}}T^{a}-\delta_{\mathbf{xy}}W_{\mathbf{z}}^{\dagger}W_{\mathbf{y}}T^{a}\right)^{ab},\\
 & =-ig_{s}\delta_{\mathbf{xy}}\left(W_{\mathbf{z}}^{\dagger}W_{\mathbf{y}}T^{a}\right)^{ab},
\end{aligned}
\end{equation}
which disappears in combination with the dipole kernel $\mathcal{M}_{\mathbf{xyz}}$.

To conclude this chapter, let us show that the JIMWLK equation is
equal to the Balitsky hierarchy of equations, when one choses to evolve
a dipole in the fundamental representation. We have that:
\begin{equation}
\begin{aligned} & \frac{\delta}{\delta\alpha_{Y}^{a}\left(\mathbf{x}\right)}\frac{\delta}{\delta\alpha_{Y}^{b}\left(\mathbf{y}\right)}D\left(\mathbf{v}-\mathbf{w}\right)=\frac{1}{N_{c}}\frac{\delta}{\delta\alpha_{Y}^{a}\left(\mathbf{x}\right)}\frac{\delta}{\delta\alpha_{Y}^{b}\left(\mathbf{y}\right)}\mathrm{Tr}\Bigl\langle U^{\dagger}\left(\mathbf{v}\right)U\left(\mathbf{w}\right)\Bigr\rangle_{Y}\\
 & =\frac{ig_{s}}{N_{c}}\frac{\delta}{\delta\alpha_{Y}^{a}\left(\mathbf{x}\right)}\mathrm{Tr}\Bigl\langle-t^{b}\delta_{\mathbf{vy}}U^{\dagger}\left(\mathbf{v}\right)U\left(\mathbf{w}\right)+\delta_{\mathbf{wy}}t^{b}U^{\dagger}\left(\mathbf{v}\right)U\left(\mathbf{w}\right)\Bigr\rangle_{Y},\\
 & =\frac{\left(ig_{s}\right)^{2}}{N_{c}}\mathrm{Tr}\Bigl\langle\Bigl(\delta_{\mathbf{vy}}\delta_{\mathbf{vx}}t^{b}t^{a}-\delta_{\mathbf{xw}}\delta_{\mathbf{vy}}t^{a}t^{b}-\delta_{\mathbf{vx}}\delta_{\mathbf{wy}}t^{b}t^{a}+\delta_{\mathbf{wx}}\delta_{\mathbf{wy}}t^{a}t^{b}\Bigr)U^{\dagger}\left(\mathbf{v}\right)U\left(\mathbf{w}\right)\Bigr\rangle_{Y},\\
 & =-\frac{g_{s}^{2}}{N_{c}}\left(\delta_{\mathbf{vy}}-\delta_{\mathbf{wy}}\right)\left(\delta_{\mathbf{vx}}\mathrm{Tr}\Bigl\langle t^{b}t^{a}U_{\mathbf{v}}^{\dagger}U_{\mathbf{w}}\Bigr\rangle_{Y}-\delta_{\mathbf{xw}}\mathrm{Tr}\Bigl\langle t^{a}t^{b}U_{\mathbf{v}}^{\dagger}U_{\mathbf{w}}\Bigr\rangle_{Y}\right).
\end{aligned}
\label{eq:BKttst0}
\end{equation}
Together with the dipole kernel, the contributions $\delta_{\mathbf{vy}}\delta_{\mathbf{vx}}$
and $\delta_{\mathbf{wy}}\delta_{\mathbf{wx}}$ disappear. The other
two terms both yield the same kernel $\mathcal{M}_{\mathbf{vwz}}$,
and their combination with the Wilson line structure can be calculated
as follows: 
\begin{equation}
\begin{aligned} & \left(1+W_{\mathbf{x}}^{\dagger}W_{\mathbf{y}}-W_{\mathbf{x}}^{\dagger}W_{\mathbf{z}}-W_{\mathbf{z}}^{\dagger}W_{\mathbf{y}}\right)^{ab}\delta_{\mathbf{wy}}\delta_{\mathbf{vx}}\mathrm{Tr}\Bigl\langle t^{b}t^{a}U_{\mathbf{v}}^{\dagger}U_{\mathbf{w}}\Bigr\rangle_{Y}\\
 & =\left(1+W_{\mathbf{v}}^{\dagger}W_{\mathbf{w}}-W_{\mathbf{v}}^{\dagger}W_{\mathbf{z}}-W_{\mathbf{z}}^{\dagger}W_{\mathbf{w}}\right)^{ab}\mathrm{Tr}\Bigl\langle t^{b}t^{a}U_{\mathbf{v}}^{\dagger}U_{\mathbf{w}}\Bigr\rangle_{Y},\\
 & =\biggl(\mathrm{Tr}\Bigl\langle t^{a}t^{a}U_{\mathbf{v}}^{\dagger}U_{\mathbf{w}}\Bigr\rangle_{Y}+\mathrm{Tr}\Bigl\langle W_{\mathbf{w}}^{cb}t^{b}W_{\mathbf{v}}^{ca}t^{a}U_{\mathbf{v}}^{\dagger}U_{\mathbf{w}}\Bigr\rangle_{Y}\\
 & -\mathrm{Tr}\Bigl\langle W_{\mathbf{z}}^{cb}t^{b}W_{\mathbf{v}}^{ca}t^{a}U_{\mathbf{v}}^{\dagger}U_{\mathbf{w}}\Bigr\rangle_{Y}-\mathrm{Tr}\Bigl\langle W_{\mathbf{w}}^{cb}t^{b}W_{\mathbf{z}}^{ca}t^{a}U_{\mathbf{v}}^{\dagger}U_{\mathbf{w}}\Bigr\rangle_{Y}\biggr).
\end{aligned}
\label{eq:BKttst}
\end{equation}
With the help of the Lie algebra identities Eqs. (\ref{eq:U2W}) and
(\ref{eq:Fierz}), we find that:
\begin{equation}
\begin{aligned}\mathrm{Tr}\Bigl\langle t^{a}t^{a}U_{\mathbf{v}}^{\dagger}U_{\mathbf{w}}\Bigr\rangle_{Y} & =\Bigl\langle t_{ij}^{a}t_{jk}^{a}U_{\mathbf{v}}^{\dagger kl}U_{\mathbf{w}}^{li}\Bigr\rangle_{Y},\\
 & =\frac{1}{2}\delta_{jj}\Bigl\langle U_{\mathbf{v}}^{\dagger il}U_{\mathbf{w}}^{li}\Bigr\rangle_{Y}-\frac{1}{2N_{c}}\Bigl\langle U_{\mathbf{v}}^{\dagger jl}U_{\mathbf{w}}^{lj}\Bigr\rangle_{Y},\\
 & =C_{F}\mathrm{Tr}\Bigl\langle U_{\mathbf{v}}^{\dagger}U_{\mathbf{w}}\Bigr\rangle_{Y},
\end{aligned}
\end{equation}
and:
\begin{equation}
\begin{aligned}\mathrm{Tr}\Bigl\langle W_{\mathbf{z}}^{cb}t^{b}W_{\mathbf{v}}^{ca}t^{a}U_{\mathbf{v}}^{\dagger}U_{\mathbf{w}}\Bigr\rangle_{Y} & =\mathrm{Tr}\Bigl\langle U_{\mathbf{z}}^{\dagger}t^{c}U_{\mathbf{z}}U_{\mathbf{v}}^{\dagger}t^{c}U_{\mathbf{w}}\Bigr\rangle_{Y},\\
 & =\frac{1}{2}\mathrm{Tr}\Bigl\langle U_{\mathbf{z}}^{\dagger}U_{\mathbf{w}}\Bigr\rangle_{Y}\mathrm{Tr}\Bigl\langle U_{\mathbf{v}}U_{\mathbf{z}}^{\dagger}\Bigr\rangle_{Y}-\frac{1}{2N_{c}}\mathrm{Tr}\Bigl\langle U_{\mathbf{v}}^{\dagger}U_{\mathbf{w}}\Bigr\rangle_{Y}.
\end{aligned}
\end{equation}
Similarly, one obtains:
\begin{equation}
\begin{aligned}\mathrm{Tr}\Bigl\langle W_{\mathbf{w}}^{cb}t^{b}W_{\mathbf{v}}^{ca}t^{a}U_{\mathbf{v}}^{\dagger}U_{\mathbf{w}}\Bigr\rangle_{Y} & =C_{F}\mathrm{Tr}\Bigl\langle U_{\mathbf{v}}^{\dagger}U_{\mathbf{w}}\Bigr\rangle_{Y},\end{aligned}
\end{equation}
and
\begin{equation}
\begin{aligned}\mathrm{Tr}\Bigl\langle W_{\mathbf{w}}^{cb}t^{b}W_{\mathbf{z}}^{ca}t^{a}U_{\mathbf{v}}^{\dagger}U_{\mathbf{w}}\Bigr\rangle_{Y} & =\frac{1}{2}\mathrm{Tr}\Bigl\langle U_{\mathbf{v}}^{\dagger}U_{\mathbf{z}}\Bigr\rangle_{Y}\mathrm{Tr}\Bigl\langle U_{\mathbf{w}}U_{\mathbf{z}}^{\dagger}\Bigr\rangle_{Y}-\frac{1}{2N_{c}}\mathrm{Tr}\Bigl\langle U_{\mathbf{v}}^{\dagger}U_{\mathbf{w}}\Bigr\rangle_{Y},\end{aligned}
\end{equation}
such that Eq. (\ref{eq:BKttst}) becomes:
\begin{equation}
\begin{aligned} & \left(1+W_{\mathbf{x}}^{\dagger}W_{\mathbf{y}}-W_{\mathbf{x}}^{\dagger}W_{\mathbf{z}}-W_{\mathbf{z}}^{\dagger}W_{\mathbf{y}}\right)^{ab}\delta_{\mathbf{wy}}\delta_{\mathbf{vx}}\mathrm{Tr}\Bigl\langle t^{b}t^{a}U_{\mathbf{v}}^{\dagger}U_{\mathbf{w}}\Bigr\rangle_{Y}\\
 & =N_{c}^{2}\left(D\left(\mathbf{v}-\mathbf{w}\right)-\frac{1}{N_{c}^{2}}\Bigl\langle\mathrm{Tr}\left(U_{\mathbf{z}}^{\dagger}U_{\mathbf{w}}\right)\mathrm{Tr}\left(U_{\mathbf{v}}U_{\mathbf{z}}^{\dagger}\right)\Bigr\rangle_{Y}\right).
\end{aligned}
\end{equation}
The calculation for the $\delta_{\mathbf{xw}}\delta_{\mathbf{vy}}$
term in Eq. (\ref{eq:BKttst0}) is completely analogous, and yields
exactly the same result as above. Since the dipole kernel $\mathcal{M}_{\mathbf{xyz}}$
is symmetric in $\mathbf{x}$ and \textbf{$\mathbf{y}$}, we end up
with: 
\begin{equation}
\begin{aligned}\frac{\partial}{\partial Y}D_{Y}\left(\mathbf{x}-\mathbf{y}\right) & =-\frac{\bar{\alpha}}{2\pi^{2}}\mathcal{M}_{\mathbf{xyz}}\int_{\mathbf{z}}\left(D_{Y}\left(\mathbf{x}-\mathbf{y}\right)-\frac{1}{N_{c}^{2}}\Bigl\langle\mathrm{Tr}\left(U_{\mathbf{x}}U_{\mathbf{z}}^{\dagger}\right)\mathrm{Tr}\left(U_{\mathbf{z}}^{\dagger}U_{\mathbf{y}}\right)\Bigr\rangle_{Y}\right),\end{aligned}
\label{eq:BK}
\end{equation}
which we recognize as the first equation of the Balitsky hierarchy,
Eq. (\ref{eq:Balitksy}). From the CGC theory for the saturated target,
we thus reproduced the Balitsky-Kovchegov and BFKL equations which
we derived in Sec. \ref{sec:BKBFKL} from the point of view of the
projectile.

\newpage{}

\thispagestyle{simple}

\part{\label{part:projectCyrille}Role of gluon polarization in forward
heavy quark production in proton-nucleus collisions}

\section{Introduction}

One of the most important concepts in QCD is factorization (see \protect\cite{CSS1989,cteq,collins}
for some standard references). Due to the fact that the running coupling
$\alpha_{s}\left(Q^{2}\right)$ is large for small momentum transfers
$Q^{2}$, many of the ingredients of a QCD cross section cannot be
calculated in perturbation theory. Factorization provides a consistent
framework to separate the perturbatively calculable \textquoteleft hard
part' of the cross section, from the nonperturbative or \textquoteleft soft
part'. In particular, most of the information on the hadron structure
is part of this nonperturbative domain.

A famous example is deep-inelastic scattering, which we already studied
in detail in Part \ref{part:DIS}. The leading-order cross section
is given by (see Eqs. (\ref{eq:photonprotoncrosssection}), (\ref{eq:partonmodel})):
\begin{equation}
\begin{aligned}\sigma_{\gamma^{*}p} & =\underset{\sigma_{\mathrm{hard}}\left(x,Q^{2}\right)}{\underbrace{\frac{4\pi^{2}\alpha_{\mathrm{em}}}{Q^{2}}\sum_{f}e_{q_{f}}^{2}}}\underset{\mathrm{DGLAP}\left(\mu^{2}\to Q^{2}\right)\otimes q_{f}\left(x,\mu^{2}\right)}{\underbrace{\biggl(xq_{f}\left(x,Q^{2}\right)+x\bar{q}_{f}\left(x,Q^{2}\right)\biggr)}},\end{aligned}
\end{equation}
in which the perturbatively calculable partonic cross section $\sigma_{\mathrm{hard}}\left(x,Q^{2}\right)$
is separated from the nonperturbative PDFs. The DGLAP evolution equations,
also perturbative, can be viewed as the connection between both, bringing
the PDFs, measured at a certain (perturbative) scale $\mu^{2}$, up
to the \textquoteleft hard scale' $Q^{2}$, which has to be large
enough for the coupling $\alpha_{s}\left(Q^{2}\right)$ to be small.
Schematically:
\begin{equation}
\sigma_{\mathrm{coll}}\left(x,Q^{2}\right)=\sigma_{\mathrm{hard}}\left(x,Q^{2}\right)\otimes\mathrm{DGLAP}\left(\mu^{2}\to Q^{2}\right)\otimes\mathrm{PDFs}\left(x,\mu^{2}\right).\label{eq:DISfactorization}
\end{equation}
This is an example of what is known as collinear factorization (Ref.
\protect\cite{Collins1988}). A central feature of this framework is the property
that the PDFs are universal. This means that, once they are measured
in a certain experiment, the PDFs can be plugged into any other cross
section in collinear factorization, and evolved with DGLAP to the
desired hard scale $Q^{2}$. 

It is well known that, for less inclusive processes characterized
by a second transverse scale, collinear factorization breaks down.
An important example is the differential cross section of the Drell-Yan
process $pp\rightarrow\gamma^{*}\rightarrow ll$, in which both the
invariant mass $Q^{2}$ of the lepton pair, and its total transverse
momentum $q_{\perp}^{2}$ play a role. As long as $q_{\perp}^{2}\sim Q^{2}$,
a collinear factorization formula reminiscent of Eq. (\ref{eq:DISfactorization})
holds (Ref. \protect\cite{CSSDY,Qiu}):
\begin{equation}
\frac{\mathrm{d}\sigma_{\mathrm{DY}}}{\mathrm{d}^{4}q}=\sum_{f}\int\mathrm{d}x_{A}\int\mathrm{d}x_{B}\,q_{f/A}\left(x_{A},Q^{2}\right)\bar{q}_{\bar{f}/B}\left(x_{B},Q^{2}\right)\,\hat{\sigma}_{0},
\end{equation}
where $q_{f/A}\left(x_{A},Q^{2}\right)$ and $\bar{q}_{\bar{f}/B}\left(x_{B},Q^{2}\right)$
are the (anti)quark PDFs of the two protons, and where $\hat{\sigma}_{0}$
is the partonic cross section for the process $q\bar{q}\to\gamma^{*}$.
However, when $q_{\perp}^{2}\ll Q^{2}$, large corrections from initial-state
radiation become important and the above expression diverges as $q_{\perp}\to0$.
It should be replaced by (Ref. \protect\cite{CSSDY,Qiu}):
\begin{equation}
\begin{aligned}\frac{\mathrm{d}\sigma_{\mathrm{DY}}}{\mathrm{d}^{4}q} & =\sum_{f}\int\mathrm{d}^{2}\mathbf{k}_{\perp A}\mathrm{d}^{2}\mathbf{k}_{\perp B}\delta^{\left(2\right)}\left(\mathbf{k}_{\perp A}+\mathbf{k}_{\perp B}-\mathbf{q}_{\perp}\right)\\
 & \times\mathcal{F}_{f/A}\left(x_{A},k_{\perp A}\right)\mathcal{F}_{\bar{f}/B}\left(x_{B},k_{\perp B}\right)\,\hat{\sigma}_{0}.
\end{aligned}
\end{equation}
The above formula is an example of transverse-momentum dependent (TMD)
factorization, in which one works with transverse-momentum dependent
PDFs $\mathcal{F}_{f/A}\left(x_{A},k_{\perp A}\right)$ (TMDs for
short) which, just like the unintegrated PDFs, contain a three-dimensional
picture of the hadron by including the $k_{\perp}$-dependence of
the parton (for some contemporary reviews, see Refs. \protect\cite{collins,REF2015,Rogers2016,Signori2016,Boer2017}).
For example, TMDs can provide detailed information on the correlation
between transverse momentum and spin, thus explaining many phenomena
in which the spin degree of freedom is taken into account.

Though widely used, and successful in describing many data, TMD factorization
suffers from several difficulties. First, it is very hard to establish
formal factorization theorems that are valid to all orders. Such theorems
nowadays only exist for the Drell-Yan process we mentioned above (Refs.
\protect\cite{CSSDY,Ji2004,Echevarria2012}), for semi-inclusive deep-inelastic
scattering (Ref. \protect\cite{Ji2005}), and for $e^{+}e^{-}$ annihilation
into two hadrons (see Ref. \protect\cite{collins}). For many other processes,
one assumes TMD factorization and works within this framework, even
when a formal proof is not available. 

Moreover, TMD parton distributions are more complicated than the usual,
collinear PDFs, in that they are process dependent rather than universal
(Refs. \protect\cite{CSS1983,Boer2000,Mulders2001,Brodsky2002bis,Collins2002,Belitsky2003}).
Indeed, in Sec. \ref{sec:GluonTMDs} we studied the example of the
unintegrated gluon distribution, and argued that many different gluon
TMDs arise, depending on the precise Wilson-line configuration that
is used to render the operator structure gauge invariant. In general,
the physical process to which the TMD factorization is applied  determines
the appropriate Wilson-line structure. A well-known example is the
Sivers function (Ref. \protect\cite{Sivers}), which is the TMD that describes
the azimuthal distribution of unpolarized partons inside a nucleon
that is transversely polarized with respect to its direction of motion.
The Sivers function can be probed in the study of single spin asymmetries
(SSAs, see e.g. Ref. \protect\cite{Anselmino,dalesio2007}) in polarized $p^{\uparrow}p$
collisions, defined as the ratio: 
\begin{equation}
\begin{aligned}A_{N} & \equiv\frac{\mathrm{d}\sigma^{\uparrow}-\mathrm{d}\sigma^{\downarrow}}{\mathrm{d}\sigma^{\uparrow}+\mathrm{d}\sigma^{\downarrow}},\end{aligned}
\end{equation}
where the arrow indicates the polarization of the proton, which is
transverse w.r.t. the beam line. It is widely predicted (see e.g.
Refs. \protect\cite{Collins2002,Brodsky2002,Boer2003,Belitsky2003,SSA2009})
that the quark Sivers function probed in the SSA in the Drell-Yan
process has the opposite sign of the one probed in SIDIS. This sign
difference, and hence process-dependence, can be explained as the
difference between the Wilson-line configuration that takes the effect
of the initial-state interactions in the Drell-Yan process into account,
and the one that encodes the final-state interactions in SIDIS, resulting
in two different quark Sivers TMDs.

Processes that are sensitive to TMDs are in general difficult to measure,
and as a consequence there are relatively few data on TMDs available
(see Ref. \protect\cite{Bachetta2017} for a recent extraction of unpolarized
quark TMDs.) Data on gluon TMDs are even more scarce due to the high
energies needed to access them. 

In the small-$x$ limit and at large enough transverse momentum ($Q^{2}\gg Q_{s}^{2}$),
there is yet another factorization scheme known as $k_{\perp}$-factorization
or high-energy factorization (HEF) (Ref. \protect\cite{Catani1990,Catani1991,Catani1993,Catani1994,Forshaw1997}).
The unintegrated PDF $f\left(x,k_{\perp}\right)$ can, with the help
of BFKL, be evolved in rapidity to match the hard parts of the cross
section, which are commonly called impact factors. Since the gluons
from the unintegrated PDF carry transverse momentum and are in general
off shell, these impact factors are calculated with off-shell incoming
partons. This is in contrast to TMD factorization, in which the incoming
virtualities in the matrix elements are neglected and all the information
on the transverse kinematics is solely encoded in the Wilson lines
(Ref. \protect\cite{collins}). 

In this part of the thesis, we study forward dijet production in proton-nucleus
collisions (Ref. \protect\cite{Kotko2015}), a process which is sensitive
to the large-$x$ content of the proton, and to the small-$x$ content
of the nucleus. This process is characterized by three momentum scales:
$\tilde{P}_{\perp}$, the typical transverse momentum of one of the
jets and always the largest scale; $q_{\perp}$, the transverse-momentum
imbalance between both jets, which is a measure of the $k_{\perp}$
of the gluons coming from the target; and the saturation scale $Q_{s}$
of the nucleus which is always the softest scale. The value of $q_{\perp}$
with respect to $Q_{s}$ and $\tilde{P}_{\perp}$ governs which factorization
scheme is valid. Indeed, when $q_{\perp}\sim Q_{s}\ll\tilde{P}_{\perp}$,
there are effectively two strongly ordered scales $q_{\perp}$ and
$\tilde{P}_{\perp}$ in the problem and TMD factorization applies
(Refs. \protect\cite{Boer2000,Boer2003,Belitsky2003,Bomhof2006,Collins2007,Vogelsang2007,Rogers2010,Xiao2010}).
In the other regime: $Q_{s}\ll q_{\perp}\sim\tilde{P}_{\perp}$, $q_{\perp}$
and $\tilde{P}_{\perp}$ are of the same order and far above the saturation
scale, hence high-energy factorization is applicable (Refs. \protect\cite{Deak2009,KutakSapeta,vanHameren2013,vanHameren,vanHameren2014}).
In both cases, the large-$x$ parton coming from the proton is described
by an integrated PDF, just like in collinear factorization.

However, since forward dijet production in $pA$ collisions is sensitive
to small-$x$ dynamics, it can also be computed in the CGC formalism
(Refs. \protect\cite{zhou,Cyrille,vanHameren2014,Kotko2015,vanHameren2016}),
again using a hybrid approach (see Refs. \protect\cite{Dumitru2005,Altinoluk2011,Chirilli2012})
in which the parton from the proton is described by an integrated
PDF. Since the CGC provides a weakly-coupled and hence perturbatively
calculable theory for a proton or nucleus in the high-energy limit,
one can match the result with the one from the TMD point of view,
such that one obtains analytical or numerical expressions for the
\textendash in principle\textendash{} nonperturbative TMDs from the
nucleus. One can then exploit the full CGC machinery to compute these
small-$x$ TMDs explicitly within the McLerran-Venugopalan model,
and evolve them in rapidity with the help of the JIMWLK equation.
This research program was developed in recent years, and indeed, both
the TMD and the HEF (taking the limit $Q_{s}\ll q_{\perp}\sim\tilde{P}_{\perp}$)
results can be reproduced from the CGC calculation, see for instance
Refs. \protect\cite{Dominguez2009,Dominguez2011,fabio,Metz2011,fabioqiu,zhou,Laidet2013,Kotko2015,vanHameren2016,Cyrille}.
We should mention that it is also possible to generalize the TMD factorization
formula for dijet production by keeping the small-$x$ gluon in the
matrix elements off shell. This approach, dubbed the \textquoteleft Improved
TMD factorization' (Refs. \protect\cite{Kotko2015,vanHameren2016}), allows
one to interpolate between the HEF and the TMD frameworks without
resorting to the CGC, and has the advantage that it is more applicable
to phenomenology. 

Specifically, in this part we build further on the work done in Ref.
\protect\cite{Cyrille} by studying the forward production of two heavy quarks,
performing the CGC calculation while keeping track of the masses.
As already observed earlier for the case of $ep\rightarrow q\bar{q}$
(see for instance Ref. \protect\cite{Mulders2001,Boer2011,Metz2011,fabioqiu}),
by adding masses the cross section becomes sensitive to additional
TMDs which describe the linearly polarized gluon content of the unpolarized
proton, or in our case, nucleus. In our calculation, this means that
the three gluon TMDs that describe the gluon channel $gA\rightarrow q\bar{q}$
will be accompanied by three \textquoteleft polarized' partners. We
compute the gluon TMDs analytically in both the MV and the GBW model,
after which they are numerically evaluated and evolved in rapidity
with the help of a lattice implementation of the JIMWLK equation.
Finally, we compare our results with the existing literature.

Before starting with the calculation, let us discuss the relevant
kinematics. As already explained, we study forward dijet or heavy
quark production in $pA$ collisions. We choose the forward case because
this implies that the partons from the proton have a generic Bjorken-$x$
of the order $x_{p}\sim1$, while on the nucleus side mainly gluons
with $x\ll1$ are involved. We therefore expect the transverse momentum
of the parton coming from the proton to be of the order of $\Lambda_{\mathrm{QCD}}$,
which is much smaller than the average transverse momentum of the
gluons in the target, which should be around the saturation scale
$Q_{s}$. Furthermore, the requirement that the two outgoing particles
are almost back-to-back, provides us with the two strongly ordered
scales required by TMD factorization. Indeed, while the momenta $k_{1\perp}\sim k_{2\perp}\sim\tilde{P}_{\perp}$
of the individual jets can be fairly large, the total transverse momentum
$\mathbf{q}_{\perp}=\mathbf{k}_{1\perp}+\mathbf{k}_{2\perp}$ of the
dijet system is small. Moreover, since the imbalance between the jets
is caused by the interaction with the nucleus, we expect the total
transverse momentum to be of the order of the saturation scale: $q_{\perp}\sim Q_{s}$.
To summarize, we have parametrically:
\begin{equation}
\begin{aligned}\Lambda_{\mathrm{QCD}}^{2} & \ll Q_{s}^{2}\ll\tilde{P}_{\perp}^{2}\ll\hat{s},\\
0.04\,\mathrm{GeV}^{2} & \ll4\,\mathrm{GeV}^{2}\ll400\,\mathrm{GeV^{2}}\ll40\,\mathrm{TeV}^{2},
\end{aligned}
\end{equation}
and we reach values of Bjorken-$x$ (of the nucleus) down to approximately:
\begin{equation}
x\simeq\frac{\tilde{P}_{\perp}^{2}}{\hat{s}}\sim10^{-5}.
\end{equation}

\section{Kinematics}

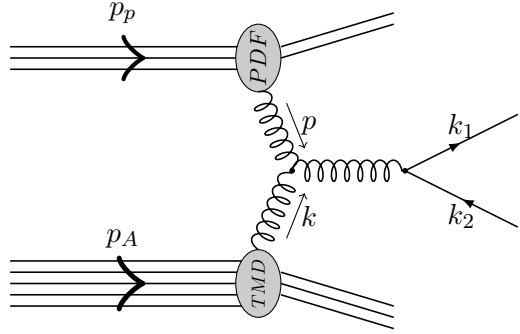
\begin{figure}[t]
\begin{centering}
\begin{tikzpicture}[scale=1.5] 

\tikzset{photon/.style={decorate,decoration={snake}},
		electron/.style={ postaction={decorate},decoration={markings,mark=at position .5 with {\arrow[draw]{latex}}}},      	gluon/.style={decorate,decoration={coil,amplitude=4pt, segment length=5pt}}}

\draw[semithick] (0,0) --++ (2.05,0);
\draw[semithick,postaction={decorate},decoration={markings,mark=at position .6 with {\arrow[thin,scale=5]{to}}}] (0,-.1) --++ (2,0);
\draw[semithick] (0,-.2) --++ (2.05,0);
\node at (1,.3) {$p_p$};
\filldraw [fill=black!20] (2.2,-.1) ellipse (.2 and .3);
\node [rotate=90,scale=.8] at (2.2,-.1) {$PDF$};
\draw[semithick] (2.4,0) --++ (1,.3);
\draw[semithick] (2.4,-.1) --++ (1,.3);

\draw[semithick] (0,-1.9) --++ (2.1,0);
\draw[semithick] (0,-2) --++ (2.05,0);
\draw[semithick,postaction={decorate},decoration={markings,mark=at position .6 with {\arrow[thin,scale=6]{to}}}] (0,-2.1) --++ (2,0);
\draw[semithick] (0,-2.2) --++ (2.05,0);
\draw[semithick] (0,-2.3) --++ (2.1,0);
\node at (1,-1.7) {$p_A$};
\filldraw [fill=black!20] (2.2,-2.1) ellipse (.2 and .3);
\node [rotate=90,scale=.65] at (2.2,-2.1) {$TMD$};
\draw[semithick] (2.4,-2) --++ (1,-.3);
\draw[semithick] (2.4,-2.1) --++ (1,-.3);
\draw[semithick] (2.4,-2.2) --++ (1,-.3);

\draw[gluon,semithick] (2.2,-.4) --++ (.3,-.7);
\draw[->] (2.45,-.5) --++(.16,-.4);
\node at (2.65,-.7) {$p$};
\draw[gluon,semithick] (2.2,-1.8) --++ (.3,.7);
\draw[->] (2.45,-1.7) --++(.16,.4);
\node at (2.65,-1.5) {$k$};
\filldraw[black] (2.5,-1.1) circle(.02);
\draw[gluon,semithick] (2.5,-1.1) --++ (1,0);
\filldraw[black] (3.5,-1.1) circle(.02);
\draw[electron,semithick] (3.5,-1.1) --++ (1,.5);
\node at (4,-.7) {$k_1$};
\draw[electron,semithick] (4.5,-1.6) --++ (-1,.5); 
\node at (4,-1.5) {$k_2$};
\end{tikzpicture} 
\par\end{centering}
\caption{\label{fig:pA2qq}One of the leading order diagrams for inclusive
heavy dijet production in $pA$ collisions. }
\end{figure}

Let us study the kinematics of our process in some more detail, and
confirm the claims that we made above. Although we will compute the
$pA\to q\bar{q}X$ cross section in the CGC, it is arguably more  illuminating
to study the kinematics from the TMD point of view. At leading order,
the reaction is depicted in Fig. \ref{fig:pA2qq}. In light-cone coordinates,
the momenta of the proton and the nucleus are:
\begin{equation}
p_{p}=\sqrt{\frac{s}{2}}\left(1,0,\mathbf{0}\right),\qquad p_{A}=\sqrt{\frac{s}{2}}\left(0,1,\mathbf{0}\right),
\end{equation}
where $s=\left(p_{p}+p_{A}\right)^{2}$ is the center of mass energy
squared. On the level of the partons, we have $g\left(p\right)+g\left(k\right)\rightarrow q\left(k_{1}\right)+\bar{q}\left(k_{2}\right)$,
where $p$ is the momentum of the gluon coming from the proton, $k$
the momentum of the gluon from the target, and where $k_{1}$ and
$k_{2}$ are the momenta of the outgoing quark and antiquark. As already
stated in the introduction, we assume that the transverse momentum
$p_{\perp}\sim\Lambda_{\mathrm{QCD}}$ of the gluon from the proton
can be neglected with respect to the transverse momentum $k_{\perp}\sim Q_{s}$
of the gluon from the target. We have, in the eikonal approximation:
\begin{equation}
\begin{aligned}p & =\left(p^{+},0^{-},\mathbf{0}_{\perp}\right),\\
k & =\left(0^{+},k_{1}^{-}+k_{2}^{-},\mathbf{k}_{1\perp}+\mathbf{k}_{2\perp}\right),\\
k_{1} & =\left(k_{1}^{+},k_{1}^{-},\mathbf{k}_{1\perp}\right)=\left(k_{1}^{+},\frac{m^{2}+k_{1\perp}^{2}}{2k_{1}^{+}},\mathbf{k}_{1\perp}\right),\\
k_{2} & =\left(k_{2}^{+},k_{2}^{-},\mathbf{k}_{2\perp}\right)=\left(k_{2}^{+},\frac{m^{2}+k_{2\perp}^{2}}{2k_{2}^{+}},\mathbf{k}_{2\perp}\right).
\end{aligned}
\end{equation}
The Bjorken-$x$ of the gluon from the target is given by:
\begin{align}
x & \equiv\frac{k^{-}}{p_{A}^{-}}=\sqrt{\frac{2}{s}}\left(k_{1}^{-}+k_{2}^{-}\right)=\frac{1}{\sqrt{s}}\left(k_{1\perp}e^{-y_{1}}+k_{2\perp}e^{-y_{2}}\right),
\end{align}
which confirms our claim in the introduction, namely that in forward
dijet production, and therefore at large rapidities $y_{1}$ and $y_{2}$,
the small-$x$ gluon content of the nucleus is probed. The energy
fraction carried by the gluon from the proton is:
\begin{align}
x_{p} & \equiv\frac{p^{+}}{p_{p}^{+}}=\frac{1}{\sqrt{s}}\left(k_{1\perp}e^{y_{1}}+k_{2\perp}e^{y_{2}}\right),
\end{align}
hence we expect $x_{p}\sim1$.

For later reference, it is useful to compare with the three main parameters
that will turn up in the CGC cross section, namely: $z$, the fraction
of the energy of the gluon from the proton carried by the outgoing
jet or heavy quark, $\tilde{\mathbf{P}}_{\perp}$, which is approximately
equal to the transverse momentum of both of the jets, and $\mathbf{q}_{\perp}$,
the vector sum of the transverse momenta of the two jets. In formulas:
\begin{equation}
\begin{aligned}z & =\frac{k_{1}^{+}}{p^{+}},\qquad\tilde{\mathbf{P}}_{\perp}=\left(1-z\right)\mathbf{k}_{1\perp}-z\mathbf{k}_{2\perp},\qquad\mathbf{q}_{\perp}=\mathbf{k}_{1\perp}+\mathbf{k}_{2\perp}.\end{aligned}
\end{equation}
It is straightforward to show that in terms of the variables above,
the Mandelstam variables on the parton level:
\begin{equation}
\begin{aligned}\hat{s} & =\left(p+k\right)^{2}=\left(k_{1}+k_{2}\right)^{2},\\
\hat{t} & =\left(p-k_{1}\right)^{2}=\left(k-k_{2}\right)^{2},\\
\hat{u} & =\left(p-k_{2}\right)^{2}=\left(k-k_{1}\right)^{2},
\end{aligned}
\end{equation}
can be written as follows:

\begin{equation}
\begin{aligned}\hat{s} & =\frac{m^{2}+\tilde{P}_{\perp}^{2}}{z\left(1-z\right)},\quad\hat{t}=-\frac{\left(1-z\right)m^{2}+k_{1\perp}^{2}}{z},\quad\hat{u}=-\frac{zm^{2}+k_{2\perp}^{2}}{1-z},\end{aligned}
\label{eq:Mandelstam}
\end{equation}
with:
\begin{equation}
\hat{s}+\hat{t}+\hat{u}=2m^{2}-q_{\perp}^{2}.
\end{equation}

\section{Color Glass Condensate approach}

\subsection{\label{subsec:CGCcrosss}CGC calculation of the $gA\rightarrow q\bar{q}X$
cross section}

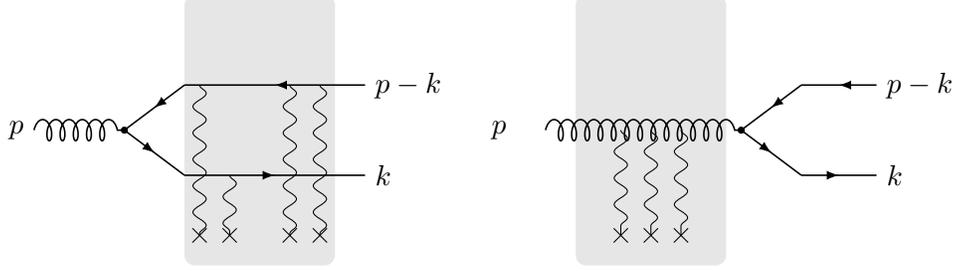
\begin{figure}[t]
\begin{centering}
\begin{tikzpicture}[scale=2] 

\tikzset{photon/.style={decorate,decoration={snake}},
		electron/.style={ postaction={decorate},decoration={markings,mark=at position .5 with {\arrow[draw]{latex}}}},      	gluon/.style={decorate,decoration={coil,amplitude=4pt, segment length=5pt}}}

\fill[black!10!white, rounded corners] (0,-.9) rectangle(1,.9);
\draw[gluon,semithick] (-1,0) node[left]{$p$}-- (-.4,0);
\filldraw[black] (-.4,0) circle(.02);
\draw[electron,semithick] (-.4,0) --++ (.4,-.3);
\draw[electron,semithick] (0,-.3)  -- ++ (1.2,0) node[right]{$k$};
\draw[electron,semithick] (0,.3) -- (-.4,0);
\draw[electron,semithick] (1.2,.3) node[right]{$p-k$}  -- ++ (-1.2,0);

\draw[photon] (0.1,.3) --++ (0,-1) node [at end, cross out, draw, solid,  inner sep=2.5 pt]{};
\draw[photon] (0.3,-.3) --++ (0,-.4) node [at end, cross out, draw, solid, inner sep=2.5 pt]{};

\draw[photon] (0.7,.3) --++ (0,-1) node [at end, cross out, draw, solid,  inner sep=2.5 pt]{};
\draw[photon] (.9,.3) --++ (0,-1) node [at end, cross out, draw, solid, inner sep=2.5 pt]{};

\fill[black!10!white, rounded corners] (2.6,-.9) rectangle(3.6,.9);
\draw[gluon,semithick] (2.4,0) node[left]{$p\quad$}--++ (1.3,0);
\filldraw[black] (3.7,0) circle(.02);
\draw[electron,semithick] (3.7,0) --++ (.4,-.3);
\draw[electron,semithick] (4.1,-.3)  -- ++ (.5,0) node[right]{$k$};
\draw[electron,semithick] (4.1,.3) --++ (-.4,-.3);
\draw[electron,semithick] (4.6,.3) node[right]{$p-k$}  -- ++ (-.5,0);

\draw[photon] (2.9,0) --++ (0,-.7) node [at end, cross out, draw, solid,  inner sep=2.5 pt]{};
\draw[photon] (3.1,0) --++ (0,-.7) node [at end, cross out, draw, solid, inner sep=2.5 pt]{};
\draw[photon] (3.3,0) --++ (0,-.7) node [at end, cross out, draw, solid,  inner sep=2.5 pt]{};
\end{tikzpicture} 
\par\end{centering}
\caption{\label{fig:CGCgA2qqX}The two amplitudes in which the gluon interacts
with the nucleus before (right) or after (left) it has fluctuated
in a quark-antiquark pair.}
\end{figure}

After these preliminaries, let us now derive the cross section for
massive quark production in $pA$ collisions, within the framework
of the Color Glass Condensate. We will closely follow Ref. \protect\cite{Marquet2007},
in which dijet production in the $qA\rightarrow qg$ channel is studied.
Although, since we study a different channel, the details of our calculation
will be different, the line of reasoning and the structure of the
calculation in this reference equally applies to our case. 

The main idea of the CGC calculation is that the outgoing quark-antiquark
pair is a fluctuation stemming from the gluon coming from the projectile.
This $g\to q\bar{q}$ splitting takes place either before or after
the interaction with the target, as illustrated in Fig. \ref{fig:CGCgA2qqX}.
Since the target is a heavily contracted shockwave, the case in which
the splitting takes place during the interaction can be neglected.
An essential assumption is that the gluon or the quark pair scatters
off the nucleus through multiple eikonal interactions with the soft
gluon fields from the target. These interactions are resummed into
Wilson lines, through which the final cross section will depend on
the target. To be more specific: the gauge links are sensitive to
the two-point correlators of the classical gluon fields in the target,
which subsequently can be computed in models such as the McLerran-Venugopalan
or Golec-Biernat-Wüsthoff model, and evolved in rapidity using the
JIMWLK equation.

Let us start with examining the gluon coming from the proton, for
which we can write the following first order Fock-space decomposition:
\begin{equation}
\begin{aligned}|\vec{p},c,\lambda\rangle & =Z|\vec{p},c,\lambda\rangle_{0}+\sum_{\alpha\beta ij}\int\mathrm{d}^{3}k\,g_{s}t_{ji}^{c}\psi_{\alpha\beta}^{T\lambda}\left(\vec{p},\vec{k},Q_{p}^{2}\right)|\left(\vec{k},i,\beta\right);\left(\vec{p}-\vec{k},j,\alpha\right)\rangle_{0}.\end{aligned}
\label{eq:Fockdressedgluon}
\end{equation}
In the above expression, $\vec{p}$ is the three-momentum, $c$ the
color, $\lambda$ the polarization, and $Q_{p}^{2}=p_{\mu}p^{\mu}$
the virtuality of the gluon. The dressed gluon state, in the left
hand side of the equation, is decomposed into a bare state, with corresponding
renormalization factor $Z$, and a the second term which describes
the gluon being split into a quark and an antiquark, with spins $\beta$
and $\alpha$, colors $i$ and $j$, and momenta $\vec{k}$ and $\vec{p}-\vec{k}$.
The wave function $\psi_{\alpha\beta}^{T\lambda}\left(\vec{p},\vec{k},Q_{p}^{2}\right)$
encodes the probability for a transversely polarized gluon to split
up and is given by (the explicit calculation is performed in the appendix,
Sec. \ref{subsec:gqqwave}):
\begin{equation}
\begin{aligned}\psi_{\alpha\beta}^{T\lambda}\left(\vec{p},\vec{k},Q_{p}^{2}\right) & =\sqrt{\frac{2}{p^{+}}}\frac{1}{\left(\mathbf{k}_{\perp}-z\mathbf{p}_{\perp}\right)^{2}+m^{2}}\\
 & \times\begin{cases}
\left(z\delta_{\alpha+}\delta_{\beta+}-\left(1-z\right)\delta_{\alpha-}\delta_{\beta-}\right)\left(\mathbf{k}_{\perp}-z\mathbf{p}_{\perp}\right)\cdot\boldsymbol{\epsilon}_{\perp}^{1}+\frac{m}{\sqrt{2}}\delta_{\alpha+}\delta_{\beta-} & \lambda=1\\
\left(z\delta_{\alpha-}\delta_{\beta-}-\left(1-z\right)\delta_{\alpha+}\delta_{\beta+}\right)\left(\mathbf{k}_{\perp}-z\mathbf{p}_{\perp}\right)\cdot\boldsymbol{\epsilon}_{\perp}^{2}-\frac{m}{\sqrt{2}}\delta_{\alpha-}\delta_{\beta+} & \lambda=2
\end{cases}.
\end{aligned}
\end{equation}
As already mentioned earlier, the transverse momentum $p_{\perp}$
and hence the virtuality $Q_{p}$ of the incoming gluon is expected
to be of the order $\Lambda_{\mathrm{QCD}}$, which is negligible
with respect to the virtualities $\sim Q_{s}$ of the gluons in the
target. We can therefore neglect the case in which the incoming gluon
is longitudinally polarized, and in what follows we suppress the index
$T$ and the dependence on $Q_{p}^{2}$, and simply write $\psi_{\alpha\beta}^{\lambda}\left(\vec{p},\vec{k}\right)$. 

Let us now transform Eq. (\ref{eq:Fockdressedgluon}) to \textquoteleft mixed
Fourier space', in which only the transverse momenta are Fourier transformed.
Using the conventions
\begin{equation}
\begin{aligned}|\vec{p},c,\lambda\rangle & =\int\frac{\mathrm{d}^{2}\mathbf{b}}{\left(2\pi\right)^{2}}e^{i\mathbf{p}_{\perp}\cdot\mathbf{b}}|p^{+},\mathbf{b},c,\lambda\rangle,\\
\psi_{\alpha\beta}^{\lambda}\left(\vec{p},\vec{k}\right) & =\int\frac{\mathrm{d}^{2}\mathbf{x}}{\left(2\pi\right)^{2}}e^{-i\mathbf{k}_{\perp}\cdot\mathbf{x}}\phi_{\alpha\beta}^{\lambda}\left(p^{+},z,\mathbf{x}\right),
\end{aligned}
\end{equation}
where $\phi_{\alpha\beta}^{\lambda}\left(p^{+},z,\mathbf{r}\right)$
is the $g\to q\bar{q}$ wave function in mixed Fourier space (we use
the notation $r=|\mathbf{r}|$):
\begin{equation}
\begin{aligned}\phi_{\alpha\beta}^{\lambda}\left(p^{+},z,\mathbf{r}\right) & =2\pi\sqrt{\frac{2}{p^{+}}}e^{iz\mathbf{p}_{\perp}\cdot\mathbf{r}}\\
 & \times\begin{cases}
imK_{1}\left(mr\right)\frac{\mathbf{r}\cdot\boldsymbol{\epsilon}_{\perp}^{1}}{r}\left(z\delta_{\alpha-}\delta_{\beta-}-\left(1-z\right)\delta_{\alpha+}\delta_{\beta+}\right)+\frac{m}{\sqrt{2}}K_{0}\left(mr\right)\delta_{\alpha+}\delta_{\beta-} & \lambda=1\\
imK_{1}\left(mr\right)\frac{\mathbf{r}\cdot\boldsymbol{\epsilon}_{\perp}^{2}}{r}\left(z\delta_{\alpha+}\delta_{\beta+}-\left(1-z\right)\delta_{\alpha-}\delta_{\beta-}\right)-\frac{m}{\sqrt{2}}K_{0}\left(mr\right)\delta_{\alpha-}\delta_{\beta+} & \lambda=2
\end{cases},
\end{aligned}
\label{eq:wavegqqMFS}
\end{equation}
we obtain:
\begin{equation}
\begin{aligned}|p^{+},\mathbf{b},c,\lambda\rangle & =Z|p^{+},\mathbf{b},c,\lambda\rangle_{0}\\
 & +g_{s}\int\mathrm{d}^{2}\mathbf{p}_{\perp}e^{-i\mathbf{p}_{\perp}\cdot\mathbf{b}}\sum_{\alpha\beta ij}t_{ji}^{c}\int\mathrm{d}^{3}k\psi_{\alpha\beta}^{\lambda}\left(\vec{p},\vec{k}\right)|\left(\vec{k},i,\beta\right);\left(\vec{p}-\vec{k},j,\alpha\right)\rangle_{0},\\
 & =Z|p^{+},\mathbf{b},c,\lambda\rangle_{0}+g_{s}\sum_{\alpha\beta ij}t_{ji}^{c}\int\mathrm{d}^{3}k\int\mathrm{d}^{2}\mathbf{p}_{\perp}\int\frac{\mathrm{d}^{2}\mathbf{y}}{\left(2\pi\right)^{2}}\frac{\mathrm{d}^{2}\mathbf{x}}{\left(2\pi\right)^{2}}\frac{\mathrm{d}^{2}\mathbf{b}'}{\left(2\pi\right)^{2}}e^{-i\mathbf{p}_{\perp}\cdot\mathbf{b}}\\
 & \times e^{-i\mathbf{k}_{\perp}\cdot\mathbf{y}}e^{i\mathbf{k}_{\perp}\cdot\mathbf{x}}e^{i\left(\mathbf{p}_{\perp}-\mathbf{k}_{\perp}\right)\cdot\mathbf{b}'}\phi_{\alpha\beta}^{\lambda}\left(\vec{p},z,\mathbf{y}\right)|\left(k^{+},\mathbf{x},i,\beta\right);\left(p^{+}-k^{+},\mathbf{b}'j,\alpha\right)\rangle_{0},\\
 & =Z|p^{+},\mathbf{b},c,\lambda\rangle_{0}+g_{s}\sum_{\alpha\beta ij}t_{ji}^{c}\int\mathrm{d}k^{+}\frac{\mathrm{d}^{2}\mathbf{x}}{\left(2\pi\right)^{2}}\frac{\mathrm{d}^{2}\mathbf{b}'}{\left(2\pi\right)^{2}}\left(2\pi\right)^{2}\delta^{\left(2\right)}\left(\mathbf{b}-z\mathbf{x}-\left(1-z\right)\mathbf{b}'\right)\\
 & \times\tilde{\phi}_{\alpha\beta}^{\lambda}\left(p^{+},z,\mathbf{x}-\mathbf{b}'\right)|\left(k^{+},\mathbf{x},i,\beta\right);\left(p^{+}-k^{+},\mathbf{b}'j,\alpha\right)\rangle_{0}.
\end{aligned}
\label{eq:FockdressedgluonMFS}
\end{equation}
A few remarks are in order: the particular delta function comes from
integrating over $\mathbf{p}_{\perp}$, over, amongst others, the
factor $\exp\left(iz\mathbf{p}_{\perp}\cdot\mathbf{y}\right)$ inside
the wave function. The last line of the above expression is hence
not dependent on $\mathbf{p}_{\perp}$ anymore, and we introduced
the $\mathbf{p}_{\perp}$-independent wave function:
\begin{equation}
\tilde{\phi}_{\alpha\beta}^{\lambda}\left(p^{+},z,\mathbf{x}-\mathbf{b}'\right)\equiv e^{-iz\mathbf{p}_{\perp}\cdot\left(\mathbf{x}-\mathbf{b}'\right)}\phi_{\alpha\beta}^{\lambda}\left(p,z,\mathbf{x}-\mathbf{b}'\right).
\end{equation}
As usual, the interaction of the gluon or the quark-antiquark pair
with the target, is quantified in the action of the $S$-matrix on
the ingoing states. One of the powers of the present formalism is
that, in mixed Fourier space, this action diagonalizes, as will become
apparent shortly. First, we need the in- and outgoing states of the
dressed gluon, which are given by:
\begin{equation}
\begin{aligned}|\Psi_{\mathrm{in}}\rangle & =|\vec{p},c,\lambda\rangle\otimes|\mathcal{T}\rangle,\qquad|\Psi_{\mathrm{out}}\rangle=S|\Psi_{\mathrm{in}}\rangle.\end{aligned}
\label{eq:ingoingoutgoingstate}
\end{equation}
In the above formula, the dressed gluon state is convolved with the
state ket of the target $|\mathcal{T}\rangle$, which in the CGC consists
of many classical color gauge fields, distributed according to the
CGC wave function $\Phi_{x}\left[\mathcal{A}\right]$:
\begin{equation}
|\mathcal{T}\rangle=\int\mathcal{D}\left[\mathcal{A}\right]\Phi_{x}\left[\mathcal{A}\right]|\mathcal{A}\rangle.\label{eq:targetket}
\end{equation}
We should stress that the quantum mechanical wave function $\Phi_{x}\left[\mathcal{A}\right]$
of the target is not known, and that we use it in a purely symbolic
fashion. However, in our calculation, we will only need to evaluate
its absolute value squared $|\Phi_{x}\left[\mathcal{A}\right]|^{2}$,
and this is naturally associated with the Gaussian distribution $\mathcal{W}_{x}\left[\mathcal{A}\right]$
which we known from the CGC.

As already alluded to above, acting with the $S$-matrix on the dressed
gluon state in mixed Fourier space is simple: it merely amounts to
multiplying the bare gluon state with an eikonal Wilson line $W\left(\mathbf{x}\right)$
in the adjoint representation, while the quark and the antiquark have
to be multiplied with an eikonal Wilson line $U\left(\mathbf{x}\right)$
or $U^{\dagger}\left(\mathbf{x}\right)$ in the fundamental representation.\footnote{Note that care should be taken with the conventions. In contrast to
Part II, the target here is a left-mover. Choosing a covariant gauge
with $A^{+}=0$, its multiple interactions with a right-moving quark
are resummed as $U\left(\mathbf{x}\right)=\mathcal{P}\exp\left(ig_{s}\int\mathrm{d}x^{+}A\left(x^{+},\mathbf{x}\right)\right)$.} The outgoing state thus becomes:
\begin{equation}
\begin{aligned}|\Psi_{\mathrm{out}}\rangle & =\int\mathcal{D}\left[\mathcal{A}\right]\Phi_{x}\left[\mathcal{A}\right]\int\frac{\mathrm{d}^{2}\mathbf{b}}{\left(2\pi\right)^{2}}e^{i\mathbf{p}_{\perp}\cdot\mathbf{b}}\biggl(Z\,W\left(\mathbf{b}\right)|p^{+},\mathbf{b},c,\lambda\rangle_{0}\otimes|\mathcal{A}\rangle+g_{s}\sum_{\alpha\beta ij}\int\mathrm{d}k^{+}\int\frac{\mathrm{d}^{2}\mathbf{x}}{\left(2\pi\right)^{2}}\\
 & \times\tilde{\phi}_{\alpha\beta}^{\lambda}\left(p^{+},k^{+},\mathbf{x}-\mathbf{b}\right)U_{jk}^{\dagger}\left(\mathbf{b}\right)t_{kl}^{c}U_{li}\left(\mathbf{x}\right)|\left(k^{+},\mathbf{x},i,\beta\right);\left(p^{+}-k^{+},\mathbf{b},j,\alpha\right)\rangle_{0}\otimes|\mathcal{A}\rangle\biggr).
\end{aligned}
\label{eq:CGCpAPsi out}
\end{equation}
It becomes now clear what the definitions (\ref{eq:ingoingoutgoingstate})
and (\ref{eq:targetket}) mean: the gluon and the quark-antiquark
pair are created from the QCD vacuum $|0\rangle$, while the target
state $|\mathcal{T}\rangle$ only pertains to the classical gluon
fields of the CGC. The latter are resummed in the Wilson lines, which
therefore effectively contain all the information on the structure
of the nucleus. According to the CGC, one has to take the statistical
average over all possible classical gauge field configurations, and
hence we expect the cross section to take the following form:
\begin{equation}
\sigma=\sigma_{0}\int\mathcal{D}\left[\mathcal{A}\right]\left|\Phi_{x}\left[\mathcal{A}\right]\right|^{2}\langle\mathcal{A}|\mathrm{Wilson\,lines|\mathcal{A}\rangle,}
\end{equation}
with, as explained earlier, $|\Phi_{x}\left[\mathcal{A}\right]|^{2}=\mathcal{W}_{x}\left[\mathcal{A}\right]$.

We mentioned earlier that the scattering of the projectile off the
target is described by two amplitudes: the one in which the gluon
interacts with the nucleus and then splits in the quark pair, and
the amplitude in which the splitting takes place earlier and therefore
the quark-antiquark pair interacts. In the expression for the outgoing
state, Eq. (\ref{eq:CGCpAPsi out}), only the latter case is apparent,
as the first term in the formula describes a bare gluon. However,
the next step in our calculation comes from the observation that the
former case is in fact contained in this bare gluon term, which becomes
apparent if we invert Eq. (\ref{eq:Fockdressedgluon}):
\begin{equation}
\begin{aligned}Z|\vec{p},c,\lambda\rangle_{0} & =|\vec{p},c,\lambda\rangle-\sum_{\alpha\beta ij}\int\mathrm{d}^{3}k\,g_{s}t_{ji}^{c}\psi_{\alpha\beta}^{\lambda}\left(\vec{p},\vec{k}\right)|\left(\vec{k},i,\beta\right);\left(\vec{p}-\vec{k},j,\alpha\right)\rangle_{0}.\end{aligned}
\end{equation}
Plugging the Fourier transform, Eq. (\ref{eq:FockdressedgluonMFS}),
of this formula in Eq. (\ref{eq:CGCpAPsi out}), and throw away the
term containing $|\vec{p},c,\lambda\rangle$ since the dressed gluon
does not contribute to the production of the dijet, we find:
\begin{equation}
\begin{aligned}|\Psi_{\mathrm{out}}\rangle & =g_{s}\int\mathcal{D}\left[\mathcal{A}\right]\Phi_{x}\left[\mathcal{A}\right]\sum_{\alpha\beta ij}\int\mathrm{d}k^{+}\frac{\mathrm{d}^{2}\mathbf{x}}{\left(2\pi\right)^{2}}\frac{\mathrm{d}^{2}\mathbf{b}}{\left(2\pi\right)^{2}}e^{i\mathbf{p}_{\perp}\cdot\mathbf{b}}\\
 & \times\Phi_{\alpha\beta ij}^{\lambda}\left(p^{+},k^{+},\mathbf{x}-\mathbf{b}\right)|\left(k^{+},\mathbf{x},i,\beta\right);\left(p^{+}-k^{+},\mathbf{b},j,\alpha\right)\rangle_{0}\otimes|\mathcal{A}\rangle,
\end{aligned}
\label{eq:outstate}
\end{equation}
where we defined:
\begin{equation}
\begin{aligned} & \Phi_{\alpha\beta ij}^{\lambda}\left(p^{+},k^{+},\mathbf{x}-\mathbf{b}\right)\\
 & \equiv\left(U_{jk}^{\dagger}\left(\mathbf{b}\right)t_{kl}^{c}U_{li}\left(\mathbf{x}\right)-W_{cd}\left(z\mathbf{x}+\left(1-z\right)\mathbf{b}\right)t_{ji}^{d}\right)\tilde{\phi}_{\alpha\beta}^{\lambda}\left(p^{+},k^{+},\mathbf{x}-\mathbf{b}\right).
\end{aligned}
\label{eq:CGCinteraction}
\end{equation}
Indeed, in the above expression, both amplitudes of Fig. \ref{fig:CGCgA2qqX}:
the one in which the gluon fluctuates in a quark-antiquark pair that
subsequently interacts with the nucleus, and the other one in which
this fluctuation takes place after the interaction of the gluon with
the medium, are now explicit. 

We have finally gathered all the ingredients to compute the $gA\rightarrow q\bar{q}X$
cross section, which is given formally by the following expression:
\begin{equation}
\begin{aligned}\frac{\mathrm{d}\sigma^{gA\rightarrow q\bar{q}X}}{\mathrm{d}^{3}k_{1}\mathrm{d}^{3}k_{2}} & =\frac{1}{2}\frac{1}{N_{c}^{2}-1}\sum_{\lambda c}\langle\Psi_{\mathrm{out}}|N_{q}\left(\vec{k}_{1}\right)N_{\bar{q}}\left(\vec{k}_{2}\right)|\Psi_{\mathrm{out}}\rangle,\end{aligned}
\label{eq:gqqcrosssection0}
\end{equation}
where $N_{q}$ and $N_{\bar{q}}$ are counting operators defined as:
\begin{equation}
\begin{aligned}N_{q}\left(\vec{k}\right) & \equiv\sum_{is}b_{i,s}^{\dagger}\left(\vec{k}\right)b_{i,s}\left(\vec{k}\right),\\
N_{\bar{q}}\left(\vec{k}\right) & \equiv\sum_{js'}d_{j,s'}^{\dagger}\left(\vec{k}\right)d_{j,s'}\left(\vec{k}\right).
\end{aligned}
\end{equation}
In the above definitions, $b_{i,s}^{\dagger}\left(\vec{k}\right)$,
$b_{i,s}\left(\vec{k}\right)$ and $d_{j,s'}^{\dagger}\left(\vec{k}\right)$,
$d_{j,s'}\left(\vec{k}\right)$ are the creation and annihilation
operators for quarks and antiquarks, respectively, satisfying the
following relations in the mixed Fourier space:
\begin{equation}
\begin{aligned}b_{i,s}^{\dagger}\left(k^{+},\mathbf{x}\right)|0\rangle & =|k^{+},\mathbf{x},i,s\rangle_{0},\\
b_{i,s}\left(k^{+},\mathbf{x}\right)|0\rangle & =0,\\
\left\{ b_{i,s}\left(k^{+},\mathbf{x}\right),b_{j,s'}^{\dagger}\left(p^{+},\mathbf{y}\right)\right\}  & =\left(2\pi\right)^{2}\delta_{ij}\delta_{ss'}\delta\left(k^{+}-p^{+}\right)\delta^{\left(2\right)}\left(\mathbf{x}-\mathbf{y}\right),
\end{aligned}
\end{equation}
where the same relations hold for the antiquark operators. Plugging
in our results for the $|\Psi_{\mathrm{out}}\rangle$ state, Eq. (\ref{eq:outstate}),
into the expression for the cross section, Eq. (\ref{eq:gqqcrosssection0}),
we obtain:
\begin{equation}
\begin{aligned}\frac{\mathrm{d}\sigma^{gA\rightarrow q\bar{q}X}}{\mathrm{d}^{3}k_{1}\mathrm{d}^{3}k_{2}} & =\frac{1}{2}\frac{1}{N_{c}^{2}-1}\int\frac{\mathrm{d}^{2}\mathbf{v}}{\left(2\pi\right)^{2}}\frac{\mathrm{d}^{2}\mathbf{v}'}{\left(2\pi\right)^{2}}\frac{\mathrm{d}^{2}\mathbf{u}}{\left(2\pi\right)^{2}}\frac{\mathrm{d}^{2}\mathbf{u}'}{\left(2\pi\right)^{2}}e^{i\mathbf{k}_{1\perp}\cdot\left(\mathbf{v}-\mathbf{v}'\right)}e^{i\mathbf{k}_{2\perp}\cdot\left(\mathbf{u}-\mathbf{u}'\right)}\\
 & \sum_{\lambda c}\sum_{ijss'}\langle\Psi_{\mathrm{out}}|b_{k,s}^{\dagger}\left(k_{1}^{+},\mathbf{v}\right)d_{l,s'}^{\dagger}\left(k_{2}^{+},\mathbf{u}\right)d_{l,s'}\left(k_{2}^{+},\mathbf{u}'\right)b_{k,s}\left(k_{1}^{+},\mathbf{v}'\right)|\Psi_{\mathrm{out}}\rangle.
\end{aligned}
\end{equation}
The action of the annihilation operators on the $|\psi_{\mathrm{out}}\rangle$
state yields:
\begin{equation}
\begin{aligned} & d_{l,s'}\left(k_{2}^{+},\mathbf{u}'\right)b_{k,s}\left(k_{1}^{+},\mathbf{v}'\right)|\Psi_{\mathrm{out}}\rangle\\
 & =g_{s}\int\mathcal{D}\left[\mathcal{A}\right]\Phi_{x}\left[\mathcal{A}\right]\sum_{\alpha\beta ij}\int\mathrm{d}k^{+}\frac{\mathrm{d}^{2}\mathbf{x}}{\left(2\pi\right)^{2}}\frac{\mathrm{d}^{2}\mathbf{b}}{\left(2\pi\right)^{2}}e^{i\mathbf{p}_{\perp}\cdot\mathbf{b}}\Phi_{\alpha\beta ij}^{\lambda}\left(p^{+},l^{+},\mathbf{x}-\mathbf{b}\right)\\
 & \times d_{l,s'}\left(k_{2}^{+},\mathbf{u}'\right)b_{k,s}\left(k_{1}^{+},\mathbf{v}'\right)b_{i,\beta}^{\dagger}\left(k^{+},\mathbf{x}\right)d_{j,\alpha}^{\dagger}\left(p^{+}-k^{+},\mathbf{b}\right)|0\rangle\otimes|\mathcal{A}\rangle,\\
 & =g_{s}\int\mathcal{D}\left[\mathcal{A}\right]\Phi_{x}\left[\mathcal{A}\right]\int\mathrm{d}k^{+}\frac{\mathrm{d}^{2}\mathbf{x}}{\left(2\pi\right)^{2}}\frac{\mathrm{d}^{2}\mathbf{b}}{\left(2\pi\right)^{2}}e^{i\mathbf{p}_{\perp}\cdot\mathbf{b}}\Phi_{s'slk}^{\lambda}\left(p^{+},k^{+},\mathbf{x}-\mathbf{b}\right)\\
 & \times\left(2\pi\right)^{4}\delta\left(k_{2}^{+}+k^{+}-p^{+}\right)\delta\left(k_{1}^{+}-k^{+}\right)\delta^{\left(2\right)}\left(\mathbf{u}'-\mathbf{b}\right)\delta^{\left(2\right)}\left(\mathbf{v}'-\mathbf{x}\right)|0\rangle\otimes|\mathcal{A}\rangle,\\
 & =g_{s}\delta\left(k_{2}^{+}+k_{1}^{+}-p^{+}\right)\int\mathcal{D}\left[\mathcal{A}\right]\Phi_{x}\left[\mathcal{A}\right]e^{i\mathbf{p}_{\perp}\cdot\mathbf{u}'}\Phi_{s'slk}^{\lambda}\left(p^{+},k_{1}^{+},\mathbf{u}'-\mathbf{v}'\right)|0\rangle\otimes|\mathcal{A}\rangle,
\end{aligned}
\end{equation}
while we find for the action of the creation operators on the $\langle\Psi_{\mathrm{out}}|$
state:
\begin{equation}
\begin{aligned}\langle\Psi_{\mathrm{out}}|b_{k,s}^{\dagger}\left(k_{1}^{+},\mathbf{v}\right)d_{l,s'}^{\dagger}\left(k_{2}^{+},\mathbf{u}\right) & =g_{s}\delta\left(p^{+}-k_{2}^{+}-k_{1}^{+}\right)\\
 & \times\int\mathcal{D}\left[\mathcal{A}\right]\Phi_{x}^{\dagger}\left[\mathcal{A}\right]\langle\mathcal{A}|\otimes\langle0|e^{-i\mathbf{p}_{\perp}\cdot\mathbf{u}}\Phi_{s'slk}^{\lambda\dagger}\left(p^{+},k_{1}^{+},\mathbf{u}-\mathbf{v}\right).
\end{aligned}
\end{equation}
We finally obtain the following result for the $gA\rightarrow q\bar{q}X$
cross section:
\begin{equation}
\begin{aligned} & \frac{\mathrm{d}\sigma^{gA\rightarrow q\bar{q}X}}{\mathrm{d}^{3}k_{1}\mathrm{d}^{3}k_{2}}\\
 & =\frac{\alpha_{s}}{N_{c}^{2}-1}\delta\left(k_{1}^{+}+k_{2}^{+}-p^{+}\right)\sum_{ijss'}\int\frac{\mathrm{d}^{2}\mathbf{v}}{\left(2\pi\right)^{2}}\frac{\mathrm{d}^{2}\mathbf{v}'}{\left(2\pi\right)^{2}}\frac{\mathrm{d}^{2}\mathbf{u}}{\left(2\pi\right)^{2}}\frac{\mathrm{d}^{2}\mathbf{u}'}{\left(2\pi\right)^{2}}e^{i\mathbf{k}_{1\perp}\cdot\left(\mathbf{v}-\mathbf{v}'\right)}e^{i\left(\mathbf{k}_{2\perp}-\mathbf{p}_{\perp}\right)\cdot\left(\mathbf{u}-\mathbf{u}'\right)}\\
 & \times\int\mathcal{D}\left[\mathcal{A}\right]\left|\Phi_{x}\left[\mathcal{A}\right]\right|^{2}\langle\mathcal{A}|\Phi_{s'sij}^{\lambda\dagger}\left(p^{+},k_{1}^{+},\mathbf{u}-\mathbf{v}\right)\Phi_{s'sij}^{\lambda}\left(p^{+},k_{1}^{+},\mathbf{u}'-\mathbf{v}'\right)|\mathcal{A}\rangle,
\end{aligned}
\end{equation}
where we divided by a factor $2\pi\delta\left(k_{1}^{+}+k_{2}^{+}-p^{+}\right)$
to remove the spurious divergency related to the use of plane waves
instead of wave packets (see Ref. \protect\cite{Marquet2007}). The interaction
part becomes:
\begin{equation}
\begin{aligned} & \int\mathcal{D}\left[\mathcal{A}\right]\left|\Phi_{x}\left[\mathcal{A}\right]\right|^{2}\langle\mathcal{A}|\left(U^{\dagger}\left(\mathbf{u}\right)t^{c}U\left(\mathbf{v}\right)-W_{cd}\left(z\mathbf{u}+\left(1-z\right)\mathbf{v}\right)t^{d}\right)^{\dagger}\tilde{\phi}_{ss'}^{\lambda*}\left(p^{+},k_{1}^{+},\mathbf{u}-\mathbf{v}\right)\\
 & \times\left(U^{\dagger}\left(\mathbf{u}'\right)t^{c}U\left(\mathbf{v}'\right)-W_{ce}\left(z\mathbf{u}'+\left(1-z\right)\mathbf{v}'\right)t^{e}\right)\tilde{\phi}_{ss'}^{\lambda}\left(p^{+},k_{1}^{+},\mathbf{u}'-\mathbf{v}'\right)|\mathcal{A}\rangle\\
 & =\tilde{\phi}_{ss'}^{\lambda*}\left(p^{+},k_{1}^{+},\mathbf{u}-\mathbf{v}\right)\tilde{\phi}_{ss'}^{\lambda}\left(p^{+},k_{1}^{+},\mathbf{u}'-\mathbf{v}'\right)\\
 & \times\biggl(\mathrm{Tr}\Bigl\langle U^{\dagger}\left(\mathbf{v}\right)t^{c}U\left(\mathbf{u}\right)U^{\dagger}\left(\mathbf{u}'\right)t^{c}U\left(\mathbf{v}'\right)\Bigr\rangle_{x}+\mathrm{Tr}\Bigl\langle W_{dc}\left(z\mathbf{u}+\left(1-z\right)\mathbf{v}\right)t^{d}W_{ce}\left(z\mathbf{u}'+\left(1-z\right)\mathbf{v}'\right)t^{e}\Bigr\rangle_{x}\\
 & -\mathrm{Tr}\Bigl\langle U^{\dagger}\left(\mathbf{v}\right)t^{c}U\left(\mathbf{u}\right)W_{ce}\left(z\mathbf{u}'+\left(1-z\right)\mathbf{v}'\right)t^{e}\Bigr\rangle_{x}-\mathrm{Tr}\Bigl\langle W_{cd}\left(z\mathbf{u}+\left(1-z\right)\mathbf{v}\right)t^{d}U^{\dagger}\left(\mathbf{u}'\right)t^{c}U\left(\mathbf{v}'\right)\Bigr\rangle_{x}\biggr),
\end{aligned}
\label{eq:CGCinteractionsquared}
\end{equation}
where we introduced the notation: 
\begin{equation}
\langle\mathcal{O}\rangle_{x}=\int\mathcal{D}\left[\mathcal{A}\right]\mathcal{W}_{x}\left[\mathcal{A}\right]\langle\mathcal{A}|\mathcal{O}|\mathcal{A}\rangle
\end{equation}
to indicate an operator being evaluated in the Color Glass Condensate
at a certain Bjorken-$x$. Using the Lie algebra and Wilson line identities
Eq. (\ref{eq:Wabreal}) and Eq. (\ref{eq:tracetatb}), we finally
obtain the following result for the cross section: (for future convenience,
we changed the notation for the transverse coordinates $\mathbf{x}\leftrightarrow\mathbf{u}$
and $\mathbf{y}\leftrightarrow\mathbf{v}$)

\begin{equation}
\begin{aligned} & \frac{\mathrm{d}\sigma^{gA\rightarrow q\bar{q}X}}{\mathrm{d}^{3}k_{1}\mathrm{d}^{3}k_{2}}=\frac{\alpha_{s}}{2}\frac{\delta\left(\left(k_{1}^{+}+k_{2}^{+}\right)/p^{+}-1\right)}{p^{+}}\\
 & \times\int\frac{\mathrm{d}^{2}\mathbf{x}}{\left(2\pi\right)^{2}}\frac{\mathrm{d}^{2}\mathbf{y}}{\left(2\pi\right)^{2}}\frac{\mathrm{d}^{2}\mathbf{x}'}{\left(2\pi\right)^{2}}\frac{\mathrm{d}^{2}\mathbf{y}'}{\left(2\pi\right)^{2}}e^{i\mathbf{k}_{1\perp}\cdot\left(\mathbf{y}-\mathbf{y}'\right)}e^{i\left(\mathbf{k}_{2\perp}-\mathbf{p}_{\perp}\right)\cdot\left(\mathbf{x}-\mathbf{x}'\right)}\\
 & \times\sum_{\lambda ss'}\tilde{\phi}_{s's}^{\lambda*}\left(p^{+},k_{1}^{+},\mathbf{x}-\mathbf{y}\right)\tilde{\phi}_{s's}^{\lambda}\left(p^{+},k_{1}^{+},\mathbf{x}'-\mathbf{y}'\right)\\
 & \times\biggl(C\left(\mathbf{x},\mathbf{y},\mathbf{x}',\mathbf{y}'\right)+D_{A}\left(z\mathbf{x}+\left(1-z\right)\mathbf{y},z\mathbf{x}'+\left(1-z\right)\mathbf{y}'\right)\\
 & -S^{\left(3\right)}\left(\mathbf{x},z\mathbf{x}'+\left(1-z\right)\mathbf{y}',\mathbf{y}\right)-S^{\left(3\right)}\left(\mathbf{y}',z\mathbf{x}+\left(1-z\right)\mathbf{y},\mathbf{x}'\right)\biggr),
\end{aligned}
\label{eq:sigmagqq}
\end{equation}
where we defined the following CGC averages over Wilson lines:
\begin{equation}
\begin{aligned}C\left(\mathbf{x},\mathbf{y},\mathbf{x}',\mathbf{y}'\right) & \equiv\frac{1}{C_{F}N_{c}}\mathrm{Tr}\Bigl\langle U^{\dagger}\left(\mathbf{y}\right)t^{c}U\left(\mathbf{x}\right)U^{\dagger}\left(\mathbf{x}'\right)t^{c}U\left(\mathbf{y}'\right)\Bigr\rangle_{x},\\
D_{A}\left(\mathbf{x},\mathbf{y}\right) & \equiv\frac{1}{N_{c}^{2}-1}\mathrm{Tr}\Bigl\langle W\left(\mathbf{x}\right)W^{\dagger}\left(\mathbf{y}\right)\Bigr\rangle_{x},\\
S^{\left(3\right)}\left(\mathbf{y},\mathbf{z},\mathbf{x}\right) & \equiv\frac{1}{C_{F}N_{c}}\Bigl\langle\mathrm{Tr}\left(U^{\dagger}\left(\mathbf{x}\right)t^{c}U\left(\mathbf{y}\right)t^{a}\right)W_{ca}\left(\mathbf{z}\right)\Bigr\rangle_{x}.
\end{aligned}
\label{eq:correlatordefinitions}
\end{equation}
This result is the same as the one quoted in, for example, Refs. \protect\cite{fabio,Cyrille}. 

\subsection{Correlation limit}

According to the hybrid dilute-dense factorization formalism, described
in Refs. \protect\cite{Kotko2015,vanHameren2016}, the full $pA$ cross section
is obtained as the convolution of our result for $gA\rightarrow q\bar{q}X$,
Eq. (\ref{eq:sigmagqq}), with the gluon PDF $x_{p}\mathcal{G}\left(x_{p},\mu^{2}\right)$.
Furthermore, in order to make contact with the TMD approach, we have
to take the so-called correlation limit, in which the two jets are
almost back-to-back. This way, we recover the two strongly ordered
transverse momentum scales, necessary to be able to use the TMD formalism. 

We first introduce the following transverse coordinates, which will
facilitate taking the correlation limit later:
\begin{equation}
\begin{aligned}\mathbf{u} & \equiv\mathbf{x}-\mathbf{y},\quad\quad\mathbf{u}'\equiv\mathbf{x}'-\mathbf{y}',\\
\mathbf{v} & \equiv z\mathbf{x}+\left(1-z\right)\mathbf{y},\quad\mathbf{v}'\equiv z\mathbf{x}'+\left(1-z\right)\mathbf{y}',
\end{aligned}
\end{equation}
or:
\begin{equation}
\begin{aligned}\mathbf{x} & =\mathbf{v}+\left(1-z\right)\mathbf{u},\qquad\mathbf{x}'=\mathbf{v}'+\left(1-z\right)\mathbf{u}',\\
\mathbf{y} & =\mathbf{v}-z\mathbf{u},\quad\quad\mathbf{y}'=\mathbf{v}'-z\mathbf{u}',
\end{aligned}
\end{equation}
as well as their conjugate momenta:
\begin{equation}
\begin{aligned}\tilde{\mathbf{P}}_{\perp} & \equiv z\mathbf{k}_{1\perp}-\left(1-z\right)\mathbf{k}_{2\perp},\\
\mathbf{q}_{\perp} & \equiv\mathbf{k}_{1\perp}+\mathbf{k}_{2\perp}.
\end{aligned}
\end{equation}
Furthermore we make use of the notation:
\begin{equation}
\begin{aligned}\mathrm{d}\mathcal{P}.\mathcal{S}. & \equiv\mathrm{d}y_{1}\mathrm{d}y_{2}\mathrm{d}^{2}\mathbf{q}_{\perp}\mathrm{d}^{2}\tilde{\mathbf{P}}_{\perp},\\
 & =\frac{1}{k_{1}^{+}k_{2}^{+}}\mathrm{d}^{3}k_{1}\mathrm{d}^{3}k_{2},
\end{aligned}
\end{equation}
and from Eq. (\ref{eq:sigmagqq}) (neglecting $\mathbf{p}_{\perp}$,
since $p_{\perp}\ll k_{\perp}$, as we already mentioned earlier)
we obtain the $pA\to q\bar{q}X$ cross section:
\begin{equation}
\begin{aligned}\frac{\mathrm{d}\sigma^{pA\rightarrow q\bar{q}X}}{\mathrm{d}\mathcal{P}.\mathcal{S}.} & =\frac{\alpha_{s}}{2}\frac{k_{1}^{+}k_{2}^{+}}{p^{+}}x_{p}\mathcal{G}\left(x_{p},\mu^{2}\right)\int\frac{\mathrm{d}^{2}\mathbf{v}}{\left(2\pi\right)^{2}}\frac{\mathrm{d}^{2}\mathbf{u}}{\left(2\pi\right)^{2}}\frac{\mathrm{d}^{2}\mathbf{v}'}{\left(2\pi\right)^{2}}\frac{\mathrm{d}^{2}\mathbf{u}'}{\left(2\pi\right)^{2}}e^{i\mathbf{q}_{\perp}\cdot\left(\mathbf{v}-\mathbf{v}'\right)}e^{-i\tilde{\mathbf{P}}_{\perp}\cdot\left(\mathbf{u}-\mathbf{u}'\right)}\\
 & \times\sum_{\lambda ss'}\tilde{\phi}_{s's}^{\lambda*}\left(p^{+},k_{1}^{+},\mathbf{u}\right)\tilde{\phi}_{s's}^{\lambda}\left(p^{+},k_{1}^{+},\mathbf{u}'\right)\\
 & \times\biggl(C\left(\mathbf{x},\mathbf{y},\mathbf{x}',\mathbf{y}'\right)+D_{A}\left(\mathbf{v},\mathbf{v}'\right)-S^{\left(3\right)}\left(\mathbf{v}+\left(1-z\right)\mathbf{u},\mathbf{v}',\mathbf{v}-z\mathbf{u}\right)\\
 & -S^{\left(3\right)}\left(\mathbf{v}'-z\mathbf{u}',\mathbf{v},\mathbf{v}'+\left(1-z\right)\mathbf{u}'\right)\biggr).
\end{aligned}
\label{eq:sigmapAqqX}
\end{equation}
A last manipulation before we take the correlation limit, is to take
a closer look at Wilson line correlators in the above cross section.
In order to compare with the TMD formalism, they have to be rewritten
in terms of only the fundamental representations. From the definitions
Eq. (\ref{eq:correlatordefinitions}), we have:
\begin{equation}
\begin{aligned}D_{A}\left(\mathbf{v},\mathbf{v}'\right) & =\frac{1}{N_{c}^{2}-1}\Bigl\langle W_{ab}\left(\mathbf{v}\right)\delta^{bc}W_{ca}^{\dagger}\left(\mathbf{v}'\right)\Bigr\rangle_{x}=\frac{2}{N_{c}^{2}-1}\Bigl\langle W_{ab}\left(\mathbf{v}\right)t_{ij}^{b}t_{ji}^{c}W_{ca}^{\dagger}\left(\mathbf{v}'\right)\Bigr\rangle_{x},\\
 & =\frac{2}{N_{c}^{2}-1}\mathrm{Tr}\Bigl\langle U\left(\mathbf{v}\right)t^{a}U^{\dagger}\left(\mathbf{v}\right)U\left(\mathbf{v}'\right)t^{a}U^{\dagger}\left(\mathbf{v}'\right)\Bigr\rangle_{x},\\
 & =\frac{1}{N_{c}^{2}-1}\left(\Bigl\langle\mathrm{Tr}\left(U\left(\mathbf{v}\right)U^{\dagger}\left(\mathbf{v}'\right)\right)\mathrm{Tr}\left(U\left(\mathbf{v}'\right)U^{\dagger}\left(\mathbf{v}\right)\right)\Bigr\rangle_{x}-1\right).
\end{aligned}
\label{eq:WW2UUUU}
\end{equation}

In the second equality of the calculation above, we used the trace
of two generators of the fundamental representation $SU\left(N_{c}\right)$,
Eq. (\ref{eq:tracetatb}), after which, in the third equality, we
made use of Eq. (\ref{eq:U2W}) to transform the Wilson lines in the
adjoint representation to the fundamental representation. Writing
the color indices explicitly, we used the Fierz identity, Eq. (\ref{eq:Fierz}),
in the fourth equality, after which the second term in the result
could be simplified, since two Wilson lines on the same transverse
position, but in the opposite direction, cancel:
\begin{equation}
\left(U\left(\mathbf{v}\right)U^{\dagger}\left(\mathbf{v}\right)\right)_{ij}=\delta_{ij}.\label{eq:UUCancel}
\end{equation}
Using the same manipulations, it is easy to show that
\begin{equation}
\begin{aligned}S^{\left(3\right)}\left(\mathbf{x},\mathbf{z},\mathbf{y}\right) & =\frac{1}{2C_{F}N_{c}}\Bigl\langle\mathrm{Tr}\left(U\left(\mathbf{x}\right)U^{\dagger}\left(\mathbf{z}\right)\right)\mathrm{Tr}\left(U\left(\mathbf{z}\right)U^{\dagger}\left(\mathbf{y}\right)\right)-\frac{1}{N_{c}}\mathrm{Tr}\left(U\left(\mathbf{x}\right)U^{\dagger}\left(\mathbf{y}\right)\right)\Bigr\rangle_{x},\end{aligned}
\end{equation}
and finally, for the quadrupole operator
\begin{equation}
\begin{aligned}C\left(\mathbf{x},\mathbf{y},\mathbf{x}',\mathbf{y}'\right) & =\frac{1}{2C_{F}N_{c}}\Bigl\langle\mathrm{Tr}\left(U\left(\mathbf{x}\right)U^{\dagger}\left(\mathbf{x}'\right)\right)\mathrm{Tr}\left(U\left(\mathbf{y}'\right)U^{\dagger}\left(\mathbf{y}\right)\right)\\
 & -\frac{1}{N_{c}}\mathrm{Tr}\left(U\left(\mathbf{x}\right)U^{\dagger}\left(\mathbf{x}'\right)U\left(\mathbf{y}'\right)U^{\dagger}\left(\mathbf{y}\right)\right)\Bigr\rangle_{x}.
\end{aligned}
\label{eq:Cfund}
\end{equation}
From the above expressions, making use of Eq. (\ref{eq:UUCancel}),
it is very straightforward to derive the following relations between
the Wilson line structures:
\begin{equation}
\begin{aligned}C\left(\mathbf{v},\mathbf{v},\mathbf{w},\mathbf{w}\right) & =D_{A}\left(\mathbf{x},\mathbf{y}\right),\\
C\left(\mathbf{x},\mathbf{y},\mathbf{w},\mathbf{w}\right) & =S^{\left(3\right)}\left(\mathbf{x},\mathbf{w},\mathbf{y}\right),\\
C\left(\mathbf{w},\mathbf{w},\mathbf{x},\mathbf{y}\right) & =S^{\left(3\right)}\left(\mathbf{y},\mathbf{w},\mathbf{x}\right),\\
S^{\left(3\right)}\left(\mathbf{x},\mathbf{y},\mathbf{x}\right) & =D_{A}\left(\mathbf{x},\mathbf{y}\right).
\end{aligned}
\label{eq:WLrelations}
\end{equation}

With these preliminaries, we are finally ready to take the correlation
limit of Eq. (\ref{eq:sigmapAqqX}), which amounts to the requirement
that:
\begin{equation}
\tilde{P}_{\perp}^{2}\gg q_{\perp}^{2},\label{eq:CorrelationLimit}
\end{equation}
implying that the outgoing jets or heavy quarks are almost back-to-back.
Indeed, $\tilde{P}_{\perp}\sim k_{1\perp}\sim k_{2\perp}$ is of the
order of the transverse momenta of the individual jets, while $\mathbf{q}_{\perp}=\mathbf{k}_{1\perp}+\mathbf{k}_{2\perp}$
is the vector sum of their transverse momenta. The latter therefore
quantifies the transverse momentum imbalance of the dijet, and since
this imbalance is caused by the multiple scatterings off the target,
we have that $q_{\perp}\sim Q_{s}$, since the saturation scale $Q_{s}$
is approximately equal to the average transverse momentum of the gluons
in the nucleus. Thus, in the correlation limit, we obtain two strongly
ordered scales, necessary for the TMD approach to be valid. Note also
that, as stated earlier, the transverse momentum $\mathbf{p}_{\perp}$
of the incoming gluon is in the ballpark of $\Lambda_{\mathrm{QCD}}$,
and therefore very small in comparison with $\mathbf{q}_{\perp}$
and $\tilde{\mathbf{P}}_{\perp}$. We can therefore set $\mathbf{p}_{\perp}\simeq0$.

The correlation limit, Eq. (\ref{eq:CorrelationLimit}), is tantamount
to the requirement
\begin{equation}
\mathbf{u},\,\mathbf{u}'\ll\mathbf{v},\,\mathbf{v}'.
\end{equation}
In particular, the Wilson line correlators in the cross section Eq.
(\ref{eq:sigmapAqqX}) can be Taylor expanded in $\mathbf{u}$ and
\textbf{$\mathbf{u}'$}, yielding, with the help of the relations
in Eq. (\ref{eq:WLrelations}):
\begin{equation}
\begin{aligned} & C\left(\mathbf{x},\mathbf{y},\mathbf{x}',\mathbf{y}'\right)+D_{A}\left(\mathbf{v},\mathbf{v}'\right)-S^{\left(3\right)}\left(\mathbf{x},\mathbf{v}',\mathbf{y}\right)-S^{\left(3\right)}\left(\mathbf{y}',\mathbf{v},\mathbf{x}'\right)\\
 & \simeq-z\left(1-z\right)u_{i}u_{j}'\frac{\partial^{2}}{\partial x_{i}\partial y_{j}'}C\left(\mathbf{x},\mathbf{v},\mathbf{v}',\mathbf{y}'\right)-z\left(1-z\right)u_{i}u_{j}'\frac{\partial^{2}}{\partial y_{i}\partial x'_{j}}C\left(\mathbf{v},\mathbf{y},\mathbf{x}',\mathbf{v}'\right)\\
 & +z^{2}u_{i}u_{j}'\frac{\partial^{2}}{\partial y_{i}\partial y_{j}'}C\left(\mathbf{v},\mathbf{y},\mathbf{v}',\mathbf{y}'\right)+\left(1-z\right)^{2}u_{i}u_{j}'\frac{\partial^{2}}{\partial x_{i}\partial x_{j}'}C\left(\mathbf{x},\mathbf{v},\mathbf{x}',\mathbf{v}'\right).
\end{aligned}
\label{eq:quadrdipexp}
\end{equation}
Using the expression for the quadrupole in terms of Wilson lines in
the fundamental representation, Eq. (\ref{eq:Cfund}), as well as
the fact that we can change derivatives around as follows:
\begin{equation}
\begin{aligned}\left[\partial_{i}U^{\dagger}\left(\mathbf{w}\right)\right]U\left(\mathbf{w}\right) & =-U^{\dagger}\left(\mathbf{w}\right)\left[\partial_{i}U\left(\mathbf{w}\right)\right],\end{aligned}
\end{equation}
it is straightforward to see that the four terms in (\ref{eq:quadrdipexp})
combine to give
\begin{equation}
\begin{aligned} & C\left(\mathbf{x},\mathbf{y},\mathbf{x}',\mathbf{y}'\right)+D_{A}\left(\mathbf{v},\mathbf{v}'\right)-S^{\left(3\right)}\left(\mathbf{x},\mathbf{v}',\mathbf{y}\right)-S^{\left(3\right)}\left(\mathbf{y}',\mathbf{v},\mathbf{x}'\right)\\
 & \simeq u_{i}u_{j}'\frac{1}{N_{c}^{2}-1}\mathcal{WL}_{ij}\left(\mathbf{v},\mathbf{v}'\right),
\end{aligned}
\label{eq:WilsonPartDef}
\end{equation}
where we defined:
\begin{equation}
\begin{aligned}\mathcal{WL}_{ij}\left(\mathbf{v},\mathbf{v}'\right) & \equiv\left[-2z\left(1-z\right)\mathrm{Re}\Bigl\langle\mathrm{Tr}\left(\left[\partial_{i}U\left(\mathbf{v}\right)\right]U^{\dagger}\left(\mathbf{v}'\right)\right)\mathrm{Tr}\left(\left[\partial_{j}U\left(\mathbf{v}'\right)\right]U^{\dagger}\left(\mathbf{v}\right)\right)\Bigr\rangle_{x}\right.\\
 & +z^{2}\Bigl\langle\mathrm{Tr}\left(\left[\partial_{i}U\left(\mathbf{v}\right)\right]\left[\partial_{j}U^{\dagger}\left(\mathbf{v}'\right)\right]\right)\mathrm{Tr}\left(U\left(\mathbf{v}'\right)U^{\dagger}\left(\mathbf{v}\right)\right)\Bigr\rangle_{x}^{*}\\
 & +\left(1-z\right)^{2}\Bigl\langle\mathrm{Tr}\left(\left[\partial_{i}U\left(\mathbf{v}\right)\right]\left[\partial_{j}U^{\dagger}\left(\mathbf{v}'\right)\right]\right)\mathrm{Tr}\left(U\left(\mathbf{v}'\right)U^{\dagger}\left(\mathbf{v}\right)\right)\Bigr\rangle_{x}\\
 & \left.+\Bigl\langle\frac{1}{N_{c}}\mathrm{Tr}\left(\left[\partial_{i}U\left(\mathbf{v}\right)\right]U^{\dagger}\left(\mathbf{v}'\right)\left[\partial_{j}U\left(\mathbf{v}'\right)\right]U^{\dagger}\left(\mathbf{v}\right)\right)\Bigr\rangle_{x}\right].
\end{aligned}
\label{eq:CGCWL}
\end{equation}
In the correlation limit, the cross section, Eq. (\ref{eq:sigmapAqqX}),
therefore becomes:

\begin{equation}
\begin{aligned}\frac{\mathrm{d}\sigma^{pA\rightarrow q\bar{q}X}}{\mathrm{d}\mathcal{P}.\mathcal{S}.} & =\frac{\alpha_{s}}{2\left(N_{c}^{2}-1\right)}\frac{k_{1}^{+}k_{2}^{+}}{p^{+}}x_{p}\mathcal{G}\left(x_{p},\mu^{2}\right)\int\frac{\mathrm{d}^{2}\mathbf{u}}{\left(2\pi\right)^{2}}\frac{\mathrm{d}^{2}\mathbf{u}'}{\left(2\pi\right)^{2}}e^{-i\tilde{\mathbf{P}}_{\perp}\cdot\left(\mathbf{u}-\mathbf{u}'\right)}u_{i}u_{j}'\\
 & \times\sum_{\lambda ss'}\tilde{\phi}_{s's}^{\lambda*}\left(p^{+},k_{1}^{+},\mathbf{u}\right)\tilde{\phi}_{s's}^{\lambda}\left(p^{+},k_{1}^{+},\mathbf{u}'\right)\int\frac{\mathrm{d}^{2}\mathbf{v}}{\left(2\pi\right)^{2}}\frac{\mathrm{d}^{2}\mathbf{v}'}{\left(2\pi\right)^{2}}e^{i\mathbf{q}_{\perp}\cdot\left(\mathbf{v}-\mathbf{v}'\right)}\mathcal{WL}_{ij}\left(\mathbf{v},\mathbf{v}'\right).
\end{aligned}
\label{eq:crosssectioncorrlimit}
\end{equation}
This expression can be simplified further by performing the $\mathbf{u}$
and $\mathbf{u}'$ integrations explicitly. The absolute value squared
of the wave functions, summed (and not averaged) over the two different
polarizations of the gluon as well as over the quark spins, is given
by:
\begin{equation}
\begin{aligned} & \left|\tilde{\phi}^{g\rightarrow q\bar{q}}\left(p^{+},z,\mathbf{u},\mathbf{u}'\right)\right|^{2}\\
 & =\frac{8\pi^{2}}{p^{+}}\left(2m^{2}K_{1}\left(mu\right)K_{1}\left(mu'\right)\frac{\mathbf{u}\cdot\mathbf{u}'}{uu'}P_{qg}\left(z\right)+m^{2}K_{0}\left(mu\right)K_{0}\left(mu'\right)\right),
\end{aligned}
\end{equation}
where
\begin{equation}
\begin{aligned}P_{qg}\left(z\right) & =\frac{z^{2}+\left(1-z\right)^{2}}{2}.\end{aligned}
\end{equation}
Using the identity:
\begin{equation}
\begin{aligned}\frac{\partial}{\partial u_{i}}K_{0}\left(mu\right) & =-m\frac{u_{i}}{u}K_{1}\left(mu\right),\end{aligned}
\end{equation}
the integrals over $\mathbf{u}$ and $\mathbf{u}'$ in Eq. (\ref{eq:crosssectioncorrlimit})
can be performed:
\begin{equation}
\begin{aligned} & \int\frac{\mathrm{d}^{2}\mathbf{u}}{\left(2\pi\right)^{2}}\frac{\mathrm{d}^{2}\mathbf{u}'}{\left(2\pi\right)^{2}}e^{-i\tilde{\mathbf{P}}_{\perp}\cdot\left(\mathbf{u}-\mathbf{u}'\right)}u_{i}u'_{j}\left|\tilde{\phi}^{g\rightarrow q\bar{q}}\left(p^{+},z,\mathbf{u},\mathbf{u}'\right)\right|^{2}\\
 & =\frac{2}{p^{+}}\Biggl[2\Bigl(\frac{\delta_{ij}}{(\tilde{P}_{\perp}^{2}+m^{2})^{2}}-\frac{4m^{2}\tilde{P}_{i}\tilde{P}_{j}}{(\tilde{P}_{\perp}^{2}+m^{2})^{4}}\Bigr)P_{qg}\left(z\right)+\frac{4m^{2}\tilde{P}_{i}\tilde{P}_{j}}{(\tilde{P}_{\perp}^{2}+m^{2})^{4}}\Biggr].
\end{aligned}
\end{equation}
We finally arrive at the following CGC result for the $pA\to q\bar{q}X$
cross section in the correlation limit:
\begin{equation}
\begin{aligned}\frac{\mathrm{d}\sigma^{pA\rightarrow q\bar{q}X}}{\mathrm{d}\mathcal{P}.\mathcal{S}.} & =\frac{2\alpha_{s}}{\left(N_{c}^{2}-1\right)}z\left(1-z\right)x_{p}\mathcal{G}\left(x_{p},\mu^{2}\right)\\
 & \times\left[\Bigl(\frac{\delta_{ij}}{(\tilde{P}_{\perp}^{2}+m^{2})^{2}}-\frac{4m^{2}\tilde{P}_{i}\tilde{P}_{j}}{(\tilde{P}_{\perp}^{2}+m^{2})^{4}}\Bigr)P_{qg}\left(z\right)+\frac{2m^{2}\tilde{P}_{i}\tilde{P}_{j}}{(\tilde{P}_{\perp}^{2}+m^{2})^{4}}\right]\\
 & \times\int\frac{\mathrm{d}^{2}\mathbf{v}}{\left(2\pi\right)^{2}}\frac{\mathrm{d}^{2}\mathbf{v}'}{\left(2\pi\right)^{2}}e^{i\mathbf{q}_{\perp}\cdot\left(\mathbf{v}-\mathbf{v}'\right)}\mathcal{WL}_{ij}\left(\mathbf{v},\mathbf{v}'\right).
\end{aligned}
\label{eq:transverseresult}
\end{equation}

\subsection{Identifying the gluon TMDs}

We will now demonstrate that the Wilson line correlators, Eq. (\ref{eq:CGCWL}),
that appear in the CGC cross section, can in fact be identified as
the small-$x$ limit of gluon TMDs. Indeed, when Fourier transforming
the correlators, each of them can be decomposed into two parts by
projecting the transverse Lorentz indices: a part that is linear in
$\delta^{ij}$, and another one that is traceless. For example:

\begin{equation}
\begin{aligned} & \frac{4}{g_{s}^{2}}\frac{1}{N_{c}}\int\frac{\mathrm{d}^{2}\mathbf{v}\mathrm{d}^{2}\mathbf{v}'}{\left(2\pi\right)^{3}}e^{-i\mathbf{q}_{\perp}\cdot\left(\mathbf{v}-\mathbf{v}'\right)}\Bigl\langle\mathrm{Tr}\left(\left[\partial_{i}U\left(\mathbf{v}\right)\right]\left[\partial_{j}U^{\dagger}\left(\mathbf{v}'\right)\right]\right)\mathrm{Tr}\left(U\left(\mathbf{v}'\right)U^{\dagger}\left(\mathbf{v}\right)\right)\Bigr\rangle_{x}\\
 & =\frac{1}{2}\delta^{ij}\mathcal{F}_{gg}^{\left(1\right)}\left(x,q_{\perp}\right)+\frac{1}{2}\left(\frac{2q_{\perp}^{i}q_{\perp}^{j}}{q_{\perp}^{2}}-\delta^{ij}\right)\mathcal{H}_{gg}^{\left(1\right)}\left(x,q_{\perp}\right).
\end{aligned}
\label{eq:FH1inverted}
\end{equation}
Both parts of the projection turn out to be gluon TMDs, as we will
demonstrate in this subsection. Interestingly, the traceless parts
\textendash in the above formula $\mathcal{H}_{gg}^{\left(1\right)}$\textendash{}
are the TMDs that correspond to the linearly polarized gluons inside
the unpolarized nucleus (Refs. \protect\cite{Mulders2001,Boer2011,Metz2011,fabioqiu}).
Gluon polarization hence does play a role in forward heavy quark production
in dilute-dense collisions, even when the proton and the nucleus themselves
are not polarized.

Indeed, performing the decomposition for the whole Wilson-line structure
$\mathcal{WL}_{ij}$, Eq. (\ref{eq:CGCWL}), we find:

\begin{equation}
\begin{aligned} & \int\frac{\mathrm{d}^{2}\mathbf{v}}{\left(2\pi\right)^{2}}\frac{\mathrm{d}^{2}\mathbf{v}'}{\left(2\pi\right)^{2}}e^{-i\mathbf{q}_{\perp}\cdot\left(\mathbf{v}-\mathbf{v}'\right)}\mathcal{WL}_{ij}\left(\mathbf{v},\mathbf{v}'\right)=\\
 & \frac{\alpha_{s}N_{c}}{2}\Biggl[2z\left(1-z\right)\left(\frac{1}{2}\delta^{ij}\mathcal{F}_{gg}^{\left(2\right)}\left(x,q_{\perp}\right)+\frac{1}{2}\left(\frac{2q_{\perp}^{i}q_{\perp}^{j}}{q_{\perp}^{2}}-\delta^{ij}\right)\mathcal{H}_{gg}^{\left(2\right)}\left(x,q_{\perp}\right)\right)\\
 & +z^{2}\left(\frac{1}{2}\delta^{ij}\mathcal{F}_{gg}^{\left(1\right)}\left(x,q_{\perp}\right)+\frac{1}{2}\left(\frac{2q_{\perp}^{i}q_{\perp}^{j}}{q_{\perp}^{2}}-\delta^{ij}\right)\mathcal{H}_{gg}^{\left(1\right)}\left(x,q_{\perp}\right)\right)\\
 & +\left(1-z\right)^{2}\left(\frac{1}{2}\delta^{ij}\mathcal{F}_{gg}^{\left(1\right)}\left(x,q_{\perp}\right)+\frac{1}{2}\left(\frac{2q_{\perp}^{i}q_{\perp}^{j}}{q_{\perp}^{2}}-\delta^{ij}\right)\mathcal{H}_{gg}^{\left(1\right)}\left(x,q_{\perp}\right)\right)\\
 & -\frac{1}{N_{c}^{2}}\left(\frac{1}{2}\delta^{ij}\mathcal{F}_{gg}^{\left(3\right)}\left(x,q_{\perp}\right)+\frac{1}{2}\left(\frac{2q_{\perp}^{i}q_{\perp}^{j}}{q_{\perp}^{2}}-\delta^{ij}\right)\mathcal{H}_{gg}^{\left(3\right)}\left(x,q_{\perp}\right)\right)\Biggr],
\end{aligned}
\label{eq:WLdecomposition}
\end{equation}
in which we identify six different gluon TMDs: $\mathcal{F}_{gg}^{\left(1\right)},$
$\mathcal{F}_{gg}^{\left(2\right)}$, and $\mathcal{F}_{gg}^{\left(3\right)}$,
defined as:
\begin{equation}
\begin{aligned}\mathcal{F}_{gg}^{\left(1\right)}\left(x,q_{\perp}\right) & \equiv\frac{4}{g_{s}^{2}}\frac{1}{N_{c}}\int\frac{\mathrm{d}^{2}\mathbf{v}\mathrm{d}^{2}\mathbf{v}'}{\left(2\pi\right)^{3}}e^{-i\mathbf{q}_{\perp}\cdot\left(\mathbf{v}-\mathbf{v}'\right)}\Bigl\langle\mathrm{Tr}\left(\left[\partial_{i}U\left(\mathbf{v}\right)\right]\left[\partial_{i}U^{\dagger}\left(\mathbf{v}'\right)\right]\right)\mathrm{Tr}\left(U\left(\mathbf{v}'\right)U^{\dagger}\left(\mathbf{v}\right)\right)\Bigr\rangle_{x},\\
\mathcal{F}_{gg}^{\left(2\right)}\left(x,q_{\perp}\right) & \equiv-\frac{4}{g_{s}^{2}}\frac{1}{N_{c}}\int\frac{\mathrm{d}^{2}\mathbf{v}\mathrm{d}^{2}\mathbf{v}'}{\left(2\pi\right)^{3}}e^{-i\mathbf{q}_{\perp}\cdot\left(\mathbf{v}-\mathbf{v}'\right)}\mathrm{Re}\Bigl\langle\mathrm{Tr}\left(\left[\partial_{i}U\left(\mathbf{v}\right)\right]U^{\dagger}\left(\mathbf{v}'\right)\right)\mathrm{Tr}\left(\left[\partial_{i}U\left(\mathbf{v}'\right)\right]U^{\dagger}\left(\mathbf{v}\right)\right)\Bigr\rangle_{x},\\
\mathcal{F}_{gg}^{\left(3\right)}\left(x,q_{\perp}\right) & \equiv-\frac{4}{g_{s}^{2}}\int\frac{\mathrm{d}^{2}\mathbf{v}\mathrm{d}^{2}\mathbf{v}'}{\left(2\pi\right)^{3}}e^{-i\mathbf{q}_{\perp}\cdot\left(\mathbf{v}-\mathbf{v}'\right)}\mathrm{Tr}\Bigl\langle\left[\partial_{i}U\left(\mathbf{v}\right)\right]U^{\dagger}\left(\mathbf{v}'\right)\left[\partial_{i}U\left(\mathbf{v}'\right)\right]U^{\dagger}\left(\mathbf{v}\right)\Bigr\rangle_{x},
\end{aligned}
\label{eq:F123}
\end{equation}
as well as their \textquoteleft polarized partners' $\mathcal{H}_{gg}^{\left(1\right)}$,
$\mathcal{H}_{gg}^{\left(2\right)}$, and $\mathcal{H}_{gg}^{\left(3\right)}$,
given by 
\begin{equation}
\begin{aligned}\mathcal{H}_{gg}^{\left(1\right)}\left(x,q_{\perp}\right) & \equiv\left(\frac{2q_{\perp}^{i}q_{\perp}^{j}}{q_{\perp}^{2}}-\delta^{ij}\right)\frac{4}{g_{s}^{2}}\frac{1}{N_{c}}\int\frac{\mathrm{d}^{2}\mathbf{v}\mathrm{d}^{2}\mathbf{v}'}{\left(2\pi\right)^{3}}e^{-i\mathbf{q}_{\perp}\cdot\left(\mathbf{v}-\mathbf{v}'\right)}\\
 & \Bigl\langle\mathrm{Tr}\left(\left[\partial_{i}U\left(\mathbf{v}\right)\right]\left[\partial_{j}U^{\dagger}\left(\mathbf{v}'\right)\right]\right)\mathrm{Tr}\left(U\left(\mathbf{v}'\right)U^{\dagger}\left(\mathbf{v}\right)\right)\Bigr\rangle_{x},\\
\mathcal{H}_{gg}^{\left(2\right)}\left(x,q_{\perp}\right) & \equiv\left(\frac{2q_{\perp}^{i}q_{\perp}^{j}}{q_{\perp}^{2}}-\delta^{ij}\right)\left(-\frac{4}{g_{s}^{2}}\right)\frac{1}{N_{c}}\int\frac{\mathrm{d}^{2}\mathbf{v}\mathrm{d}^{2}\mathbf{v}'}{\left(2\pi\right)^{3}}e^{-i\mathbf{q}_{\perp}\cdot\left(\mathbf{v}-\mathbf{v}'\right)}\\
 & \mathrm{Re}\Bigl\langle\mathrm{Tr}\left(\left[\partial_{i}U\left(\mathbf{v}\right)\right]U^{\dagger}\left(\mathbf{v}'\right)\right)\mathrm{Tr}\left(\left[\partial_{j}U\left(\mathbf{v}'\right)\right]U^{\dagger}\left(\mathbf{v}\right)\right)\Bigr\rangle_{x},\\
\mathcal{H}_{gg}^{\left(3\right)}\left(x,q_{\perp}\right) & \equiv\left(\frac{2q_{\perp}^{i}q_{\perp}^{j}}{q_{\perp}^{2}}-\delta^{ij}\right)\left(-\frac{4}{g_{s}^{2}}\right)\int\frac{\mathrm{d}^{2}\mathbf{v}\mathrm{d}^{2}\mathbf{v}'}{\left(2\pi\right)^{3}}e^{-i\mathbf{q}_{\perp}\cdot\left(\mathbf{v}-\mathbf{v}'\right)}\\
 & \mathrm{Tr}\Bigl\langle\left[\partial_{i}U\left(\mathbf{v}\right)\right]U^{\dagger}\left(\mathbf{v}'\right)\left[\partial_{j}U\left(\mathbf{v}'\right)\right]U^{\dagger}\left(\mathbf{v}\right)\Bigr\rangle_{x}.
\end{aligned}
\label{eq:H123}
\end{equation}
On a technical note: in Eq. (\ref{eq:CGCWL}), the Wilson line correlator
for $\mathcal{F}_{gg}^{\left(1\right)}$ and $\mathcal{H}_{gg}^{\left(1\right)}$
that is multiplied by $z^{2}$, is complex conjugated with respect
to Eq. (\ref{eq:FH1inverted}). However, since $\mathcal{F}_{gg}^{\left(1\right)}$
and $\mathcal{H}_{gg}^{\left(1\right)}$ (and the other TMDs) are
a function of $q_{\perp}=|\mathbf{q_{\perp}}|$ and are real, it is
easy to show that the result is the same as the one in definitions
(\ref{eq:F123}) and (\ref{eq:H123}).

Introducing the angle $\phi$ between $\mathbf{\tilde{P}}_{\perp}$
and $\mathbf{q}_{\perp}$:
\begin{equation}
\begin{aligned}\frac{1}{2}\left(\frac{2\tilde{P}_{i}q_{\perp}^{i}\tilde{P}_{j}q_{\perp}^{j}}{q_{\perp}^{2}}-\tilde{P}_{\perp}^{2}\right) & =\frac{1}{2}\tilde{P}_{\perp}^{2}\cos\left(2\phi\right),\end{aligned}
\end{equation}
we obtain the following result for the cross section:
\begin{equation}
\begin{aligned}\frac{\mathrm{d}\sigma^{pA\rightarrow q\bar{q}X}}{\mathrm{d}\mathcal{P}.\mathcal{S}.} & =\frac{\alpha_{s}^{2}}{2C_{F}}\frac{z\left(1-z\right)}{(\tilde{P}_{\perp}^{2}+m^{2})^{2}}x_{p}\mathcal{G}\left(x_{p},\mu^{2}\right)\Biggl\{\biggl(P_{qg}\left(z\right)+z\left(1-z\right)\frac{2m^{2}\tilde{P}_{\perp}^{2}}{(\tilde{P}_{\perp}^{2}+m^{2})^{2}}\biggr)\\
 & \times\left(\left(z^{2}+\left(1-z\right)^{2}\right)\mathcal{F}_{gg}^{\left(1\right)}\left(x,q_{\perp}\right)+2z\left(1-z\right)\mathcal{F}_{gg}^{\left(2\right)}\left(x,q_{\perp}\right)-\frac{1}{N_{c}^{2}}\mathcal{F}_{gg}^{\left(3\right)}\left(x,q_{\perp}\right)\right)\\
 & +z\left(1-z\right)\frac{2m^{2}\tilde{P}_{\perp}^{2}}{(\tilde{P}_{\perp}^{2}+m^{2})^{2}}\cos\left(2\phi\right)\\
 & \times\left(\left(z^{2}+\left(1-z\right)^{2}\right)\mathcal{H}_{gg}^{\left(1\right)}\left(x,q_{\perp}\right)+2z\left(1-z\right)\mathcal{H}_{gg}^{\left(2\right)}\left(x,q_{\perp}\right)-\frac{1}{N_{c}^{2}}\mathcal{H}_{gg}^{\left(3\right)}\left(x,q_{\perp}\right)\right)\Biggr\}.
\end{aligned}
\label{eq:finalCGCcrosssection}
\end{equation}
As is clear from the above formula, the information on the gluon polarization,
encoded in $\mathcal{H}_{gg}^{\left(1,2,3\right)}$, couples to the
mass $m$ of the heavy quarks, and exhibits an angular dependence
$\cos\left(2\phi\right)$, where $\phi$ is the angle between the
transverse momentum of one of the jets, and the transverse-momentum
imbalance of the two jets. 

We will now show that the cross section Eq. (\ref{eq:finalCGCcrosssection})
is the same as the one obtained from the TMD factorization approach.
This involves, as announced, identifying the structures $\mathcal{F}_{gg}^{\left(1,2,3\right)}$
and $\mathcal{H}_{gg}^{\left(1,2,3\right)}$ as being the small-$x$
limits of gluon TMDs. A well-known reference work which provides a
\textquoteleft library' of the different TMDs and their operator definitions
is Ref. \protect\cite{Boer2003}. In order to compare with this work, let
us introduce the following notation for a Wilson line with generic
integration limits: 
\begin{equation}
U\left(a^{+},b^{+};\mathbf{x}\right)\equiv\mathcal{P}\exp\left(ig\int_{a^{+}}^{b^{+}}\mathrm{d}x^{+}A_{c}^{-}\left(x^{+},\mathbf{x}\right)t^{c}\right).
\end{equation}
Note that in our conventions, the path ordering operator sets the
color matrices from left to right in the order they appear along the
path: the further towards the integration limit $b^{+}$, the more
to the right. With this convention, we can easily prove the following
identity:
\begin{equation}
\begin{aligned}\partial_{i}U\left(\mathbf{x}\right) & =\partial_{i}\mathcal{P}\exp\left(ig\int_{-\infty}^{\infty}\mathrm{d}x^{+}A_{c}^{-}\left(x^{+},\mathbf{x}\right)t^{c}\right),\\
 & =ig\int_{-\infty}^{\infty}\mathrm{d}z^{+}U\left(-\infty,z^{+};\mathbf{x}\right)\left[\partial_{i}A_{c}^{-}\left(z^{+},\mathbf{x}\right)t^{c}\right]U\left(z^{+},+\infty;\mathbf{x}\right),\\
 & =ig\int_{-\infty}^{\infty}\mathrm{d}z^{+}U\left(-\infty,z^{+};\mathbf{x}\right)F^{i-}\left(z^{+},\mathbf{x}\right)U\left(z^{+},+\infty;\mathbf{x}\right).
\end{aligned}
\end{equation}
In the last line of the above expression, we used the fact that, choosing
a covariant gauge with $A^{i},A^{+}=0$, the field strength takes
the following form:
\begin{equation}
\begin{aligned}F^{i-}\left(x^{+},\mathbf{x}\right) & =\partial_{i}A_{c}^{-}\left(x^{+},\mathbf{x}\right)t^{c}.\end{aligned}
\end{equation}
As an example, let us rewrite $\mathcal{F}_{gg}^{\left(3\right)}\left(x,q_{\perp}\right)$:
\begin{equation}
\begin{aligned} & \mathcal{F}_{gg}^{\left(3\right)}\left(x,q_{\perp}\right)\equiv-\frac{4}{g_{s}^{2}}\int\frac{\mathrm{d}^{2}\mathbf{v}\mathrm{d}^{2}\mathbf{v}'}{\left(2\pi\right)^{3}}e^{-i\mathbf{q}_{\perp}\cdot\left(\mathbf{v}-\mathbf{v}'\right)}\mathrm{Tr}\Bigl\langle\left[\partial_{i}U\left(\mathbf{v}\right)\right]U^{\dagger}\left(\mathbf{v}'\right)\left[\partial_{i}U\left(\mathbf{v}'\right)\right]U^{\dagger}\left(\mathbf{v}\right)\Bigr\rangle_{x},\\
 & =-\frac{4}{g_{s}^{2}}\left(ig_{s}\right)^{2}\int\frac{\mathrm{d}^{3}v\mathrm{d}^{3}v'}{\left(2\pi\right)^{3}}e^{-i\mathbf{q}_{\perp}\cdot\left(\mathbf{v}-\mathbf{v}'\right)}\mathrm{Tr}\Bigl\langle U\left(-\infty,v^{+};\mathbf{v}\right)F^{i-}\left(v^{+},\mathbf{v}\right)\\
 & \times U\left(v^{+},+\infty;\mathbf{v}\right)U^{\dagger}\left(\mathbf{v}'\right)U\left(-\infty,v'^{+};\mathbf{v}'\right)F^{i-}\left(v'^{+},\mathbf{v}'\right)U\left(v'^{+},+\infty;\mathbf{v}'\right)U^{\dagger}\left(\mathbf{v}\right)\Bigr\rangle_{x}.
\end{aligned}
\label{eq:F3tmdintermediate}
\end{equation}
Using the fact that overlapping Wilson lines that are oriented in
the opposite direction cancel:
\begin{equation}
\begin{aligned}U^{\dagger}\left(\mathbf{x}\right)U\left(-\infty,x{}^{+};\mathbf{x}\right) & =U\left(+\infty,x{}^{+};\mathbf{x}\right)U\left(x^{+},-\infty;\mathbf{x}\right)U\left(-\infty,x{}^{+};\mathbf{x}\right),\\
 & =U\left(+\infty,x{}^{+};\mathbf{x}\right),
\end{aligned}
\end{equation}
expression (\ref{eq:F3tmdintermediate}) becomes:
\begin{equation}
\begin{aligned}\mathcal{F}_{gg}^{\left(3\right)}\left(x,q_{\perp}\right) & =4\int\frac{\mathrm{d}^{3}v\mathrm{d}^{3}v'}{\left(2\pi\right)^{3}}e^{-i\mathbf{q}_{\perp}\cdot\left(\mathbf{v}-\mathbf{v}'\right)}\mathrm{Tr}\Bigl\langle F^{i-}\left(v^{+},\mathbf{v}\right)U\left(v^{+},+\infty;\mathbf{v}\right)U\left(+\infty,v'^{+};\mathbf{v}'\right)\\
 & \times F^{i-}\left(v'^{+},\mathbf{v}'\right)U\left(v'^{+},+\infty;\mathbf{v}'\right)U\left(+\infty,v^{+};\mathbf{v}\right)\Bigr\rangle_{x}.
\end{aligned}
\end{equation}
With the techniques explained in Sec. \ref{sec:GluonTMDs} (invariance
of the CGC under translations in the transverse plane, writing the
CGC average in terms of the hadron states, Eq. (\ref{eq:Fock2hadronic})),
we recover the Weizsäcker-Williams gluon distribution, which we introduced
in Sec. \ref{sec:GluonTMDs} and calculated in the MV model in Sec.
\ref{sec:McLerran-Venugopalan-model}:
\begin{equation}
\begin{aligned}\mathcal{F}_{gg}^{\left(3\right)}\left(x,q_{\perp}\right) & =2\int\frac{\mathrm{d}^{3}\xi}{\left(2\pi\right)^{3}p_{A}^{-}}e^{ixp_{A}^{-}\xi^{+}}e^{-i\mathbf{q}_{\perp}\cdot\boldsymbol{\xi}}\mathrm{Tr}\Bigl\langle A\Bigr|F^{i-}\left(\vec{\xi}\right)U^{\left[+\right]\dagger}F^{i-}\left(\vec{0}\right)U^{\left[+\right]}\Bigl|A\Bigr\rangle.\end{aligned}
\label{eq:F3asTMD}
\end{equation}
In the above expression, we introduced the Bjorken-$x$-dependent
exponential $e^{ixp_{A}^{-}\xi^{+}}$, which in the small-$x$ limit
was set equal to one. Note this exponential is the only way in which
the TMDs depend on $x$ in the TMD factorization approach, while in
the CGC-description of the TMDs: Eqs. (\ref{eq:F123}) and (\ref{eq:H123}),
the $x$-dependence is hidden within the CGC averages $\langle\hat{\mathcal{O}}\rangle_{x}$.

Expression (\ref{eq:F3asTMD}) proves our claim that $\mathcal{F}_{gg}^{\left(3\right)}\left(x,q_{\perp}\right)$,
introduced as the Wilson line correlator in the decomposition Eq.
(\ref{eq:WLdecomposition}), is indeed a gluon TMD in the small-$x$
limit. Using the same techniques, we can do the same for the other
structures defined in Eqs. (\ref{eq:F123}) and (\ref{eq:H123}):
\begin{equation}
\begin{aligned}\mathcal{F}_{gg}^{\left(1\right)}\left(x,q_{\perp}\right) & =\frac{2}{N_{c}}\int\frac{\mathrm{d}^{3}\xi}{\left(2\pi\right)^{3}p_{A}^{-}}e^{ixp_{A}^{-}\xi^{+}}e^{-i\mathbf{q}_{\perp}\cdot\boldsymbol{\xi}}\\
 & \Bigl\langle A\Bigr|\mathrm{Tr}\left(F^{i-}\left(\vec{\xi}\right)U^{\left[-\right]\dagger}F^{i-}\left(\vec{0}\right)U^{\left[+\right]}\right)\mathrm{Tr}\left(U^{\left[-\right]}U^{\left[+\right]\dagger}\right)\Bigl|A\Bigr\rangle,\\
\mathcal{F}_{gg}^{\left(2\right)}\left(x,q_{\perp}\right) & =\frac{2}{N_{c}}\int\frac{\mathrm{d}^{3}\xi}{\left(2\pi\right)^{3}p_{A}^{-}}e^{ixp_{A}^{-}\xi^{+}}e^{-i\mathbf{q}_{\perp}\cdot\boldsymbol{\xi}}\\
 & \mathrm{Re}\Bigl\langle A\Bigr|\mathrm{Tr}\left(F^{i-}\left(\vec{\xi}\right)U^{\left[+\right]\dagger}U^{\left[-\right]}\right)\mathrm{Tr}\left(F^{i-}\left(\vec{0}\right)U^{\left[+\right]}U^{\left[-\right]\dagger}\right)\Bigl|A\Bigr\rangle.
\end{aligned}
\end{equation}
As we already mentioned, the TMDs are required to be real, as well
as symmetric in $\mathbf{q}_{\perp}$, and hence one can write equivalently:
\begin{equation}
\begin{aligned}\mathcal{F}_{gg}^{\left(1\right)}\left(x,q_{\perp}\right) & =\frac{2}{N_{c}}\int\frac{\mathrm{d}^{3}\xi}{\left(2\pi\right)^{3}p_{A}^{-}}e^{-ixp_{A}^{-}\xi^{+}}e^{-i\mathbf{q}_{\perp}\cdot\boldsymbol{\xi}}\\
 & \Bigl\langle A\Bigr|\mathrm{Tr}\left(F^{i-}\left(\vec{\xi}\right)U^{\left[+\right]\dagger}F^{i-}\left(\vec{0}\right)U^{\left[-\right]}\right)\mathrm{Tr}\left(U^{\left[+\right]}U^{\left[-\right]\dagger}\right)\Bigl|A\Bigr\rangle,\\
\mathcal{F}_{gg}^{\left(2\right)}\left(x,q_{\perp}\right) & =\frac{2}{N_{c}}\int\frac{\mathrm{d}^{3}\xi}{\left(2\pi\right)^{3}p_{A}^{-}}e^{-ixp_{A}^{-}\xi^{+}}e^{-i\mathbf{q}_{\perp}\cdot\boldsymbol{\xi}}\\
 & \mathrm{Re}\Bigl\langle A\Bigr|\mathrm{Tr}\left(F^{i-}\left(\vec{\xi}\right)U^{\left[-\right]\dagger}U^{\left[+\right]}\right)\mathrm{Tr}\left(F^{i-}\left(\vec{0}\right)U^{\left[-\right]}U^{\left[+\right]\dagger}\right)\Bigl|A\Bigr\rangle.
\end{aligned}
\end{equation}
Likewise, we find the following expressions for the \textquoteleft polarized'
gluon TMDs $\mathcal{H}_{gg}^{\left(1\right)}$, $\mathcal{H}_{gg}^{\left(2\right)}$
and $\mathcal{H}_{gg}^{\left(3\right)}$:
\begin{equation}
\begin{aligned}\mathcal{H}_{gg}^{\left(1\right)}\left(x,q_{\perp}\right) & \equiv\left(\frac{2q_{\perp}^{i}q_{\perp}^{j}}{q_{\perp}^{2}}-\delta^{ij}\right)\frac{2}{N_{c}}\int\frac{\mathrm{d}^{3}\xi}{\left(2\pi\right)^{3}p_{A}^{-}}e^{ixp_{A}^{-}\xi^{+}}e^{-i\mathbf{q}_{\perp}\cdot\boldsymbol{\xi}}\\
 & \Bigl\langle A\Bigr|\mathrm{Tr}\left(F^{i-}\left(\vec{\xi}\right)U^{\left[-\right]\dagger}F^{j-}\left(\vec{0}\right)U^{\left[+\right]}\right)\mathrm{Tr}\left(U^{\left[-\right]}U^{\left[+\right]\dagger}\right)\Bigl|A\Bigr\rangle,\\
\mathcal{H}_{gg}^{\left(2\right)}\left(x,q_{\perp}\right) & \equiv\left(\frac{2q_{\perp}^{i}q_{\perp}^{j}}{q_{\perp}^{2}}-\delta^{ij}\right)\frac{2}{N_{c}}\int\frac{\mathrm{d}^{3}\xi}{\left(2\pi\right)^{3}p_{A}^{-}}e^{ixp_{A}^{-}\xi^{+}}e^{-i\mathbf{q}_{\perp}\cdot\boldsymbol{\xi}}\\
 & \mathrm{Re}\Bigl\langle A\Bigr|\mathrm{Tr}\left(F^{i-}\left(\vec{\xi}\right)U^{\left[+\right]\dagger}U^{\left[-\right]}\right)\mathrm{Tr}\left(F^{i-}\left(\vec{0}\right)U^{\left[+\right]}U^{\left[-\right]\dagger}\right)\Bigl|A\Bigr\rangle,\\
\mathcal{H}_{gg}^{\left(3\right)}\left(x,q_{\perp}\right) & \equiv\left(\frac{2q_{\perp}^{i}q_{\perp}^{j}}{q_{\perp}^{2}}-\delta^{ij}\right)\\
 & \times2\int\frac{\mathrm{d}^{3}\xi}{\left(2\pi\right)^{3}p_{A}^{-}}e^{ixp_{A}^{-}\xi^{+}}e^{-i\mathbf{q}_{\perp}\cdot\boldsymbol{\xi}}\mathrm{Tr}\Bigl\langle A\Bigr|F^{i-}\left(\vec{\xi}\right)U^{\left[+\right]\dagger}F^{i-}\left(\vec{0}\right)U^{\left[+\right]}\Bigl|A\Bigr\rangle.
\end{aligned}
\end{equation}
It is worth noting that $\mathcal{F}_{gg}^{\left(1\right)}$, $\mathcal{F}_{gg}^{\left(2\right)}$,
$\mathcal{H}_{gg}^{\left(1\right)}$, and $\mathcal{H}_{gg}^{\left(2\right)}$
are not independent but can be related to each other with the help
of the dipole distribution in the adjoint representation, $xG_{A}^{\left(2\right)}\left(x,q_{\perp}\right)$,
defined as:
\begin{equation}
\begin{aligned}xG_{A}^{\left(2\right)}\left(x,q_{\perp}\right) & \equiv4\frac{C_{F}}{g_{s}^{2}}\int\frac{\mathrm{d}^{2}\mathbf{v}\mathrm{d}^{2}\mathbf{v}'}{\left(2\pi\right)^{3}}e^{-i\mathbf{q}_{\perp}\left(\mathbf{v}-\mathbf{v}'\right)}\left.\frac{\partial^{2}}{\partial x^{i}\partial y^{i}}D_{A}\left(\mathbf{x}-\mathbf{y}\right)\right|_{\mathbf{x}=\mathbf{v},\,\mathbf{y}=\mathbf{v}'},\end{aligned}
\label{eq:xG2A}
\end{equation}
where:
\begin{equation}
D_{A}\left(\mathbf{x}-\mathbf{y}\right)\equiv\frac{1}{N_{c}^{2}-1}\mathrm{Tr}\Bigl\langle W\left(\mathbf{x}\right)W^{\dagger}\left(\mathbf{y}\right)\Bigr\rangle_{x}.
\end{equation}
We already showed in Eq. (\ref{eq:WW2UUUU}) that:
\begin{equation}
\begin{aligned}D_{A}\left(\mathbf{x}-\mathbf{y}\right) & =\frac{1}{N_{c}^{2}-1}\left(\Bigl\langle\mathrm{Tr}\left(U\left(\mathbf{x}\right)U^{\dagger}\left(\mathbf{y}\right)\right)\mathrm{Tr}\left(U\left(\mathbf{y}\right)U^{\dagger}\left(\mathbf{x}\right)\right)\Bigr\rangle_{x}-1\right).\end{aligned}
\end{equation}
Taking the derivatives, one then obtains:
\begin{equation}
\begin{aligned} & \left.\frac{\partial^{2}}{\partial x^{i}\partial y^{j}}D_{A}\left(\mathbf{x}-\mathbf{y}\right)\right|_{\mathbf{x}=\mathbf{v},\,\mathbf{y}=\mathbf{v}'}\\
 & =\frac{1}{N_{c}^{2}-1}\Biggl\{\Bigl\langle\mathrm{Tr}\left(U\left(\mathbf{v}\right)U^{\dagger}\left(\mathbf{v}'\right)\right)\mathrm{Tr}\left(\left[\partial_{j}U\left(\mathbf{v}'\right)\right]\left[\partial_{i}U^{\dagger}\left(\mathbf{v}\right)\right]\right)\Bigr\rangle_{x}\\
 & +\Bigl\langle\mathrm{Tr}\left(U\left(\mathbf{v}'\right)U^{\dagger}\left(\mathbf{v}\right)\right)\mathrm{Tr}\left(\left[\partial_{i}U\left(\mathbf{v}\right)\right]\left[\partial_{j}U^{\dagger}\left(\mathbf{v}'\right)\right]\right)\Bigr\rangle_{x}\\
 & +2\mathrm{Re}\Bigl\langle\mathrm{Tr}\left(\left[\partial_{i}U\left(\mathbf{v}\right)\right]U^{\dagger}\left(\mathbf{v}'\right)\right)\mathrm{Tr}\left(\left[\partial_{j}U\left(\mathbf{v}'\right)\right]U^{\dagger}\left(\mathbf{v}\right)\right)\Bigr\rangle_{x}\Biggr\}.
\end{aligned}
\label{eq:DA2F1F2}
\end{equation}
In the definitions (\ref{eq:F123}) and (\ref{eq:H123}) of the gluon
TMDs, $\mathbf{v}$ and $\mathbf{v}'$ can be interchanged, owing
to the fact that the gluon TMDs are functions of $\left|q_{\perp}\right|$
only. Therefore, if we multiply Eq. (\ref{eq:DA2F1F2}) with $\delta_{ij}$
and plug it into the definition of $xG_{A}^{\left(2\right)}\left(x,q_{\perp}\right)$,
Eq. (\ref{eq:xG2A}), the last term of the above equation becomes
equal to $-\mathcal{F}_{gg}^{\left(2\right)}\left(x,q_{\perp}\right)$,
whereas both the first and the second correlator contribute to $\mathcal{F}_{gg}^{\left(1\right)}\left(x,q_{\perp}\right)$,
yielding the following sum rule:
\begin{equation}
\begin{aligned}\mathcal{F}_{gg}^{\left(1\right)}\left(x,q_{\perp}\right)-\mathcal{F}_{gg}^{\left(2\right)}\left(x,q_{\perp}\right) & =xG_{A}^{\left(2\right)}\left(x,q_{\perp}\right).\end{aligned}
\label{eq:F1-F2general}
\end{equation}
Using the same argument, in combination with the fact that:
\begin{equation}
\begin{aligned} & xG_{A}^{\left(2\right)}\left(x,q_{\perp}\right)\\
 & =4\frac{C_{F}}{g_{s}^{2}}\int\frac{\mathrm{d}^{2}\mathbf{v}\mathrm{d}^{2}\mathbf{v}'}{\left(2\pi\right)^{3}}q_{\perp}^{2}e^{-i\mathbf{q}_{\perp}\left(\mathbf{v}-\mathbf{v}'\right)}D_{A}\left(\mathbf{v}-\mathbf{v}'\right),\\
 & =4\frac{C_{F}}{g_{s}^{2}}\int\frac{\mathrm{d}^{2}\mathbf{v}\mathrm{d}^{2}\mathbf{v}'}{\left(2\pi\right)^{3}}e^{-i\mathbf{q}_{\perp}\left(\mathbf{v}-\mathbf{v}'\right)}\left(\frac{2q_{i}q_{j}}{q_{\perp}^{2}}-\delta_{ij}\right)\frac{\partial^{2}}{\partial v^{i}\partial v'^{j}}D_{A}\left(\mathbf{v}-\mathbf{v}'\right),
\end{aligned}
\label{eq:PIGA}
\end{equation}
we immediately obtain as well:
\begin{equation}
\begin{aligned}\mathcal{H}_{gg}^{\left(1\right)}\left(x,q_{\perp}\right)-\mathcal{H}_{gg}^{\left(2\right)}\left(x,q_{\perp}\right) & =xG_{A}^{\left(2\right)}\left(x,q_{\perp}\right).\end{aligned}
\label{eq:H1-H2general}
\end{equation}

The TMD $\mathcal{F}_{gg}^{\left(3\right)}\left(x,q_{\perp}\right)$
is the very well-known Weizsäcker-Williams distribution, which we
introduced already in Sec. \ref{sec:GluonTMDs}. It is the only gluon
TMD that allows, in the light-cone gauge, for an interpretation as
the number density of gluons.

\section{\label{sec:ComparisonTMD}Matching the CGC approach with the TMD
calculations}

In the previous section, we identified the individual TMDs that appear
in the CGC cross section, Eq. (\ref{eq:finalCGCcrosssection}). The
next step is now to compare the CGC cross section itself with the
results that are obtained from the calculation in the TMD approach.
In order to do this, we have to rewrite our cross section in function
of the Mandelstam variables Eq. (\ref{eq:Mandelstam}). In the correlation
limit, we have $k_{1\perp}^{2}\simeq k_{2\perp}^{2}\simeq\tilde{P}_{\perp}^{2}$,
and we can easily express $z$ and $\tilde{P}_{\perp}^{2}$ as functions
of $\hat{s}$, $\hat{t}$ and $\hat{u}$:

\begin{equation}
z=\frac{m^{2}-\hat{u}}{\hat{s}},\quad1-z=\frac{m^{2}-\hat{t}}{\hat{s}},\quad\tilde{P}_{\perp}^{2}=\frac{\hat{u}\hat{t}-m^{4}}{\hat{s}}.\label{eq:Mandelstamrelations}
\end{equation}
With the help of the above relations, the CGC cross section, Eq. (\ref{eq:finalCGCcrosssection}),
becomes:
\begin{equation}
\begin{aligned} & \frac{\mathrm{d}\sigma^{pA\rightarrow q\bar{q}X}}{\mathrm{d}\mathcal{P}.\mathcal{S}.}=\frac{\alpha_{s}^{2}}{2C_{F}}\frac{\left(m^{2}-\hat{t}\right)\left(m^{2}-\hat{u}\right)}{\left(\hat{u}\hat{t}+m^{2}\hat{s}-m^{4}\right)^{2}}x_{p}\mathcal{G}\left(x_{p},\mu^{2}\right)\\
 & \Biggl\{\left(\frac{\left(m^{2}-\hat{t}\right)^{2}+\left(m^{2}-\hat{u}\right)^{2}}{2\hat{s}^{2}}+2m^{2}\frac{\hat{u}\hat{t}-m^{4}}{\hat{s}}\frac{\left(m^{2}-\hat{t}\right)\left(m^{2}-\hat{u}\right)}{\left(\hat{u}\hat{t}+m^{2}\hat{s}-m^{4}\right)^{2}}\right)\\
 & \times\left(\frac{\left(m^{2}-\hat{t}\right)^{2}+\left(m^{2}-\hat{u}\right)^{2}}{\hat{s}^{2}}\mathcal{F}_{gg}^{\left(1\right)}\left(x,q_{\perp}\right)+2\frac{\left(m^{2}-\hat{t}\right)\left(m^{2}-\hat{u}\right)}{\hat{s}^{2}}\mathcal{F}_{gg}^{\left(2\right)}\left(x,q_{\perp}\right)-\frac{1}{N_{c}^{2}}\mathcal{F}_{gg}^{\left(3\right)}\left(x,q_{\perp}\right)\right)\\
 & +2m^{2}\frac{\hat{u}\hat{t}-m^{4}}{\hat{s}}\frac{\left(m^{2}-\hat{t}\right)\left(m^{2}-\hat{u}\right)}{\left(\hat{u}\hat{t}+m^{2}\hat{s}-m^{4}\right)^{2}}\cos\left(2\phi\right)\\
 & \times\left(\frac{\left(m^{2}-\hat{t}\right)^{2}+\left(m^{2}-\hat{u}\right)^{2}}{\hat{s}^{2}}\mathcal{H}_{gg}^{\left(1\right)}\left(x,q_{\perp}\right)+2\frac{\left(m^{2}-\hat{t}\right)\left(m^{2}-\hat{u}\right)}{\hat{s}^{2}}\mathcal{H}_{gg}^{\left(2\right)}\left(x,q_{\perp}\right)-\frac{1}{N_{c}^{2}}\mathcal{H}_{gg}^{\left(3\right)}\left(x,q_{\perp}\right)\right)\Biggr\}.
\end{aligned}
\label{eq:finalCGCcrosssectionMandelstam}
\end{equation}
In the massless limit, the angular dependence is lost, and the cross
section is reduced to the following expression:
\begin{equation}
\begin{aligned}\left.\frac{\mathrm{d}\sigma^{pA\rightarrow q\bar{q}X}}{\mathrm{d}\mathcal{P}.\mathcal{S}.}\right|_{m=0} & =\frac{\alpha_{s}^{2}}{2C_{F}}\frac{1}{\hat{s}^{2}}x_{p}\mathcal{G}\left(x_{p},\mu^{2}\right)\frac{\hat{t}^{2}+\hat{u}^{2}}{2\hat{u}\hat{t}}\\
 & \times\left\{ \frac{\hat{t}^{2}+\hat{u}^{2}}{\hat{s}^{2}}\mathcal{F}_{gg}^{\left(1\right)}\left(x,q_{\perp}\right)+\frac{2\hat{u}\hat{t}}{\hat{s}^{2}}\mathcal{F}_{gg}^{\left(2\right)}\left(x,q_{\perp}\right)-\frac{1}{N_{c}^{2}}\mathcal{F}_{gg}^{\left(3\right)}\left(x,q_{\perp}\right)\right\} .
\end{aligned}
\label{eq:TMDcrosssectionunp}
\end{equation}
which is the same as the one obtained within the TMD framework (Eq.
(65) of \protect\cite{fabio} in the large-$N_{c}$ limit, Eq. (4.14) in \protect\cite{Kotko2015}
at finite-$N_{c}$).

Likewise, we find for the polarized part of the cross section (at
leading order in $m^{2}/\tilde{P}_{\perp}^{2}$):
\begin{equation}
\begin{aligned}\left.\frac{\mathrm{d}\sigma^{pA\rightarrow q\bar{q}X}}{\mathrm{d}\mathcal{P}.\mathcal{S}.}\right|_{\phi} & =\frac{\alpha_{s}^{2}}{2C_{F}}\frac{1}{\hat{s}^{2}}x_{p}\mathcal{G}\left(x_{p},\mu^{2}\right)\frac{2m^{2}}{\tilde{P}_{\perp}^{2}}\cos\left(2\phi\right)\\
 & \times\left\{ \frac{\hat{t}^{2}+\hat{u}^{2}}{\hat{s}^{2}}\mathcal{H}_{gg}^{\left(1\right)}\left(x,q_{\perp}\right)+\frac{2\hat{u}\hat{t}}{\hat{s}^{2}}\mathcal{H}_{gg}^{\left(2\right)}\left(x,q_{\perp}\right)-\frac{1}{N_{c}^{2}}\mathcal{H}_{gg}^{\left(3\right)}\left(x,q_{\perp}\right)\right\} ,
\end{aligned}
\label{eq:resultcrosssectionpol}
\end{equation}
Using the sum rules of Eqs. (\ref{eq:F1-F2general}) and (\ref{eq:H1-H2general}),
$\mathcal{F}_{gg}^{\left(2\right)}\left(x,q_{\perp}\right)$ and $\mathcal{H}_{gg}^{\left(2\right)}\left(x,q_{\perp}\right)$
can be eliminated in favor of the better-known adjoint dipole TMD,
yielding, again in leading order of $m^{2}/\tilde{P}_{\perp}^{2}$:
\begin{equation}
\begin{aligned}\left.\frac{\mathrm{d}\sigma^{pA\rightarrow q\bar{q}X}}{\mathrm{d}\mathcal{P}.\mathcal{S}.}\right|_{m=0} & =\frac{\alpha_{s}^{2}}{2C_{F}}\frac{1}{\hat{s}^{2}}x_{p}\mathcal{G}\left(x_{p},\mu^{2}\right)\frac{\hat{t}^{2}+\hat{u}^{2}}{2\hat{u}\hat{t}}\\
 & \times\left\{ \mathcal{F}_{gg}^{\left(1\right)}\left(x,q_{\perp}\right)-\frac{2\hat{u}\hat{t}}{\hat{s}^{2}}xG_{A}^{\left(2\right)}\left(x,q_{\perp}\right)-\frac{1}{N_{c}^{2}}\mathcal{F}_{gg}^{\left(3\right)}\left(x,q_{\perp}\right)\right\} ,
\end{aligned}
\end{equation}
and:
\begin{equation}
\begin{aligned}\left.\frac{\mathrm{d}\sigma^{pA\rightarrow q\bar{q}X}}{\mathrm{d}\mathcal{P}.\mathcal{S}.}\right|_{\phi} & =\frac{\alpha_{s}^{2}}{2C_{F}}\frac{1}{\hat{s}^{2}}x_{p}\mathcal{G}\left(x_{p},\mu^{2}\right)\frac{2m^{2}}{\tilde{P}_{\perp}^{2}}\cos\left(2\phi\right)\\
 & \times\left\{ \mathcal{H}_{gg}^{\left(1\right)}\left(x,q_{\perp}\right)-\frac{2\hat{u}\hat{t}}{\hat{s}^{2}}xG_{A}^{\left(2\right)}\left(x,q_{\perp}\right)-\frac{1}{N_{c}^{2}}\mathcal{H}_{gg}^{\left(3\right)}\left(x,q_{\perp}\right)\right\} .
\end{aligned}
\label{eq:resultcrosssectionpolDP}
\end{equation}
Expressions (\ref{eq:resultcrosssectionpol}) and (\ref{eq:resultcrosssectionpolDP})
for the polarized cross section are the main analytical results of
this work. In an independent study, Ref. \protect\cite{zhou}, the authors
follow a similar approach, but limit themselves to the GBW model for
the gluon TMDs. While being equally simple as the one found in Ref.
\protect\cite{zhou}, our result is more general, since it is model-independent.
We will demonstrate that, when explicitly evaluating the TMDs in the
GBW model, Eq. (\ref{eq:resultcrosssectionpolDP}) indeed coincides
with the findings of our colleagues.

$\mathcal{F}_{gg}^{\left(1,2,3\right)}$ and $\mathcal{H}_{gg}^{\left(1,2,3\right)}$
can be evaluated analytically in the McLerran-Venugopalan model, at
finite-$N_{c}$. The results are:
\begin{equation}
\begin{aligned}\mathcal{F}_{gg}^{\left(1\right)}\left(x,q_{\perp}\right) & =\frac{S_{\perp}}{\alpha_{s}}\frac{C_{F}}{N_{c}^{2}}\frac{1}{32\pi^{3}}\int\mathrm{d}r\frac{J_{0}\left(q_{\perp}r\right)}{r}e^{-\frac{N_{c}}{2}\Gamma\left(r\right)}\Biggl[64\left(e^{\frac{N_{c}}{2}\Gamma\left(r\right)}-1\right)\\
 & -\alpha_{s}^{2}N_{c}^{4}\mu_{A}^{2}r^{4}\left(1-2\gamma_{E}+\ln4+\ln\frac{1}{r^{2}\Lambda^{2}}\right)^{2}+8\alpha_{s}N_{c}\left(N_{c}^{2}-2\right)\mu_{A}r^{2}\ln\frac{1}{r^{2}\Lambda^{2}}\Biggr)\Biggr]
\end{aligned}
\label{eq:F1finiteNc-1}
\end{equation}
\begin{equation}
\begin{aligned}\mathcal{H}_{gg}^{\left(1\right)}\left(x,q_{\perp}\right) & =\frac{S_{\perp}}{\alpha_{s}}\frac{C_{F}}{N_{c}^{2}}\frac{1}{32\pi^{3}}\int\mathrm{d}r\frac{J_{2}\left(q_{\perp}r\right)}{r}e^{-\frac{N_{c}}{2}\Gamma\left(r\right)}\Biggl[64\frac{1}{\ln\frac{1}{r^{2}\Lambda^{2}}}\left(e^{\frac{N_{c}}{2}\Gamma\left(r\right)}-1\right)\\
 & +\alpha_{s}^{2}N_{c}^{4}\mu_{A}^{2}r^{4}\left(1-2\gamma_{E}+\ln4+\ln\frac{1}{r^{2}\Lambda^{2}}\right)^{2}+8\alpha_{s}\mu_{A}N_{c}\left(N_{c}^{2}-2\right)r^{2}\Biggr]
\end{aligned}
\label{eq:H1finiteNc-1}
\end{equation}
\begin{equation}
\begin{aligned}\mathcal{F}_{gg}^{\left(2\right)}\left(x,q_{\perp}\right) & =\frac{S_{\perp}}{\alpha_{s}}\frac{C_{F}}{N_{c}^{2}}\frac{1}{32\pi^{3}}\int\mathrm{d}r\frac{J_{0}\left(q_{\perp}r\right)}{r}e^{-\frac{N_{c}}{2}\Gamma\left(r\right)}\Biggl[64\left(e^{\frac{N_{c}}{2}\Gamma\left(r\right)}-1\right)\\
 & +N_{c}^{4}\mu_{A}^{2}\alpha_{s}^{2}r^{4}\left(1-2\gamma_{E}+\ln4+\ln\frac{1}{r^{2}\Lambda^{2}}\right)^{2}-16\alpha_{s}N_{c}\mu_{A}r^{2}\ln\frac{1}{r^{2}\Lambda^{2}}\Biggr]
\end{aligned}
\label{eq:F2finiteNc-1}
\end{equation}
\begin{equation}
\begin{aligned} & \mathcal{H}_{gg}^{\left(2\right)}\left(x,q_{\perp}\right)=\frac{S_{\perp}}{\alpha_{s}}\frac{C_{F}}{N_{c}^{2}}\frac{1}{32\pi^{3}}\int\mathrm{d}r\frac{J_{2}\left(q_{\perp}r\right)}{r}e^{-\frac{N_{c}}{2}\Gamma\left(r\right)}\\
 & \Biggl[64\left(e^{\frac{N_{c}}{2}\Gamma\left(r\right)}-1\right)\frac{1}{\ln\frac{1}{r^{2}\Lambda^{2}}}-\alpha_{s}^{2}N_{c}^{4}\mu_{A}^{2}r^{4}\left(1-2\gamma_{E}+\ln4+\ln\frac{1}{r^{2}\Lambda^{2}}\right)^{2}-16\alpha_{s}N_{c}\mu_{A}r^{2}\Biggr]
\end{aligned}
\label{eq:H2finiteNc-1}
\end{equation}
The expressions for the Weizsäcker-Williams gluon TMD, $\mathcal{F}_{gg}^{\left(3\right)}\left(x,q_{\perp}\right)$,
and its polarized partner $\mathcal{H}_{gg}^{\left(3\right)}\left(x,q_{\perp}\right)$
in the MV model (see Eqs. (\ref{eq:WWfiniteNc}) and (\ref{eq:H3finiteNc}))
were already established in the literature, see for instance \protect\cite{Dominguez2011,Metz2011,fabioqiu}. 

We are now ready to turn to the numerical part of this work, in which
we implement our gluon TMDs on the lattice and evolve them with JIMWLK.
The following section is devoted to casting expressions (\ref{eq:F123})
and (\ref{eq:H123}) in a more appropriate form. 

\section{Implementation on the lattice}

In this short chapter we derive, for completeness, the particular
forms in which we cast the gluon TMDs in order to implement them numerically.
Our starting points are the definitions Eqs. (\ref{eq:F123}), (\ref{eq:H123}).
For example, $\mathcal{F}_{gg}^{\left(1\right)}\left(x,q_{\perp}\right)$
can be written as:

\begin{equation}
\begin{aligned}\mathcal{F}_{gg}^{\left(1\right)}\left(x,q_{\perp}\right) & =\frac{4}{g_{s}^{2}}\frac{1}{N_{c}}\int\frac{\mathrm{d}^{2}\mathbf{v}\mathrm{d}^{2}\mathbf{v}'}{\left(2\pi\right)^{3}}e^{-i\mathbf{q}_{\perp}\cdot\left(\mathbf{v}-\mathbf{v}'\right)}\Bigl\langle\mathrm{Tr}\left(\left[\partial_{i}U\left(\mathbf{v}\right)\right]\left[\partial_{i}U^{\dagger}\left(\mathbf{v}'\right)\right]\right)\mathrm{Tr}\left(U\left(\mathbf{v}'\right)U^{\dagger}\left(\mathbf{v}\right)\right)\Bigr\rangle_{x},\\
 & =\frac{4}{g_{s}^{2}}\frac{1}{N_{c}}\int\frac{\mathrm{d}^{2}\mathbf{v}\mathrm{d}^{2}\mathbf{v}'}{\left(2\pi\right)^{3}}e^{-i\mathbf{q}_{\perp}\cdot\left(\mathbf{v}-\mathbf{v}'\right)}\Bigl\langle\left[\partial_{i}U\left(\mathbf{v}\right)\right]_{ij}\left[\partial_{i}U^{\dagger}\left(\mathbf{v}'\right)\right]_{ji}U\left(\mathbf{v}'\right)_{kl}U^{\dagger}\left(\mathbf{v}\right)_{lk}\Bigr\rangle_{x},\\
 & =\frac{4}{g_{s}^{2}}\frac{1}{N_{c}}\int\frac{\mathrm{d}^{2}\mathbf{v}\mathrm{d}^{2}\mathbf{v}'}{\left(2\pi\right)^{3}}e^{-i\mathbf{q}_{\perp}\cdot\left(\mathbf{v}-\mathbf{v}'\right)}\Bigl\langle\left[\partial_{i}U\left(\mathbf{v}\right)\right]_{ij}U^{\dagger}\left(\mathbf{v}\right)_{lk}\left[\partial_{i}U^{\dagger}\left(\mathbf{v}'\right)\right]_{ji}U\left(\mathbf{v}'\right)_{kl}\Bigr\rangle_{x},\\
 & =\frac{4}{g_{s}^{2}}\frac{1}{N_{c}}\int\frac{\mathrm{d}^{2}\mathbf{v}\mathrm{d}^{2}\mathbf{v}'}{\left(2\pi\right)^{3}}e^{-i\mathbf{q}_{\perp}\cdot\left(\mathbf{v}-\mathbf{v}'\right)}\Bigl\langle\left[\partial_{i}U\left(\mathbf{v}\right)\right]U^{\dagger}\left(\mathbf{v}\right)_{lk}\left(\left[\partial_{i}U\left(\mathbf{v}'\right)\right]_{ij}U^{\dagger}\left(\mathbf{v}'\right)_{lk}\right)^{*}\Bigr\rangle_{x},\\
 & =\frac{8\pi}{g_{s}^{2}}\frac{1}{N_{c}}\sum_{i=1}^{2}\sum_{ij=1}^{N_{c}^{2}}\sum_{lk=1}^{N_{c}^{2}}\Bigl\langle\Bigl|\int\frac{\mathrm{d}^{2}\mathbf{v}}{\left(2\pi\right)^{2}}e^{-i\mathbf{q}_{\perp}\cdot\mathbf{v}}\left[\partial_{i}U\left(\mathbf{v}\right)\right]_{ij}U^{\dagger}\left(\mathbf{v}\right)_{lk}\Bigr|^{2}\Bigr\rangle_{x}.
\end{aligned}
\end{equation}
where we used the fact that:
\begin{equation}
\begin{aligned}\left[\partial_{i}U^{\dagger}\left(\mathbf{x}\right)\right]_{ij} & =\left[\partial_{i}\mathcal{P}\exp\left(ig\int_{+\infty}^{-\infty}\mathrm{d}x^{+}A_{c}^{-}\left(x^{+},\mathbf{x}\right)t^{c}\right)\right]_{ij},\\
 & =ig\int_{+\infty}^{-\infty}\mathrm{d}z^{+}U_{ik}\left(+\infty,z^{+};\mathbf{x}\right)\left[\partial_{i}A_{c}^{-}\left(z^{+},\mathbf{x}\right)t_{kl}^{c}\right]U_{lj}\left(z^{+},-\infty;\mathbf{x}\right),\\
 & =\biggl(-ig\int_{+\infty}^{-\infty}\mathrm{d}z^{+}U_{ki}\left(z^{+},+\infty;\mathbf{x}\right)\left[\partial_{i}A_{c}^{-}\left(z^{+},\mathbf{x}\right)t_{lk}^{c}\right]U_{jl}\left(-\infty,z^{+};\mathbf{x}\right)\biggr)^{*},\\
 & =\left[\partial_{i}U\left(\mathbf{x}\right)\right]_{ji}^{*}.
\end{aligned}
\end{equation}
In a similar calculation, we can show that:
\begin{equation}
\begin{aligned}\mathcal{F}_{gg}^{\left(2\right)}\left(x,q_{\perp}\right) & =-\frac{8\pi}{g_{s}^{2}}\frac{1}{N_{c}}\sum_{ij=1}^{N_{c}^{2}}\sum_{lk=1}^{N_{c}^{2}}\mathrm{Re}\Bigl\langle\biggl(\int\frac{\mathrm{d}^{2}\mathbf{v}}{\left(2\pi\right)^{2}}e^{-i\mathbf{q}_{\perp}\cdot\mathbf{v}}\left[\partial_{i}U\left(\mathbf{v}\right)\right]_{ij}U^{\dagger}\left(\mathbf{v}\right)_{lk}\biggr)\\
 & \times\biggl(\int\frac{\mathrm{d}^{2}\mathbf{v}'}{\left(2\pi\right)^{2}}e^{-i\mathbf{q}_{\perp}\cdot\mathbf{v}'}U\left(\mathbf{v}'\right){}_{ij}\left[\partial_{i}U\left(\mathbf{v}'\right)\right]_{kl}^{*}\biggr)^{*}\Bigr\rangle_{x},\\
\mathcal{F}_{gg}^{\left(3\right)}\left(x,q_{\perp}\right) & =\frac{8\pi}{g_{s}^{2}}\sum_{i=1}^{2}\sum_{ij=1}^{N_{c}^{2}}\Bigl\langle\biggl|\int\frac{\mathrm{d}^{2}\mathbf{v}}{\left(2\pi\right)^{2}}e^{-i\mathbf{q}_{\perp}\cdot\mathbf{v}}U^{\dagger}\left(\mathbf{v}\right)\partial_{i}U\left(\mathbf{v}\right)\biggr|_{ij}^{2}\Bigr\rangle_{x}.
\end{aligned}
\end{equation}
For $\mathcal{H}_{gg}^{\left(1\right)}\left(x,q_{\perp}\right)$,
$\mathcal{H}_{gg}^{\left(2\right)}\left(x,q_{\perp}\right)$ and $\mathcal{H}_{gg}^{\left(3\right)}\left(x,q_{\perp}\right)$,
we find the following formulas:
\begin{equation}
\begin{aligned}\mathcal{H}_{gg}^{\left(1\right)}\left(x,q_{\perp}\right) & =\frac{16\pi}{g_{s}^{2}}\frac{1}{N_{c}}\sum_{ij=1}^{N_{c}^{2}}\sum_{lk=1}^{N_{c}^{2}}\Bigl\langle\biggr|\frac{q_{\perp}^{i}}{q_{\perp}}\int\frac{\mathrm{d}^{2}\mathbf{v}}{\left(2\pi\right)^{2}}e^{-i\mathbf{q}_{\perp}\cdot\mathbf{v}}\left[\partial_{i}U\left(\mathbf{v}\right)\right]_{ij}U^{\dagger}\left(\mathbf{v}\right)_{lk}\biggr|^{2}\Bigr\rangle_{x}-\mathcal{F}_{gg}^{\left(1\right)}\left(x,q_{\perp}\right).\\
\mathcal{H}_{gg}^{\left(2\right)}\left(x,q_{\perp}\right) & =-\frac{16\pi}{g_{s}^{2}}\frac{1}{N_{c}}\sum_{ij=1}^{N_{c}^{2}}\sum_{lk=1}^{N_{c}^{2}}\mathrm{Re}\Bigl\langle\Bigl(\frac{q_{\perp}^{i}}{\left|q_{\perp}\right|}\int\frac{\mathrm{d}^{2}\mathbf{v}}{\left(2\pi\right)^{2}}e^{-i\mathbf{q}_{\perp}\cdot\mathbf{v}}\left[\partial_{i}U\left(\mathbf{v}\right)\right]_{ij}U^{\dagger}\left(\mathbf{v}\right)_{lk}\Bigr)\\
 & \times\Bigl(\frac{q_{\perp}^{j}}{\left|q_{\perp}\right|}\int\frac{\mathrm{d}^{2}\mathbf{v}'}{\left(2\pi\right)^{2}}e^{-i\mathbf{q}_{\perp}\mathbf{v}'}U\left(\mathbf{v}'\right){}_{ij}\left[\partial_{j}U\left(\mathbf{v}'\right)\right]_{kl}^{*}\Bigr)^{*}\Bigr\rangle_{x}-\mathcal{F}_{gg}^{\left(2\right)}\left(x,q_{\perp}\right),\\
\mathcal{H}_{gg}^{\left(3\right)}\left(x,q_{\perp}\right) & =\frac{16\pi}{g_{s}^{2}}\sum_{ij=1}^{N_{c}^{2}}\Bigl\langle\biggr|\frac{q_{\perp}^{i}}{\left|q_{\perp}\right|}\int\frac{\mathrm{d}^{2}\mathbf{v}}{\left(2\pi\right)^{2}}e^{-i\mathbf{q}_{\perp}\cdot\mathbf{v}}U^{\dagger}\left(\mathbf{v}\right)\partial_{i}U\left(\mathbf{v}\right)\biggr|_{ij}^{2}\Bigr\rangle_{x}-\mathcal{F}_{gg}^{\left(3\right)}\left(x,q_{\perp}\right).
\end{aligned}
\end{equation}

\section{Results from QCD lattice JIMWLK evolution}

The JIMWLK evolution equation can be simulated on a two dimensional
transverse QCD lattice. The numerical code with which we perform the
evolution  has been written by Claude Roiesnel, and was used previously
in Ref. \protect\cite{Cyrille}. It is primarily based on the algorithms described
in Refs. \protect\cite{Rummukainen2004,Lappi2013}. The plots we show here
are the results of $50$ independent simulations with a lattice size
$L=1024$. 

For the sake of comparison, let us assign an approximate scale to
the results presented here. The saturation scale on the lattice is
defined via the correlation length:
\begin{equation}
R_{s}\equiv\frac{\sqrt{2}}{Q_{s}},\label{eq:CorrelationlengthvsSat}
\end{equation}
which in turn is determined from the numerical values of the dipole
correlator, requiring:
\begin{equation}
D\left(R_{s}\right)=e^{-1/2}.\label{eq:CorrelationlengthDP}
\end{equation}
From the numerical analysis, the optimal value of the correlation
length was determined to be:
\begin{equation}
R_{s}\simeq66\,a.
\end{equation}
If we assume a starting Bjorken-$x$ of $x_{0}=10^{-2}$, with associated
saturation scale $Q_{s}^{2}\left(x_{0}\right)=0.2\,\mathrm{GeV}^{2}$,
we can restore the lattice spacing $a$ in the above formula and find:
\begin{equation}
a=\frac{\sqrt{2}}{66\times\sqrt{0.2}}\mathrm{GeV}^{-1}\simeq0.05\,\mathrm{GeV}^{-1}.
\end{equation}
It is worth keeping in mind that, due to lattice and discretization
effects, the lattice results are only dependable up to: 
\begin{equation}
q_{\perp}a\apprle\frac{\pi}{4},
\end{equation}
which corresponds to $q_{\perp}\simeq16\,\mathrm{GeV}$. The rapidity
evolution is expressed in terms of 
\begin{equation}
\frac{\alpha_{s}}{\pi^{2}}y=\frac{\alpha_{s}}{\pi^{2}}\ln\frac{x_{0}}{x},
\end{equation}
where the numerical code evolves the TMDs in steps of $\left(\alpha_{s}/\pi^{2}\right)\delta y=10^{-4}$.
Choosing $\alpha_{s}=0.15$, a value of $\left(\alpha_{s}/\pi^{2}\right)y=0.1$
corresponds to $x=10^{-5}$. The different values of $\left(\alpha_{s}/\pi^{2}\right)y$,
for which we plot the numerical results, are listed in Table \ref{tab:SvsX},
along with the corresponding value of Bjorken-$x$ as well as the
approximate value of the saturation scale.

Our results for $\mathcal{F}_{gg}^{\left(1\right)}\left(x,q_{\perp}\right)$,
$\mathcal{F}_{gg}^{\left(2\right)}\left(x,q_{\perp}\right)$ and $\mathcal{F}_{gg}^{\left(3\right)}\left(x,q_{\perp}\right)$
match indeed with the earlier work in Ref. \protect\cite{Cyrille}. Furthermore,
we reproduce, at least quantitatively, the numerical results for $\mathcal{F}_{gg}^{\left(3\right)}\left(x,q_{\perp}\right)$
and $\mathcal{H}_{gg}^{\left(3\right)}\left(x,q_{\perp}\right)$ in
Ref. \protect\cite{Lappi}.

\begin{figure}[H]
\begin{centering}
\includegraphics[scale=0.4]{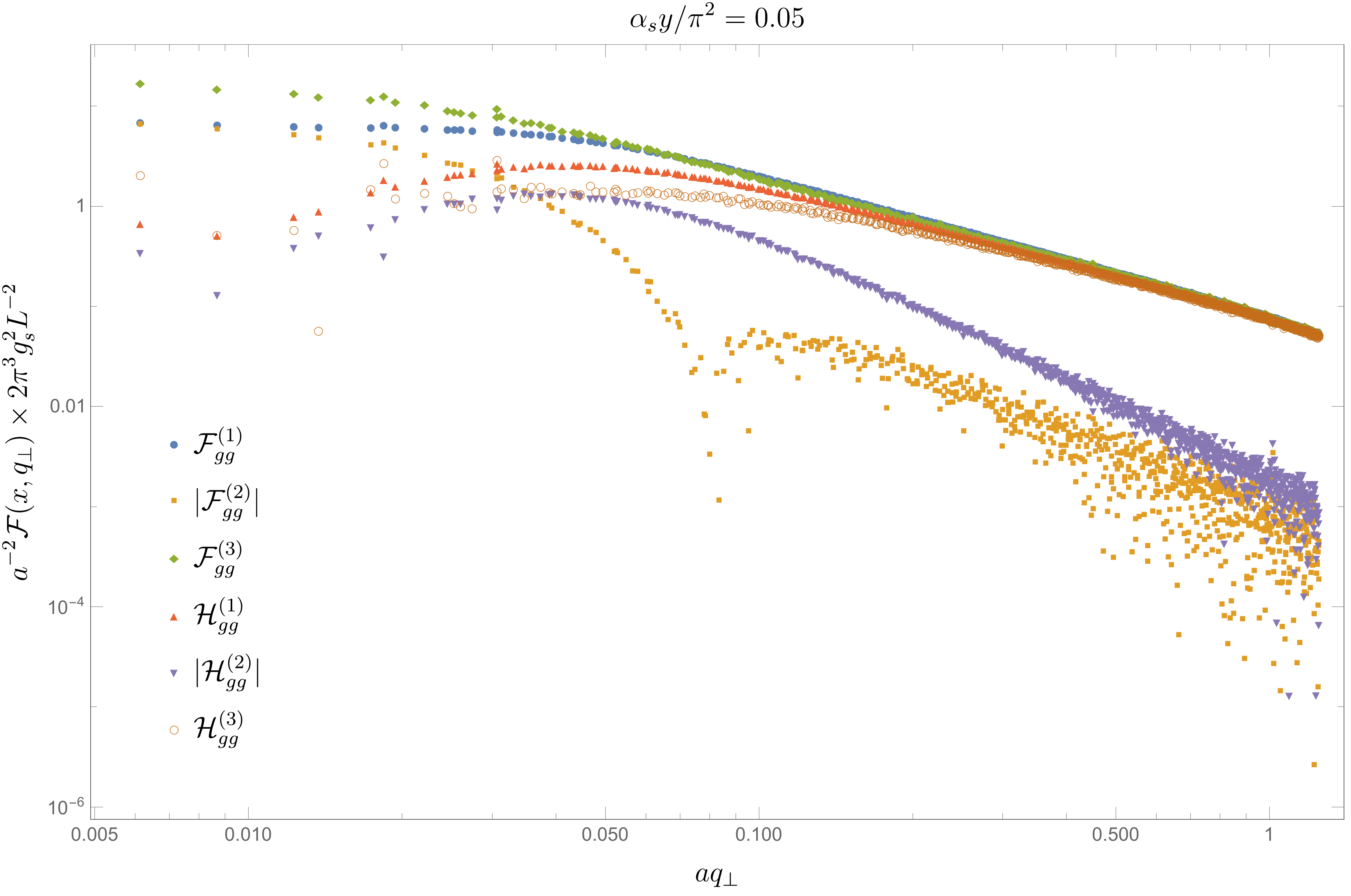}
\par\end{centering}
\caption{\label{fig:AllgTMDslog}The gluon TMDs after 500 steps in the rapidity
evolution with a logarithmic $y$-axis. The convergence of $\mathcal{F}_{gg}^{\left(2\right)}$
and $\mathcal{H}_{gg}^{\left(2\right)}$ to zero (their absolute value
is depicted), for large values of $q_{\perp}$, is now apparent, while
the other gluon TMDs exhibit the same power tail. }
\end{figure}

A first important observation to be made is that the data confirms
the claims we made earlier in the introduction, namely, the property
that in the limit of large $q_{\perp}$ the HEF or $k_{\perp}$-factorization
regime is recovered, in which the TMDs, set apart by their intricate
gauge link structure, converge to a common unintegrated PDF. The sole
exceptions are $\mathcal{F}_{gg}^{\left(2\right)}\left(x,q_{\perp}\right)$
and $\mathcal{H}_{gg}^{\left(2\right)}\left(x,q_{\perp}\right)$,
which disappear (Fig. \ref{fig:AllgTMDslog}). 

Furthermore, the TMDs shift towards larger values of $q_{\perp}$
if we evolve towards smaller values of $x$. This is of course to
be expected from the fact that the distributions follow the saturation
scale $Q_{s}^{2}\left(x\right)$, which grows when Bjorken-$x$ becomes
smaller.

\begin{figure}[H]
\begin{centering}
\includegraphics[scale=0.4]{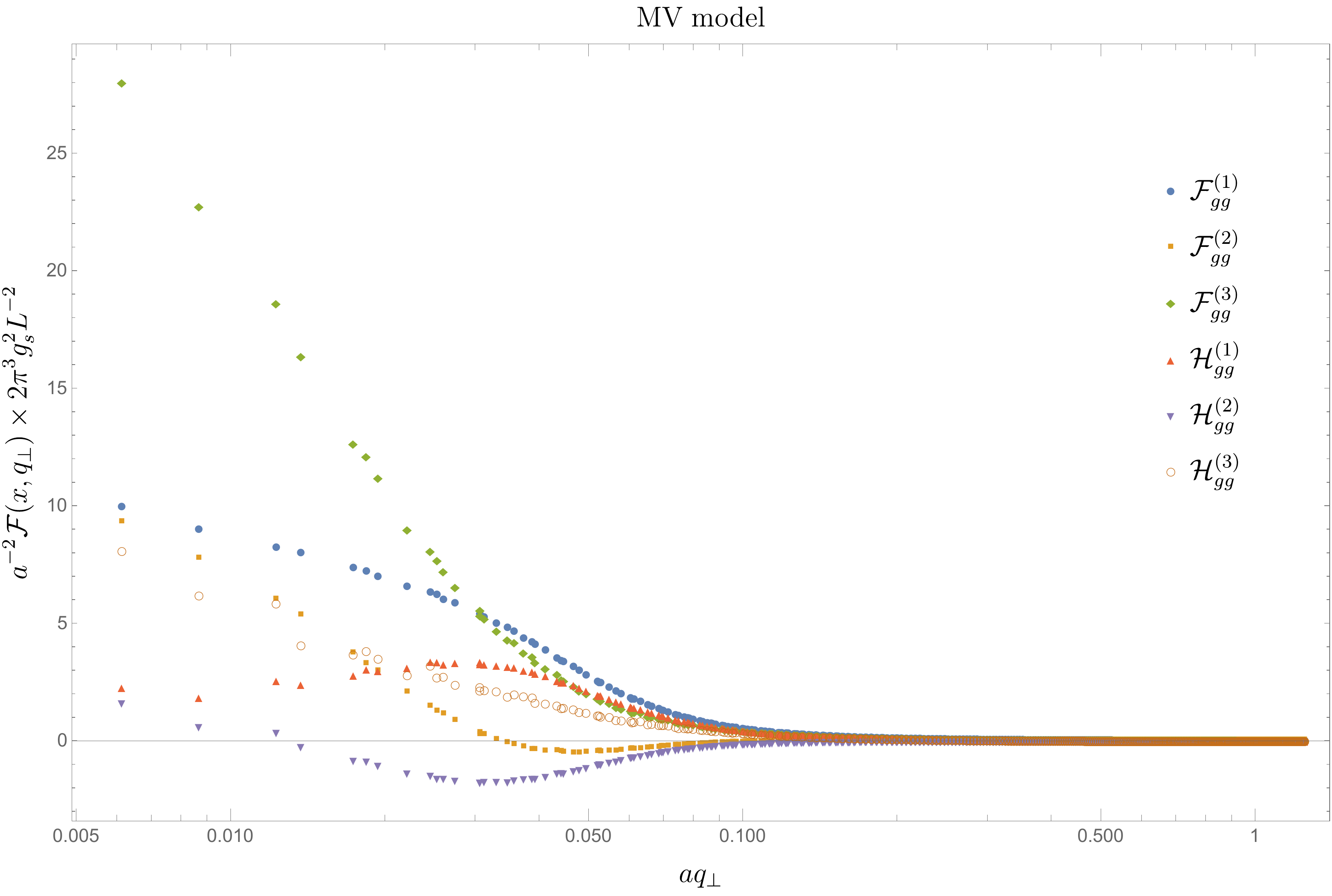}
\par\end{centering}
\bigskip{}
\begin{centering}
\includegraphics[scale=0.4]{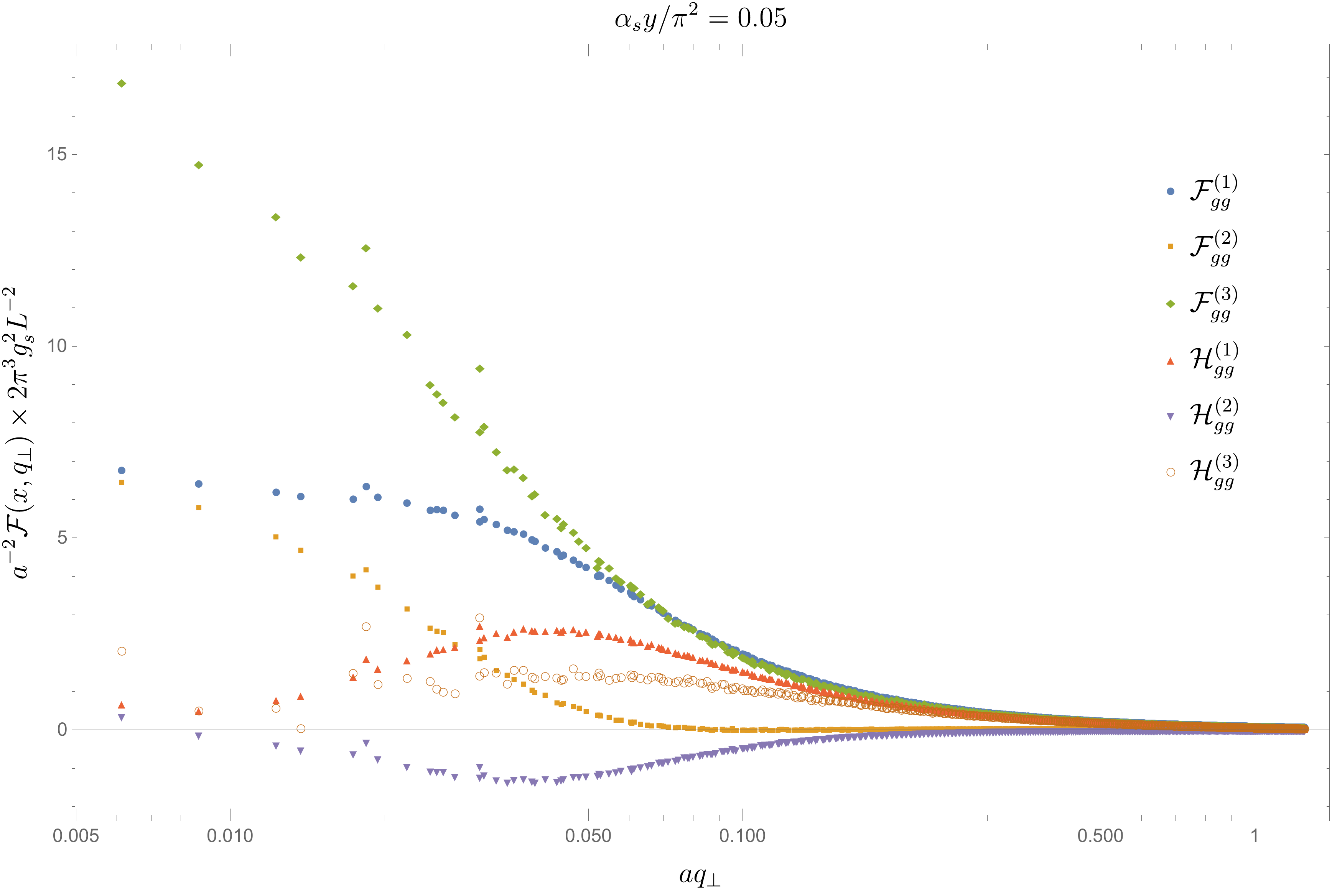}
\par\end{centering}
\caption{\label{fig:AllgTMDs0}Above: the initial conditions (MV model) for
all six gluon TMDs on the QCD lattice. Below: the gluon TMDs after
500 steps in the rapidity evolution.}
\end{figure}

The information on the gluon polarization is washed out by the evolution
fairly fast: $\mathcal{H}_{gg}^{\left(1\right)}$ and $\mathcal{H}_{gg}^{\left(2\right)}$
are small to begin with, and $\mathcal{H}_{gg}^{\left(3\right)}$
is strongly suppressed after the first steps in the evolution (cf.
(\ref{fig:AllgTMDs0})).

Finally, for very small values of $q_{\perp}$, $\mathcal{F}_{gg}^{\left(1\right)}\left(x,q_{\perp}\right)$
is equal to $\mathcal{F}_{gg}^{\left(2\right)}\left(x,q_{\perp}\right)$
and $\mathcal{H}_{gg}^{\left(1\right)}\left(x,q_{\perp}\right)$ is
equal to $\mathcal{H}_{gg}^{\left(2\right)}\left(x,q_{\perp}\right)$.
This is a consequence of the identities Eqs. (\ref{eq:F1-F2general})
and (\ref{eq:H1-H2general}), in combination with the fact that the
adjoint dipole TMD $xG_{A}^{\left(2\right)}\left(x,q_{\perp}\right)$
disappears for $q_{\perp}=0$, which follows directly from Eq. (\ref{eq:PIGA}).

\begin{figure}[H]
\begin{centering}
\includegraphics[scale=0.4]{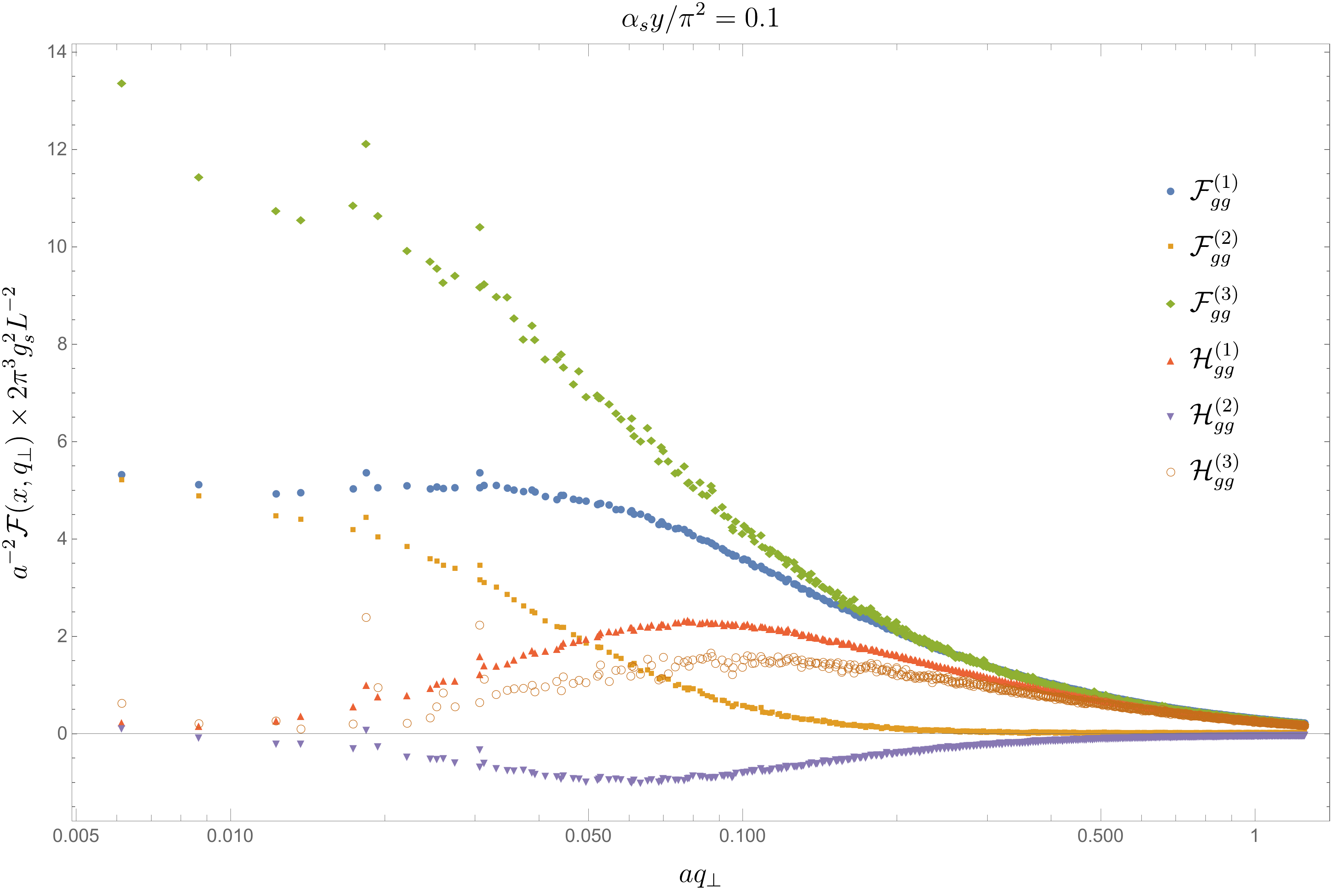}
\par\end{centering}
\bigskip{}
\begin{centering}
\includegraphics[scale=0.4]{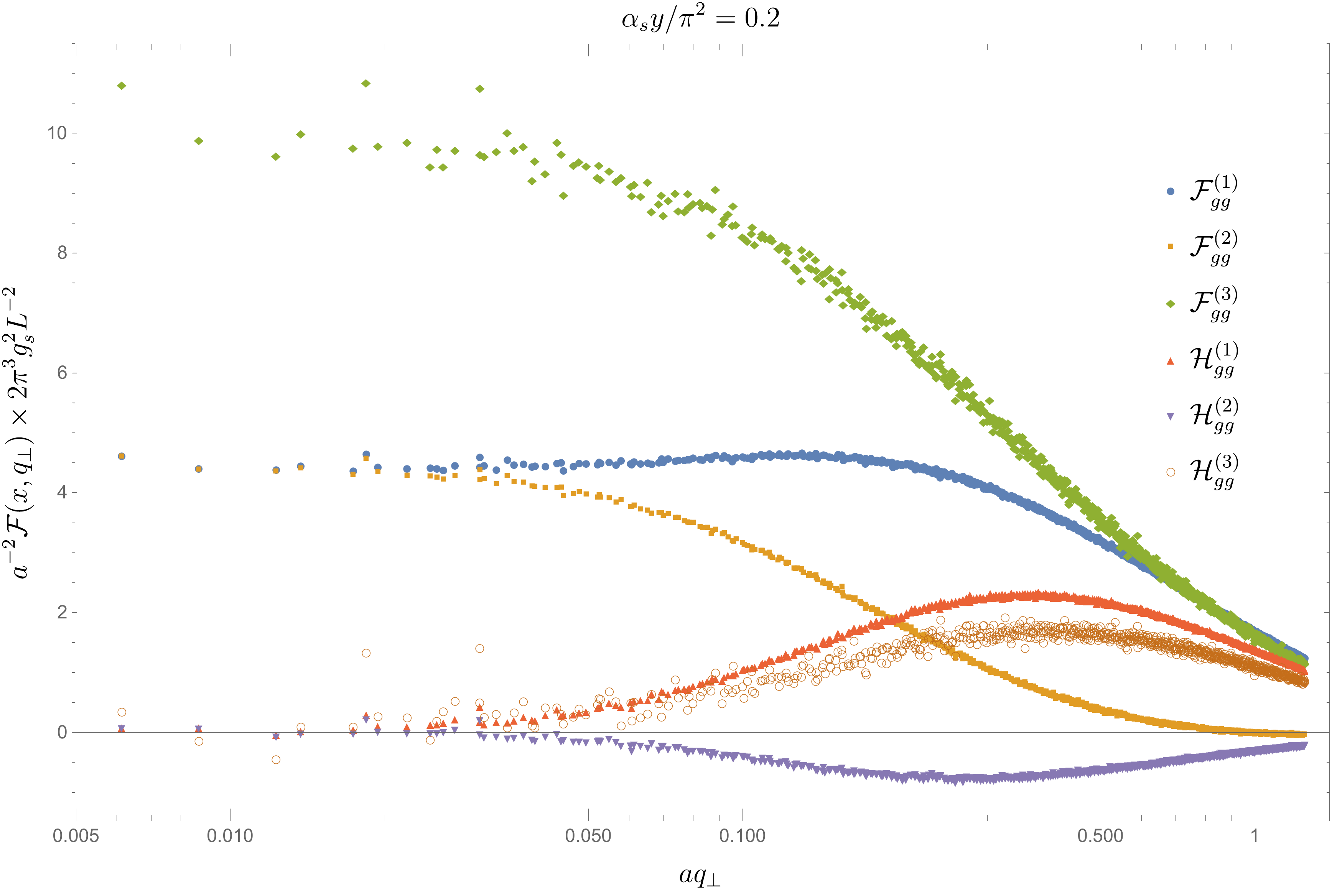}
\par\end{centering}
\caption{\label{fig:AllgTMDs1000}The gluon TMDs after 1000 (above) and 2000
(below) steps in the rapidity evolution.}
\end{figure}
\begin{table}[H]
\begin{centering}
\begin{tabular}{|c|c|c|}
\hline 
$\left(\alpha_{s}/\pi^{2}\right)y$ & $x$ & $Q_{s}$\tabularnewline
\hline 
\hline 
$0$ & $x_{0}=10^{-2}$ & $0.4\,\mathrm{\mathrm{GeV}}$\tabularnewline
\hline 
$0.05$ & $4\cdot10^{-4}$ & $0.7\,\mathrm{\mathrm{GeV}}$\tabularnewline
\hline 
$0.1$ & $10^{-5}$ & $1\,\mathrm{\mathrm{GeV}}$\tabularnewline
\hline 
$0.2$ & $2\cdot10^{-8}$ & $6\,\mathrm{\mathrm{GeV}}$\tabularnewline
\hline 
\end{tabular}
\par\end{centering}
\caption{\label{tab:SvsX}The values of $\left(\alpha_{s}/\pi^{2}\right)y$
versus Bjorken-$x$, and the corresponding value of the saturation
scale, calculated from the dipole using Eqs. (\ref{eq:CorrelationlengthDP}),
(\ref{eq:CorrelationlengthvsSat}).}
\end{table}

\section{$\gamma A\rightarrow q\bar{q}X$}

Before turning to the conclusions, we should remark that from our
present study of dijet production in proton-nucleus collisions, the
results for the photoproduction case $\gamma A\rightarrow q\bar{q}X$
appear as a simple corollary. 

Indeed, when computing the $\gamma^{*}A\rightarrow q\bar{q}X$ cross
section within the CGC, Sec. \ref{subsec:CGCcrosss}, the only difference
with the case in which a photon interacts with the nucleus (for example
in DIS) is the different coupling constant and the lack of color charge.
Therefore, the equivalent of the term Eq. (\ref{eq:CGCinteraction}),
which encodes the interaction of the projectile with the CGC, is given
by:
\begin{equation}
\begin{aligned}\Phi_{ss',ij}^{\lambda}\left(p^{+},k_{1}^{+},\mathbf{x}-\mathbf{b}\right) & \equiv\left(U_{il}^{\dagger}\left(\mathbf{b}\right)U_{lj}\left(\mathbf{x}\right)-1_{ij}\right)\phi_{s's}^{\lambda}\left(p,k_{1}^{+},\mathbf{x}-\mathbf{b}\right).\end{aligned}
\end{equation}
When taking the absolute value squared, Eq. (\ref{eq:CGCinteractionsquared}),
we now obtain the much simpler expression:
\begin{equation}
\begin{aligned} & \int\mathcal{D}\left[\mathcal{A}\right]\left|\Phi\left(\mathcal{A}\right)\right|^{2}\,_{\mathcal{A}}\langle0|\left(U_{il}^{\dagger}\left(\mathbf{u}\right)U_{lj}\left(\mathbf{v}\right)-1_{ij}\right)^{\text{*}}\phi_{ss'}^{\lambda*}\left(p,k_{1}^{+},\mathbf{u}-\mathbf{v}\right)\\
 & \times\left(U_{il}^{\dagger}\left(\mathbf{u}'\right)U_{lj}\left(\mathbf{v}'\right)-1_{ij}\right)\phi_{ss'}^{\lambda}\left(p,k_{1}^{+},\mathbf{u}'-\mathbf{v}'\right)|0\rangle_{\mathcal{A}}\\
 & =\phi_{ss'}^{\lambda*}\left(p,k_{1}^{+},\mathbf{u}-\mathbf{v}\right)\phi_{ss'}^{\lambda}\left(p,k_{1}^{+},\mathbf{u}'-\mathbf{v}'\right)\biggl(\mathrm{Tr}\Bigl\langle U\left(\mathbf{u}\right)U^{\dagger}\left(\mathbf{u}'\right)U\left(\mathbf{v}'\right)U^{\dagger}\left(\mathbf{v}\right)\Bigr\rangle_{x}\\
 & -\mathrm{Tr}\Bigl\langle U\left(\mathbf{v}\right)U^{\dagger}\left(\mathbf{u}\right)\Bigr\rangle_{x}-\mathrm{Tr}\Bigl\langle U\left(\mathbf{v}'\right)U^{\dagger}\left(\mathbf{u}'\right)\Bigr\rangle_{x}+N_{c}\biggr).
\end{aligned}
\end{equation}
The cross section for the $\gamma A\rightarrow q\bar{q}X$ finally
becomes, changing $\alpha_{s}\to\alpha_{em}e_{q}^{2}$:
\begin{equation}
\begin{aligned}\frac{\mathrm{d}\sigma^{\gamma A\rightarrow q\bar{q}X}}{\mathrm{d}^{3}k_{1}\mathrm{d}^{3}k_{2}} & =\alpha_{em}e_{q}^{2}N_{c}\delta\left(k_{1}^{+}+k_{2}^{+}-p^{+}\right)\int\frac{\mathrm{d}^{2}\mathbf{x}}{\left(2\pi\right)^{2}}\frac{\mathrm{d}^{2}\mathbf{y}}{\left(2\pi\right)^{2}}\frac{\mathrm{d}^{2}\mathbf{x}'}{\left(2\pi\right)^{2}}\frac{\mathrm{d}^{2}\mathbf{y}'}{\left(2\pi\right)^{2}}e^{i\mathbf{k}_{1\perp}\cdot\left(\mathbf{y}-\mathbf{y}'\right)}e^{i\left(\mathbf{k}_{2\perp}-\mathbf{p}_{\perp}\right)\cdot\left(\mathbf{x}-\mathbf{x}'\right)}\\
 & \times\sum_{\lambda ss'}\phi_{s's}^{\lambda*}\left(p^{+},k_{1}^{+},\mathbf{x}-\mathbf{y}\right)\phi_{s's}^{\lambda}\left(p^{+},k_{1}^{+},\mathbf{x}'-\mathbf{y}'\right)\\
 & \times\left(S^{\left(4\right)}\left(\mathbf{x},\mathbf{y},\mathbf{y}',\mathbf{x}'\right)-D\left(\mathbf{y},\mathbf{x}\right)-D\left(\mathbf{y}',\mathbf{x}'\right)+1\right),
\end{aligned}
\label{eq:DIS2qqcrosssection}
\end{equation}
in accordance with the result quoted in Ref. \protect\cite{fabio}. In the
above expression, the familiar dipole appears, along with the gauge
link structure:
\begin{equation}
\begin{aligned}S^{\left(4\right)}\left(\mathbf{x},\mathbf{y},\mathbf{y}',\mathbf{x}'\right) & \equiv\frac{1}{N_{c}}\mathrm{Tr}\Bigl\langle U\left(\mathbf{x}\right)U^{\dagger}\left(\mathbf{x}'\right)U\left(\mathbf{y}'\right)U^{\dagger}\left(\mathbf{y}\right)\Bigr\rangle_{x}.\end{aligned}
\end{equation}
Taking the correlation limit, we obtain:

\begin{equation}
\begin{aligned}\frac{\mathrm{d}\sigma^{\gamma A\rightarrow q\bar{q}X}}{\mathrm{d}\mathcal{P}.\mathcal{S}.} & =\alpha_{s}\alpha_{em}e_{q}^{2}\delta\left(\frac{k_{1}^{+}+k_{2}^{+}}{p^{+}}-1\right)z\left(1-z\right)\frac{1}{(\tilde{P}_{\perp}^{2}+m^{2})^{4}}\\
 & \times\Biggl\{\left((\tilde{P}_{\perp}^{4}+m^{4})(z^{2}+(1-z)^{2})+2m^{2}\tilde{P}_{\perp}^{2}\right)\mathcal{F}_{gg}^{\left(3\right)}\left(x,q_{\perp}\right)\\
 & +z\left(1-z\right)4m^{2}\tilde{P}_{\perp}^{2}\cos\left(2\phi\right)\mathcal{H}_{gg}^{\left(3\right)}\left(x,q_{\perp}\right)\Biggr\}.
\end{aligned}
\end{equation}
In terms of the Mandelstam variables, we find, keeping only the leading
terms in $m^{2}/\tilde{P}_{\perp}^{2}$:
\begin{equation}
\begin{aligned}\left.\frac{\mathrm{d}\sigma^{\gamma A\rightarrow q\bar{q}X}}{\mathrm{d}\mathcal{P}.\mathcal{S}.}\right|_{m=0} & =\alpha_{s}\alpha_{em}e_{q}^{2}\delta\left(\frac{k_{1}^{+}+k_{2}^{+}}{p^{+}}-1\right)\frac{\hat{t}^{2}+\hat{u}^{2}}{\hat{s}^{3}}\mathcal{F}_{gg}^{\left(3\right)}\left(x,q_{\perp}\right),\end{aligned}
\end{equation}
and
\begin{equation}
\begin{aligned}\left.\frac{\mathrm{d}\sigma^{\gamma A\rightarrow q\bar{q}X}}{\mathrm{d}\mathcal{P}.\mathcal{S}.}\right|_{\phi} & =2\alpha_{s}\alpha_{em}e_{q}^{2}\delta\left(\frac{k_{1}^{+}+k_{2}^{+}}{p^{+}}-1\right)\frac{\hat{u}\hat{t}}{\hat{s}^{3}}\frac{2m^{2}}{\tilde{P}_{\perp}^{2}}\cos\left(2\phi\right)\mathcal{H}_{gg}^{\left(3\right)}\left(x,q_{\perp}\right).\end{aligned}
\end{equation}
Since the incoming photon doesn't produce initial state radiation,
as compared to the gluon in the process $pA\to q\bar{q}X$, the corresponding
cross section is much simpler, and only the Weizsäcker-Williams gluon
TMD $\mathcal{F}_{gg}^{\left(3\right)}\left(x,q_{\perp}\right)$ and
its polarized partner, $\mathcal{H}_{gg}^{\left(3\right)}\left(x,q_{\perp}\right)$,
play a role.

Incidentally, from Eq. (\ref{eq:DIS2qqcrosssection}), restoring the
virtuality of the photon (hence we have to take both the longitudinal
and the transverse $\gamma\to q\bar{q}$ wave function into account)
and integrating over $k_{1}$ and $k_{2}$, one recovers the dipole
DIS cross section, given by Eqs. (\ref{eq:DISdipolecrosssection})
and (\ref{eq:dipolecrossection}).

\section{Conclusion and outlook}

In this part of the thesis, we used the Color Glass Condensate to
compute the cross section for the forward production of two heavy
quarks in dilute-dense, i.e. proton-nucleon collisions. In the result,
given by Eqs. (\ref{eq:finalCGCcrosssection}) or (\ref{eq:finalCGCcrosssectionMandelstam}),
we could discern six different Wilson line correlators, which, by
comparing with TMD calculation, could be identified as the small-$x$
limits of six gluon TMDs: three unpolarized ones, and their three
partners which describe the linearly polarized gluons in the unpolarized
nucleus. We obtained analytical expressions (\ref{eq:F1finiteNc-1}),
(\ref{eq:F2finiteNc-1}), (\ref{eq:H1finiteNc-1}), (\ref{eq:H2finiteNc-1}),
(\ref{eq:WWfiniteNc}) and (\ref{eq:H3finiteNc}) for the TMDs in
the McLerran-Venugopalan model at finite-$N_{c}$, and showed that,
when limiting ourself to the simple Golec-Biernat and Wüsthoff model,
the result of Ref. \protect\cite{zhou} is recovered. Furthermore, the gluon
TMDs were implemented on the QCD lattice, and the nonlinear JIMWLK
evolution in rapidity was carried out numerically, as shown in Figs.
\ref{fig:AllgTMDs0} and \ref{fig:AllgTMDs1000}. As expected, the
distributions we obtained are centered around the saturation momentum
$Q_{s}\left(x\right)$, and either disappear or coincide in the limit
$k_{\perp}\gg Q_{s}$. 

At least in principle, our predictions could be measured at the LHC.
Moreover, we showed that in the simpler process $\gamma A\rightarrow q\bar{q}X$,
only two of the gluon TMDs appear: $\mathcal{F}_{gg}^{\left(3\right)}\left(x,q_{\perp}\right)$
and $\mathcal{H}_{gg}^{\left(3\right)}\left(x,q_{\perp}\right)$.
These could be extracted from measurements in a future electron-ion
collider (Ref. \protect\cite{Boer2011}), or in ultraperipheral collisions
at RHIC (Ref. \protect\cite{Kotko2017}), and be then used to constrain the
four other TMDs that play a role in the $pA\to q\bar{q}X$. In addition,
it should be possible to apply our calculation to proton-proton collisions
at very high energies, taking care of the proton's smaller transverse
area as compared to a nucleus by taking the impact-parameter dependence
into account. 

Finally, for completeness we should note that there are other possible
sources for the azimuthal asymmetry of the heavy quarks, which are
not included in our CGC analysis, see e.g. Refs. \protect\cite{Boer2009,Boer2011,Metz2011,Pisano2013}.

\newpage{}

\thispagestyle{simple}

\part{\label{part:edmond}Renormalization of the jet quenching parameter}

\section{Introduction}

In high-energy collisions of two heavy nuclei, an extremely dense
and hot state of QCD matter is produced known as the quark-gluon plasma
(QGP), see for instance Ref. \protect\cite{dEnterria2007} and references
therein. Hard jets that were produced during the scattering, are attenuated
by this plasma when traveling through on their way to the detector.
This phenomenon and the resulting set of observables, such as dijet
asymmetries, angular broadening or a suppression of the hadron yield,
are collectively known as \textquoteleft jet quenching' (see, e.g.
Refs. \protect\cite{Salgadolectures,dEnterria2010}), first theorized by J.
Bjorken (Ref. \protect\cite{BjorkenQGP}). Jet quenching is one of the principal
probes to resolve the properties of the QGP. 

In this part of the thesis, we concentrate on one aspect of jet quenching:
the transverse momentum broadening of a hard parton traveling through
the medium, caused by the \textendash in general multiple\textendash{}
scatterings off the plasma's constituents (Refs. \protect\cite{BDMPS3,Mueller2016}).
The aim is to employ small-$x$ techniques to study the radiative
corrections, due to the emission of soft gluons, to this broadening.
Since the hard parton interacts with the nuclear medium during its
whole lifetime, rather than during a very short time interval, the
problem differs significantly from what we encountered earlier in
proton-nucleus collisions, for which the CGC approach was applicable.
In particular, the medium cannot be regarded as a shockwave anymore,
and as a result, to be applicable, the JIMWLK evolution equation needs
to be extended beyond the eikonal approximation.

Interestingly, theoretical studies indicate that transverse momentum
broadening is intimately related to jet energy loss through induced
gluon radiation (Refs. \protect\cite{Gyulassy1994,Wang1995,Zakharov1996,Zakharov1997,BDMPS1,BDMPS2,BDMPS3,BDMPS4,BDMPS5,BDMPS6,Baier2000,Wiedemann2000},
or Refs. \protect\cite{Salgado2003,Casalderrey2008,Arnold2009,Arnold2009-2,CaronHuot2009,Taels2014,Blaizot2013,MehtarTani2013,Al,Blaizot2014,Blaizot2014RG,Wu2014,Iancu,Mueller2016}
for some recent developments). Both only depend on the medium via
the so-called jet quenching parameter $\hat{q}$ (see \protect\cite{JETCollaboration2014}
for a recent extraction), which can be regarded as a transport coefficient
of the medium (Refs. \protect\cite{AMY1,AMY2}). As it was first pointed out
in \protect\cite{Blaizot2014RG}, it turns out that the radiative corrections
to the transverse momentum broadening can be absorbed into a renormalization
of $\hat{q}$.

We start this part by introducing the non-eikonal (in the sense that
the transverse coordinates of the soft gluon are allowed to vary during
the multiple scatterings off the medium) generalization of the JIMWLK
Hamiltonian, constructed in Ref. \protect\cite{Iancu}, and demonstrate its
reduction to the regular JIMWLK evolution in the limit in which the
medium is a shockwave. We then carefully analyze its structure, in
particular the cancellation of divergences through the interplay of
real and virtual contributions. 

We then turn to the problem of transverse momentum broadening, and
show how our generalized high-energy evolution equation can be applied
to resum the large logarithms, associated with soft gluon radiation,
which appear when calculating the radiative corrections to the broadening.
To double leading logarithmic accuracy, the evolution equation can
be solved analytically, resulting in a renormalization of the jet
quenching parameter. Interestingly, the problem of gluon radiation
in an extended medium is drastically different from the one in a shockwave,
in particular, the medium introduces a mutual dependence between the
longitudinal and transverse phase space of the gluon fluctuation.
As a result, the double logarithmic evolution is not the same as the
one in the vacuum, which we encountered in Sec. \ref{subsec:The-soft-Bremsstrahlung}. 

Finally, we give an outline of the calculation in Ref. \protect\cite{Al},
in which the authors were able to compute a single iteration of the
evolution to single logarithmic accuracy. During my internship with
Dr. Iancu, it was the aim to try to solve evolution equation Eq. (\ref{eq:qhatevolutionequation}),
derived in Ref. \protect\cite{Iancu}, which in principle contains all the
leading logarithmic contributions. Solving this equation would therefore
be tantamount to a full resummation of the single leading logarithms.
At the present moment, however, it is not clear whether such a resummation
is possible, or even meaningful, beyond DLA accuracy. Rather, we closely
follow Ref. \protect\cite{Iancu} and set up the approach to the problem,
elucidating on the small-$x$ physics we encountered earlier by studying
the more difficult example of in-medium radiation. As such, this chapter
should be seen as a case-study of a theoretical problem which is interesting
in its own right, even if, ultimately, we didn't reach our original
goal.

\section{\label{sec:generalJIMWLK}A general high-energy evolution equation}

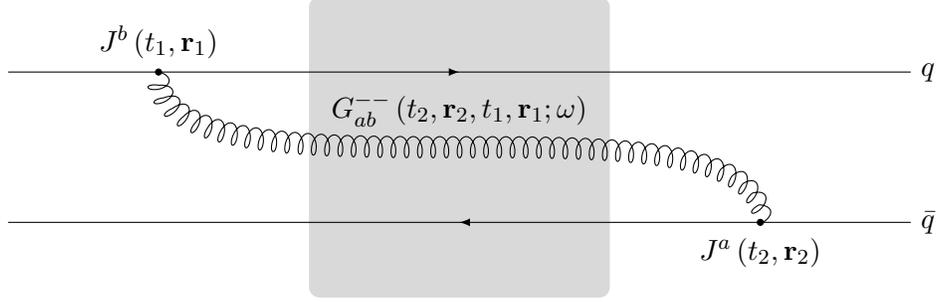
\begin{figure}[t]
\begin{centering}
\begin{tikzpicture}[scale=2] 

\tikzset{photon/.style={decorate,decoration={snake}},
		electron/.style={ postaction={decorate},decoration={markings,mark=at position .5 with {\arrow[draw]{latex}}}},      	gluon/.style={decorate,decoration={coil,amplitude=4pt, segment length=5pt}}}

\fill[black!15!white, rounded corners] (-1,-1) rectangle(1,1);
\draw[electron] (-3,.5) --++ (6,0) node[right]{$q$};
\draw[electron] (3,-.5) node[right]{$\bar{q}$}--++(-6,0);
\draw[gluon] (-2,.5) .. controls (-2,-0.5) and (2,.5) .. (2,-0.5);
\node at (-2,0.7) {$J^{b}\left(t_{1},\mathbf{r}_{1}\right)$};
\node at (2,-0.7) {$J^{a}\left(t_{2},\mathbf{r}_{2}\right)$};
\node at (0,.25) {$G_{ab}^{--}\left(t_{2},\mathbf{r}_{2},t_{1},\mathbf{r}_{1};\omega\right)$};
\filldraw[fill=black] (-2,0.5) circle(.02);
\filldraw[fill=black] (2,-0.5) circle(.02);
\end{tikzpicture}
\par\end{centering}
\caption{\label{fig:HEE}One step in the non-eikonal high-energy evolution
of a dipole.}
\end{figure}

The Hamiltonian $H$ for the high-energy evolution of a right-moving
projectile through a dense medium is given by (see Fig. \ref{fig:HEE}).
\begin{equation}
H=\frac{\omega}{2\pi}\int\mathrm{d}t_{2}\int^{t_{2}}\mathrm{d}t_{1}\int\mathrm{d}^{2}\mathbf{r}_{2}\int\mathrm{d}^{2}\mathbf{r}_{1}G_{ab}^{--}\left(t_{2},\mathbf{r}_{2},t_{1},\mathbf{r}_{1};\omega\right)J^{a}\left(t_{2},\mathbf{r}_{2}\right)J^{b}\left(t_{1},\mathbf{r}_{1}\right),\label{eq:GEE}
\end{equation}
corresponding to the following evolution equation for the $S$-matrix
$\mathcal{O}$ of the projectile\textendash medium interaction:
\begin{equation}
\frac{\partial\mathcal{O}}{\partial Y}=H\mathcal{O}.\label{eq:GEEeq}
\end{equation}
As usual, we assume that the gluon fluctuation, like its parent quark
or gluon, has a longitudinal momentum component $p^{+}=\omega$ that
is much larger than its transverse momentum. It therefore only couples
to the \textquoteleft minus' field $\alpha_{a}\equiv A_{a}^{-}$ of
the (left-moving) medium. The operator $J^{a}\left(t,\mathbf{r}\right)$,
defined as:
\begin{equation}
J^{a}\left(t,\mathbf{r}\right)\equiv\frac{\delta}{\delta\alpha_{a}\left(t,\mathbf{r}\right)},
\end{equation}
generates the emission or absorption of a soft gluon from the projectile,
which is required to be built from Wilson lines:
\begin{equation}
U\left(\mathbf{x}\right)=\mathcal{P}e^{ig_{s}\int\mathrm{d}z^{+}\alpha_{a}\left(z^{+},\mathbf{r}\right)t^{a}}.
\end{equation}
The propagation of the soft gluon, with energy $\omega$, through
the medium, is described by the in-medium propagator $G_{ab}^{--}$,
given by (see e.g. Refs. \protect\cite{JIMWLK4,Iancu}):
\begin{equation}
G_{ab}^{--}\left(t_{2},\mathbf{r}_{2},t_{1},\mathbf{r}_{1};\omega\right)=\frac{1}{\omega^{2}}\mathbf{\partial}_{\mathbf{r}_{1}}^{i}\mathbf{\partial}_{\mathbf{r}_{2}}^{i}G_{ab}\left(t_{2},\mathbf{r}_{2},t_{1},\mathbf{r}_{1};\omega\right)+\frac{i}{\omega^{2}}\delta_{ab}\delta\left(t_{2}-t_{1}\right)\delta^{\left(2\right)}\left(\mathbf{r}_{2}-\mathbf{r}_{1}\right).\label{eq:inmediumpropagator}
\end{equation}
The first term in the above expression is referred to as the radiation
part of the propagator, while the second term, which is instantaneous
and local, is called the Coulomb part. The object $G_{ab}\left(t_{2},\mathbf{r}_{2},t_{1},\mathbf{r}_{1};\omega\right)$
is known as the scalar propagator, and is given by the non-relativistic
path integral for a particle with an effective mass $\omega$ in two
spatial dimensions, in a color background field:
\begin{equation}
G_{ab}\left(t_{2},\mathbf{r}_{2},t_{1},\mathbf{r}_{1};\omega\right)=\frac{1}{2\omega}\int\left[\mathcal{D}\mathbf{r}\left(t\right)\right]\exp\left\{ i\frac{\omega}{2}\int_{t_{1}}^{t_{2}}\mathrm{d}t\,\dot{\mathbf{r}}^{2}\left(t\right)\right\} W_{ab}^{\dagger}\left(t_{2},t_{1},\mathbf{r}\left(t\right)\right).\label{eq:scalarpropagator}
\end{equation}
The non-eikonal, in the sense that the transverse coordinates are
allowed to vary, interaction of the soft gluon with the color fields
in the medium is encoded in $W_{ab}^{\dagger}$, which is the generalization
of a Wilson line to a functional of the gluon's transverse trajectory
$\mathbf{r}\left(t\right)$:
\begin{equation}
\begin{aligned}W_{ab}^{\dagger}\left(t_{2},t_{1},\mathbf{r}\left(t\right)\right) & \equiv\mathcal{\bar{P}}\exp\left(ig_{s}\int_{t_{1}}^{t_{2}}\mathrm{d}t\,\alpha_{a}\left(t,\mathbf{r}\left(t\right)\right)T^{a}\right),\end{aligned}
\label{eq:WilsonFunctional}
\end{equation}

\subsection{Recovering the JIMWLK equation}

To motivate the expression (\ref{eq:GEE}) for the evolution Hamiltonian,
let us show that it is reduced to the JIMWLK Hamiltonian, Eq. (\ref{eq:JIMWLKH}),
in the limit in which the medium is a shockwave.

Indeed, taking the shockwave to be centered around $z^{+}=t=0$, the
action of the operators $J^{a}\left(t,\mathbf{r}\right)$ on the Wilson
lines inside the $S$-matrix $\mathcal{O}$ yields:
\begin{equation}
\begin{aligned}J^{a}\left(t,\mathbf{r}\right)W_{\mathbf{z}}^{\dagger} & =-ig_{s}\delta^{\left(2\right)}\left(\mathbf{r}-\mathbf{z}\right)W_{\mathbf{z}}^{\dagger}\left(\infty,t\right)T^{a}W_{\mathbf{z}}^{\dagger}\left(t,-\infty\right).\end{aligned}
\label{eq:J2LR0}
\end{equation}
In the vacuum, i.e. when their path doesn't cross the shockwave, the
Wilson lines become equal to one, as there is no background field
for the gluon to interact with. We can then make use of the transitivity
rule Eq. (\ref{eq:Wilsonlinesum}) to write:
\begin{equation}
\begin{aligned}t<0: & \quad W_{\mathbf{z}}^{\dagger}=W_{\mathbf{z}}^{\dagger}\left(\infty,t\right)W_{\mathbf{z}}^{\dagger}\left(t,-\infty\right)=W_{\mathbf{z}}^{\dagger}\left(\infty,t\right),\\
t>0: & \quad W_{\mathbf{z}}^{\dagger}=W_{\mathbf{z}}^{\dagger}\left(\infty,t\right)W_{\mathbf{z}}^{\dagger}\left(t,-\infty\right)=W_{\mathbf{z}}^{\dagger}\left(t,-\infty\right),
\end{aligned}
\end{equation}
using which Eq. (\ref{eq:J2LR0}) can be simplified as follows:
\begin{equation}
\begin{aligned}t<0: & \quad J^{a}\left(t,\mathbf{r}\right)W_{\mathbf{z}}^{\dagger}=R_{\mathbf{r}}^{a}W_{\mathbf{z}}^{\dagger},\\
t>0: & \quad J^{a}\left(t,\mathbf{r}\right)W_{\mathbf{z}}^{\dagger}=L_{\mathbf{r}}^{a}W_{\mathbf{z}}^{\dagger}.
\end{aligned}
\label{eq:J2LR}
\end{equation}
In the above formula, we introduce the \textquoteleft right' and \textquoteleft left'
operators $R_{\mathbf{r}}^{a}$ and $L_{\mathbf{r}}^{a}$, defined
by their action on a Wilson line:
\begin{equation}
\begin{aligned}R_{\mathbf{r}}^{a}W_{\mathbf{z}}^{\dagger} & \equiv-ig_{s}\delta^{\left(2\right)}\left(\mathbf{r}-\mathbf{z}\right)W_{\mathbf{z}}^{\dagger}T^{a},\\
L_{\mathbf{r}}^{a}W_{\mathbf{z}}^{\dagger} & \equiv-ig_{s}\delta^{\left(2\right)}\left(\mathbf{r}-\mathbf{z}\right)T^{a}W_{\mathbf{z}}^{\dagger},
\end{aligned}
\label{eq:defRL}
\end{equation}
and:
\begin{equation}
\begin{aligned}R_{\mathbf{r}}^{a}W_{\mathbf{z}} & \equiv ig_{s}\delta^{\left(2\right)}\left(\mathbf{r}-\mathbf{z}\right)T^{a}W_{\mathbf{z}},\\
L_{\mathbf{r}}^{a}W_{\mathbf{z}} & \equiv ig_{s}\delta^{\left(2\right)}\left(\mathbf{r}-\mathbf{z}\right)W_{\mathbf{z}}T^{a}.
\end{aligned}
\label{eq:defLR}
\end{equation}
The operators $J^{a}(t,\mathbf{r})$ thus generate a gluon vertex
either before or after the shockwave. An important feature of Eq.
(\ref{eq:J2LR}) is that it is independent of time. Therefore, assuming
that the $S$-matrix $\mathcal{O}$ is built from Wilson lines and
is time-independent as well, so is the action of $J\left(t,\mathbf{r}\right)$
on $\mathcal{O}$. As a consequence, the problem of the full time
integration in Eq. (\ref{eq:GEE}) is reduced to the much simpler
problem of evaluating the integrals over $t_{1}$ and $t_{2}$ over
the in-medium propagator Eq. (\ref{eq:inmediumpropagator}) in the
shockwave limit:
\begin{equation}
\int\mathrm{d}t_{2}\int^{t_{2}}\mathrm{d}t_{1}G_{ab}^{--}\left(t_{2},\mathbf{r}_{2},t_{1},\mathbf{r}_{1};\omega\right).\label{eq:threetimeintegrals}
\end{equation}
One distinguishes three different cases, corresponding to three different
regions for the time integrations in Eq. (\ref{eq:GEE}). In the first
two cases:
\begin{equation}
-\infty<t_{1}\leq t_{2}<0\quad\mathrm{or}\quad0<t_{1}\leq t_{2}<\infty,
\end{equation}
the gluon is emitted and reabsorbed prior to, or after, the interaction
with the shockwave (similar to diagrams (a) and (c) in the dipole-shockwave
scattering in Fig. \ref{fig:BK}). The gluon fluctuation thus takes
place in the vacuum, and the Wilson line functional Eq. (\ref{eq:WilsonFunctional}),
encoding the gluon's interaction with the color background, becomes
simply: 
\begin{equation}
W_{ab}^{\dagger}\left(t_{2},t_{1},\mathbf{r}\left(t\right)\right)\overset{\mathrm{vacuum}}{\to}\delta_{ab}.
\end{equation}
Correspondingly, the scalar propagator, Eq. (\ref{eq:scalarpropagator}),
reduces to the non-relativistic two-dimensional propagator, familiar
from quantum mechanics:
\begin{equation}
\begin{aligned}G_{ab}\left(t_{2},\mathbf{r}_{2},t_{1},\mathbf{r}_{1};\omega\right) & \overset{\mathrm{vacuum}}{\to}\frac{\delta_{ab}}{2\omega}G_{0}\left(t_{2},\mathbf{r}_{2},t_{1},\mathbf{r}_{1};\omega\right),\\
 & \equiv\frac{\delta_{ab}}{2\omega}\int\left[\mathcal{D}\mathbf{r}\left(t\right)\right]\exp\left\{ i\frac{\omega}{2}\int_{t_{1}}^{t_{2}}\mathrm{d}t\,\dot{\mathbf{r}}^{2}\left(t\right)\right\} ,\\
 & =\frac{\delta_{ab}}{2\omega}\int\frac{\mathrm{d}^{2}\mathbf{k}_{\perp}}{\left(2\pi\right)^{2}}e^{i\mathbf{k}_{\perp}\left(\mathbf{r}_{2}-\mathbf{r}_{1}\right)}e^{-i\frac{k_{\perp}^{2}}{2\omega}\left(t_{2}-t_{1}\right)}.
\end{aligned}
\label{eq:vacuumpropagator}
\end{equation}
To proceed, let us focus on the case in which the gluon fluctuation
takes place before the collision: $-\infty<t_{1}\leq t_{2}<0$. The
time integrations over the radiation piece of the in-medium propagator
can be performed as follows: 
\begin{equation}
\begin{aligned} & \int_{-\infty}^{0}\mathrm{d}t_{2}\int_{-\infty}^{t_{2}}\mathrm{d}t_{1}G_{ab,0,\mathrm{rad}}^{--}\left(t_{2},\mathbf{r}_{2},t_{1},\mathbf{r}_{1};\omega\right),\\
 & =\frac{\delta_{ab}}{2\omega^{3}}\int_{-\infty}^{0}\mathrm{d}t_{2}\int_{-\infty}^{t_{2}}\mathrm{d}t_{1}\mathbf{\partial}_{\mathbf{r}_{1}}^{i}\mathbf{\partial}_{\mathbf{r}_{2}}^{i}\int\frac{\mathrm{d}^{2}\mathbf{k}_{\perp}}{\left(2\pi\right)^{2}}e^{i\mathbf{k}_{\perp}\left(\mathbf{r}_{2}-\mathbf{r}_{1}\right)}e^{-i\frac{k_{\perp}^{2}}{2\omega}\left(t_{2}-t_{1}\right)}e^{\epsilon\left(t_{2}+t_{1}\right)},\\
 & =\frac{\delta_{ab}}{2\omega^{3}}\int_{-\infty}^{0}\mathrm{d}t_{2}\int\frac{\mathrm{d}^{2}\mathbf{k}_{\perp}}{\left(2\pi\right)^{2}}k_{\perp}^{2}e^{i\mathbf{k}_{\perp}\left(\mathbf{r}_{2}-\mathbf{r}_{1}\right)}\frac{e^{2\epsilon t_{2}}}{ik^{-}+\epsilon},\\
 & =\frac{\delta_{ab}}{2\omega^{3}}\int\frac{\mathrm{d}^{2}\mathbf{k}_{\perp}}{\left(2\pi\right)^{2}}k_{\perp}^{2}e^{i\mathbf{k}_{\perp}\left(\mathbf{r}_{2}-\mathbf{r}_{1}\right)}\left(\frac{1}{2\epsilon}\frac{1}{ik^{-}}+\frac{1}{2}\frac{1}{\left(k^{-}\right)^{2}}+\mathcal{O}\left(\epsilon\right)\right),
\end{aligned}
\label{eq:timeintradvac}
\end{equation}
where we wrote $k^{-}=k_{\perp}^{2}/2\omega$. In the calculation
above, we introduced an infinitesimal constant $\epsilon>0$ to parametrize
the divergences stemming from gluon fluctuations in the far past.
The finite (in $\epsilon$) part of Eq. (\ref{eq:timeintradvac})
yields (using Eq. (\ref{eq:fourier2Kxyz})):
\begin{equation}
\frac{\delta_{ab}}{\omega}\int\frac{\mathrm{d}^{2}\mathbf{k}_{\perp}}{\left(2\pi\right)^{2}}\frac{1}{k_{\perp}^{2}}e^{i\mathbf{k}_{\perp}\left(\mathbf{r}_{2}-\mathbf{r}_{1}\right)}=\frac{\delta_{ab}}{\omega}\frac{1}{\left(2\pi\right)^{2}}\int\mathrm{d}^{2}\mathbf{z}\,\mathcal{K}_{\mathbf{r}_{1}\mathbf{r}_{2}\mathbf{z}},
\end{equation}
while we obtain for the singular part:
\begin{equation}
-\frac{i}{2\epsilon}\frac{\delta_{ab}}{\omega^{2}}\int\frac{\mathrm{d}^{2}\mathbf{k}_{\perp}}{\left(2\pi\right)^{2}}e^{i\mathbf{k}_{\perp}\left(\mathbf{r}_{2}-\mathbf{r}_{1}\right)}=-\frac{i}{2\epsilon}\frac{\delta_{ab}}{\omega^{2}}\delta^{\left(2\right)}\left(\mathbf{r}_{2}-\mathbf{r}_{1}\right).\label{eq:GEEsingular}
\end{equation}
The \textquoteleft large-time' divergency when $\epsilon\to0$ (see
Fig. \ref{fig:HEEdivergences}) is not physical, and has to cancel
with the similar large-time divergency coming from the Coulomb part
of the propagator. Indeed, we have:
\begin{equation}
\begin{aligned} & \int_{-\infty}^{0}\mathrm{d}t_{2}\int_{-\infty}^{t_{2}}\mathrm{d}t_{1}G_{ab,0,\mathrm{Coul}}^{--}\left(t_{2},\mathbf{r}_{2},t_{1},\mathbf{r}_{1};\omega\right)\\
 & =\frac{1}{2}\int_{-\infty}^{0}\mathrm{d}t_{2}\int_{-\infty}^{0}\mathrm{d}t_{1}\frac{i}{\omega^{2}}\delta_{ab}\delta\left(t_{2}-t_{1}\right)\delta^{\left(2\right)}\left(\mathbf{r}_{2}-\mathbf{r}_{1}\right)e^{\epsilon\left(t_{2}+t_{1}\right)},\\
 & =\frac{i}{2\epsilon}\frac{1}{\omega^{2}}\delta_{ab}\delta^{\left(2\right)}\left(\mathbf{r}_{2}-\mathbf{r}_{1}\right),
\end{aligned}
\label{eq:GEECoulombsingular}
\end{equation}
which is exactly the opposite of Eq. (\ref{eq:GEEsingular}). 

In conclusion, performing the time integrations over the in-medium
propagator, corresponding to a gluon fluctuation before the shockwave,
yields:
\begin{equation}
\int_{-\infty}^{0}\mathrm{d}t_{2}\int_{-\infty}^{t_{2}}\mathrm{d}t_{1}G_{ab}^{--}\left(t_{2},\mathbf{r}_{2},t_{1},\mathbf{r}_{1};\omega\right)=\frac{\delta_{ab}}{\omega}\frac{1}{\left(2\pi\right)^{2}}\int_{\mathbf{z}}\mathcal{K}_{\mathbf{r}_{1}\mathbf{r}_{2}\mathbf{z}}.\label{eq:swnoncrossG}
\end{equation}
When the gluon fluctuation takes place after the scattering: $0<t_{1}\leq t_{2}<\infty$,
exactly the same result is found. This time, the large-time divergences
correspond to fluctuations in the far future (see Fig. \ref{fig:HEEdivergences}),
and are once again cancelled between the radiative and the Coulomb
part of the propagator. 

Let us now study the case where the soft gluon crosses the shockwave:
\begin{equation}
-\infty<t_{1}<0<t_{2}<\infty.
\end{equation}
The Coulomb part of the propagator (\ref{eq:GEE}) disappears, since
it is local in time, hence we only have to compute the radiative term.
Because the medium is a shockwave, we can make the usual assumption
that the transverse position of the gluon doesn't change during the
scattering. As a result, the Wilson line functional $W_{ab}^{\dagger}\left(t_{2},t_{1},\mathbf{r}\left(t\right)\right)$
reduces to the familiar, eikonal, Wilson line $W_{ab}^{\dagger}\left(t_{2},t_{1},\mathbf{r}\left(0\right)\right)$,
and the scalar propagator, Eq. (\ref{eq:scalarpropagator}), can be
simplified as follows:
\begin{equation}
\begin{aligned} & G_{ab}\left(t_{2},\mathbf{r}_{2},t_{1},\mathbf{r}_{1};\omega\right)=\frac{1}{2\omega}\int\left[\mathcal{D}\mathbf{r}\left(t\right)\right]\exp\left\{ i\frac{\omega}{2}\int_{t_{1}}^{t_{2}}\mathrm{d}t\dot{\mathbf{r}}^{2}\left(t\right)\right\} W_{ab}^{\dagger}\left(t_{2},t_{1},\mathbf{r}\left(t\right)\right),\\
 & =\frac{1}{2\omega}\int\left[\mathcal{D}\mathbf{r}\left(t\right)\right]\exp\left\{ i\frac{\omega}{2}\int_{t_{1}}^{t_{2}}\mathrm{d}t\mathbf{\dot{r}}^{2}\left(t\right)\right\} \int\mathrm{d}^{2}\mathbf{z}\,\delta^{\left(2\right)}\left(\mathbf{z}-\mathbf{r}\left(0\right)\right)W_{ab}^{\dagger}\left(t_{2},t_{1},\mathbf{z}\right),\\
 & =\frac{1}{2\omega}\int\mathrm{d}^{2}\mathbf{z}\,G_{0}\left(t_{2},\mathbf{r}_{2}-\mathbf{z};\omega\right)G_{0}\left(-t_{1},\mathbf{z}-\mathbf{r}_{1};\omega\right)W_{ab}^{\dagger}\left(\mathbf{z}\right),
\end{aligned}
\label{eq:propagatorSSA}
\end{equation}
where, in the last line, we used that:
\begin{equation}
W_{ab}^{\dagger}\left(t_{1}<0<t_{2},\mathbf{z}\right)=W_{ab}^{\dagger}\left(\infty,-\infty,\mathbf{z}\right)=W_{ab}^{\dagger}\left(\mathbf{z}\right).
\end{equation}
The time integrations in Eq. (\ref{eq:threetimeintegrals}) can now
be performed separately:
\begin{equation}
\begin{aligned} & \frac{1}{2\omega}\int_{-\infty}^{0}\mathrm{d}t_{1}\partial_{\mathbf{r}_{1}}^{i}G_{0}\left(-t_{1},\mathbf{z}-\mathbf{r}_{1};\omega\right)\\
 & =\frac{1}{2\omega}\int_{-\infty}^{0}\mathrm{d}t_{1}\partial_{\mathbf{r}_{1}}^{i}\int\frac{\mathrm{d}^{2}\mathbf{k}_{\perp}}{\left(2\pi\right)^{2}}e^{i\mathbf{k}_{\perp}\left(\mathbf{z}-\mathbf{r}_{1}\right)}e^{i\frac{k_{\perp}^{2}}{2\omega}t_{1}}e^{\epsilon t_{1}},\\
 & =-\frac{i}{2\omega}\int\frac{\mathrm{d}^{2}\mathbf{k}_{\perp}}{\left(2\pi\right)^{2}}k_{\perp}^{i}e^{i\mathbf{k}_{\perp}\left(\mathbf{z}-\mathbf{r}_{1}\right)}\frac{1}{ik^{-}+\epsilon},\\
 & =-\int\frac{\mathrm{d}^{2}\mathbf{k}_{\perp}}{\left(2\pi\right)^{2}}\frac{k_{\perp}^{i}}{k_{\perp}^{2}}e^{i\mathbf{k}_{\perp}\left(\mathbf{z}-\mathbf{r}_{1}\right)}=\frac{i}{2\pi}\frac{\mathbf{r}_{1}^{i}-\mathbf{z}^{i}}{\left(\mathbf{r}_{1}-\mathbf{z}\right)^{2}},
\end{aligned}
\label{eq:G2K}
\end{equation}
and analogously 
\begin{equation}
\begin{aligned}\frac{1}{2\omega}\int_{0}^{\infty}\mathrm{d}t_{2}\partial_{\mathbf{r}_{2}}^{i}G_{0}\left(t_{2},\mathbf{r}_{2}-\mathbf{z};p^{+}\right) & =\frac{i}{2\pi}\frac{\mathbf{r}_{2}^{i}-\mathbf{z}^{i}}{\left(\mathbf{r}_{2}^{i}-\mathbf{z}^{i}\right)^{2}}.\end{aligned}
\label{eq:G2K2}
\end{equation}
Combining these two results, we find:
\begin{equation}
\begin{aligned}\int_{-\infty}^{0}\mathrm{d}t_{1}\int_{0}^{\infty}\mathrm{d}t_{2}G_{ab}^{--}\left(t_{2},\mathbf{r}_{2},t_{1},\mathbf{r}_{1};\omega\right) & =-\frac{1}{\left(2\pi\right)^{2}}\frac{2}{\omega}\int_{\mathbf{z}}\mathcal{K}_{\mathbf{r}_{1}\mathbf{r}_{2}\mathbf{z}}W_{ab}^{\dagger}\left(\mathbf{z}\right).\end{aligned}
\label{eq:SWcrossG}
\end{equation}

We evaluated the time integrals over the in-medium propagator for
three different configurations, corresponding to a gluon fluctuation
before and after the interaction of the dipole with the shockwave,
and to a fluctuation that participates in the scattering. Combining
the results (\ref{eq:swnoncrossG}) and (\ref{eq:SWcrossG}) with
the action Eq. (\ref{eq:J2LR}) of the functional derivatives $J\left(t,\mathbf{r}\right)$
on the operator $\mathcal{O}$, we can finally calculate the r.h.s.
of the generalized high-energy evolution equation (\ref{eq:GEEeq}):
\begin{equation}
\begin{aligned}H\mathcal{O}\overset{\mathrm{shockwave}}{\to} & \frac{1}{\left(2\pi\right)^{3}}\int_{\mathbf{x}\mathbf{y}\mathbf{z}}\mathcal{K}_{\mathbf{x}\mathbf{y}\mathbf{z}}\left(R_{\mathbf{x}}^{a}R_{\mathbf{y}}^{a}+L_{\mathbf{x}}^{a}L_{\mathbf{y}}^{a}-2L_{\mathbf{x}}^{a}W_{\mathbf{z}}^{\dagger ab}R_{\mathbf{y}}^{b}\right)\mathcal{O}\\
 & =-H_{\mathrm{JIMWLK}}.
\end{aligned}
\label{eq:JIMWLKLR}
\end{equation}
This is indeed another formulation of the JIMWLK Hamiltonian, Eq.
(\ref{eq:JIMWLKH}):
\begin{equation}
\begin{aligned}H_{\mathrm{JIMWLK}} & \equiv-\frac{1}{\left(2\pi\right)^{3}}\int_{\mathbf{x}\mathbf{y}\mathbf{z}}\mathcal{K}_{\mathbf{xyz}}\frac{\delta}{\delta\alpha_{Y}^{a}\left(\mathbf{x}\right)}\left(1+W_{\mathbf{x}}^{\dagger}W_{\mathbf{y}}-W_{\mathbf{x}}^{\dagger}W_{\mathbf{z}}-W_{\mathbf{z}}^{\dagger}W_{\mathbf{y}}\right)^{ab}\frac{\delta}{\delta\alpha_{Y}^{b}\left(\mathbf{y}\right)},\end{aligned}
\label{eq:JIMWLKH-1}
\end{equation}
which can be proven with the help of definitions (\ref{eq:defRL})
and (\ref{eq:defLR}). To do so, let us observe that, for a generic
operator $\mathcal{O}$ that is built from Wilson lines, the action
of $L_{\mathbf{x}}^{a}$ and $R_{\mathbf{x}}^{a}$ is equivalent to:
\begin{equation}
\begin{aligned}L_{\mathbf{x}}^{a}\mathcal{O} & =\frac{\delta}{\delta\alpha_{Y}^{a}\left(\mathbf{x}\right)}\mathcal{O},\\
R_{\mathbf{x}}^{a}\mathcal{O} & =\frac{\delta}{\delta\alpha_{Y}^{b}\left(\mathbf{x}\right)}W_{\mathbf{x}}^{\dagger bc}\mathcal{O},
\end{aligned}
\end{equation}
from which it follows immediately that:
\begin{equation}
\left(L_{\mathbf{x}}^{c}L_{\mathbf{y}}^{c}+R_{\mathbf{x}}^{c}R_{\mathbf{y}}^{c}\right)\mathcal{O}=\frac{\delta}{\delta\alpha_{Y}^{a}\left(\mathbf{x}\right)}\left(1+W_{\mathbf{x}}^{\dagger}W_{\mathbf{y}}\right)^{ab}\frac{\delta}{\delta\alpha_{Y}^{b}\left(\mathbf{y}\right)}\mathcal{O}.
\end{equation}
Furthermore, to prove the equality:
\begin{equation}
\begin{aligned}2L_{\mathbf{x}}^{a}W_{\mathbf{z}}^{\dagger ab}R_{\mathbf{y}}^{b}\mathcal{O} & =\frac{\delta}{\delta\alpha_{Y}^{a}\left(\mathbf{x}\right)}\left(W_{\mathbf{z}}^{\dagger}W_{\mathbf{y}}W_{\mathbf{x}}^{\dagger}W_{\mathbf{z}}\right)^{ab}\frac{\delta}{\delta\alpha_{Y}^{b}\left(\mathbf{y}\right)}\mathcal{O},\end{aligned}
\label{eq:leftrightdefjimwlk}
\end{equation}
we should show that:
\begin{equation}
L_{\mathbf{x}}^{a}W_{\mathbf{z}}^{\dagger ab}R_{\mathbf{y}}^{b}\mathcal{O}=L_{\mathbf{x}}^{a}W_{\mathbf{x}}^{\dagger ab}R_{\mathbf{z}}^{b}\mathcal{O}.\label{eq:lrjimwlkequality}
\end{equation}
Choosing an auxiliary Wilson line $W_{\mathbf{u}}^{\dagger}$, we
find that the l.h.s. of the above equation is equal to (using the
short-hand notation $\delta_{\mathbf{xy}}=\delta^{\left(2\right)}\left(\mathbf{x}-\mathbf{y}\right)$):
\begin{equation}
\begin{aligned}L_{\mathbf{x}}^{a}W_{\mathbf{z}}^{\dagger ac}W_{\mathbf{y}}^{cb}L_{\mathbf{y}}^{b}W_{\mathbf{u}}^{\dagger} & =-ig_{s}\delta_{\mathbf{yu}}L_{\mathbf{x}}^{a}W_{\mathbf{z}}^{\dagger ac}W_{\mathbf{y}}^{cb}T^{b}W_{\mathbf{u}}^{\dagger},\\
 & =g_{s}^{2}\delta_{\mathbf{yu}}\delta_{\mathbf{xy}}W_{\mathbf{z}}^{\dagger ac}W_{\mathbf{y}}^{cd}T_{db}^{a}T^{b}W_{\mathbf{u}}^{\dagger}-g_{s}^{2}\delta_{\mathbf{yu}}\delta_{\mathbf{xu}}W_{\mathbf{z}}^{\dagger ac}W_{\mathbf{y}}^{cb}T^{b}T^{a}W_{\mathbf{u}}^{\dagger},\\
 & =-g_{s}^{2}\delta_{\mathbf{yu}}\delta_{\mathbf{xy}}W_{\mathbf{z}}^{\dagger ac}W_{\mathbf{y}}^{cb}T^{a}T^{b}W_{\mathbf{u}}^{\dagger},
\end{aligned}
\label{eq:leftjimwlkequality}
\end{equation}
where we used that $L_{\mathbf{x}}^{a}W_{\mathbf{z}}^{\dagger ac}\propto T_{ac}^{a}=0$,
as well as:
\begin{equation}
\left[T^{a},T^{b}\right]=-T_{bc}^{a}T^{c}.
\end{equation}
The r.h.s. of Eq. (\ref{eq:lrjimwlkequality}) gives:
\begin{equation}
\begin{aligned}L_{\mathbf{x}}^{a}W_{\mathbf{x}}^{\dagger ac}W_{\mathbf{z}}^{cb}L_{\mathbf{y}}^{b}W_{\mathbf{u}}^{\dagger} & =-ig_{s}\delta_{\mathbf{yu}}L_{\mathbf{x}}^{a}W_{\mathbf{x}}^{\dagger ac}W_{\mathbf{z}}^{cb}T^{b}W_{\mathbf{u}}^{\dagger},\\
 & =g_{s}^{2}\delta_{\mathbf{yu}}\delta_{\mathbf{xz}}W_{\mathbf{x}}^{\dagger ac}W_{\mathbf{z}}^{cd}T_{db}^{a}T^{b}W_{\mathbf{u}}^{\dagger}-g_{s}^{2}\delta_{\mathbf{yu}}\delta_{\mathbf{xu}}W_{\mathbf{x}}^{\dagger ac}W_{\mathbf{z}}^{cb}T^{b}T^{a}W_{\mathbf{u}}^{\dagger},\\
 & =-g_{s}^{2}\delta_{\mathbf{yu}}\delta_{\mathbf{xu}}W_{\mathbf{z}}^{\dagger bc}W_{\mathbf{x}}^{ca}T^{b}T^{a}W_{\mathbf{u}}^{\dagger},
\end{aligned}
\end{equation}
which is equal to Eq. (\ref{eq:leftjimwlkequality}), thus proving
relation (\ref{eq:lrjimwlkequality}). It is then easy to check that
Eq. (\ref{eq:leftrightdefjimwlk}) follows from Eq. (\ref{eq:lrjimwlkequality}),
concluding our proof of the equivalence of the Hamiltonians (\ref{eq:JIMWLKH-1})
and (\ref{eq:JIMWLKLR}).

\subsection{Local form of the evolution Hamiltonian}

We showed that, in the limit where the medium is a shockwave, the
high-energy evolution Hamiltonian, Eq. (\ref{eq:GEE}) is reduced
to the familiar JIMWLK Hamiltonian. The next step in our analysis
is to study our Hamiltonian in the general case of an extended medium,
scrutinizing the possible divergences, and explicitly verifying their
cancellation. 

First, notice that when deriving the JIMWLK equation (cf. Sec. \ref{sec:JIMWLK}),
we were never bothered by large-time divergences, such as the ones
we encountered in the previous paragraph. The reason for this is that
in the case of JIMWLK, the gluon fluctuation was always confined to
the rapidity bin $Y+\mathrm{d}Y\simeq\ln x^{+}+\mathrm{d}x^{+}/x^{+}$
under consideration, and hence to a fixed interval of the longitudinal
position, centered around the position of the shockwave. In the present
case, however, time is a variable over which we integrate. Therefore,
the operators $J^{a}\left(t,\mathbf{r}\right)$ need to be supplemented
with an exponential cutoff, which parametrizes the spurious large-time
divergences:
\begin{equation}
J^{a}\left(t,\mathbf{r}\right)\to J^{a}\left(t,\mathbf{r}\right)e^{-\epsilon\left|t\right|},
\end{equation}
just like we did in the calculations in the previous paragraph. 

With this adjustment, let us study the evolution Hamiltonian (\ref{eq:GEE})
in the vacuum, which we can do by setting the gluon fields in the
target equal to zero: $\alpha_{a}=0$. Since, in this case, the action
of the functional derivatives $J^{a}\left(t,\mathbf{x}\right)$ becomes
time-independent, we can once again explicitly perform the time integrations
over the in-medium propagator Eq. (\ref{eq:inmediumpropagator}):
\begin{equation}
G_{0,ab}^{--}\left(t_{2},\mathbf{r}_{2},t_{1},\mathbf{r}_{1};\omega\right)=\frac{\delta_{ab}}{2\omega^{3}}\mathbf{\partial}_{\mathbf{r}_{1}}^{i}\mathbf{\partial}_{\mathbf{r}_{2}}^{i}G_{0}\left(t_{2},\mathbf{r}_{2},t_{1},\mathbf{r}_{1};\omega\right)+\frac{i}{\omega^{2}}\delta_{ab}\delta\left(t_{2}-t_{1}\right)\delta^{\left(2\right)}\left(\mathbf{r}_{2}-\mathbf{r}_{1}\right).
\end{equation}
The integrations over the radiation part yield, plugging in Eq. (\ref{eq:vacuumpropagator})
for the scalar propagator in Fourier space and in the vacuum:
\begin{equation}
\begin{aligned} & \int\mathrm{d}t_{2}\int\mathrm{d}t_{1}G_{0,\mathrm{rad}}^{--}\left(t_{2},t_{1},\mathbf{k}_{\perp};\omega\right)e^{-\epsilon\left|t_{2}\right|}e^{-\epsilon\left|t_{1}\right|}\\
 & =\frac{k_{\perp}^{2}}{2\omega^{3}}\int_{-\infty}^{+\infty}\mathrm{d}t_{2}\int_{-\infty}^{t_{2}}\mathrm{d}t_{1}e^{-ik^{-}\left(t_{2}-t_{1}\right)}e^{-\epsilon\left|t_{2}\right|}e^{-\epsilon\left|t_{1}\right|},\\
 & =\frac{k_{\perp}^{2}}{2\omega^{3}}\int_{0}^{+\infty}\mathrm{d}t_{2}\biggl(\left.\frac{e^{-ik^{-}\left(t_{2}-t_{1}\right)}e^{-\epsilon t_{1}}}{ik^{-}-\epsilon}\right|_{0}^{t_{2}}+\left.\frac{e^{-ik^{-}\left(t_{2}-t_{1}\right)}e^{\epsilon t_{1}}}{ik^{-}+\epsilon}\right|_{-\infty}^{0}\biggr)e^{-\epsilon t_{2}}\\
 & +\frac{k_{\perp}^{2}}{2\omega^{3}}\int_{-\infty}^{0}\mathrm{d}t_{2}\left.\frac{e^{-ik^{-}\left(t_{2}-t_{1}\right)}e^{\epsilon t_{1}}}{ik^{-}+\epsilon}\right|_{-\infty}^{t_{2}}e^{\epsilon t_{2}}=-\frac{i}{\omega^{2}}\frac{1}{\epsilon},
\end{aligned}
\end{equation}
which cancels precisely with the time integrations over the Coulomb
part of the propagator:
\begin{equation}
\begin{aligned} & \int\mathrm{d}t_{2}\int\mathrm{d}t_{1}G_{0,\mathrm{Coul}}^{--}\left(t_{2},t_{1},\mathbf{k}_{\perp};\omega\right)e^{-\epsilon\left|t_{2}\right|}e^{-\epsilon\left|t_{1}\right|}\\
 & =\frac{i}{\omega^{2}}\int_{-\infty}^{+\infty}\mathrm{d}t_{2}\int_{-\infty}^{t_{2}}\mathrm{d}t_{1}\delta\left(t_{2}-t_{1}\right)e^{-\epsilon\left|t_{2}\right|}e^{-\epsilon\left|t_{1}\right|},\\
 & =\frac{i}{\omega^{2}}\int_{-\infty}^{+\infty}\mathrm{d}t_{2}e^{-2\epsilon\left|t_{2}\right|}=\frac{i}{\omega^{2}}\frac{1}{\epsilon}.
\end{aligned}
\end{equation}
As a result, integrating the in-medium propagator over time, in the
absence of interactions with the medium, yields zero:
\begin{equation}
\int\mathrm{d}t_{2}\int\mathrm{d}t_{1}G_{0,ab}^{--}\left(t_{2},t_{1},\mathbf{k}_{\perp};\omega\right)e^{-\epsilon\left|t_{2}\right|}e^{-\epsilon\left|t_{1}\right|}=0,
\end{equation}
or equivalently:
\begin{equation}
\left.H\mathcal{O}\right|_{\alpha=0}=0.\label{eq:HOnull}
\end{equation}
Rather than being a mathematical coincidence, the above equality is
required to be fulfilled on physical grounds. Indeed, in the absence
of scattering, whatever fluctuations take place in the projectile,
they cannot influence the evolution of the $S$-matrix. In the vacuum,
the action of the evolution Hamiltonian on an $S$-matrix therefore
has to disappear. 

\begin{figure}[t]
\begin{centering}

\begin{tikzpicture}[scale=2] 

\tikzset{photon/.style={decorate,decoration={snake}},
		electron/.style={ postaction={decorate},decoration={markings,mark=at position .5 with {\arrow[draw]{latex}}}},      	gluon/.style={decorate,decoration={coil,amplitude=4pt, segment length=5pt}}}

\fill[black!15!white, rounded corners] (-1,-1) rectangle(1,1);
\draw[electron] (-3,.5) --++ (6,0) node[right]{$q$};
\draw[electron] (3,-.5) node[right]{$\bar{q}$}--++(-6,0);
\draw[gluon] (-2.5,.5) --++ (0,-1);
\draw[gluon] (2.5,.5) --++ (0,-1);
\node at (-3.2,0) {$\mathrm{large\:time}\leftarrow $};
\node at (3.2,0) {$\rightarrow\mathrm{large\:time} $};

\draw[gluon] (-1.5,.5) .. controls (-2,1.1) and (-1,1.1) .. (-1.5,.5);
\draw[gluon] (-1.5,-.5) .. controls (-2,0.1) and (-1,0.1) .. (-1.5,-.5);
\filldraw[fill=black] (-1.5,-.5) circle(.02);
\filldraw[fill=black] (-1.5,.5) circle(.02);
\draw[gluon] (1.5,.5) .. controls (1,1.1) and (2,1.1) .. (1.5,.5);
\draw[gluon] (1.5,-.5) .. controls (1,0.1) and (2,0.1) .. (1.5,-.5);
\filldraw[fill=black] (1.5,-.5) circle(.02);
\filldraw[fill=black] (1.5,.5) circle(.02);
\filldraw[fill=black] (-2.5,.5) circle(.02);
\filldraw[fill=black] (-2.5,-.5) circle(.02);
\filldraw[fill=black] (2.5,.5) circle(.02);
\filldraw[fill=black] (2.5,-.5) circle(.02);
\node at (-1.87,-.7) {$\mathrm{tadpoles}\;\uparrow$};
\node at (1.87,-.7) {$\uparrow\;\mathrm{tadpoles}$};

\end{tikzpicture}
\par\end{centering}
\caption{\label{fig:HEEdivergences}The divergences appearing in the high-energy
evolution of a dipole using the Hamiltonian Eq. (\ref{eq:GEE}).}
\end{figure}
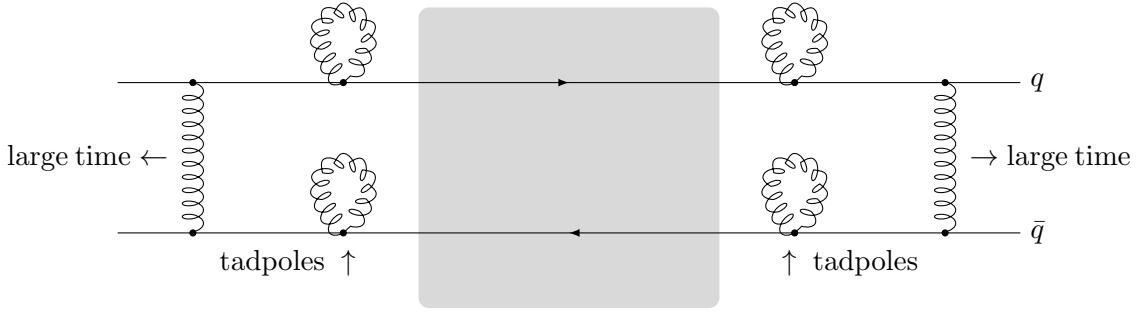

Remarkably, property (\ref{eq:HOnull}) can be exploited to rewrite
the action of the Hamiltonian on an observable in an alternative way,
in which the various cancellations of infinities take place locally
in time. Returning to the case of an extended medium, that is $\alpha_{a}\neq0$,
the action of the two functional derivatives on a dipole $D_{Y}\left(\mathbf{x}-\mathbf{y}\right)\equiv\langle S_{\mathbf{xy}}\rangle_{Y}$
is:
\begin{equation}
\begin{aligned} & J^{b}\left(t_{2},\mathbf{r}_{2}\right)J^{a}\left(t_{1},\mathbf{r}_{1}\right)D_{Y}\left(\mathbf{x}-\mathbf{y}\right)\\
 & =-g_{s}^{2}\delta_{\mathbf{xr}_{1}}\delta_{\mathbf{xr}_{2}}\frac{1}{N_{c}}\mathrm{Tr}\Bigl\langle U_{\mathbf{x}}^{\dagger}\left(\infty,t_{2}\right)t^{b}U_{\mathbf{x}}^{\dagger}\left(t_{2},t_{1}\right)t^{a}U_{\mathbf{x}}^{\dagger}\left(t_{1},-\infty\right)U_{\mathbf{y}}\Bigr\rangle_{Y}\\
 & +g_{s}^{2}\delta_{\mathbf{xr}_{1}}\delta_{\mathbf{yr}_{2}}\frac{1}{N_{c}}\mathrm{Tr}\Bigl\langle U_{\mathbf{x}}^{\dagger}\left(\infty,t_{1}\right)t^{a}U_{\mathbf{x}}^{\dagger}\left(t_{1},-\infty\right)U_{\mathbf{y}}\left(-\infty,t_{2}\right)t^{b}U_{\mathbf{y}}\left(t_{2},\infty\right)\Bigr\rangle_{Y}\\
 & +g_{s}^{2}\delta_{\mathbf{yr}_{1}}\delta_{\mathbf{xr}_{2}}\frac{1}{N_{c}}\mathrm{Tr}\Bigl\langle U_{\mathbf{x}}^{\dagger}\left(\infty,t_{2}\right)t^{b}U_{\mathbf{x}}^{\dagger}\left(t_{2},-\infty\right)U_{\mathbf{y}}\left(-\infty,t_{1}\right)t^{a}U_{\mathbf{y}}\left(t_{1},\infty\right)\Bigr\rangle_{Y}\\
 & -g_{s}^{2}\delta_{\mathbf{yr}_{1}}\delta_{\mathbf{yr}_{2}}\frac{1}{N_{c}}\mathrm{Tr}\Bigl\langle U_{\mathbf{x}}^{\dagger}U_{\mathbf{y}}\left(-\infty,t_{1}\right)t^{a}U_{\mathbf{y}}\left(t_{1},t_{2}\right)t^{b}U_{\mathbf{y}}\left(t_{2},\infty\right)\Bigr\rangle_{Y}.
\end{aligned}
\end{equation}
Assuming that the medium averages are local in time, in the spirit
of the CGC (see Eq. (\ref{eq:MV2point})), they can be factorized,
like we did when computing the dipole in the MV model, cf. Eq. (\ref{eq:dipolecalcu}).
This yields:
\begin{equation}
\begin{aligned} & J^{b}\left(t_{2},\mathbf{r}_{2}\right)J^{a}\left(t_{1},\mathbf{r}_{1}\right)D_{Y}\left(\mathbf{x}-\mathbf{y}\right)\\
 & =g_{s}^{2}\left(\delta_{\mathbf{xr}_{1}}\delta_{\mathbf{yr}_{2}}+\delta_{\mathbf{yr}_{1}}\delta_{\mathbf{xr}_{2}}-\delta_{\mathbf{xr}_{1}}\delta_{\mathbf{xr}_{2}}-\delta_{\mathbf{yr}_{1}}\delta_{\mathbf{yr}_{2}}\right)\\
 & \times\Bigl\langle S_{\mathbf{xy}}\left(t_{2},\infty\right)\Bigr\rangle_{Y}\Bigl\langle S_{\mathbf{xy}}\left(-\infty,t_{1}\right)\Bigr\rangle_{Y}\frac{1}{N_{c}}\mathrm{Tr}\Bigl\langle t^{b}U_{\mathbf{x}}^{\dagger}\left(t_{2},t_{1}\right)t^{a}U_{\mathbf{y}}\left(t_{1},t_{2}\right)\Bigr\rangle_{Y}.
\end{aligned}
\label{eq:JJD}
\end{equation}
The action of the Coulomb part of the Hamiltonian on the dipole then
gives:
\begin{equation}
\begin{aligned} & H_{\mathrm{Coul}}D_{Y}\left(\mathbf{x}-\mathbf{y}\right)=\frac{i}{4\pi}\frac{\delta_{ab}}{\omega}\int_{t_{2}t_{1}}\int_{\mathbf{r}_{2}\mathbf{r}_{1}}\delta\left(t_{2}-t_{1}\right)\delta^{\left(2\right)}\left(\mathbf{r}_{2}-\mathbf{r}_{1}\right)J^{b}\left(t_{2},\mathbf{r}_{2}\right)J^{a}\left(t_{1},\mathbf{r}_{1}\right)D_{Y}\left(\mathbf{x}-\mathbf{y}\right),\\
 & =i\alpha_{s}\frac{C_{F}}{\omega}D_{Y}\left(\mathbf{x}-\mathbf{y}\right)\int_{t}e^{-2\epsilon\left|t\right|}\int_{\mathbf{r}_{2}\mathbf{r}_{1}}\delta^{\left(2\right)}\left(\mathbf{r}_{2}-\mathbf{r}_{1}\right)\left(\delta_{\mathbf{xr}_{1}}\delta_{\mathbf{yr}_{2}}+\delta_{\mathbf{yr}_{1}}\delta_{\mathbf{xr}_{2}}-\delta_{\mathbf{xr}_{1}}\delta_{\mathbf{xr}_{2}}-\delta_{\mathbf{yr}_{1}}\delta_{\mathbf{yr}_{2}}\right),\\
 & =-\delta_{\mathbf{xx}}i\alpha_{s}\frac{C_{F}}{\omega}\frac{1}{\epsilon}D_{Y}\left(\mathbf{x}-\mathbf{y}\right).
\end{aligned}
\label{eq:HcoulD}
\end{equation}
Clearly, the above result is plagued by several divergences, which
are illustrated in Fig. \ref{fig:HEEdivergences}. As we already explained,
$\epsilon$ parametrizes the large-time singularities $\epsilon\to0$.
Furthermore, there is a soft divergence when $\omega\to0$, and finally,
due to the local nature of $H_{\mathrm{Coul}}$, there are transverse
tadpole divergences $\delta^{\left(2\right)}\left(\mathbf{0}\right)$.
Since the divergent eigenvalue of the dipole in Eq. (\ref{eq:HcoulD})
is independent of the gluon fields in the target, we have that:
\begin{equation}
\begin{aligned}H_{\mathrm{Coul}}D_{Y}\left(\mathbf{x}-\mathbf{y}\right) & =-\delta_{\mathbf{xx}}i\alpha_{s}\frac{C_{F}}{\omega}\frac{1}{\epsilon}D_{Y}\left(\mathbf{x}-\mathbf{y}\right),\\
 & =\left.H_{\mathrm{Coul}}D_{Y}\left(\mathbf{x}-\mathbf{y}\right)\right|_{\alpha=0}D_{Y}\left(\mathbf{x}-\mathbf{y}\right).
\end{aligned}
\end{equation}
Therefore, by virtue of Eq. (\ref{eq:HOnull}), the large-time and
tadpole divergences will be cancelled by the action of $H_{\mathrm{rad}}$
on the dipole, like we checked explicitly in the case of a shockwave
(cf. Eqs. (\ref{eq:timeintradvac}), (\ref{eq:GEEsingular}), and
(\ref{eq:GEECoulombsingular})). Using Eq. (\ref{eq:HOnull}), the
above identity allows us to write the action of the Hamiltonian as
follows:
\begin{equation}
\begin{aligned}HD_{Y}\left(\mathbf{x}-\mathbf{y}\right) & =\Bigl(H_{\mathrm{rad}}-\left.H_{\mathrm{rad}}D_{\tau}\left(\mathbf{x}-\mathbf{y}\right)\right|_{\alpha=0}\Bigr)D_{Y}\left(\mathbf{x}-\mathbf{y}\right),\end{aligned}
\end{equation}
which is now formulated only in terms of the radiation part of the
Hamiltonian. Although we took the example of a dipole operator to
prove the above identity, the arguments we used were generic, and
therefore we expect it to hold for any gauge-invariant observable
$\mathcal{O}$ built from Wilson lines: 
\begin{equation}
\begin{aligned}H\mathcal{O} & =\Bigl(H_{\mathrm{rad}}-\left.H_{\mathrm{rad}}\mathcal{O}\right|_{\alpha=0}\Bigr)\mathcal{O},\end{aligned}
\end{equation}
or, more explicitly:
\begin{equation}
\begin{aligned}H\mathcal{O} & =\frac{\omega}{2\pi}\int_{-\infty}^{+\infty}\mathrm{d}t_{2}\int_{-\infty}^{t_{2}}\mathrm{d}t_{1}\int\mathrm{d}^{2}\mathbf{r}_{2}\int\mathrm{d}^{2}\mathbf{r}_{1}\Bigl(\mathcal{H}_{\mathrm{rad}}-\left.\mathcal{H}_{\mathrm{rad}}\mathcal{O}\right|_{\alpha=0}\Bigr)\mathcal{O},\end{aligned}
\label{eq:alternativeGEE}
\end{equation}
where we defined:
\begin{equation}
\mathcal{H}_{\mathrm{rad}}\equiv G_{ab,\mathrm{rad}}^{--}\left(t_{2},\mathbf{r}_{2},t_{1},\mathbf{r}_{1};\omega\right)J^{a}\left(t_{2},\mathbf{r}_{2}\right)J^{b}\left(t_{1},\mathbf{r}_{1}\right).
\end{equation}
Evaluating equation (\ref{eq:alternativeGEE}) in the vacuum, using
$\left.\mathcal{O}\right|_{\alpha=0}=1$, we see that the property
(\ref{eq:HOnull}) is now satisfied locally in time and space:
\begin{equation}
\left.\Bigl(\mathcal{H}_{\mathrm{rad}}-\left.\mathcal{H}_{\mathrm{rad}}\mathcal{O}\right|_{\alpha=0}\Bigr)\mathcal{O}\right|_{\alpha=0}=0,\label{eq:localHOnull}
\end{equation}
implying that the tadpole and large-time divergences are cancelled
locally, before the evaluation of the integrals. When they are harmless
enough, i.e. when they are logarithmic, the soft divergences $\omega\to0$
will of course be resummed by solving the evolution equation. Moreover,
since the contributions of the terms in which the gluon fluctuation
doesn't cross the medium, are always independent of the background
fields $\alpha$, we have that:
\begin{equation}
\mathcal{H}_{\mathrm{rad}}^{\mathrm{non-cross}}\mathcal{O}=\left.\mathcal{H}_{\mathrm{rad}}^{\mathrm{non-cross}}\mathcal{O}\right|_{\alpha=0}\mathcal{O},
\end{equation}
and therefore
\begin{equation}
H^{\mathrm{non-cross}}\mathcal{O}=0.\label{eq:noncrossingtermsdisappear}
\end{equation}
Not only does this imply that only the time orderings in which the
gluon fluctuation crosses the medium play a role in Eq. (\ref{eq:alternativeGEE}),
the above property also assures that the evolution equation in this
form is free from large-time divergences, since these are precisely
generated by the non-crossing terms. 

In conclusion, Eq. (\ref{eq:alternativeGEE}) is a more robust form
of the action of the Hamiltonian, in particular in cases where it
is difficult to perform the time integrations explicitly. Furthermore,
cast in this form, the Hamiltonian allows a probabilistic interpretation,
in which the \textquoteleft real term' $\mathcal{H}_{\mathrm{rad}}\mathcal{O}$
describes the change in the $S$-matrix due to a real emission, while
the \textquoteleft virtual term\textquoteleft{} $\left.\mathcal{H}_{\mathrm{rad}}\mathcal{O}\right|_{\alpha=0}\mathcal{O}$
takes care of the conservation of probability.

As an example, let us calculate the action of Eq. (\ref{eq:alternativeGEE})
on the dipole $S$-matrix. From Eq. (\ref{eq:JJD}), in the case in
which the gluon fluctuation doesn't cross the medium, the functional
derivatives yield: 
\begin{equation}
\begin{aligned} & J^{b}\left(t_{2},\mathbf{r}_{2}\right)J^{a}\left(t_{1},\mathbf{r}_{1}\right)D_{Y}\left(\mathbf{x}-\mathbf{y}\right)\\
 & \overset{\mathrm{non-cross.}}{=}g_{s}^{2}C_{F}\left(\delta_{\mathbf{xr}_{1}}\delta_{\mathbf{yr}_{2}}+\delta_{\mathbf{yr}_{1}}\delta_{\mathbf{xr}_{2}}-\delta_{\mathbf{xr}_{1}}\delta_{\mathbf{xr}_{2}}-\delta_{\mathbf{yr}_{1}}\delta_{\mathbf{yr}_{2}}\right)D_{Y}\left(\mathbf{x}-\mathbf{y}\right),\\
 & =\left.J^{b}\left(t_{2},\mathbf{r}_{2}\right)J^{a}\left(t_{1},\mathbf{r}_{1}\right)D_{Y}\left(\mathbf{x}-\mathbf{y}\right)\right|_{\alpha=0}D_{Y}\left(\mathbf{x}-\mathbf{y}\right),
\end{aligned}
\end{equation}
confirming our assertion (\ref{eq:noncrossingtermsdisappear}) that
only the crossing terms play a role. For these terms, we have:
\begin{equation}
\begin{aligned} & H_{\mathrm{rad}}^{\mathrm{cross}}D_{Y}\left(\mathbf{x}-\mathbf{y}\right)\\
 & =\frac{1}{4\pi}\frac{1}{\omega}\int_{t_{2}t_{1}}^{\mathrm{cross}}\int_{\mathbf{r}_{2}\mathbf{r}_{1}}\mathbf{\partial}_{\mathbf{r}_{1}}^{i}\mathbf{\partial}_{\mathbf{r}_{2}}^{i}\frac{1}{2\omega}\int\left[\mathcal{D}\mathbf{r}\left(t\right)\right]\exp\left\{ i\frac{\omega}{2}\int_{t_{1}}^{t_{2}}\mathrm{d}t\dot{\mathbf{r}}^{2}\left(t\right)\right\} \\
 & \times W_{ab}^{\dagger}\left(t_{2},t_{1},\mathbf{r}\left(t\right)\right)J^{b}\left(t_{2},\mathbf{r}_{2}\right)J^{a}\left(t_{1},\mathbf{r}_{1}\right)D_{Y}\left(\mathbf{x}-\mathbf{y}\right),\\
 & =-\frac{\alpha_{s}}{\omega}\int_{t_{2}t_{1}}^{\mathrm{cross}}\mathbf{\partial}_{\mathbf{r}_{1}}^{i}\mathbf{\partial}_{\mathbf{r}_{2}}^{i}\frac{1}{2\omega}\int\left[\mathcal{D}\mathbf{r}\left(t\right)\right]\exp\left\{ i\frac{\omega}{2}\int_{t_{1}}^{t_{2}}\mathrm{d}t\dot{\mathbf{r}}^{2}\left(t\right)\right\} \\
 & \times\Bigl\langle S_{\mathbf{xy}}\left(t_{2},\infty\right)\Bigr\rangle_{Y}\Bigl\langle S_{\mathbf{xy}}\left(-\infty,t_{1}\right)\Bigr\rangle_{Y}\\
 & \times\left(\frac{N_{c}}{2}\Bigl\langle S_{\mathbf{yr}\left(t\right)}\left(t_{1},t_{2}\right)S_{\mathbf{r}\left(t\right)\mathbf{x}}\left(t_{1},t_{2}\right)\Bigr\rangle_{Y}-\frac{1}{2N_{c}}\Bigl\langle S_{\mathbf{xy}}\left(t_{1},t_{2}\right)\Bigr\rangle_{Y}\right)\biggr|_{\mathbf{r}_{1}=\mathbf{y}}^{\mathbf{r}_{1}=\mathbf{x}}\biggr|_{\mathbf{r}_{2}=\mathbf{y}}^{\mathbf{r}_{2}=\mathbf{x}}.
\end{aligned}
\end{equation}
To obtain the last equality in the calculation above, we plugged the
Wilson functional $W_{ab}^{\dagger}\left(t_{2},t_{1},\mathbf{r}\left(t\right)\right)$
into the medium average in Eq. (\ref{eq:JJD}), after which we made
use of Eq. (\ref{eq:U2W}). In the vacuum limit $\alpha_{a}=0$, the
above result becomes:
\begin{equation}
\begin{aligned}\left.H_{\mathrm{rad}}^{\mathrm{cross}}D_{Y}\left(\mathbf{x}-\mathbf{y}\right)\right|_{\alpha=0} & =-\frac{\alpha_{s}C_{F}}{\omega}\int_{t_{2}t_{1}}^{\mathrm{cross}}\mathbf{\partial}_{\mathbf{r}_{1}}^{i}\mathbf{\partial}_{\mathbf{r}_{2}}^{i}\\
 & \times\frac{1}{2\omega}\int\left[\mathcal{D}\mathbf{r}\left(t\right)\right]\exp\left\{ i\frac{\omega}{2}\int_{t_{1}}^{t_{2}}\mathrm{d}t\dot{\mathbf{r}}^{2}\left(t\right)\right\} \biggr|_{\mathbf{r}_{1}=\mathbf{y}}^{\mathbf{r}_{1}=\mathbf{x}}\biggr|_{\mathbf{r}_{2}=\mathbf{y}}^{\mathbf{r}_{2}=\mathbf{x}},
\end{aligned}
\label{eq:virtualcrossing}
\end{equation}
and therefore, the r.h.s. of evolution equation (\ref{eq:alternativeGEE})
gives:
\begin{equation}
\begin{aligned}HD_{Y}\left(\mathbf{x}-\mathbf{y}\right) & =\frac{\alpha_{s}}{\omega}\frac{N_{c}}{2}\int_{t_{2}t_{1}}^{\mathrm{cross}}\mathbf{\partial}_{\mathbf{r}_{1}}^{i}\mathbf{\partial}_{\mathbf{r}_{2}}^{i}\frac{1}{2\omega}\int\left[\mathcal{D}\mathbf{r}\left(t\right)\right]\exp\left\{ i\frac{\omega}{2}\int_{t_{1}}^{t_{2}}\mathrm{d}t\dot{\mathbf{r}}^{2}\left(t\right)\right\} \\
 & \times\Bigl\langle S_{\mathbf{xy}}\left(t_{2},\infty\right)\Bigr\rangle_{Y}\Bigl\langle S_{\mathbf{xy}}\left(-\infty,t_{1}\right)\Bigr\rangle_{Y}\\
 & \times\left(\Bigl\langle S_{\mathbf{xy}}\left(t_{1},t_{2}\right)\Bigr\rangle_{Y}-\Bigl\langle S_{\mathbf{yr}\left(t\right)}\left(t_{1},t_{2}\right)S_{\mathbf{r}\left(t\right)\mathbf{x}}\left(t_{1},t_{2}\right)\Bigr\rangle_{Y}\right)\biggr|_{\mathbf{r}_{1}=\mathbf{y}}^{\mathbf{r}_{1}=\mathbf{x}}\biggr|_{\mathbf{r}_{2}=\mathbf{y}}^{\mathbf{r}_{2}=\mathbf{x}}.
\end{aligned}
\label{eq:GEEdipole}
\end{equation}
The above formula, in combination with Eq. (\ref{eq:GEEeq}), is our
\textquoteleft master equation' for the non-eikonal high-energy evolution
of a dipole through a medium. It will play a major role in the next
section, where we will use it to calculate the evolution of the jet
quenching parameter.

To conclude this section, let us once again consider the limit in
which the medium is a shockwave, which allows us to explicitly evaluate
the integrals over time: 
\begin{equation}
\int_{t_{2}t_{1}}^{\mathrm{cross}}\overset{\mathrm{sw.}}{\to}2\int_{-\infty}^{0}\mathrm{d}t_{1}\int_{0}^{\infty}\mathrm{d}t_{2}.
\end{equation}
Using the intermediate results from Eqs. (\ref{eq:G2K}) and (\ref{eq:G2K2}),
Eq. (\ref{eq:GEEdipole}) becomes:
\begin{equation}
\begin{aligned}\frac{\partial}{\partial Y}D_{Y}\left(\mathbf{x}-\mathbf{y}\right) & =-\frac{\alpha_{s}C_{F}}{\pi^{2}}\int_{\mathbf{z}}\mathcal{K}_{\mathbf{r}_{1}\mathbf{r}_{2}\mathbf{z}}\left(\Bigl\langle S_{\mathbf{xy}}\Bigr\rangle_{Y}-\Bigl\langle S_{\mathbf{yz}}S_{\mathbf{z}\mathbf{x}}\Bigr\rangle_{Y}\right)\biggr|_{\mathbf{r}_{1}=\mathbf{y}}^{\mathbf{r}_{1}=\mathbf{x}}\biggr|_{\mathbf{r}_{2}=\mathbf{y}}^{\mathbf{r}_{2}=\mathbf{x}}.\\
 & =\frac{\alpha_{s}C_{F}}{\pi^{2}}\int_{\mathbf{z}}\mathcal{M}_{\mathbf{x}\mathbf{y}\mathbf{z}}\left(\Bigl\langle S_{\mathbf{yz}}S_{\mathbf{z}\mathbf{x}}\Bigr\rangle_{Y}-\Bigl\langle S_{\mathbf{xy}}\Bigr\rangle_{Y}\right),
\end{aligned}
\end{equation}
which is of course the first equation in the Balitsky hierarchy, as
expected. We should make the remark that, for a shockwave, the virtual
crossing term Eq. (\ref{eq:virtualcrossing}) is:
\begin{equation}
\begin{aligned}\left.-H_{\mathrm{rad}}^{\mathrm{cross}}D_{Y}\left(\mathbf{x}-\mathbf{y}\right)\right|_{\alpha=0}D_{Y}\left(\mathbf{x}-\mathbf{y}\right) & =-\frac{\alpha_{s}C_{F}}{\pi^{2}}\int_{\mathbf{z}}\mathcal{M}_{\mathbf{x}\mathbf{y}\mathbf{z}}D_{Y}\left(\mathbf{x}-\mathbf{y}\right).\end{aligned}
\end{equation}
Since this term measures the decrease in the probability to find the
original dipole at the time of the interaction, we expect it to be
equal to the contributions from the non-crossing gluon fluctuations,
which take place before or after the interaction. That this is indeed
the case, is apparent from the following calculation:
\begin{equation}
\begin{aligned} & H_{\mathrm{rad}}^{\mathrm{non-cross}}D_{Y}\left(\mathbf{x}-\mathbf{y}\right)=\frac{1}{4\pi}\frac{1}{\omega}\int_{t_{2}t_{1}}^{\mathrm{non-cross}}\int_{\mathbf{r}_{2}\mathbf{r}_{1}}\mathbf{\partial}_{\mathbf{r}_{1}}^{i}\mathbf{\partial}_{\mathbf{r}_{2}}^{i}\frac{1}{2\omega}\int\left[\mathcal{D}\mathbf{r}\left(t\right)\right]\exp\left\{ i\frac{\omega}{2}\int_{t_{1}}^{t_{2}}\mathrm{d}t\dot{\mathbf{r}}^{2}\left(t\right)\right\} \\
 & \times W_{ab}^{\dagger}\left(t_{2},t_{1},\mathbf{r}\left(t\right)\right)J^{b}\left(t_{2},\mathbf{r}_{2}\right)J^{a}\left(t_{1},\mathbf{r}_{1}\right)D_{Y}\left(\mathbf{x}-\mathbf{y}\right),\\
 & =-\frac{\alpha_{s}C_{F}}{\omega}\int_{t_{2}t_{1}}^{\mathrm{non-cross}}\int_{\mathbf{r}_{2}\mathbf{r}_{1}}\mathbf{\partial}_{\mathbf{r}_{1}}^{i}\mathbf{\partial}_{\mathbf{r}_{2}}^{i}\frac{1}{2\omega}\int\left[\mathcal{D}\mathbf{r}\left(t\right)\right]\exp\left\{ i\frac{\omega}{2}\int_{t_{1}}^{t_{2}}\mathrm{d}t\dot{\mathbf{r}}^{2}\left(t\right)\right\} \\
 & \times D_{Y}\left(\mathbf{x}-\mathbf{y}\right)\biggr|_{\mathbf{r}_{1}=\mathbf{y}}^{\mathbf{r}_{1}=\mathbf{x}}\biggr|_{\mathbf{r}_{2}=\mathbf{y}}^{\mathbf{r}_{2}=\mathbf{x}}\quad\overset{\mathrm{sw.}}{=}-\frac{\alpha_{s}C_{F}}{\pi^{2}}\int_{\mathbf{z}}\mathcal{M}_{\mathbf{x}\mathbf{y}\mathbf{z}}D_{Y}\left(\mathbf{x}-\mathbf{y}\right).
\end{aligned}
\end{equation}
This further supports our assertion that the evolution process can
be given a probabilistic interpretation, where $\left(\bar{\alpha}/2\pi\right)\mathcal{M}_{\mathbf{x}\mathbf{y}\mathbf{z}}\mathrm{d}Y\mathrm{d}^{2}\mathbf{z}$
is the differential probability for a dipole, with legs at the positions
$\left(\mathbf{x},\mathbf{y}\right)$, to emit a gluon with energy
$\omega$ at a transverse coordinate $\mathbf{z}$.

\section{Transverse momentum broadening}

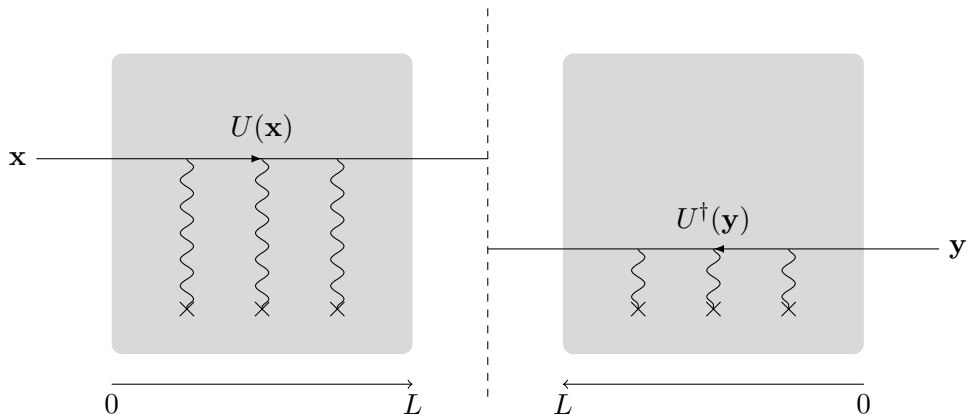
\begin{figure}[t]
\begin{centering}
\begin{tikzpicture}[scale=2] 

\tikzset{photon/.style={decorate,decoration={snake}},
		electron/.style={ postaction={decorate},decoration={markings,mark=at position .5 with {\arrow[draw]{latex}}}},      	gluon/.style={decorate,decoration={coil,amplitude=4pt, segment length=5pt}}}

\fill[black!15!white, rounded corners] (-1,-1) rectangle(1,1);
\fill[black!15!white, rounded corners] (2,-1) rectangle(4,1);
\draw[->] (-1,-1.2) node [below]{$0$} -- (1,-1.2)node [below]{$L$};
\draw[->] (4,-1.2) node [below]{$0$} -- (2,-1.2)node [below]{$L$};
\draw[electron] (-1.5,.3) node[left]{$\mathbf{x}$} --++ (3,0) ;
\draw[electron] (4.5,-.3) node[right]{$\mathbf{y}$}--++(-3,0);
\node at (0,.5) {$U(\mathbf{x})$};
\node at (3,-.1) {$U^\dagger(\mathbf{y})$};

\draw[photon] (-.5,0.3) --++ (0,-1) node [at end, cross out, draw, solid,  inner sep=2.5 pt]{};
\draw[photon] (0,0.3) --++ (0,-1) node [at end, cross out, draw, solid, inner sep=2.5 pt]{};
\draw[photon] (.5,0.3) --++ (0,-1) node [at end, cross out, draw, solid,  inner sep=2.5 pt]{};

\draw[photon] (2.5,-.3) --++ (0,-.4) node [at end, cross out, draw, solid,  inner sep=2.5 pt]{};
\draw[photon] (3,-.3) --++ (0,-.4) node [at end, cross out, draw, solid, inner sep=2.5 pt]{};
\draw[photon] (3.5,-.3) --++ (0,-.4) node [at end, cross out, draw, solid,  inner sep=2.5 pt]{};
\draw[dashed] (1.5,1.3) -- (1.5,-1.3);
\end{tikzpicture}
\par\end{centering}
\caption{\label{fig:TMB}The leading-order cross section for the transverse
momentum broadening of a parton traversing the medium.}
\end{figure}

We are now ready to introduce (one aspect of) the problem of jet quenching,
to which we aim to apply the machinery of non-eikonal high-energy
evolution, on which we elaborated in the previous section.

The transverse momentum distribution of a highly energetic parton,
obtained after traveling through, and interacting with, a nuclear
medium (see Refs. \protect\cite{BDMPS3,KovchegovTuchin,Salgadolectures,MuellerMunier2012}),
is easily calculated at tree level by multiplying the Wilson line
in the amplitude with the one in the complex conjugate amplitude,
which both resum the multiple soft scatterings of the parton with
the background field in the eikonal approximation, see Fig. \ref{fig:TMB}.
The result is:
\begin{equation}
\frac{\mathrm{d}N}{\mathrm{d}^{2}\mathbf{k}_{\perp}}=\frac{1}{\left(2\pi\right)^{2}}\int\mathrm{d}^{2}\mathbf{r}e^{-i\mathbf{k}_{\perp}\cdot\mathbf{r}}D\left(\mathbf{r}\right),\label{eq:tmd}
\end{equation}
where $\mathbf{r}=\mathbf{x}-\mathbf{y}$, and where $D\left(\mathbf{r}\right)$
is the usual dipole, see Eq. (\ref{eq:dipole}), limited to the longitudinal
extent $L$ of the medium:
\begin{equation}
\begin{aligned}D\left(\mathbf{r}\right) & =\exp\left(-\frac{g_{s}^{2}}{\mu_{A}}\frac{C_{F}}{2}\int_{0}^{L}\mathrm{d}x^{+}\lambda_{A}\left(x^{+}\right)\Gamma\left(\mathbf{r}\right)\right).\end{aligned}
\end{equation}
It should be stressed that, in contrast to e.g. DIS, the dipole in
the present formalism is not a physical one, but rather a mathematical
construction corresponding to the quark in the amplitude times its
complex conjugate. 

The second moment of the distribution Eq. (\ref{eq:tmd}) gives the
transverse momentum broadening (TMB) of the parton:
\begin{equation}
\begin{aligned}\left\langle k_{\perp}^{2}\right\rangle  & =\int\mathrm{d}^{2}\mathbf{k}_{\perp}k_{\perp}^{2}\frac{\mathrm{d}N}{\mathrm{d}^{2}\mathbf{k}_{\perp}},\\
 & =\left.-\partial_{\mathbf{r}}^{2}D\left(\mathbf{r}\right)\right|_{\mathbf{r}=\mathbf{0}}.
\end{aligned}
\label{eq:TMB}
\end{equation}
The above definitions are only useful if they come with a set of assumptions
about the nuclear medium. First, since the parton is very fast: $p^{+}=\omega\gg p_{\perp}$,
it only interacts with the \textquoteleft minus' medium gluon fields
$\alpha_{a}\left(x^{+},\mathbf{x}\right)\equiv A_{a}^{-}\left(x^{+},\mathbf{x}\right)$.
For a weakly-coupled quark-gluon plasma, which is the medium in which
we are interested, we can presume \textendash in the spirit of the
Color Glass Condensate\textendash{} that the only non-trivial gluon
correlator is the two-point function: 
\begin{equation}
\begin{aligned}\langle\alpha_{a}\left(x^{+},\mathbf{x}\right)\alpha_{b}\left(y^{+},\mathbf{y}\right)\rangle & =\frac{1}{g_{s}^{2}}\delta_{ab}\delta\left(x^{+}-y^{+}\right)\lambda_{A}\left(x^{+}\right)L\left(\mathbf{x}-\mathbf{y}\right).\end{aligned}
\end{equation}
Note that the roles of the $+$ and $-$ coordinate are reversed with
respect to the corresponding expression in the MV model (see Eq. (\ref{eq:MV2point})),
since in our case the nuclear medium is a left-mover. 

In addition, the plasma comes with a natural infrared cutoff, provided
by the Debye mass $m_{D}$, which corresponds to the screening of
the color interactions over a transverse distance $\sim1/m_{D}$.
Parametrically, we have that $m_{D}^{2}\sim\alpha_{s}N_{c}T^{2}$
and $Q_{s}^{2}\sim\alpha_{s}^{2}N_{c}^{2}T^{3}L$ (see e.g. Ref. \protect\cite{Arnold2009}),
and therefore, for a large medium that is very hot, the Debye mass
is much smaller than the saturation scale $Q_{s}$:
\begin{equation}
m_{D}\ll Q_{s}.
\end{equation}
Finally, the medium is assumed to be uniform both in the longitudinal
direction and in the transverse plane. 

When $r\ll1/m_{D}$, which, as we shall argue later, is always the
case in the regime of our interest, the dipole can be evaluated to
logarithmic accuracy in what we will call the \textquoteleft MV approximation'
(cf. Eqs. (\ref{eq:Gamma}), (\ref{eq:GammaQ}) and (\ref{eq:DipoleMVGamma})),
yielding:
\begin{equation}
\begin{aligned}D\left(\mathbf{r}\right) & \simeq\exp\left(-\frac{1}{4}Q_{s}^{2}\left(r^{2}\right)\mathbf{r}^{2}\right),\\
Q_{s}^{2}\left(r^{2}\right) & \equiv\alpha_{s}C_{F}\mu_{A}\ln\frac{1}{r^{2}m_{D}^{2}}.
\end{aligned}
\label{eq:dipoleQs}
\end{equation}
In the above formula, the transverse color charge density $\mu_{A}$
appears, which in the CGC was given by Eq. (\ref{eq:muA}):
\begin{equation}
\mu_{A}\equiv g_{s}^{2}\int\mathrm{d}z^{+}\lambda_{A}\left(z^{+}\right)=\frac{g_{s}^{2}A}{2\pi R_{A}^{2}}.\label{eq:muA2}
\end{equation}
In the present case, it is more appropriate to work with the number
density $n_{0}$ of the color charges in the medium, related to $\mu_{A}$
as follows:
\begin{equation}
\mu_{A}=g_{s}^{2}\int_{0}^{L}\mathrm{d}z^{+}\lambda_{A}\left(z^{+}\right)=g_{s}^{2}n_{0}L,\label{eq:n0}
\end{equation}
where we used the fact that the medium is homogeneous. Eq. (\ref{eq:dipoleQs})
can now be rewritten as:
\begin{equation}
\begin{aligned}D\left(\mathbf{r}\right) & \simeq\exp\left(-\frac{1}{4}\hat{q}\left(r^{2}\right)L\mathbf{r}^{2}\right),\\
\hat{q}\left(r^{2}\right) & \equiv\frac{Q_{s}^{2}\left(r^{2}\right)}{L}=4\pi\alpha_{s}^{2}C_{F}n_{0}\ln\frac{1}{r^{2}m_{D}^{2}},
\end{aligned}
\label{eq:jetquenchingparameter}
\end{equation}
where we introduced the jet quenching parameter $\hat{q}$: an intrinsic
scale of the nuclear medium whose significance will become more clear
in what follows. Parametrically, we have that $n_{0}\sim N_{c}T^{3}$
(see Ref. \protect\cite{Arnold2009}) and hence $\hat{q}\sim\alpha_{s}^{2}N_{c}^{2}T^{3}$.

With the above expression for the dipole, we are now ready to evaluate
the transverse momentum distribution, Eq. (\ref{eq:tmd}). For momenta
$m_{D}^{2}\ll k_{\perp}^{2}\lesssim Q_{s}^{2}$, the integral over
$\mathbf{r}$ is dominated by small dipole sizes $r\sim1/Q_{s}$ and
can be evaluated by neglecting the logarithmic dependence $\sim\ln r^{2}m_{D}^{2}$
in $\hat{q}$:
\begin{equation}
\begin{aligned}\frac{\mathrm{d}N}{\mathrm{d}^{2}\mathbf{k}_{\perp}} & =\frac{1}{\left(2\pi\right)^{2}}\int_{1/Q_{s}}\mathrm{d}^{2}\mathbf{r}e^{-i\mathbf{k}_{\perp}\cdot\mathbf{r}}\exp\left(-\frac{1}{4}\hat{q}L\mathbf{r}^{2}\right),\\
 & \simeq\frac{1}{\pi Q_{s}^{2}}e^{-k_{\perp}^{2}/Q_{s}^{2}}.
\end{aligned}
\label{eq:transversediffusion}
\end{equation}
From Eq. (\ref{eq:TMB}), we then obtain the following result for
the transverse momentum broadening: 
\begin{equation}
\begin{aligned}\left\langle k_{\perp}^{2}\right\rangle  & =\hat{q}L=Q_{s}^{2}.\end{aligned}
\label{eq:kisql}
\end{equation}
The above results imply that, within approximations of interest, the
transverse momentum broadening of a parton traveling through the medium
can be regarded as a diffusive process. Due to multiple soft scatterings
of the parton with the medium, the system performs a random walk in
transverse momentum space, corresponding to a Gaussian distribution
of the parton momenta. The jet quenching parameter $\hat{q}$, introduced
in Eq. (\ref{eq:jetquenchingparameter}), is found to be equal to
the average transverse momentum broadening of the parton per unit
distance. A gluon with lifetime $\tau$ thus picks up a transverse
momentum:
\begin{equation}
k_{\mathrm{br}}^{2}\left(\tau\right)=\hat{q}\tau.\label{eq:kbr}
\end{equation}
In the case where the whole transverse momentum of the gluon fluctuation
is obtained through the interaction with the medium, the above formula
implies:
\begin{equation}
k_{\mathrm{br}}^{2}\left(\omega\right)=\sqrt{\hat{q}\omega}\quad\mathrm{and}\quad\tau_{\mathrm{br}}\left(\omega\right)=\sqrt{\frac{\omega}{\hat{q}}}.\label{eq:taubrownian}
\end{equation}
After traversing the whole length $L$ of the plasma, the TMB turns
out to be equal to the medium saturation scale $Q_{s}^{2}$. 

Note that, in the above approximation, Eq. (\ref{eq:transversediffusion}),
the probability distribution is already normalized:
\begin{equation}
\int_{k_{\perp}\sim Q_{s}}\mathrm{d}^{2}\mathbf{k}\frac{\mathrm{d}N}{\mathrm{d}^{2}\mathbf{k}_{\perp}}=1.\label{eq:intdN/dk}
\end{equation}
However, the diffusion approximation we just described ignores the
fact that the \textquoteleft true' transverse momentum distribution
has a tail $\sim1/k_{\perp}^{4}$, which is caused by single hard
collisions with $k_{\perp}\gg Q_{s}$. Indeed, in such a case, the
integral in Eq. (\ref{eq:tmd}) over the dipole size is cut off at
a value $r\sim1/k_{\perp}\ll1/Q_{s}$. It is therefore more appropriate
to expand the exponential, but to keep the logarithmic $r$-dependence
of $\hat{q}$:
\begin{equation}
\begin{aligned}\frac{\mathrm{d}N}{\mathrm{d}^{2}\mathbf{k}_{\perp}} & =\frac{1}{\left(2\pi\right)^{2}}\int\mathrm{d}^{2}\mathbf{r}e^{-i\mathbf{k}_{\perp}\cdot\mathbf{r}}\exp\left(-\frac{1}{4}\hat{q}\left(r^{2}\right)L\mathbf{r}^{2}\right),\\
 & \simeq-\frac{\alpha_{s}^{2}}{4\pi}C_{F}n_{0}L\int\mathrm{d}^{2}\mathbf{r}\,\mathbf{r}^{2}e^{-i\mathbf{k}_{\perp}\cdot\mathbf{r}}\ln\frac{1}{r^{2}m_{D}^{2}},\\
 & =-\frac{\alpha_{s}^{2}}{4\pi}C_{F}n_{0}L\times\left(-\frac{16\pi}{k_{\perp}^{4}}+\mathcal{O}\left(Q_{s}\right)\right),\\
 & =\frac{1}{\pi Q_{s}^{2}\ln\left(Q_{s}^{2}/m_{D}^{2}\right)}\frac{Q_{s}^{4}}{k_{\perp}^{4}}.
\end{aligned}
\label{eq:hardtail}
\end{equation}
The above single hard collisions are rare events, suppressed by a
large logarithm $\ln\left(Q_{s}^{2}/m_{D}^{2}\right)$, which explains
why the multiple soft scatterings already approximately exhaust the
integral over the probability Eq. (\ref{eq:intdN/dk}). However, these
hard collisions render the definition (\ref{eq:TMB}) of TMB as the
second moment of $\mathrm{d}N/\mathrm{d}^{2}\mathbf{k}$ divergent
(see also the discussions in Refs. \protect\cite{BDMPS3,Arnold2009-2,Al}).
In what follows, we will therefore always focus on the typical events,
encoded in Eq. (\ref{eq:transversediffusion}), in which the transverse
momentum broadening is a diffusive process caused by multiple soft
scatterings. Accordingly, we assume throughout this work that the
fictitious dipole has a size $r\sim1/Q_{s}$.

\section{Resummation of the soft radiative corrections to the TMB}

Equipped with the machinery that we developed in section \ref{sec:generalJIMWLK},
we are ready to study the high-energy QCD corrections to the transverse
momentum broadening of a parton in a medium. Indeed, the TMB can be
written as the Fourier transform of a fictitious dipole, Eq. (\ref{eq:TMB}).
The soft gluon fluctuations, associated with this dipole, will be
resummed with the help of the BK equation (\ref{eq:GEEdipole}), generalized
beyond the eikonal approximation.

Before delving into the equations, we should stress a couple of remarkable
differences between the present problem and the calculations that
we performed in the context of the Color Glass Condensate. 

First, in the CGC, the energy $E$ of the incoming probe was always
assumed to be much larger than the characteristic energy $\omega_{c}$
of the target. Therefore, a large phase space was available for gluon
fluctuations from the dipole, with energy $\omega_{c}\ll\omega\ll E$
and accordingly very large lifetimes $\tau\gg L$. As a result, the
target could be treated as a shockwave, and the large logarithms $\alpha_{s}\ln E/\omega_{c}\simeq\alpha_{s}\ln1/x$
were resummed in the JIMWLK/BK equation. In the present context of
jet quenching, however, the energy phase space looks completely different.
Now, both the probe and the target have an energy scale of the same
order ($\sim10^{2}\,\mathrm{GeV}$ at the LHC), hence the soft gluon
fluctuations that we are interested in have an energy $\omega$ much
smaller than the medium energy scale $\omega_{c}$, and correspondingly
a short lifetime $\tau$ as compared to the medium length:
\begin{equation}
\omega\ll\omega_{c}\quad\mathrm{and}\quad\tau\ll L.\label{eq:mediumphasespace0}
\end{equation}
Thus, the large logarithms that will be resummed in the problem at
hand stem from the following region of phase space:
\begin{equation}
\omega_{0}\ll\omega\ll\omega_{c},
\end{equation}
where $\omega_{0}$ is a lower cutoff which will be specified in a
moment.

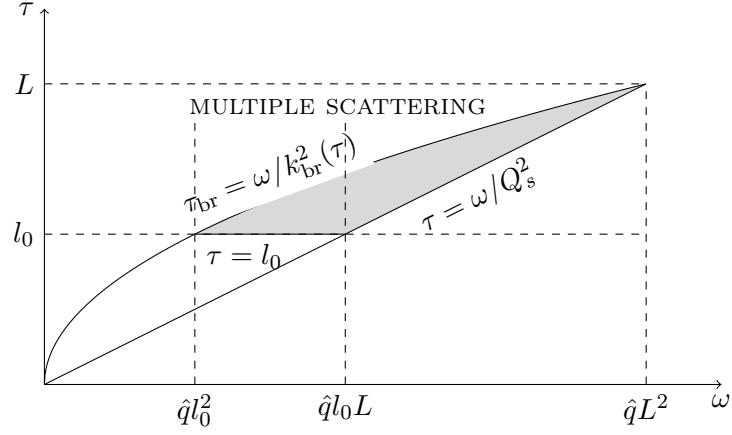
\begin{figure}[t]
\begin{centering}

\begin{tikzpicture}[scale=2] 

\tikzset{electron/.style={ postaction={decorate},decoration={markings,mark=at position .5 with {\arrow[]{>}}}},      	gluon/.style={decorate,decoration={coil,amplitude=4pt, segment length=5pt}}};
\draw[->] (0,0)  --++(4.5,0) node[below]{$\omega$};
\draw[->] (0,0)  --++(0,2.5) node[left]{$\tau$};

\fill[color=black!15!white]  plot [samples = 1000, domain=1:4] (\x,{sqrt(\x)}) -- (4,2) -- (2,1) -- cycle;
\draw[samples = 1000, domain=0:4]  plot(\x,{sqrt(\x)}) ;
\draw(0,0)  --++(4,2) node[below,sloped,pos=.7]{$\tau=\omega / Q_{s}^2$};
\draw(1,1)  --++(1,0) node[below,sloped,pos=.33]{$\tau=l_0$};
\node[rotate=20,preaction={fill, white}] at (1.5,1.4) {$\tau_\mathrm{br}=\omega / k_{\mathrm{br}}^{2}(\tau)$};
\draw[dashed] (1,0) node[below]{$\hat{q}l_0^2$} --++(0,2);
\draw[dashed] (0,1) node[left]{$l_0$} --++(4,0);
\draw[dashed] (2,0) node[below]{$\hat{q}l_0 L$} --++(0,2);
\draw[dashed] (0,2) node[left]{$L$} --++(4,0);
\draw[dashed] (4,0) node[below]{$\hat{q}L^2$} --++(0,2);
\node[preaction={fill, white}] at (1.95,1.85) {\sc{multiple scattering}}; 

\end{tikzpicture} 
\par\end{centering}
\centering{}\caption{\label{fig:gluonphasespace}The phase space for the virtual gluon
in the high-energy evolution of the jet quenching parameter. The shaded
region, bounded by the limits $k_{\mathrm{br}}^{2}\leq k_{\perp}^{2}\leq Q_{s}^{2}$
and $l_{0}\leq\tau\leq L$, corresponds to the kinematic region of
single scattering.}
\end{figure}

Related to this, there is a second, more dramatic difference with
the shockwave case: the fact that in the phase space for the gluon
fluctuation, in the approximation of our interest, the medium introduces
a strong correlation between the limits for the energy $\omega$ and
for the transverse momentum $k_{\perp}^{2}$ of the gluon. We will
come back to this in detail later, but let us already present one
consequence of this correlation, namely the precise expression for
the maximum $\omega_{c}$ of the gluon's energy. Indeed, from Eq.
(\ref{eq:mediumphasespace0}), we have that lifetime of the gluon
fluctuation is limited by the medium size, hence:
\begin{equation}
\tau_{\mathrm{max}}=L.
\end{equation}
Since $\tau=2\omega/p_{\perp}^{2}$, this means that:
\begin{equation}
\frac{2\omega_{c}}{p_{\perp,\mathrm{min}}^{2}}=L.
\end{equation}
The minimal transverse momentum $p_{\perp,\mathrm{min}}^{2}$ of the
gluon is now obtained from the argument that, for small transverse
momenta, the derivation of the momentum broadening of an eikonal parton,
in the previous chapter, equally applies to the gluon fluctuation
under consideration. Hence, during its lifetime the gluon obtains
at least a momentum $p_{\perp,\mathrm{min}}^{2}\left(\tau\right)=\hat{q}\tau$
due to multiple soft scattering, and after traveling through the medium
$p_{\perp,\mathrm{min}}^{2}\left(L\right)=\hat{q}L$ (cf. Eqs. (\ref{eq:kisql}),
(\ref{eq:kbr})), from which we obtain the following value for the
maximal gluon energy $\omega_{c}$:
\begin{equation}
\omega_{c}=Q_{s}^{2}L=\hat{q}L^{2}.
\end{equation}
Likewise, a lower cutoff $l_{0}$ on the lifetime, whose physical
origin we will explain later, yields a value $\omega_{0}=\hat{q}l_{0}^{2}$
for the lower limit on the energy. Thus, the presence of the medium
imposes restrictions on the lifetime of the gluon fluctuations, resulting
in the phase space (see Fig. \ref{fig:gluonphasespace}):
\begin{equation}
l_{0}\ll\tau\ll L\quad\leftrightarrow\quad\omega_{0}=\hat{q}l_{0}^{2}\ll\omega\ll\omega_{c}=\hat{q}L^{2}.\label{eq:qhatenergyphasespace}
\end{equation}

As we will see, a second consequence of the mutual dependency of the
transverse and longitudinal phase space is the existence of double
large logarithms in the same large ratio $\omega_{c}/\omega_{0}=L/l_{0}$.
This is in contrast to the DLA in the shockwave, in which the logarithms
$\ln Q^{2}/\Lambda_{\mathrm{QCD}}^{2}$ and $\ln1/x$ are large separately
(cf. (\ref{eq:xGDLLA})). We will analytically perform the resummation
of the double large logarithms associated with soft gluon fluctuations,
and demonstrate that they can be absorbed into a renormalization of
the jet quenching parameter.

\subsection{Evolution equation for the dipole}

In order to apply the generalized high-energy evolution equation for
a dipole, Eqs. (\ref{eq:GEEeq}) and (\ref{eq:GEEdipole}), to the
problem of transverse momentum broadening, let us perform a couple
of manipulations to cast the equation into a more useful form. First,
using
\begin{equation}
\begin{aligned}\frac{\partial D_{\omega}\left(\mathbf{x}-\mathbf{y}\right)}{\partial\omega}\end{aligned}
=D_{\omega}\left(\mathbf{x}-\mathbf{y}\right)\begin{aligned}\frac{\partial\ln D_{\omega}\left(\mathbf{x}-\mathbf{y}\right)}{\partial\omega}\end{aligned}
,
\end{equation}
the evolution equation becomes:
\begin{equation}
\begin{aligned}\omega\begin{aligned}\frac{\partial\ln D_{\omega}\left(\mathbf{x}-\mathbf{y}\right)}{\partial\omega}\end{aligned}
 & =\frac{\alpha_{s}}{\omega}\frac{N_{c}}{2}\int_{t_{2}t_{1}}^{\mathrm{cross}}\mathbf{\partial}_{\mathbf{r}_{1}}^{i}\mathbf{\partial}_{\mathbf{r}_{2}}^{i}\frac{1}{2\omega}\int\left[\mathcal{D}\mathbf{r}\left(t\right)\right]\exp\left\{ i\frac{\omega}{2}\int_{t_{1}}^{t_{2}}\mathrm{d}t\dot{\mathbf{r}}^{2}\left(t\right)\right\} \\
 & \times\left(1-\Bigl\langle S_{\mathbf{yr}\left(t\right)}\left(t_{1},t_{2}\right)S_{\mathbf{r}\left(t\right)\mathbf{x}}\left(t_{1},t_{2}\right)\Bigr\rangle_{\omega}\Bigl\langle S_{\mathbf{x}\mathbf{y}}\left(t_{1},t_{2}\right)\Bigr\rangle_{\omega}^{-1}\right)\biggr|_{\mathbf{r}_{1}=\mathbf{y}}^{\mathbf{r}_{1}=\mathbf{x}}\biggr|_{\mathbf{r}_{2}=\mathbf{y}}^{\mathbf{r}_{2}=\mathbf{x}}.
\end{aligned}
\label{eq:towardsGEEDipapplied}
\end{equation}
Note that we write the evolution in function of the energy $\omega$,
rather than the rapidity $\tau=\ln P/\omega$, in order to prevent
confusion with the gluon's lifetime, also denoted $\tau$. We are
interested in soft gluon fluctuations whose probability is enhanced
by a large logarithm in the energy $\sim\ln\omega_{c}/\omega$. Correspondingly,
we expect the lifetime $\tau\simeq\omega/k_{\perp}^{2}$ of the gluon
to be small as well, hence it is a good approximation to assume that
the fluctuation takes place deeply inside the medium, i.e. $0\ll t_{1}<t_{2}\ll L$.
The time integrations in Eq. (\ref{eq:towardsGEEDipapplied}) thus
reduce to:
\begin{equation}
\int_{t_{2}t_{1}}^{\mathrm{cross}}\to\int_{0}^{L}\mathrm{d}t_{2}\int_{0}^{t_{2}}\mathrm{d}t_{1}.
\end{equation}
Furthermore, introducing the notation:
\begin{equation}
\bar{\Gamma}_{\omega}\left(\mathbf{r}\right)\equiv\frac{n_{0}}{2\mu_{A}}\Gamma_{\omega}\left(\mathbf{r}\right),
\end{equation}
the (anti)quark-gluon dipole in the medium can be written as:
\begin{equation}
\begin{aligned}D_{\omega}\left(t_{2},t_{1},\mathbf{r}\right) & =\exp\Bigl(-g_{s}^{2}C_{F}\tau\bar{\Gamma}_{\omega}\left(\mathbf{r}\right)\Bigr),\end{aligned}
\label{eq:dipoleinmedium}
\end{equation}
where we identified $\tau=t_{2}-t_{1}$. Lastly, in the large-$N_{c}$
limit, the medium average over the two dipoles factorizes, and we
can finally write: 
\begin{equation}
\begin{aligned}\begin{aligned}L\omega\frac{\partial\bar{\Gamma}_{\omega}\left(\mathbf{x},\mathbf{y}\right)}{\partial\omega}\end{aligned}
 & =\frac{1}{4\pi\omega^{2}}\int_{0}^{L}\mathrm{d}t_{2}\int_{0}^{t_{2}}\mathrm{d}t_{1}\mathbf{\partial}_{\mathbf{r}_{1}}^{i}\mathbf{\partial}_{\mathbf{r}_{2}}^{i}\int\left[\mathcal{D}\mathbf{r}\left(t\right)\right]\exp\left\{ i\frac{\omega}{2}\int_{t_{1}}^{t_{2}}\mathrm{d}t\dot{\mathbf{r}}^{2}\left(t\right)\right\} \\
 & \times\left[\exp\left\{ -g_{s}^{2}\frac{N_{c}}{2}\int_{t_{1}}^{t_{2}}\mathrm{d}t\left(\bar{\Gamma}_{\omega}\left(\mathbf{x},\mathbf{r}\left(t\right)\right)+\bar{\Gamma}_{\omega}\left(\mathbf{y},\mathbf{r}\left(t\right)\right)-\bar{\Gamma}_{\omega}\left(\mathbf{x},\mathbf{y}\right)\right)\right\} -1\right]\biggr|_{\mathbf{r}_{1}=\mathbf{y}}^{\mathbf{r}_{1}=\mathbf{x}}\biggr|_{\mathbf{r}_{2}=\mathbf{y}}^{\mathbf{r}_{2}=\mathbf{x}}.
\end{aligned}
\label{eq:qhatevolutionequation}
\end{equation}
In what follows, we further study the above equation, scrutinize its
regions of validity, and solve it in the double leading logarithmic
approximation.

\subsection{Single scattering approximation}

It is apparent from formula (\ref{eq:dipoleinmedium}) that the dipole
$S$-matrix resums the amplitudes $g_{s}^{2}C_{F}\bar{\Gamma}_{\omega}\left(\mathbf{r}\right)$
for a single scattering per unit time, for which we established the
evolution equation (\ref{eq:qhatevolutionequation}). When the exponent
of Eq. (\ref{eq:dipoleinmedium}) is small enough, i.e. when:
\begin{equation}
g_{s}^{2}C_{F}\tau\bar{\Gamma}_{\omega}\left(\mathbf{r}\right)\ll1,\label{eq:singlescatteringrequirement}
\end{equation}
the dipole can be expanded to leading order:
\begin{equation}
\begin{aligned}D_{\omega}\left(t_{2},t_{1},\mathbf{r}\right) & \simeq1-g_{s}^{2}C_{F}\tau\bar{\Gamma}_{\omega}\left(\mathbf{r}\right),\end{aligned}
\end{equation}
approximating the full dipole-medium interaction with one single scattering
off the medium constituents.

In the previous paragraph we argued that, since we are interested
in the contributions that are amplified by large logarithms, we can
focus on gluon fluctuations in the soft limit $\omega\to0$, in which
the gluon's lifetime $\tau\simeq\omega/k_{\perp}^{2}$ is very short.
At first sight, this leading logarithmic approximation complies with
the requirement for single scattering Eq. (\ref{eq:singlescatteringrequirement}).
However, in the denominator of $\tau$, there is the transverse momentum
$k_{\perp}^{2}$ of the gluon, which, when very small, can still invalidate
Eq. (\ref{eq:singlescatteringrequirement}). This implies that the
single scattering regime is naturally associated with the double,
rather than the single, leading logarithmic approximation of evolution
equation Eq. (\ref{eq:qhatevolutionequation}). In this regime, $\omega$
is small enough and $k_{\perp}$ is large enough for requirement (\ref{eq:singlescatteringrequirement})
to be satisfied. 

To show more precisely how the DLA solution to Eq. (\ref{eq:qhatevolutionequation})
comes into play, let us simplify the evolution equation (\ref{eq:qhatevolutionequation})
within the single scattering approximation by linearizing the exponential
in the r.h.s.:
\begin{equation}
\begin{aligned}\begin{aligned}L\omega\frac{\partial\bar{\Gamma}_{\omega}\left(\mathbf{r}\right)}{\partial\omega}\end{aligned}
 & =-\frac{\alpha_{s}}{\omega^{2}}\frac{N_{c}}{2}\int_{0}^{L}\mathrm{d}t_{2}\int_{0}^{t_{2}}\mathrm{d}t_{1}\int_{t_{1}}^{t_{2}}\mathrm{d}t\mathbf{\partial}_{\mathbf{r}_{1}}^{i}\mathbf{\partial}_{\mathbf{r}_{2}}^{i}\int\left[\mathcal{D}\mathbf{r}\left(t\right)\right]\exp\left\{ i\frac{\omega}{2}\int_{t_{1}}^{t_{2}}\mathrm{d}t\dot{\mathbf{r}}^{2}\left(t\right)\right\} \\
 & \times\left(\bar{\Gamma}_{\omega}\left(\mathbf{x},\mathbf{r}\left(t\right)\right)+\bar{\Gamma}_{\omega}\left(\mathbf{y},\mathbf{r}\left(t\right)\right)-\bar{\Gamma}_{\omega}\left(\mathbf{x},\mathbf{y}\right)\right)\biggr|_{\mathbf{r}_{1}=\mathbf{y}}^{\mathbf{r}_{1}=\mathbf{x}}\biggr|_{\mathbf{r}_{2}=\mathbf{y}}^{\mathbf{r}_{2}=\mathbf{x}}.
\end{aligned}
\label{eq:qhatssa}
\end{equation}
Rewriting the time integrations as follows:
\begin{equation}
\int_{0}^{L}\mathrm{d}t_{2}\int_{0}^{t_{2}}\mathrm{d}t_{1}\int_{t_{1}}^{t_{2}}\mathrm{d}t=\int_{0}^{L}\mathrm{d}t\int_{0}^{t}\mathrm{d}t_{1}\int_{t}^{L}\mathrm{d}t_{2},
\end{equation}
and introducing $1=\int_{\mathbf{z}}\delta^{\left(2\right)}\left(\mathbf{z}-\mathbf{r}\left(t\right)\right)$,
Eq. (\ref{eq:qhatssa}) becomes:
\begin{equation}
\begin{aligned} & \begin{aligned}L\omega\frac{\partial\bar{\Gamma}_{\omega}\left(\mathbf{r}\right)}{\partial\omega}\end{aligned}
=-\frac{\alpha_{s}}{\omega^{2}}\frac{N_{c}}{2}\int_{0}^{L}\mathrm{d}t\int_{0}^{t}\mathrm{d}t_{1}\int_{t}^{L}\mathrm{d}t_{2}\mathbf{\partial}_{\mathbf{r}_{1}}^{i}\mathbf{\partial}_{\mathbf{r}_{2}}^{i}\\
 & \times\int\mathrm{d}^{2}\mathbf{z}\,G_{0}\left(t_{2},\mathbf{r}_{2},t,\mathbf{z};\omega\right)G_{0}\left(t,\mathbf{z},t_{1},\mathbf{r}_{1};\omega\right)\left(\bar{\Gamma}_{\omega}\left(\mathbf{x},\mathbf{z}\right)+\bar{\Gamma}_{\omega}\left(\mathbf{y},\mathbf{z}\right)-\bar{\Gamma}_{\omega}\left(\mathbf{x},\mathbf{y}\right)\right)\biggr|_{\mathbf{r}_{1}=\mathbf{y}}^{\mathbf{r}_{1}=\mathbf{x}}\biggr|_{\mathbf{r}_{2}=\mathbf{y}}^{\mathbf{r}_{2}=\mathbf{x}}.
\end{aligned}
\end{equation}
The time integrals can now be calculated explicitly, similar to what
we did in Eqs. (\ref{eq:G2K}) and (\ref{eq:G2K2}):
\begin{equation}
\begin{aligned} & \frac{1}{2\omega}\int_{t}^{L}\mathrm{d}t_{2}\partial_{\mathbf{r}_{2}}^{i}G_{0}\left(t_{2}-t,\mathbf{r}_{2}-\mathbf{z};\omega\right)\\
 & =-\frac{i}{2\omega}\int\frac{\mathrm{d}^{2}\mathbf{k}_{\perp}}{\left(2\pi\right)^{2}}k_{\perp}^{i}e^{i\mathbf{k}_{\perp}\left(\mathbf{r}_{2}-\mathbf{z}\right)}\frac{1}{ik^{-}}\left(e^{-ik^{-}\left(L-t\right)}-1\right),\\
 & \simeq\frac{i}{2\pi}\frac{\mathbf{r}_{2}^{i}-\mathbf{z}^{i}}{\left(\mathbf{r}_{2}-\mathbf{z}\right)^{2}},
\end{aligned}
\label{eq:G2Kmedium1}
\end{equation}
and
\begin{equation}
\begin{aligned} & \frac{1}{2\omega}\int_{0}^{t}\mathrm{d}t_{1}\partial_{\mathbf{r}_{1}}^{i}G_{0}\left(t-t_{1},\mathbf{z}-\mathbf{r}_{1};\omega\right)\\
 & =-\frac{i}{2\omega}\int\frac{\mathrm{d}^{2}\mathbf{k}_{\perp}}{\left(2\pi\right)^{2}}k_{\perp}^{i}e^{i\mathbf{k}_{\perp}\left(\mathbf{z}-\mathbf{r}_{1}\right)}\frac{1}{ik^{-}}\left(1-e^{-ik^{-}t}\right),\\
 & \simeq\frac{i}{2\pi}\frac{\mathbf{r}_{1}^{i}-\mathbf{z}^{i}}{\left(\mathbf{r}_{1}-\mathbf{z}\right)^{2}}.
\end{aligned}
\label{eq:G2Kmedium2}
\end{equation}
In the above calculations, we neglected the exponentials $e^{-ik^{-}\left(L-t\right)}\approx e^{-ik^{-}t}\approx0$
in order to obtain the familiar Weizsäcker-Williams wave functions
$\sim x^{i}/x^{2}$. These approximations are well-justified within
our LLA approach, since we both assume that the scattering takes place
deeply inside the medium $0\ll t\ll L$, and require that the gluon
lifetime $\tau$ is small, hence $t+\tau\ll L$ and $\tau\ll t$.

From the above intermediate results, we easily obtain the following
equation:
\begin{equation}
\begin{aligned}\omega\begin{aligned}\frac{\partial\bar{\Gamma}_{\omega}\left(\mathbf{x},\mathbf{y}\right)}{\partial\omega}\end{aligned}
 & =\frac{\bar{\alpha}}{2\pi}\int\mathrm{d}^{2}\mathbf{z}\mathcal{M}_{\mathbf{x}\mathbf{y}\mathbf{z}}\left(\bar{\Gamma}_{\omega}\left(\mathbf{x},\mathbf{z}\right)+\bar{\Gamma}_{\omega}\left(\mathbf{y},\mathbf{z}\right)-\bar{\Gamma}_{\omega}\left(\mathbf{x},\mathbf{y}\right)\right),\end{aligned}
\label{eq:BFKLmedium}
\end{equation}
in which finally the logarithmic enhancement $\mathrm{d}\omega/\omega$
is explicit. Although this evolution equation closely resembles the
BFKL equation for a dipole in a shockwave (see Eq. (\ref{eq:BFKL})),
it differs on some crucial points. Indeed, the regular BFKL equation
is an evolution equation for dipole amplitudes $\langle T\left(\mathbf{r}\right)\rangle_{\omega}$
in the dilute regime, valid in the limit in which these amplitudes,
and therefore the corresponding dipole sizes, are small: $\langle T\left(\mathbf{r}\right)\rangle_{\omega}\sim r^{2}\ll1/Q_{s}^{2}$
(cf. (\ref{eq:sigmadipdilute})). Eq. (\ref{eq:BFKLmedium}), however,
is formulated in function of the amplitudes per unit time $\bar{\Gamma}_{\omega}\left(\mathbf{r}\right)$,
for which the condition (\ref{eq:singlescatteringrequirement}) has
to hold. Since, from the requirement $\tau\ll L$, the (anti)quark-gluon
dipoles are very short-living in comparison with their parents, they
can become very large and still satisfy Eq. (\ref{eq:singlescatteringrequirement}):
\begin{equation}
\begin{aligned}\bar{\Gamma}_{\omega}\left(\mathbf{x},\mathbf{z}\right),\bar{\Gamma}_{\omega}\left(\mathbf{y},\mathbf{z}\right) & \gg\bar{\Gamma}_{\omega}\left(\mathbf{x},\mathbf{y}\right).\end{aligned}
\label{eq:parentneglection}
\end{equation}
In the MV approximation, 
\begin{equation}
g_{s}^{2}C_{F}\bar{\Gamma}_{\omega}\left(r\right)\simeq\frac{1}{4}\hat{q}\left(r^{2}\right)r^{2},\label{eq:MV approximation}
\end{equation}
which follows from Eqs. (\ref{eq:jetquenchingparameter}) and (\ref{eq:dipoleinmedium}),
formula (\ref{eq:parentneglection}) becomes:
\begin{equation}
\begin{aligned}\left(\mathbf{x}-\mathbf{z}\right)^{2},\left(\mathbf{y}-\mathbf{z}\right)^{2} & \gg\left(\mathbf{x}-\mathbf{y}\right)^{2}\sim1/Q_{s}^{2}.\end{aligned}
\label{eq:parentneglectionHO}
\end{equation}
This observation motivates us to write an approximation to Eq. (\ref{eq:BFKLmedium})
which \textendash as will become clear shortly\textendash{} is tantamount
to the double logarithmic approximation. Assuming that the daughter
dipoles are indeed much larger than their parent, as in Eq. (\ref{eq:parentneglection}),
the latter can be omitted from the r.h.s. of Eq. (\ref{eq:BFKLmedium}).
The difference between both daughter dipoles then becomes negligible,
hence we write $\bar{\Gamma}_{\omega}\left(\mathbf{x},\mathbf{z}\right)\simeq\bar{\Gamma}_{\omega}\left(\mathbf{y},\mathbf{z}\right)$
and $\mathbf{B}^{2}\equiv\left(\mathbf{x}-\mathbf{z}\right)^{2}\simeq\left(\mathbf{y}-\mathbf{z}\right)^{2}$,
thus obtaining:
\begin{equation}
\begin{aligned}\omega\begin{aligned}\frac{\partial\bar{\Gamma}_{\omega}\left(r\right)}{\partial\omega}\end{aligned}
 & =\bar{\alpha}r^{2}\int_{1/Q_{s}^{2}}^{2/k_{\mathrm{br}}^{2}\left(\omega\right)}\frac{\mathrm{d}B^{2}}{B^{4}}\bar{\Gamma}_{\omega}\left(B\right),\end{aligned}
\label{eq:BFKLmediumDLA}
\end{equation}
where $\mathbf{r}\equiv\mathbf{x}-\mathbf{y}$ is the size of the
parent dipole. The integration limits in the above equation are crucial,
and are obtained as follows: first, in order for the assumption (\ref{eq:parentneglection}),
which asserts that the daughter dipoles are clearly distinguishable
from their parent, to be valid, we require that:
\begin{equation}
B^{2}\gg r^{2}\sim1/Q_{s}^{2}.\label{eq:Bupper}
\end{equation}
again making use of the MV approximation Eq. (\ref{eq:MV approximation}).
Second, the upper limit in Eq. (\ref{eq:BFKLmediumDLA}) comes from
the single scattering requirement, Eq. (\ref{eq:singlescatteringrequirement}):
\begin{equation}
2g_{s}^{2}C_{F}\tau\bar{\Gamma}_{\omega}\left(\mathbf{B}\right)\ll1,
\end{equation}
where the factor two accounts for the two daughter dipoles. Using
the MV approximation to write $g_{s}^{2}C_{F}\bar{\Gamma}_{\omega}\left(\mathbf{B}\right)=\frac{1}{4}\hat{q}B^{2}$,
the above expression yields the following upper limit on the daughter
dipole size:
\begin{equation}
B^{2}\ll\frac{2}{\hat{q}\tau}=\frac{2}{k_{\mathrm{br}}^{2}\left(\omega\right)},\label{eq:Blower}
\end{equation}
where we identified $\hat{q}\tau$ with the average transverse momentum
acquired by the gluon due to multiple soft scatterings off the medium
constituents: $k_{\mathrm{br}}^{2}\left(\omega\right)=\hat{q}\tau\simeq\sqrt{\omega\hat{q}}$
(see Eq. (\ref{eq:kbr})). The onset of the multiple soft scattering
regime therefore acts as the infrared cutoff (instead of $\Lambda_{\mathrm{QCD}}^{2}$
in the vacuum) for the DLA.

Interestingly, there is yet another requirement, that we overlooked.
Indeed, we argued already (see Eq. (\ref{eq:qhatenergyphasespace}))
that the gluon fluctuations should take place deeply inside the medium,
which led to the requirement:
\begin{equation}
l_{0}\ll\tau\ll L.
\end{equation}
The upper limit is automatically satisfied Eq. (\ref{eq:Blower}):
\begin{equation}
\tau_{\mathrm{max}}\simeq\frac{\omega_{c}}{k_{\perp,\mathrm{min}}^{2}}=\frac{\omega_{c}}{k_{\mathrm{br}}^{2}\left(\omega\right)}=\frac{\hat{q}L^{2}}{\hat{q}\tau}=L.
\end{equation}
However, this is not the case for the lower limit $l_{0}\ll\tau$,
hence we have to enforce it explicitly, modifying the upper limit
$k_{\perp}^{2}\ll Q_{s}^{2}$ for the gluon's momentum (from Eq. (\ref{eq:Bupper}))
to $k_{\perp}^{2}\ll\min\left(Q_{s}^{2},2\omega/l_{0}\right)$. 

\begin{figure}[t]
\begin{centering}
\begin{tikzpicture}[scale=2] 

\tikzset{photon/.style={decorate,decoration={snake}},
		electron/.style={postaction={decorate},decoration={markings,mark=at position .5 with {\arrow[draw]{>}}}},      	gluon/.style={decorate,decoration={coil,amplitude=4pt, segment length=5pt}}};

\fill[black!15!white, rounded corners] (-1,-1) rectangle(1,1);
\draw[electron] (1.5,.4) --++(-3,0);
\draw[electron] (-1.5,.8) --++(3,0);

\draw[gluon] (-.6,.4) .. controls (-.4,-1) and (.4,-1)..  (.6,.8) ;
\draw[<->] (-1.3, .75) --++ (0,-.15) node[left]{$r\sim 1/Q_s$} --++ (0,-.15);
\draw[<->] (-1.1, .6) --++ (0,-.6) node[left]{$B$} --++ (0,-.6);
\draw[photon](-.2,-.5) --++ (0,-.4) node [at end, cross out, draw, solid, inner sep=2.5 pt]{};
\draw[photon](.2,.4) --++ (0,-1.3) node [at end, cross out, draw, solid, inner sep=2.5 pt]{};
\node at (-1,-1.1) {$0$};
\node at (1,-1.1) {$L$};

\end{tikzpicture} 
\par\end{centering}
\caption{\label{fig:DLAinmedium}One step in the evolution of a dipole through
the nuclear medium in the DLA approximation, as encoded in Eq. (\ref{eq:BFKLmediumDLA}).
The size $B$ of the $qg$ and $\bar{q}g$ daughter dipoles is much
larger than the size $r$ of the parent dipole.}
\end{figure}
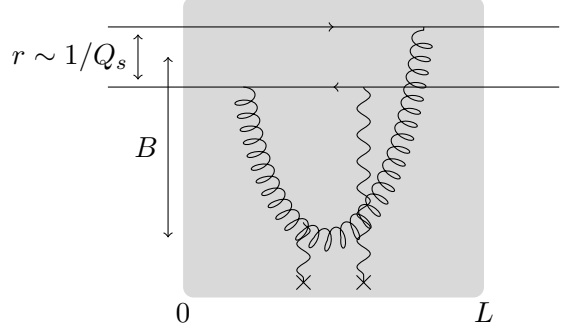

In summary, the phase space of the double-logarithmic regime is the
one restricted by the following limits on the energy and on the transverse
extent of the gluon fluctuation:
\begin{equation}
\begin{aligned} & \omega_{0}=\hat{q}l_{0}^{2}\ll\omega\ll\omega_{c}=\hat{q}L^{2},\\
 & \max\left(r^{2}\sim\frac{1}{Q_{s}^{2}},\frac{l_{0}}{2\omega}\right)\ll B^{2}\ll\frac{2}{k_{\mathrm{br}}^{2}\left(\omega\right)}.
\end{aligned}
\label{eq:DLAlimits}
\end{equation}
One step in the evolution is illustrated in Fig. \ref{fig:DLAinmedium},
and the DLA phase space is depicted in Fig. \ref{fig:gluonphasespace}
(in terms of $(\tau,\omega)$).

\section{Renormalization of the jet quenching parameter}

We will now solve evolution equation (\ref{eq:BFKLmediumDLA}) explicitly
in the double logarithmic approximation. Interestingly, the evolution
preserves the MV approximation Eq. (\ref{eq:MV approximation}). This
implies that the double-logarithmic high-energy corrections can be
absorbed into a renormalization of the jet quenching parameter $\hat{q}_{\omega}$.
Indeed, combining Eqs. (\ref{eq:MV approximation}) and (\ref{eq:BFKLmediumDLA}),
one obtains the following evolution equation for $\hat{q}_{\omega}\left(r\right)$:
\begin{equation}
\begin{aligned}\omega\begin{aligned}\frac{\partial\hat{q}_{\omega}\left(r\right)}{\partial\omega}\end{aligned}
 & =\bar{\alpha}\int\frac{\mathrm{d}B^{2}}{B^{2}}\hat{q}_{\omega}\left(B\right),\end{aligned}
\end{equation}
or, in integral form with the correct limits, Eq. (\ref{eq:DLAlimits}):
\begin{equation}
\hat{q}_{\omega}\left(r\right)=\bar{\alpha}\int_{\omega_{0}}^{\omega_{c}}\frac{\mathrm{d}\omega}{\omega}\int_{\max\left(r^{2},l_{0}/2\omega\right)}^{2/k_{\mathrm{br}}^{2}\left(\omega\right)}\frac{\mathrm{d}B^{2}}{B^{2}}\hat{q}_{\omega}\left(B\right).\label{eq:qhatdlaintegral}
\end{equation}
This equation can be solved iteratively as follows: denoting the zeroth-order
solution, which we assume to be scale-independent, as $\hat{q}^{\left(0\right)}$,
we have that:
\begin{equation}
\begin{aligned}\delta\hat{q}_{\omega}^{\left(1\right)}\left(r\right) & =\bar{\alpha}\hat{q}^{\left(0\right)}\int_{l_{0}/2r^{2}}^{\omega_{c}}\frac{\mathrm{d}\omega}{\omega}\int_{r^{2}}^{2/k_{\mathrm{br}}^{2}\left(\omega\right)}\frac{\mathrm{d}B^{2}}{B^{2}}+\bar{\alpha}\hat{q}^{\left(0\right)}\int_{\omega_{0}}^{l_{0}/2r^{2}}\frac{\mathrm{d}\omega}{\omega}\int_{l_{0}/2\omega}^{2/k_{\mathrm{br}}^{2}\left(\omega\right)}\frac{\mathrm{d}B^{2}}{B^{2}},\\
 & =\bar{\alpha}\hat{q}^{\left(0\right)}\int_{l_{0}/2r^{2}}^{\omega_{c}}\frac{\mathrm{d}\omega}{\omega}\ln\frac{2}{k_{\mathrm{br}}^{2}\left(\omega\right)r^{2}}+\bar{\alpha}\hat{q}^{\left(0\right)}\int_{\omega_{0}}^{l_{0}/2r^{2}}\frac{\mathrm{d}\omega}{\omega}\ln\frac{4\omega}{k_{\mathrm{br}}^{2}\left(\omega\right)l_{0}},\\
 & \simeq\bar{\alpha}\hat{q}^{\left(0\right)}\int_{l_{0}/2r^{2}}^{\omega_{c}}\frac{\mathrm{d}\omega}{\omega}\ln\frac{Q_{s}^{2}}{k_{\mathrm{br}}^{2}\left(\omega\right)}+\bar{\alpha}\hat{q}^{\left(0\right)}\int_{\omega_{0}}^{l_{0}/2r^{2}}\frac{\mathrm{d}\omega}{\omega}\ln\frac{4\omega}{k_{\mathrm{br}}^{2}\left(\omega\right)l_{0}}.
\end{aligned}
\end{equation}
From Eqs. (\ref{eq:qhatenergyphasespace}), (\ref{eq:kisql}) and
(\ref{eq:taubrownian}), the logarithms of the transverse momentum
scales can be written in function of the energies:
\begin{equation}
\begin{aligned}\ln\frac{Q_{s}^{2}}{k_{\mathrm{br}}^{2}\left(\omega\right)} & =\ln\sqrt{\frac{\omega_{c}}{\omega}},\\
\ln\frac{\omega}{k_{\mathrm{br}}^{2}\left(\omega\right)l_{0}} & =\ln\sqrt{\frac{\omega}{\omega_{0}}},
\end{aligned}
\end{equation}
hence we obtain:
\begin{equation}
\begin{aligned}\delta\hat{q}_{\omega}^{\left(1\right)}\left(r\right) & \simeq\frac{\bar{\alpha}}{2}\hat{q}^{\left(0\right)}\int_{l_{0}/2r^{2}}^{\omega_{c}}\frac{\mathrm{d}\omega}{\omega}\ln\frac{\omega_{c}}{\omega}+\frac{\bar{\alpha}}{2}\hat{q}^{\left(0\right)}\int_{\omega_{0}}^{l_{0}/2r^{2}}\frac{\mathrm{d}\omega}{\omega}\ln\frac{\omega}{\omega_{0}},\\
 & =\frac{\bar{\alpha}}{4}\hat{q}^{\left(0\right)}\left(\ln^{2}\frac{r^{2}\omega_{c}}{l_{0}}+\ln^{2}\frac{l_{0}}{r^{2}\omega_{0}}\right).
\end{aligned}
\label{eq:qhat1}
\end{equation}
The \textquoteleft physical' jet quenching parameter is then the one
evaluated on the point $\tau=L$ and $p_{\perp}^{2}=Q_{s}^{2}$, and
is equal to:
\begin{equation}
\begin{aligned}\hat{q}_{\omega_{c}}\left(1/Q_{s}^{2}\right) & =\hat{q}^{\left(0\right)}\left(1+\frac{\bar{\alpha}}{2}\ln^{2}\frac{L}{l_{0}}+\mathcal{O}\left(\bar{\alpha}^{2}\right)\right).\end{aligned}
\label{eq:qhat1phys}
\end{equation}
In order to calculate the next-to-leading order correction, and eventually
resum all DLA contributions, it is more convenient to work in the
phase space $\left(\tau,k_{\perp}^{2}\right)$, with $\tau$ the lifetime
of the gluon. We then do not have to make use of the \textendash somewhat
awkward to handle\textendash{} condition $B_{\min}=\max\left(r^{2}\sim\frac{1}{Q_{s}^{2}},\frac{l_{0}}{2\omega}\right)$
anymore, but instead, the DLA regime is simply defined by (writing
$k_{\perp}^{2}\sim1/B^{2}$):
\begin{equation}
\begin{aligned} & l_{0}\ll\tau\ll L,\quad\mathrm{and}\quad k_{\mathrm{br}}^{2}=\hat{q}\tau\ll k_{\perp}^{2}\ll Q_{s}^{2}.\end{aligned}
\end{equation}
The integral form of the DLA evolution equation (\ref{eq:qhatdlaintegral})
then becomes:
\begin{equation}
\hat{q}_{\tau}\left(p_{\perp}^{2}\right)=\bar{\alpha}\int_{l_{0}}^{\tau}\frac{\mathrm{d}\tau'}{\tau'}\int_{\hat{q}\tau'}^{p_{\perp}^{2}}\frac{\mathrm{d}k_{\perp}^{2}}{k_{\perp}^{2}}\hat{q}_{\tau'}\left(k_{\perp}^{2}\right).\label{eq:qhatRG}
\end{equation}
It is easy to check that the first iteration of the solution to the
above equation is the same as Eq. (\ref{eq:qhat1}), as it should.
In order to obtain the next iteration, one should observe that the
contributions that generate a large logarithm are the ones in which
each subsequent emission has a shorter lifetime than the one of its
parent, resulting in a gluon density that keeps growing (cf. the discussion
on BFKL, Sec. \ref{sec:BKBFKL}):
\begin{equation}
L\gg\tau_{1}\gg\tau_{2}\gg...\gg l_{0}.\label{eq:qhattauordering}
\end{equation}
Interestingly, however, in our case the strong ordering of the virtualities
is opposite to the one in DGLAP. Indeed, remember that when building
the DGLAP ladders, each emission is treated as if coming from an on-shell
parton: an approximation which can only be valid when the virtuality
of each radiated parton is much larger than the previous one. In our
case of in-medium radiation, due to the requirement (\ref{eq:parentneglection}),
it is precisely the other way around: each gluon that is emitted results
in a daughter dipole with a larger transverse size $B$, and hence
a smaller transverse momentum $k_{\perp}\sim1/B$, than its parent:
\begin{equation}
Q_{s}^{2}\gg k_{1\perp}^{2}\gg k_{2\perp}^{2}\gg...\gg k_{\mathrm{br}}^{2}\left(\tau\right).\label{eq:qhatkperpordering}
\end{equation}
Implementing the orderings (\ref{eq:qhattauordering}) and (\ref{eq:qhatkperpordering}),
the second iteration of Eq. (\ref{eq:qhatRG}) is:
\begin{equation}
\begin{aligned}\delta\hat{q}_{\tau}^{\left(2\right)}\left(p_{\perp}^{2}\right) & =\bar{\alpha}^{2}\hat{q}^{\left(0\right)}\int_{l_{0}}^{\tau}\frac{\mathrm{d}\tau_{1}}{\tau_{1}}\int_{\hat{q}\tau_{1}}^{p_{\perp}^{2}}\frac{\mathrm{d}k_{1\perp}^{2}}{k_{1\perp}^{2}}\int_{l_{0}}^{\tau_{1}}\frac{\mathrm{d}\tau_{2}}{\tau_{2}}\int_{\hat{q}\tau_{2}}^{k_{1\perp}^{2}}\frac{\mathrm{d}k_{2\perp}^{2}}{k_{2\perp}^{2}},\\
 & =\frac{\bar{\alpha}^{2}}{2}\hat{q}^{\left(0\right)}\int_{l_{0}}^{\tau}\frac{\mathrm{d}\tau_{1}}{\tau_{1}}\int_{\hat{q}\tau_{1}}^{p_{\perp}^{2}}\frac{\mathrm{d}k_{1\perp}^{2}}{k_{1\perp}^{2}}\ln\frac{\tau_{1}}{l_{0}}\ln\frac{k_{1\perp}^{4}}{\hat{q}^{2}\tau_{1}l_{0}},\\
 & =\frac{\bar{\alpha}^{2}}{2}\hat{q}^{\left(0\right)}\int_{l_{0}}^{\tau}\frac{\mathrm{d}\tau_{1}}{\tau_{1}}\ln\frac{\tau_{1}}{l_{0}}\ln\frac{p_{\perp}^{2}}{l_{0}\hat{q}}\ln\frac{p_{\perp}^{2}}{\hat{q}\tau_{1}},\\
 & =\frac{\bar{\alpha}^{2}}{12}\hat{q}^{\left(0\right)}\ln^{2}\frac{\tau}{l_{0}}\ln\frac{p_{\perp}^{2}}{l_{0}\hat{q}}\ln\frac{p_{\perp}^{6}}{\hat{q}^{3}\tau^{2}l_{0}},
\end{aligned}
\end{equation}
which, on the physical point $\tau=L$ and $p_{\perp}^{2}=Q_{s}^{2}$
yields:
\begin{equation}
\begin{aligned}\delta\hat{q}_{L}^{\left(2\right)}\left(Q_{s}^{2}\right) & =\hat{q}^{\left(0\right)}\frac{\bar{\alpha}^{2}}{12}\ln^{4}\frac{L}{l_{0}}.\end{aligned}
\label{eq:qhat2}
\end{equation}
As it turns out (see Refs. \protect\cite{Al,Blaizot2014RG,Iancu}), the full
solution to Eq. (\ref{eq:qhatRG}) is found to be: 
\begin{equation}
\begin{aligned}\hat{q}_{L}\left(Q_{s}^{2}\right) & =\hat{q}^{\left(0\right)}\frac{1}{\sqrt{\bar{\alpha}}\ln\frac{L}{l_{0}}}I_{1}\left(2\sqrt{\bar{\alpha}}\ln\frac{L}{l_{0}}\right),\end{aligned}
\label{eq:DLArenormalizationqhat}
\end{equation}
of which $\delta\hat{q}_{L}^{\left(1\right)}\left(Q_{s}^{2}\right)$,
Eq. (\ref{eq:qhat1phys}) and $\delta\hat{q}_{L}^{\left(2\right)}\left(Q_{s}^{2}\right)$
are indeed the first terms in the expansion in $\bar{\alpha}$. This
is our final result for the radiative corrections to the jet quenching
parameter in the double leading approximation. For a small coupling
constant $\bar{\alpha}\sim0.3$, a lower bound $l_{0}\sim1/T$ with
the temperature $T\sim300\,\mathrm{MeV}$, and a plasma with longitudinal
extent $L\sim5\,\mathrm{fermi}$, the enhancement seems to be rather
sizable, of the order:
\begin{equation}
\hat{q}_{L}\left(Q_{s}^{2}\right)\sim1.8\,\hat{q}^{\left(0\right)}.
\end{equation}

\section{Multiple soft scattering and gluon saturation}

In the previous sections, we constructed a high-energy evolution equation
for a probe in an extended nuclear medium, and applied this equation
to the fictitious dipole that describes the transverse momentum broadening
of a parton in this medium. Since the MV approximation for a dipole
turned out to be conserved by the evolution, at least to DLA accuracy,
the double large logarithms associated with soft gluon fluctuations
could be absorbed into a renormalization of the jet quenching parameter
$\hat{q}$.

In order to gain further insight into the physics of transverse momentum
broadening, and to elucidate on the similarities and differences with
the Color Glass Condensate, let us have a closer look on the above
analysis from the point of view of the target. 

First however, we should explain the physical basis for the lower
bound $l_{0}$ on the gluon's lifetime (see Ref. \protect\cite{Al}). Since
the gluon fluctuation has $k_{\perp}^{2}\ll Q_{s}^{2}$, we can treat
it as being nearly on shell and write:
\begin{equation}
\begin{aligned}k^{\mu} & =\left(\omega,k^{-}\simeq\frac{k_{\perp}^{2}}{2\omega},\mathbf{k}_{\perp}\right).\end{aligned}
\end{equation}
The gluon is emitted from a right-moving parent $P\simeq\left(P^{+},0^{-},\mathbf{0}\right)$,
from which it cannot inherit a momentum component $k^{-}$. Hence,
this component is obtained from the interactions with the medium constituents.
Since the lifetime of the gluon is $\tau\sim1/k^{-}$, this confirms
our earlier assertions that the medium imposes strong conditions on
the gluon lifetime. A typical thermal parton of the medium has, in
the medium rest frame, a momentum component $p^{-}\sim T$. Upon interacting
with the gluon fluctuation, a fraction $x=k^{-}/p^{-}\leq1$ of this
momentum is absorbed by the gluon, and we find:
\begin{equation}
\begin{aligned}\tau & \simeq\frac{1}{k^{-}}=\frac{1}{xp^{-}}.\end{aligned}
\end{equation}
In the single scattering approximation, the maximal momentum component
$k^{-}$ that the gluon can obtain from the interaction with the medium
is therefore given by $x=1$, corresponding to a lower limit $\tau_{\mathrm{min}}=l_{0}\equiv1/T$
on the gluon's lifetime.

Let us now perform a Lorentz boost to an infinite momentum frame for
the left-moving medium, in which all the soft gluons emitted by the
fictitious dipole can be regarded as partons of the medium. In the
rest frame of the plasma, gluons with energy $\omega$ and transverse
momentum $k_{\perp}$ correspond to a Lorentz factor $\gamma\simeq\omega/k_{\perp}$
(cf. (\ref{eq:DISfluctime})). Therefore, since the gluon fluctuations
that we are interested in have $k_{\perp}\gtrsim k_{\mathrm{br}}$,
the medium should be boosted with a factor
\begin{equation}
\gamma\simeq\frac{\omega_{\mathrm{max}}}{k_{\perp,\mathrm{min}}}\simeq\frac{\omega_{c}}{k_{\mathrm{br}}\left(\omega_{c}\right)}=\frac{\omega_{c}}{Q_{s}}=\sqrt{\hat{q}L^{3}},
\end{equation}
for all the relevant fluctuations to be naturally associated with
the medium, instead of with the dipole. Accordingly, the high-energy
evolution of the fictitious dipole can be viewed as the evolution
of the gluon distribution in the medium. 

Interestingly, this evolution differs from the one we encountered
previously in the case of a shockwave. This is a consequence of the
fact that, on the \textendash for the left-moving medium\textendash{}
longitudinal axis $x^{+}$, the gluon fluctuations do not necessarily
overlap with each other. This is in sharp contrast with the CGC, where
it was always assumed that for each step in the evolution, the semi-fast
gluons were well located from the point of view of the soft gluons,
and hence could be integrated out and absorbed into the color sources
for the latter. 

To quantify this better, let us revisit the gluon occupation number,
Eq. (\ref{eq:phioverlap}), for a shockwave, but now in unintegrated
form:
\begin{equation}
\begin{aligned}\varphi\left(x,k_{\perp}\right) & \equiv\frac{4\pi^{3}}{N_{c}^{2}-1}\frac{1}{\pi R^{2}}x\frac{\mathrm{d}N}{\mathrm{d}x\mathrm{d}^{2}\mathbf{k}_{\perp}},\\
 & =\frac{4\pi^{3}}{N_{c}^{2}-1}\frac{1}{\pi R^{2}}\mathcal{F}_{gg}^{\left(3\right)}\left(x,k_{\perp}\right).
\end{aligned}
\label{eq:gluonoccupationnumber}
\end{equation}
In the dilute approximation and at lowest order, the Weizsäcker-Williams
gluon distribution is equal to the amount of valence quarks in the
medium: $AN_{c}$, times the corresponding tree-level Bremsstrahlung
spectrum $\alpha_{s}C_{F}/\pi^{2}k_{\perp}^{2}$ (see Eqs. (\ref{eq:WWlargekT}),
(\ref{eq:WWdilute})): 
\begin{equation}
\begin{aligned}\varphi\left(x,k_{\perp}\right) & \simeq\frac{4\pi^{3}}{N_{c}^{2}-1}\frac{1}{\pi R^{2}}AN_{c}\frac{\alpha_{s}C_{F}}{\pi^{2}k_{\perp}^{2}},\\
 & \simeq\frac{1}{\alpha_{s}N_{c}}\frac{Q_{sg}^{2}}{k_{\perp}^{2}}.
\end{aligned}
\label{eq:phidilute}
\end{equation}
Indeed, approaching the saturation scale from the dilute regime, the
gluon occupation number reaches its maximal allowed value $\varphi\sim1/\alpha_{s}N_{c}$
when $k_{\perp}^{2}=Q_{sg}^{2}\simeq\alpha_{s}N_{c}\mu_{A}$. To extend
the validity of definition (\ref{eq:gluonoccupationnumber}) to an
extended medium, we should observe that the distribution $\varphi\left(x,k_{\perp}\right)$
is integrated in the $k^{-}$-direction, since
\begin{equation}
\varphi\left(x,k_{\perp}\right)\propto x\frac{\mathrm{d}N}{\mathrm{d}x\mathrm{d}^{2}\mathbf{k}_{\perp}}=k^{-}\frac{\mathrm{d}N}{\mathrm{d}k^{-}\mathrm{d}^{2}\mathbf{k}_{\perp}}.
\end{equation}
A truly three-dimensional gluon occupation number $\varphi_{\mathrm{med}}\left(k^{-},k_{\perp}\right)$,
appropriate for an extended medium, is therefore obtained by writing:
\begin{equation}
\begin{aligned}\varphi_{\mathrm{med}}\left(k^{-},k_{\perp}\right) & \equiv\frac{4\pi^{3}}{N_{c}^{2}-1}\frac{1}{\pi R^{2}L}\frac{\mathrm{d}N}{\mathrm{d}k^{-}\mathrm{d}^{2}\mathbf{k}_{\perp}},\\
 & =\frac{4\pi^{3}}{N_{c}^{2}-1}\frac{1}{\pi R^{2}L}\frac{\mathcal{F}_{gg}^{\left(3\right)}\left(k^{-},k_{\perp}\right)}{k^{-}}.
\end{aligned}
\label{eq:gluonoccupationnumbermedium}
\end{equation}
Again in the dilute limit and at lowest order, this yields:
\begin{equation}
\begin{aligned}\varphi_{\mathrm{med}}\left(k^{-},k_{\perp}\right) & \simeq4\pi\alpha_{s}n_{0}\frac{1}{k^{-}k_{\perp}^{2}},\end{aligned}
\label{eq:gluonoccupationnumbermediumdilute}
\end{equation}
where now the three-dimensional density $n_{0}$, Eq. (\ref{eq:n0}),
comes into play, instead of the transverse one ($\mu_{A}/g_{s}^{2}$
, see Eq. (\ref{eq:muA2}), which is encoded in $Q_{sg}^{2}$ in Eq.
(\ref{eq:phidilute})). If we now use the above quantity to estimate
the saturation scale, requiring 
\begin{equation}
\varphi_{\mathrm{med}}\left(k^{-},Q_{sg}^{2}\right)\sim\frac{1}{\alpha_{s}N_{c}},
\end{equation}
we obtain:
\begin{equation}
Q_{sg,\mathrm{med}}^{2}=4\pi\alpha_{s}^{2}N_{c}n_{0}\frac{1}{k^{-}}=Q_{sg}^{2}\times\frac{1}{Lk^{-}}.
\end{equation}
Since the lifetime of a right-mover is given by $\tau\sim1/k^{-}$,
and using Eq. (\ref{eq:jetquenchingparameter}) as well as the relation
$N_{c}Q_{s}^{2}=C_{F}Q_{sg}^{2}$, we find that:
\begin{equation}
Q_{s,\mathrm{med}}^{2}\left(\tau\right)=Q_{s}^{2}\frac{\tau}{L}=\hat{q}\tau\simeq\hat{q}\frac{T}{x}.\label{eq:saturationlinetreelevel}
\end{equation}
The average transverse momentum obtained by a gluon after multiple
soft scatterings with the medium: $k_{\mathrm{br}}^{2}\left(\tau\right)=\hat{q}\tau$,
is therefore equal to the saturation line $Q_{s,\mathrm{med}}^{2}\left(\tau\right)$
in a frame in which the target is highly relativistic. In contrast
to a shockwave, in which the $x$-dependence of the saturation scale
comes into play after considering evolution (see Sec. \ref{sec:The-dipole-picture-and-saturation}
and Refs. \protect\cite{MuellerTriantafyllopoulos2002,Iancu2002,Triantafyllopoulos2003}),
in the present case this dependence is already strongly present at
tree-level.

In the double logarithmic approximation, we found out that the evolution
led to a renormalization of $\hat{q}$. Hence, to DLA-accuracy, Eq.
(\ref{eq:gluonoccupationnumbermediumdilute}) becomes:
\begin{equation}
\begin{aligned}\varphi_{\mathrm{med}}\left(k^{-},k_{\perp}\right) & \simeq\frac{1}{\alpha_{s}N_{c}}\frac{\hat{q}_{\tau}\left(k_{\perp}\right)}{k^{-}k_{\perp}^{2}}.\end{aligned}
\end{equation}
The fact that the momentum broadening $k_{\mathrm{br}}^{2}$ due to
multiple soft scattering can be interpreted as the medium saturation
scale $Q_{s,\mathrm{med}}^{2}$ is not a coincidence. Rather, it can
be seen from the \textquoteleft complete', i.e., valid beyond the
single scattering approximation, evolution equation (\ref{eq:qhatevolutionequation})
that the multiple scattering effects become sizable when the exponent
is of order one. However, in the MV approximation, this is equal to
the requirement that:
\begin{equation}
\begin{aligned}1 & \sim g_{s}^{2}\frac{N_{c}}{2}\tau\bar{\Gamma}_{\omega}\left(\mathbf{r}\left(t\right)\right)\simeq\frac{1}{4}\tau\hat{q}_{\tau}r^{2}\sim\hat{q}_{\tau}\frac{1}{k^{-}k_{\perp}^{2}}\\
\Longleftrightarrow & \qquad\varphi_{\mathrm{med}}\left(Q_{s,\mathrm{med}}^{2}\left(\tau\right)\right)\sim\frac{1}{\alpha_{s}N_{c}}.
\end{aligned}
\end{equation}
The saturation line is thus given to DLA-accuracy by:
\begin{equation}
Q_{s,\mathrm{med}}^{2}\left(\tau\right)=\hat{q}_{\tau}\left(Q_{s,\mathrm{med}}^{2}\left(\tau\right)\right)\tau.\label{eq:dlasaturationline}
\end{equation}
Revisiting the steps leading to our result (\ref{eq:DLArenormalizationqhat}),
it is easy to see that we have:
\begin{equation}
\begin{aligned}\hat{q}\left(x\right) & =\hat{q}^{\left(0\right)}\frac{1}{\sqrt{\bar{\alpha}}\ln\frac{1}{x}}I_{1}\left(2\sqrt{\bar{\alpha}}\ln\frac{1}{x}\right).\end{aligned}
\end{equation}
In the limit $2\sqrt{\bar{\alpha}}\ln(1/x)\gg1$, this can be used
to solve Eq. (\ref{eq:dlasaturationline}), for which we find:
\begin{equation}
Q_{s,\mathrm{med}}^{2}\left(x\right)\simeq\frac{Q_{s,0}^{2}}{x^{1+\gamma_{s}}}
\end{equation}
with $\gamma_{s}=2\sqrt{\bar{\alpha}}$. Hence, the double logarithmic
contributions of the high-energy evolution add a small-$x$ anomalous
dimension $\gamma_{s}$ to the $x$-dependence of the saturation line
in a medium, which is already strong at tree-level (cf. Eq. (\ref{eq:saturationlinetreelevel})).

\section{Beyond the single scattering approximation}

In this section, we have a closer look at the full evolution equation
(\ref{eq:qhatevolutionequation}) beyond the single scattering approximation.
In particular, we argue that the limits for the gluon phase space,
which we introduced using semi-quantitative arguments, are indeed
encoded in the formalism.

To do so, let us rewrite Eq. (\ref{eq:qhatevolutionequation}) in
the harmonic approximation (writing Eq. (\ref{eq:MV approximation})
and furthermore neglecting the logarithmic $r$-dependence of $\hat{q}$):
\begin{equation}
\begin{aligned} & \begin{aligned}L\omega\frac{\partial\bar{\Gamma}_{\omega}\left(\mathbf{x},\mathbf{y}\right)}{\partial\omega}\end{aligned}
=\frac{1}{4\pi\omega^{2}}\int_{0}^{L}\mathrm{d}t_{2}\int_{0}^{t_{2}}\mathrm{d}t_{1}\mathbf{\partial}_{\mathbf{r}_{1}}^{i}\mathbf{\partial}_{\mathbf{r}_{2}}^{i}\int\left[\mathcal{D}\mathbf{r}\left(t\right)\right]\exp\left\{ i\frac{\omega}{2}\int_{t_{1}}^{t_{2}}\mathrm{d}t\dot{\mathbf{r}}^{2}\left(t\right)\right\} \\
 & \times\left[\exp\left\{ -\frac{1}{4}\hat{q}\int_{t_{1}}^{t_{2}}\mathrm{d}t\left(\left(\mathbf{x}-\mathbf{r}\left(t\right)\right)^{2}+\left(\mathbf{y}-\mathbf{r}\left(t\right)\right)^{2}-\left(\mathbf{x}-\mathbf{y}\right)^{2}\right)\right\} -1\right]\biggr|_{\mathbf{r}_{1}=\mathbf{y}}^{\mathbf{r}_{1}=\mathbf{x}}\biggr|_{\mathbf{r}_{2}=\mathbf{y}}^{\mathbf{r}_{2}=\mathbf{x}}.
\end{aligned}
\label{eq:qhatevolutionho}
\end{equation}

\begin{figure}[t]
\begin{centering}
\begin{tikzpicture}[scale=2.5] 

\tikzset{photon/.style={decorate,decoration={snake}},
		electron/.style={postaction={decorate},decoration={markings,mark=at position .5 with {\arrow[draw]{>}}}},      	gluon/.style={decorate,decoration={coil,amplitude=4pt, segment length=5pt}}};

\fill[black!15!white, rounded corners] (-1,-.7) rectangle(1,1);
\draw[electron] (1.5,.4) --++(-3,0);
\draw[electron] (-1.5,.8) --++(3,0);
\draw[photon](-.8,.6) -- (-.8,-.5) node [at end, cross out, draw, solid, inner sep=2.5 pt]{};
\draw[photon](-.6,.6) -- (-.6,-.5) node [at end, cross out, draw, solid, inner sep=2.5 pt]{};
\draw[photon](.8,.6) -- (.8,-.5) node [at end, cross out, draw, solid, inner sep=2.5 pt]{};
\draw[photon](.6,.6) -- (.6,-.5) node [at end, cross out, draw, solid, inner sep=2.5 pt]{};

\draw[gluon] (-.4,.4) .. controls (-.2,-.5) and (.2,-.5)..  (.4,.8) ;
\draw[photon](-.2,-.2) --++ (0,-.3) node [at end, cross out, draw, solid, inner sep=2.5 pt]{};
\draw[photon](.2,.6) --++ (0,-1.1) node [at end, cross out, draw, solid, inner sep=2.5 pt]{};

\draw[->] (-1,-.8) --++(2.1,0);
\draw[] (-1,-.78) --++(0,-.04);
\draw[] (-.4,-.78) --++(0,-.04);
\draw[] (.4,-.78) --++(0,-.04);
\draw[] (1,-.78) --++(0,-.04);
\node at (-1,-.9) {$0$};
\node at (-.4,-.9) {$t_1$};
\node at (.4,-.9) {$t_2$};
\node at (1,-.9) {$L$};

\draw[] (-1,1.1) -- (-.5,1.1);
\draw[] (-1,1.1) --++(0,-.05);
\draw[] (-.5,1.1) --++(0,-.05);4
\node at (-.7,1.3) {$e^{-\frac{1}{4}\hat{q}\mathbf{x}^2 t_1}$};
\node at (.8,1.3) {$e^{-\frac{1}{4}\hat{q}\mathbf{x}^2 (L-t_2)}$};
\node at (0,1.3) {$G(\mathbf{r}_2,\mathbf{r}_1,\tau)$};
\draw[] (-.45,1.1) --(.45,1.1);
\draw[] (-.45,1.1) --++(0,-.05);
\draw[] (.45,1.1) --++(0,-.05);

\draw[] (.5,1.1) -- (1,1.1);
\draw[] (.5,1.1) --++(0,-.05);
\draw[] (1,1.1) --++(0,-.05);

\end{tikzpicture} 
\par\end{centering}
\caption{\label{fig:MMevo}One step in the evolution of a dipole through the
nuclear medium, as encoded in Eq. (\ref{eq:MuellerMaster}).}
\end{figure}
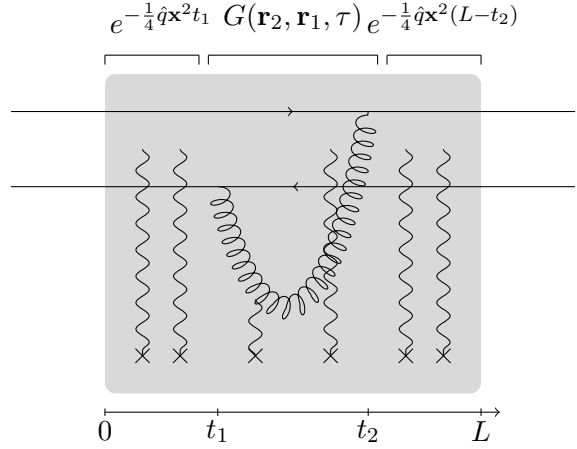

In the above equation, we can discern both the vacuum propagator of
the gluon:
\begin{equation}
G_{0}\left(\mathbf{r}_{2},\mathbf{r}_{1},\tau\right)=\int\left[\mathcal{D}\mathbf{r}\left(t\right)\right]\exp\left\{ i\frac{\omega}{2}\int_{t_{1}}^{t_{2}}\mathrm{d}t\dot{\mathbf{r}}^{2}\left(t\right)\right\} ,
\end{equation}
as well as the Green function that describes the motion of the gluon
in an imaginary harmonic potential, which describes the multiple soft
scatterings with the background:

\begin{equation}
\begin{aligned}G\left(\mathbf{r}_{2},\mathbf{r}_{1},\tau\right) & \equiv\int\left[\mathcal{D}\mathbf{r}\left(t\right)\right]\exp\left\{ i\frac{\omega}{2}\int_{t_{1}}^{t_{2}}\mathrm{d}t\dot{\mathbf{r}}^{2}\left(t\right)\right\} \\
 & \times\exp\left\{ -\frac{1}{4}\hat{q}\int_{t_{1}}^{t_{2}}\mathrm{d}t\left(\left(\mathbf{x}-\mathbf{r}\left(t\right)\right)^{2}+\left(\mathbf{y}-\mathbf{r}\left(t\right)\right)^{2}\right)\right\} .
\end{aligned}
\end{equation}
For both, we can write down an explicit expression, provided by a
standard calculation (see e.g. \protect\cite{FeynmanHibbs}):
\begin{equation}
\begin{aligned}G_{0}\left(\mathbf{r}_{2},\mathbf{r}_{1},\tau\right) & =-i\frac{\omega}{2\pi\tau}e^{i\frac{\omega}{2\tau}\left(\mathbf{r}_{2}-\mathbf{r}_{1}\right)^{2}},\end{aligned}
\end{equation}
\begin{equation}
\begin{aligned} & G\left(\mathbf{r}_{2},\mathbf{r}_{1},\tau\right)=-\frac{i}{2\pi}\frac{\omega\Omega}{\sinh\Omega\tau}\\
 & \times\exp\left\{ \frac{i}{2}\frac{\omega\Omega}{\sinh\Omega\tau}\left[\left(\left(\mathbf{r}_{2}-\mathbf{R}\right)^{2}+\left(\mathbf{r}_{1}-\mathbf{R}\right)^{2}\right)\cosh\Omega\tau-2\left(\mathbf{r}_{2}-\mathbf{R}\right)\cdot\left(\mathbf{r}_{1}-\mathbf{R}\right)\right]\right\} ,
\end{aligned}
\end{equation}
where
\begin{equation}
\mathbf{R}\equiv\frac{\mathbf{x}+\mathbf{y}}{2},\quad\mathrm{and}\quad\Omega\equiv\frac{1+i}{\sqrt{2}}\sqrt{\frac{\hat{q}}{\omega}}.\label{eq:Omega}
\end{equation}
In the notation:
\begin{equation}
\begin{aligned}N_{\omega}\left(\mathbf{x}\right) & \equiv\frac{\partial S}{\partial\ln\omega}=\omega S\frac{\partial\ln S}{\partial\omega}\end{aligned}
=-g^{2}C_{F}LS\omega\frac{\partial\bar{\Gamma}_{\omega}\left(\mathbf{x}\right)}{\partial\omega},\label{eq:Ndef}
\end{equation}
and, without loss of generality, setting $\mathbf{y}=\mathbf{0}$,
Eq. (\ref{eq:qhatevolutionho}) becomes: 
\begin{equation}
\begin{aligned}N_{\omega}\left(\mathbf{x}\right) & =\frac{-\alpha_{s}N_{c}}{2\omega^{2}}\int_{0}^{L}\mathrm{d}t_{2}\int_{0}^{t_{2}}\mathrm{d}t_{1}\partial_{\mathbf{r}_{1}}^{i}\partial_{\mathbf{r}_{2}}^{i}\\
 & \biggl(e^{-\frac{\hat{q}}{4}\left(L-t_{2}\right)\mathbf{x}^{2}}G\left(\mathbf{r}_{2},\mathbf{r}_{1},\tau\right)e^{-\frac{\hat{q}}{4}t_{1}\mathbf{x}^{2}}-e^{-\frac{\hat{q}}{4}L\mathbf{x}^{2}}G_{0}\left(\mathbf{r}_{2},\mathbf{r}_{1},\tau\right)\biggr)\Biggr|_{\mathbf{r}_{2}=\mathbf{0}}^{\mathbf{r}_{2}=\mathbf{x}}\Biggr|_{\mathbf{r}_{1}=\mathbf{0}}^{\mathbf{r}_{1}=\mathbf{x}}.
\end{aligned}
\label{eq:MuellerMaster}
\end{equation}
A nice feature of the above equation (we cast it in such a form to
match the notations in Ref. \protect\cite{Al}) is that it paints a very transparent
picture of the physics at work (see Fig. \ref{fig:MMevo}): one step
in the evolution corresponds to a gluon emission, whose motion in
the transverse plane, as well as its multiple scattering (and the
one of the parent dipole) off the medium, is encoded in the Green
function $G\left(\mathbf{r}_{2},\mathbf{r}_{1},\tau=t_{2}-t_{1}\right)$.
Before and after the gluon fluctuation, the parent dipole undergoes
multiple soft scattering, described by $\exp\left(-\frac{\hat{q}}{4}\mathbf{x}^{2}t_{1}\right)$
and $\exp\left(-\frac{\hat{q}}{4}\mathbf{x}^{2}\left(L-t_{2}\right)\right)$,
respectively. From this, the virtual contributions are subtracted,
which correspond to the free gluon propagator times the multiple scattering
factor of the parent dipole. Note that in Ref. \protect\cite{Al}, the factor
$\exp\left(-\frac{\hat{q}}{4}\mathbf{x}^{2}L\right)$ isn't included,
since the authors obtain their evolution equation in the spirit of
the BDMPS-Z formalism, in which they simply subtract a medium independent
contribution $G_{0}\left(\mathbf{r}_{2},\mathbf{r}_{1},\tau\right)$.
In this sense, the work of Iancu (Ref. \protect\cite{Iancu}), to which this
part of the thesis is devoted, can be regarded as a more formal, and
ultimately more correct, approach to the same problem. By carefully
constructing the in-medium evolution equation, we do obtain the correct
virtual term which however, as we have seen, doesn't play a role in
the double logarithmic approximation. Rather, the virtual contribution
will become important when performing the calculation to single-logarithmic
accuracy, for which the authors of Ref. \protect\cite{Al} inserted the correct
virtual corrections by hand.

Let us now show, as we already announced, that the limits to the phase
space of the gluon fluctuation are encoded in Eq. (\ref{eq:MuellerMaster}).
For this, we should first make the observation that the factor $\sqrt{\omega/\hat{q}}$,
which appears in the definition of $\Omega$, Eq. (\ref{eq:Omega}),
is equal to the lifetime $\tau_{\mathrm{br}}$ of a gluon whose transverse
momentum is solely obtained from multiple soft scatterings off the
medium. Indeed, we have:
\begin{equation}
\begin{aligned}k_{\mathrm{br}}^{2} & =\hat{q}\tau,\quad\mathrm{and}\quad\tau\simeq\frac{2\omega}{k_{\mathrm{br}}^{2}},\\
\Longleftrightarrow & \qquad\tau_{\mathrm{br}}\left(\omega\right)\simeq\sqrt{\frac{\omega}{\hat{q}}}.
\end{aligned}
\end{equation}
The region of multiple scattering is therefore the one above the $\tau_{\mathrm{br}}\left(\omega\right)$-line
in the $(\tau,\omega)$-plane (see Fig. \ref{fig:gluonphasespace}).
Accordingly, this regime corresponds to values: 

\begin{equation}
\Omega\tau=\frac{1+i}{\sqrt{2}}\frac{\tau}{\tau_{\mathrm{br}}}\gg1
\end{equation}
for which:
\begin{equation}
\begin{aligned}G\left(\mathbf{r}_{2},\mathbf{r}_{1},\tau\gg\tau_{\mathrm{br}}\right) & \simeq-\frac{i}{\pi}\omega\Omega e^{-\Omega\tau}\exp\left\{ i\frac{\omega\Omega}{2}\left(\left(\mathbf{r}_{2}-\mathbf{R}\right)^{2}+\left(\mathbf{r}_{1}-\mathbf{R}\right)^{2}\right)\right\} .\end{aligned}
\end{equation}
Gluon fluctuations with a lifetime that surpasses the characteristic
lifetime $\tau_{\mathrm{br}}\left(\omega\right)$ are thus exponentially
suppressed. The emergence of the characteristic lifetime, during which
the gluon undergoes coherent scattering with the medium, is recognized
as an aspect of the Landau-Pomeranchuk-Migdal (LPM) effect (Refs.
\protect\cite{Landau1953,Migdal1956}) in QCD (BDMPS-Z, see Refs. \protect\cite{Zakharov1996,Zakharov1997,BDMPS1,BDMPS2,BDMPS3,BDMPS4,BDMPS5,BDMPS6,Baier2000}).

Furthermore, for values $\tau\sim\tau_{\mathrm{br}}$:
\begin{equation}
\begin{aligned}G\left(\mathbf{r}_{2},\mathbf{r}_{1},\tau\sim\tau_{\mathrm{br}}\right) & \propto\exp\left\{ \frac{i}{2}\omega\Omega\left(\left(\mathbf{r}_{2}-\mathbf{R}\right)^{2}+\left(\mathbf{r}_{1}-\mathbf{R}\right)^{2}\right)\right\} ,\\
 & \propto\exp\left\{ -\frac{1}{\sqrt{2}}\frac{\omega}{\tau_{\mathrm{br}}}B^{2}\right\} ,
\end{aligned}
\end{equation}
hence the growth of the gluon fluctuation is restricted to $B^{2}\lesssim2/k_{\mathrm{br}}^{2}\left(\omega\right)$,
as we anticipated in Eqs. (\ref{eq:Blower}), (\ref{eq:DLAlimits}).

Finally, for small lifetimes $\tau\ll\tau_{\mathrm{br}}$ (but still
$\tau\gg\omega/Q_{s}^{2}$), one can recover the double logarithmic
approximation by expanding the r.h.s. of Eq. (\ref{eq:MuellerMaster})
around small sizes $\mathbf{x}$ of the parent dipole:
\begin{equation}
\begin{aligned}e^{-\frac{\hat{q}}{4}L\mathbf{x}^{2}}\partial_{\mathbf{r}_{1}}\cdot\partial_{\mathbf{r}_{2}}G_{0}\left(\mathbf{r}_{2},\mathbf{r}_{1},\tau\right)\Bigr|_{\mathbf{0}}^{\mathbf{x}}\Bigr|_{\mathbf{0}}^{\mathbf{x}} & \simeq\frac{2i\omega^{3}}{\pi\tau^{3}}x^{2},\\
e^{-\frac{\hat{q}}{4}\left(L-\tau\right)\mathbf{x}^{2}}\partial_{\mathbf{r}_{1}}\cdot\partial_{\mathbf{r}_{2}}G\left(\mathbf{r}_{2},\mathbf{r}_{1},\tau\right)\Bigr|_{\mathbf{0}}^{\mathbf{x}}\Bigr|_{\mathbf{0}}^{\mathbf{x}} & \simeq\frac{i\omega^{3}\Omega^{3}}{2\pi\sinh^{3}\Omega\tau}x^{2}\left(\sinh^{2}\Omega\tau+4\right),
\end{aligned}
\label{eq:Taylordipole}
\end{equation}
such that:
\begin{align}
N_{\omega}\left(\mathbf{x}\right) & \simeq\frac{-i\alpha_{s}N_{c}}{4\pi\omega^{2}}x^{2}\int_{0}^{L}\mathrm{d}t_{2}\int_{0}^{t_{2}}\mathrm{d}t_{1}\left(\frac{\omega}{\tau}\right)^{3}\left(\frac{\left(\Omega\tau\right)^{3}}{\sinh^{3}\Omega\tau}\left(\sinh^{2}\Omega\tau+4\right)-4\right).\label{eq:smalldipoles}
\end{align}
In Eq. (\ref{eq:Taylordipole}) we see that, as expected, the virtual
corrections do not play a role at the present DLA accuracy. Rewriting
the time integrations as follows:
\begin{equation}
\begin{aligned}\int_{0}^{L}\mathrm{d}t_{2}\int_{0}^{t_{2}}\mathrm{d}t_{1}f\left(\tau=t_{2}-t_{1}\right) & =\int_{0}^{L}\mathrm{d}t_{2}\int_{0}^{t_{2}}\mathrm{d}\tau f\left(\tau\right),\\
 & =\int_{0}^{L}\mathrm{d}t_{2}\int_{0}^{L}\mathrm{d}\tau\Theta\left(t_{2}-\tau\right)f\left(\tau\right),\\
 & =\int_{0}^{L}\mathrm{d}\tau\left(L-\tau\right)f\left(\tau\right),
\end{aligned}
\end{equation}
the equation becomes:
\begin{align}
N_{\omega}\left(\mathbf{x}\right) & =\frac{-i\alpha_{s}N_{c}\omega}{4\pi}x^{2}\int_{l_{0}}^{L}\mathrm{d}\tau\frac{L-\tau}{\tau^{3}}\left(\frac{\left(\Omega\tau\right)^{3}}{\sinh^{3}\Omega\tau}\left(\sinh^{2}\Omega\tau+4\right)-4\right).
\end{align}

Finally, taking the single scattering limit, i.e. $\Omega\tau$ small:
\begin{equation}
\begin{aligned}N_{\omega}\left(\mathbf{x}\right) & \simeq\frac{i\alpha_{s}N_{c}}{4\pi\omega^{2}}x^{2}\int_{l_{0}}^{L}\mathrm{d}\tau\left(L-\tau\right)\left(\frac{\omega}{\tau}\right)^{3}\left(\Omega\tau\right)^{2},\\
 & =\frac{-\alpha_{s}N_{c}}{4\pi}\hat{q}x^{2}\int_{l_{0}}^{L}\mathrm{d}t\frac{L-\tau}{\tau}.
\end{aligned}
\end{equation}
Combining this result with Eq. (\ref{eq:Ndef}) and with the definition
(\ref{eq:TMB}) of transverse momentum broadening, we obtain:
\begin{equation}
\begin{aligned}\left\langle k_{\perp}^{2}\right\rangle  & =\frac{\alpha_{s}N_{c}}{\pi}\hat{q}\int_{l_{0}}^{L}\mathrm{d}\tau\frac{L-\tau}{\tau}\int_{\hat{q}\tau^{2}}^{\hat{q}L\tau}\frac{\mathrm{d}\omega}{\omega},\\
 & \simeq\frac{\bar{\alpha}}{2}\hat{q}L\ln^{2}\frac{L}{l_{0}}.
\end{aligned}
\end{equation}
in agreement with our result (\ref{eq:qhat1}).

The authors of Ref. \protect\cite{Al} push the calculation even further,
to single logarithmic accuracy, that is: they manage to calculate
all logarithmic contributions deeply inside the region $l_{0}\ll\tau\ll L$
and $\omega_{0}\ll\omega\ll\omega_{c}$ (see Fig. \ref{fig:gluonphasespace}).
Let us quickly sketch their approach: first, one chooses a point $\tau_{\mathrm{br}}\ll\tau_{0}\ll\omega/Q_{s}^{2}$,
hence within the single scattering region. The integration from this
point onwards to the maximal lifetime $\tau=L$, crossing the multiple
scattering boundary $\tau=\tau_{\mathrm{br}}\left(\omega\right)$,
is then governed by Eq. (\ref{eq:smalldipoles}):
\begin{align}
N_{\omega}^{\tau_{0}\to L}\left(\mathbf{x}\right) & \simeq\frac{-i\alpha_{s}N_{c}}{4\pi\omega^{2}}x^{2}\int_{\tau_{0}}^{L}\mathrm{d}\tau\left(L-\tau\right)\left(\frac{\omega}{\tau}\right)^{3}\left(\frac{\left(\Omega\tau\right)^{3}}{\sinh^{3}\Omega\tau}\left(\sinh^{2}\Omega\tau+4\right)-4\right).\label{eq:boundary(b)}
\end{align}
That is, the original dipole size is still assumed to be small enough
for the harmonic approximation to hold ($N_{\omega}\left(\mathbf{x}\right)\propto x^{2}$),
yet one doesn't enforce single scattering: there is no expansion in
$\Omega\tau$. 

This result is then matched with the integration that crosses the
boundary $\tau=\omega/Q_{s}^{2}$, which is the regime in which the
gluon's transverse size is of the same order as the parent dipole.
The appropriate equation is now the single-scattering \textquoteleft BFKL'
equation (\ref{eq:BFKLmedium}), in which the virtual term is crucial
to guarantee ultraviolet convergence. Note that the time integrations,
that we evaluated in Eqs. (\ref{eq:G2Kmedium1}), (\ref{eq:G2Kmedium2}),
need to be restored in order to extract an integral over $\tau$ that
can be matched with the one in Eq. (\ref{eq:boundary(b)}). We have,
approximately:
\begin{equation}
\begin{aligned} & \int_{0}^{L}\mathrm{d}t\int_{0}^{t}\mathrm{d}t_{1}\int_{t}^{L}\mathrm{d}t_{2}G_{0}\left(t_{2},\mathbf{r}_{2},t,\mathbf{z};\omega\right)G_{0}\left(t,\mathbf{z},t_{1},\mathbf{r}_{1};\omega\right)\\
 & \simeq L\int_{l_{0}}^{\tau_{0}}\mathrm{d}\tau\int_{-\tau}^{0}\mathrm{d}t_{1}G_{0}\left(\tau+t_{1},\mathbf{r}_{2},0,\mathbf{z};\omega\right)G_{0}\left(0,\mathbf{z},t_{1},\mathbf{r}_{1};\omega\right),
\end{aligned}
\end{equation}
which is in accordance with the manipulations in Eqs. (\ref{eq:G2Kmedium1}),
(\ref{eq:G2Kmedium2}), in which we neglected the exponentials in
$L$ and $t$. The result is:
\begin{equation}
\begin{aligned} & N_{\omega}^{l_{0}\to\tau_{0}}\left(\mathbf{x}\right)=\frac{\alpha_{s}}{\omega^{2}}\frac{N_{c}}{8}\hat{q}e^{-\frac{1}{4}\hat{q}Lx^{2}}L\int_{l_{0}}^{\tau_{0}}\mathrm{d}\tau\int_{-\tau}^{0}\mathrm{d}t_{1}\mathbf{\partial}_{\mathbf{r}_{1}}^{i}\mathbf{\partial}_{\mathbf{r}_{2}}^{i}\\
 & \times\int\mathrm{d}^{2}\mathbf{z}\,G_{0}\left(\tau+t_{1},\mathbf{r}_{2},0,\mathbf{z};\omega\right)G_{0}\left(0,\mathbf{z},t_{1},\mathbf{r}_{1};\omega\right)\left(\left(\mathbf{x}-\mathbf{z}\right)^{2}+z^{2}-x^{2}\right)\biggr|_{\mathbf{r}_{1}=\mathbf{0}}^{\mathbf{r}_{1}=\mathbf{x}}\biggr|_{\mathbf{r}_{2}=\mathbf{0}}^{\mathbf{r}_{2}=\mathbf{x}}.
\end{aligned}
\label{eq:boundary(a)}
\end{equation}
Finally, near the boundary $\tau=l_{0}$, the gluon distribution of
the nucleon at large values of $x$ is probed, since $\tau\simeq l_{0}/x$
(see the discussion in the previous section). The approximation that
$\hat{q}$ is a constant is only valid in small-$x$, and therefore
breaks down. In Ref. \protect\cite{Al} this problem is circumvented by calculating
directly the induced gluon radiation of a highly-energetic quark and
setting $\left\langle k_{\perp}^{2}\right\rangle \simeq k_{\perp}^{2}$,
where $k_{\perp}$ is the transverse momentum of the radiated gluon.
The result is:
\begin{equation}
\left\langle k_{\perp}^{2}\right\rangle \simeq\bar{\alpha}\int\frac{\mathrm{d}\omega}{\omega}\hat{q}L\left(\ln\frac{t_{0}}{L}+\gamma_{E}\right).\label{eq:boundary(c)}
\end{equation}

Matching results (\ref{eq:boundary(b)}) with (\ref{eq:boundary(c)})
for $\omega<\hat{q}l_{0}L$, and (\ref{eq:boundary(b)}) with (\ref{eq:boundary(a)})
for $\omega>\hat{q}l_{0}L$, one obtains the final result for the
leading-order correction to the transverse momentum broadening to
single leading logarithmic accuracy:
\begin{equation}
\left\langle k_{\perp}^{2}\right\rangle \simeq\bar{\alpha}\hat{q}L\left[\left(\ln2-\frac{1}{3}\right)\ln\frac{L}{l_{0}}+\frac{1}{2}\ln^{2}\frac{L}{l_{0}}\right]+C,
\end{equation}
where $C$ is constant that cannot be calculated in the present formalism.

\section{Conclusion}

After constructing a non-eikonal generalization of the JIMWLK equation,
we were well-equipped to attack the problem of the radiative corrections
to the transverse momentum broadening of a hard parton that travels
through a nuclear medium. To double leading logarithmic accuracy,
these corrections could be absorbed into a renormalization of the
jet quenching parameter, recovering the results in Refs. \protect\cite{Al,Blaizot2014RG}.
Moreover, from our more formal evolution point of view, we derived
the complete in-medium evolution equation for a dipole, Eq. (\ref{eq:qhatevolutionequation}),
which contains all the virtual terms necessary to guarantee ultraviolet
convergence. In principle, solving the equation would be tantamount
to the resummation of all the leading logarithmic corrections to $\hat{q}$,
the single large logarithms included. In practice, however, we do
not know how to solve such a functional integro-differential equation,
which is complicated by the mixing of the longitudinal and transverse
phase space due to the influence of the medium. Furthermore, most
probably the LLA corrections cannot be associated with a renormalization
of $\hat{q}$ anymore, but one rather has to renormalize the transverse
momentum broadening $\langle k_{\perp}^{2}\rangle$ itself.

\newpage{}

\thispagestyle{simple}

\appendix

\part{Appendices}

\section{Lie algebra and Wilson line relations}

The representations of $SU\left(3\right)$ are always Hermitian: $t^{a}=t^{a\dagger}$,
hence:
\begin{equation}
\begin{aligned}t_{ij}^{a} & =t_{ji}^{a*}.\end{aligned}
\end{equation}
The adjoint representation of $SU\left(3\right)$ is generated by
the matrices:
\begin{equation}
\begin{aligned}T_{bc}^{a} & =-if^{abc},\end{aligned}
\end{equation}
where $f^{abc}$ are the structure constants, given by:
\begin{equation}
\begin{aligned}\left[l^{a},l^{b}\right] & =if^{abc}l^{c},\end{aligned}
\end{equation}
and where $l^{a}$ are the generators of\emph{ }any\emph{ }$SU\left(3\right)$
representation. The structure constants $f^{abc}$ are real and antisymmetric,
hence:
\begin{equation}
\begin{aligned}T_{bc}^{a} & =-T_{bc}^{a*}.\end{aligned}
\label{eq:Tstar}
\end{equation}
In our conventions, we define a Wilson line in the fundamental and
adjoint representation, respectively, as follows:
\begin{equation}
\begin{aligned}U\left(\mathbf{x}\right) & \equiv\mathcal{P}\exp\left(ig_{s}\int\mathrm{\mathrm{d}}x^{+}A_{c}^{-}\left(\mathbf{x}\right)t^{c}\right),\\
W\left(\mathbf{x}\right) & \equiv\mathcal{P}\exp\left(ig_{s}\int\mathrm{\mathrm{d}}x^{+}A_{c}^{-}\left(\mathbf{x}\right)T^{c}\right),
\end{aligned}
\end{equation}
where the path ordering operator $\mathcal{P}$ orders the color matrices
from left to right, such that the rightmost matrix is the one that
appears last along the integration path. The Hermitian conjugate of
a Wilson line is then equal to:
\begin{equation}
\begin{aligned}U^{\dagger}\left(\mathbf{x}\right) & =\left(\mathcal{P}\exp\left(ig_{s}\int\mathrm{\mathrm{d}}x^{+}A_{c}^{-}\left(\mathbf{x}\right)t^{c}\right)\right)^{\dagger}=\bar{\mathcal{P}}\exp\left(\left(ig_{s}\int\mathrm{\mathrm{d}}x^{+}A_{c}^{-}\left(\mathbf{x}\right)t^{c}\right)^{\dagger}\right),\\
 & =\bar{\mathcal{P}}\exp\left(-ig_{s}\int\mathrm{\mathrm{d}}x^{+}A_{c}^{-}\left(\mathbf{x}\right)t^{c}\right)=U_{\leftrightarrow}\left(\mathbf{x}\right),
\end{aligned}
\end{equation}
where $\bar{\mathcal{P}}$ is the anti-path ordering operator, ordering
the color matrices from right to left according to their appearance
along the path. Hermitian conjugation is therefore tantamount to inverting
the direction along the integration path of the Wilson line.

Using the notation $U_{\mathbf{x}}\left(a,b\right)$ for a Wilson
line over a finite length:
\begin{equation}
\begin{aligned}U_{\mathbf{x}}\left(a,b\right) & \equiv\mathcal{P}\exp\left(ig_{s}\int_{a}^{b}\mathrm{\mathrm{d}}x^{+}A_{c}^{-}\left(\mathbf{x}\right)t^{c}\right),\end{aligned}
\end{equation}
it is clear that the following rule holds:
\begin{equation}
\begin{aligned}U_{\mathbf{x}}\left(a,b\right) & =U_{\mathbf{x}}\left(a,c\right)U_{\mathbf{x}}\left(c,b\right).\end{aligned}
\label{eq:Wilsonlinesum}
\end{equation}

From Eq. (\ref{eq:Tstar}), one has that:
\begin{equation}
\begin{aligned}W_{ab}^{*}\left(\mathbf{x}\right) & =\left(\mathcal{P}\exp\left(ig_{s}\int\mathrm{\mathrm{d}}x^{+}A_{c}^{-}T^{c}\right)^{*}\right)_{ab},\\
 & =\left(\mathcal{P}\exp\left(-ig_{s}\int\mathrm{\mathrm{d}}x^{+}A_{c}^{-}T^{c*}\right)\right)_{ab},\\
 & =W_{ab}\left(\mathbf{x}\right),
\end{aligned}
\label{eq:Wabreal}
\end{equation}
and hence a Wilson line in the adjoint representation is real.

Using the expansion of the Wilson lines around unity, we find the
following important identity:
\begin{equation}
\begin{aligned}U^{\dagger}t^{a}U & =W_{ac}t^{c}.\end{aligned}
\label{eq:U2W}
\end{equation}
The same identity holds for Wilson lines in the adjoint representation:
\begin{equation}
\begin{aligned}W^{\dagger}t^{a}W & =W_{ac}t^{c}.\end{aligned}
\label{eq:adjoint2adjoint}
\end{equation}
We will frequently take the traces over the representation matrices
\begin{equation}
\begin{aligned}\mathrm{Tr}\left(t^{a}t^{b}\right) & =\frac{1}{2}\delta_{ab},\\
\mathrm{Tr}\left(T^{a}T^{b}\right) & =N_{c}\delta_{ab},
\end{aligned}
\label{eq:tracetatb}
\end{equation}
and use the very important Fierz identity:
\begin{equation}
\begin{aligned}t_{ij}^{a}t_{kl}^{a} & =\frac{1}{2}\delta_{il}\delta_{jk}-\frac{1}{2N_{c}}\delta_{ij}\delta_{kl}.\end{aligned}
\label{eq:Fierz}
\end{equation}
The Casimir operators of the fundamental and adjoint representation
are:
\begin{equation}
t_{ik}^{a}t_{kj}^{a}=C_{F}\delta_{ij},\quad\mathrm{with}\quad C_{F}\equiv\frac{N_{c}^{2}-1}{2N_{c}},\label{eq:Casimirfund}
\end{equation}
and
\begin{equation}
T_{ac}^{d}T_{cb}^{d}=N_{c}\delta_{ab}.\label{eq:Casimiradj}
\end{equation}

\section{Some useful integrals}

The following integral, which is the Green function of the two-dimensional
Laplace equation, will often play a role in our calculations:
\begin{equation}
\begin{aligned}\int\mathrm{d}^{2}\mathbf{k}_{\perp}\frac{e^{i\mathbf{k}_{\perp}\mathbf{r}}}{k_{\perp}^{2}} & =2\pi\int_{\Lambda}^{\infty}\frac{\mathrm{d}k_{\perp}}{k_{\perp}}J_{0}\left(k_{\perp}r\right),\\
 & =2\pi\ln\frac{1}{r\Lambda}.
\end{aligned}
\label{eq:fourier2log}
\end{equation}
Moreover, one can show that:
\begin{equation}
\int\mathrm{d}^{2}\mathbf{k}_{\perp}\frac{e^{i\mathbf{k}_{\perp}\mathbf{r}}}{k_{\perp}^{2}}=\int\mathrm{d}^{2}\mathbf{z}\frac{z^{i}\left(r+z\right)^{i}}{z^{2}\left(\mathbf{r}+\mathbf{z}\right)^{2}},\label{eq:fourier2Kxyz}
\end{equation}
and
\begin{equation}
\int\mathrm{d}^{2}\mathbf{k}_{\perp}\frac{k_{\perp}^{i}}{k_{\perp}^{2}}e^{i\mathbf{k}_{\perp}\mathbf{x}}=2\pi i\frac{x^{i}}{x^{2}}.\label{eq:WWfieldrelation}
\end{equation}

\section{Light-cone perturbation theory conventions\label{subsec:Conventions}}

We follow the so-called \textquoteleft Kogut-Soper' conventions \protect\cite{Brodsky1998},
in which the transverse polarization vectors of the gluon are written
as follows: 
\begin{equation}
\epsilon_{\lambda}^{\mu}\left(k\right)=\left(0,\frac{k_{\perp}\cdot\epsilon_{\perp}^{\lambda}}{k^{+}},\boldsymbol{\epsilon}_{\perp}^{\lambda}\right),
\end{equation}
and the longitudinal polarization vector (with $Q^{2}=k^{2}$):
\begin{equation}
\epsilon_{L}^{\mu}\left(k\right)=\left(0,\frac{Q}{k^{+}},0\right).
\end{equation}
One chooses circular polarization, in which 
\begin{equation}
\boldsymbol{\epsilon}_{\perp}^{1}=\frac{1}{\sqrt{2}}\left(\begin{array}{c}
1\\
i
\end{array}\right),\qquad\boldsymbol{\epsilon}_{\perp}^{2}=\frac{1}{\sqrt{2}}\left(\begin{array}{c}
1\\
-i
\end{array}\right),
\end{equation}
with $\boldsymbol{\epsilon}_{\perp}^{\lambda\dagger}\cdot\boldsymbol{\epsilon}_{\perp}^{\lambda'}=\delta^{\lambda\lambda'}$.
We work in the chiral representation of the gamma matrices:
\begin{equation}
\gamma^{0}=\left(\begin{array}{cc}
0 & I\\
I & 0
\end{array}\right),\quad\gamma^{i}=\left(\begin{array}{cc}
0 & -\sigma^{i}\\
\sigma^{i} & 0
\end{array}\right),
\end{equation}
\begin{align}
\gamma^{+} & =\left(\begin{array}{cccc}
0 & 0 & 0 & 0\\
0 & 0 & 0 & \sqrt{2}\\
\sqrt{2} & 0 & 0 & 0\\
0 & 0 & 0 & 0
\end{array}\right),\quad\gamma^{1}=\left(\begin{array}{cccc}
0 & 0 & 0 & -1\\
0 & 0 & -1 & 0\\
0 & 1 & 0 & 0\\
1 & 0 & 0 & 0
\end{array}\right),\quad\gamma^{2}=\left(\begin{array}{cccc}
0 & 0 & 0 & i\\
0 & 0 & -i & 0\\
0 & -i & 0 & 0\\
i & 0 & 0 & 0
\end{array}\right),
\end{align}
with the usual Pauli matrices:
\begin{align}
\sigma^{1} & =\left(\begin{array}{cc}
0 & 1\\
1 & 0
\end{array}\right),\qquad\sigma^{2}=\left(\begin{array}{cc}
0 & -i\\
i & 0
\end{array}\right),\qquad\sigma^{3}=\left(\begin{array}{cc}
1 & 0\\
0 & -1
\end{array}\right).
\end{align}
With these conventions, the Dirac spinors are given by:
\begin{equation}
u_{+}\left(k\right)=\frac{1}{2^{1/4}\sqrt{k^{+}}}\left(\begin{array}{c}
\sqrt{2}k^{+}\\
k_{1}+ik_{2}\\
m\\
0
\end{array}\right),\qquad u_{-}\left(k\right)=\frac{1}{2^{1/4}\sqrt{k^{+}}}\left(\begin{array}{c}
0\\
m\\
-k_{1}+ik_{2}\\
\sqrt{2}k^{+}
\end{array}\right),
\end{equation}
and
\begin{equation}
v_{+}\left(k\right)=\frac{1}{2^{1/4}\sqrt{k^{+}}}\left(\begin{array}{c}
0\\
-m\\
-k_{1}+ik_{2}\\
\sqrt{2}k^{+}
\end{array}\right),\qquad v_{-}\left(k\right)=\frac{1}{2^{1/4}\sqrt{k^{+}}}\left(\begin{array}{c}
\sqrt{2}k^{+}\\
k_{1}+ik_{2}\\
-m\\
0
\end{array}\right).
\end{equation}

\section{The $g\rightarrow q\bar{q}$ wave function\label{subsec:gqqwave}}
\begin{center}
\begin{tikzpicture}[scale=2] 
\tikzset{photon/.style={semithick,decorate,decoration={snake}}, electron/.style={ postaction={decorate},decoration={markings,mark=at position .5 with {\arrow[]{latex}}}},	positron/.style={ postaction={decorate},decoration={markings,mark=at position .5 with {\arrow[]{latex reversed}}}},      	gluon/.style={decorate,decoration={coil,amplitude=4pt, segment length=5pt}}}

\draw[semithick,gluon] (0.5,-1.5) node [left]{$p,\lambda$} --(1,-1.5);
\draw[semithick,electron] (1,-1.5).. controls (1.2,-1.3) and (1.8,-1.1)  .. (2,-1.1) node [right]{$k,\beta$};
\draw[semithick,positron] (1,-1.5).. controls (1.2,-1.7) and (1.8,-1.9)  .. (2,-1.9) node [right]{$p-k,\alpha$};

\end{tikzpicture} 
\par\end{center}

Choosing $p^{\mu}=\left(p^{+},\left(p_{\perp}^{2}-Q^{2}\right)/2p^{+},\mathbf{p}_{\perp}\right)$
to be the momentum of the gluon, and $k$ and $q\equiv p-k$ the momenta
of the quark and antiquark, respectively, the wave function will have
the form:

\begin{equation}
\psi_{\alpha\beta}^{\lambda,L}\left(p,k\right)=\frac{1}{\sqrt{8\left(p-k\right)^{+}p^{+}k^{+}}}\frac{\bar{u}_{\beta}\left(k\right)\gamma_{\mu}\epsilon_{\lambda,L}^{\mu}\left(p\right)v_{\alpha}\left(p-k\right)}{\left(p-k\right)^{-}+k^{-}-p^{-}},
\end{equation}
where:
\begin{align}
\gamma_{\mu}\epsilon_{\lambda}^{\mu}\left(k\right) & =\gamma^{+}\frac{\mathbf{k}_{\perp}\cdot\boldsymbol{\mathbf{\epsilon}}_{\perp}^{\lambda}}{k^{+}}-\gamma_{\perp}\cdot\epsilon_{\perp}^{\lambda},\\
 & =\left(\begin{array}{cccc}
0 & 0 & 0 & \epsilon_{1}^{\lambda}-i\epsilon_{2}^{\lambda}\\
0 & 0 & \epsilon_{1}^{\lambda}+i\epsilon_{2}^{\lambda} & \sqrt{2}\frac{\mathbf{k}_{\perp}\cdot\boldsymbol{\mathbf{\epsilon}}_{\perp}^{\lambda}}{k^{+}}\\
\sqrt{2}\frac{\mathbf{k}_{\perp}\cdot\boldsymbol{\mathbf{\epsilon}}_{\perp}^{\lambda}}{k^{+}} & -\epsilon_{1}^{\lambda}+i\epsilon_{2}^{\lambda} & 0 & 0\\
-\epsilon_{1}^{\lambda}-i\epsilon_{2}^{\lambda} & 0 & 0 & 0
\end{array}\right),
\end{align}
and
\begin{equation}
\begin{aligned}\gamma_{\mu}\epsilon_{L}^{\mu}\left(k\right) & =\gamma^{+}\epsilon_{L}^{-}\left(k\right)=\frac{Q}{p^{+}}\left(\begin{array}{cccc}
0 & 0 & 0 & 0\\
0 & 0 & 0 & \sqrt{2}\\
\sqrt{2} & 0 & 0 & 0\\
0 & 0 & 0 & 0
\end{array}\right).\end{aligned}
\end{equation}
The energy denominator is the same for each spin configuration, and
can be calculated as follows:
\begin{equation}
\begin{aligned}\left(p-k\right)^{-}+k^{-}-p^{-} & =\frac{m^{2}+\left(\mathbf{p}_{\perp}-\mathbf{k}_{\perp}\right)^{2}}{2\left(p^{+}-k^{+}\right)}+\frac{k_{\perp}^{2}+m^{2}}{2k^{+}}-\frac{p_{\perp}^{2}-Q^{2}}{2p^{+}},\\
 & =\frac{\left(\mathbf{k}_{\perp}-z\mathbf{p}_{\perp}\right)^{2}+m^{2}+\left(1-z\right)zQ^{2}}{2\left(p^{+}-k^{+}\right)z}.
\end{aligned}
\end{equation}

\subsubsection*{Helicity minus - minus}

In the case of a longitudinal polarized gluon, the numerator reads:
\begin{equation}
\begin{aligned}\bar{u}_{-}\left(k\right)\gamma_{\mu}\epsilon_{L}^{\mu}\left(p\right)v_{+}\left(q\right) & =\frac{1}{\sqrt{2k^{+}\left(p-k\right)^{+}}}\frac{Q}{p^{+}}\left(\begin{array}{cccc}
-k_{1}-ik_{2}, & \sqrt{2}k^{+}, & 0, & m\end{array}\right)\\
 & \times\left(\begin{array}{cccc}
0 & 0 & 0 & 0\\
0 & 0 & 0 & \sqrt{2}\\
\sqrt{2} & 0 & 0 & 0\\
0 & 0 & 0 & 0
\end{array}\right)\left(\begin{array}{c}
0\\
-m\\
-q_{1}+iq_{2}\\
\sqrt{2}q^{+}
\end{array}\right)=\frac{2zQq^{+}}{\sqrt{k^{+}\left(p-k\right)^{+}}},
\end{aligned}
\end{equation}
while we have for the transverse case:
\begin{equation}
\begin{aligned} & \bar{u}_{-}\left(k\right)\gamma_{\mu}\epsilon_{\lambda}^{\mu}\left(p\right)v_{+}\left(q\right)\\
 & =\frac{1}{\sqrt{2k^{+}\left(p-k\right)^{+}}}\Biggl(\sqrt{2}q^{+}\left(\epsilon_{1}^{\lambda}-i\epsilon_{2}^{\lambda}\right)\left(-k_{1}-ik_{2}\right)\\
 & +\sqrt{2}k^{+}\left(\left(-q_{1}+iq_{2}\right)\left(\epsilon_{1}^{\lambda}+i\epsilon_{2}^{\lambda}\right)+2q^{+}\frac{\mathbf{p}_{\perp}\cdot\boldsymbol{\mathbf{\epsilon}}_{\perp}^{\lambda}}{p^{+}}\right)\Biggr),\\
 & =\frac{1}{\sqrt{k^{+}\left(p-k\right)^{+}}}\left(2q^{+}\delta_{\lambda1}\left(-k_{1}-ik_{2}\right)+2k^{+}\left(-q_{1}+iq_{2}\right)\delta_{\lambda2}+2zq^{+}\mathbf{p}_{\perp}\cdot\boldsymbol{\mathbf{\epsilon}}_{\perp}^{\lambda}\right),\\
 & =\frac{1}{\sqrt{k^{+}\left(p-k\right)^{+}}}\left(-2q^{+}\delta_{\lambda1}\mathbf{k}_{\perp}\cdot\boldsymbol{\mathbf{\epsilon}}_{\perp}^{1}-2k^{+}\delta_{\lambda2}\mathbf{q}_{\perp}\cdot\boldsymbol{\mathbf{\epsilon}}_{\perp}^{2}+2zq^{+}\mathbf{p}_{\perp}\cdot\boldsymbol{\mathbf{\epsilon}}_{\perp}^{\lambda}\right),\\
 & =\frac{1}{\sqrt{k^{+}\left(p-k\right)^{+}}}\left(-2q^{+}\delta_{\lambda1}\left(\mathbf{k}_{\perp}-z\mathbf{p}_{\perp}\right)\cdot\boldsymbol{\mathbf{\epsilon}}_{\perp}^{1}+2k^{+}\delta_{\lambda2}\left(\mathbf{k}_{\perp}-z\mathbf{p}_{\perp}\right)\cdot\boldsymbol{\mathbf{\epsilon}}_{\perp}^{2}\right),
\end{aligned}
\end{equation}
where in the last line we used the definition $\mathbf{q}_{\perp}=\mathbf{p}_{\perp}-\mathbf{k}_{\perp}$.
Finally, one obtains the following expressions for the wave functions:
\begin{equation}
\begin{aligned}\psi_{--}^{L}\left(p,k\right) & =\frac{1}{\sqrt{8\left(p-k\right)^{+}p^{+}k^{+}}}\frac{2\left(p^{+}-k^{+}\right)z}{\left(\mathbf{k}_{\perp}-z\mathbf{p}_{\perp}\right)^{2}+m^{2}+\left(1-z\right)zQ^{2}}\frac{2zQq^{+}}{\sqrt{k^{+}\left(p-k\right)^{+}}},\\
 & =\sqrt{\frac{2}{p^{+}}}\frac{Qz\left(1-z\right)}{\left(\mathbf{k}_{\perp}-z\mathbf{p}_{\perp}\right)^{2}+m^{2}+\left(1-z\right)zQ^{2}},
\end{aligned}
\end{equation}
\begin{equation}
\begin{aligned}\psi_{--}^{T\lambda}\left(p,k\right) & =\frac{1}{\sqrt{8\left(p-k\right)^{+}p^{+}k^{+}}}\frac{2\left(p^{+}-k^{+}\right)z}{\left(\mathbf{k}_{\perp}-z\mathbf{p}_{\perp}\right)^{2}+m^{2}+\left(1-z\right)zQ^{2}}\\
 & \times\frac{1}{\sqrt{k^{+}\left(p-k\right)^{+}}}\left(-2q^{+}\delta_{\lambda1}\left(\mathbf{k}_{\perp}-z\mathbf{p}_{\perp}\right)\cdot\boldsymbol{\mathbf{\epsilon}}_{\perp}^{1}+2k^{+}\delta_{\lambda2}\left(\mathbf{k}_{\perp}-z\mathbf{p}_{\perp}\right)\cdot\boldsymbol{\mathbf{\epsilon}}_{\perp}^{2}\right),\\
 & =\sqrt{\frac{2}{p^{+}}}\frac{-\left(1-z\right)\delta_{\lambda1}\left(\mathbf{k}_{\perp}-z\mathbf{p}_{\perp}\right)\cdot\boldsymbol{\mathbf{\epsilon}}_{\perp}^{1}+z\delta_{\lambda2}\left(\mathbf{k}_{\perp}-z\mathbf{p}_{\perp}\right)\cdot\boldsymbol{\mathbf{\epsilon}}_{\perp}^{2}}{\left(\mathbf{k}_{\perp}-z\mathbf{p}_{\perp}\right)^{2}+m^{2}+\left(1-z\right)zQ^{2}}.
\end{aligned}
\end{equation}

\subsubsection*{Helicity plus - plus}

For the longitudinal polarized gluon, the numerator reads:
\begin{equation}
\begin{aligned}\bar{u}_{+}\left(k\right)\gamma_{\mu}\epsilon_{L}^{\mu}\left(p\right)v_{-}\left(q\right) & =\frac{2zQq^{+}}{\sqrt{k^{+}\left(p-k\right)^{+}}},\end{aligned}
\end{equation}
and in the transverse case, we have:

\begin{equation}
\begin{aligned} & \bar{u}_{+}\left(k\right)\gamma_{\mu}\epsilon_{\lambda}^{\mu}\left(p\right)v_{-}\left(q\right)\\
 & =\frac{1}{\sqrt{k^{+}\left(p-k\right)^{+}}}\left(2k^{+}\left(\mathbf{k}_{\perp}-z\mathbf{p}_{\perp}\right)\cdot\boldsymbol{\epsilon}_{\perp}^{1}\delta_{\lambda1}-2q^{+}\delta_{\lambda2}\left(\mathbf{k}_{\perp}-z\mathbf{p}_{\perp}\right)\cdot\boldsymbol{\epsilon}_{\perp}^{2}\right).
\end{aligned}
\end{equation}
The wave functions finally read:
\begin{equation}
\begin{aligned}\psi_{++}^{L}\left(p,k\right) & =\frac{1}{\sqrt{8\left(p-k\right)^{+}p^{+}k^{+}}}\frac{2\left(p^{+}-k^{+}\right)z}{\left(\mathbf{k}_{\perp}-z\mathbf{p}_{\perp}\right)^{2}+m^{2}+\left(1-z\right)zQ^{2}}\frac{2zQq^{+}}{\sqrt{k^{+}\left(p-k\right)^{+}}},\\
 & =\sqrt{\frac{2}{p^{+}}}\frac{Q\left(1-z\right)z}{\left(\mathbf{k}_{\perp}-z\mathbf{p}_{\perp}\right)^{2}+m^{2}+\left(1-z\right)zQ^{2}},
\end{aligned}
\end{equation}
and
\begin{equation}
\begin{aligned}\psi_{++}^{\lambda}\left(p,k\right) & =\frac{1}{\sqrt{8\left(p-k\right)^{+}p^{+}k^{+}}}\frac{2\left(p^{+}-k^{+}\right)z}{\left(\mathbf{k}_{\perp}-z\mathbf{p}_{\perp}\right)^{2}+m^{2}+\left(1-z\right)zQ^{2}}\\
 & \times\frac{1}{\sqrt{k^{+}\left(p-k\right)^{+}}}\left(2k^{+}\left(\mathbf{k}_{\perp}-z\mathbf{p}_{\perp}\right)\cdot\boldsymbol{\epsilon}_{\perp}^{1}\delta_{\lambda1}-2q^{+}\delta_{\lambda2}\left(\mathbf{k}_{\perp}-z\mathbf{p}_{\perp}\right)\cdot\boldsymbol{\epsilon}_{\perp}^{2}\right),\\
 & =\sqrt{\frac{2}{p^{+}}}\frac{z\delta_{\lambda1}\left(\mathbf{k}_{\perp}-z\mathbf{p}_{\perp}\right)\cdot\boldsymbol{\epsilon}_{\perp}^{1}-\left(1-z\right)\delta_{\lambda2}\left(\mathbf{k}_{\perp}-z\mathbf{p}_{\perp}\right)\cdot\boldsymbol{\epsilon}_{\perp}^{2}}{\left(\mathbf{k}_{\perp}-z\mathbf{p}_{\perp}\right)^{2}+m^{2}+\left(1-z\right)zQ^{2}}.
\end{aligned}
\end{equation}

\subsubsection*{Helicity-flip cases}

For the longitudinal polarization, we have:

\begin{equation}
\begin{aligned}\bar{u}_{+}\left(k\right)\gamma_{\mu}\epsilon_{L}^{\mu}\left(p\right)v_{+}\left(q\right) & =0,\\
\bar{u}_{-}\left(k\right)\gamma_{\mu}\epsilon_{L}^{\mu}\left(p\right)v_{-}\left(q\right) & =0.
\end{aligned}
\end{equation}
The numerators for the transversely polarized gluon are given by:

\begin{equation}
\begin{aligned}\bar{u}_{+}\left(k\right)\gamma_{\mu}\epsilon_{\lambda}^{\mu}\left(p\right)v_{+}\left(q\right) & =\frac{\sqrt{2}mp^{+}}{\sqrt{k^{+}\left(p-k\right)^{+}}}\delta_{\lambda1},\\
\bar{u}_{-}\left(k\right)\gamma_{\mu}\epsilon_{\lambda}^{\mu}\left(p\right)v_{-}\left(q\right) & =\frac{-\sqrt{2}mp^{+}}{\sqrt{k^{+}\left(p-k\right)^{+}}}\delta_{\lambda2}.
\end{aligned}
\end{equation}
The wave functions read:
\begin{equation}
\begin{aligned}\psi_{+-}^{L}\left(p,k\right) & =\psi_{-+}^{L}\left(p,k\right)=0,\end{aligned}
\end{equation}

\begin{equation}
\begin{aligned}\psi_{+-}^{\lambda}\left(p,k\right) & =\frac{1}{\sqrt{8\left(p-k\right)^{+}p^{+}k^{+}}}\frac{2\left(p^{+}-k^{+}\right)z}{\left(\mathbf{k}_{\perp}-z\mathbf{p}_{\perp}\right)^{2}+m^{2}+\left(1-z\right)zQ^{2}}\frac{mp^{+}}{\sqrt{k^{+}\left(p-k\right)^{+}}}\sqrt{2}\delta_{\lambda1},\\
 & =\sqrt{\frac{2}{p^{+}}}\frac{1}{\left(\mathbf{k}_{\perp}-z\mathbf{p}_{\perp}\right)^{2}+m^{2}+\left(1-z\right)zQ^{2}}\frac{m}{\sqrt{2}}\delta_{\lambda1},
\end{aligned}
\end{equation}
\begin{equation}
\begin{aligned}\psi_{-+}^{\lambda}\left(p,k\right) & =\frac{1}{\sqrt{8\left(p-k\right)^{+}p^{+}k^{+}}}\frac{2\left(p^{+}-k^{+}\right)z}{\left(\mathbf{k}_{\perp}-z\mathbf{p}_{\perp}\right)^{2}+m^{2}+\left(1-z\right)zQ^{2}}\frac{-p^{+}m}{\sqrt{k^{+}\left(p-k\right)^{+}}}\sqrt{2}\delta_{\lambda2}\\
 & =\sqrt{\frac{2}{p^{+}}}\frac{1}{\left(\mathbf{k}_{\perp}-z\mathbf{p}_{\perp}\right)^{2}+m^{2}+\left(1-z\right)zQ^{2}}\frac{-m}{\sqrt{2}}\delta_{\lambda2}.
\end{aligned}
\end{equation}

\subsubsection*{A general expression for the wave function}

The wave functions for the specific helicity cases can be written
in full generality as follows:
\begin{equation}
\begin{aligned}\psi_{\alpha\beta}^{T\lambda}\left(p,k\right) & =\sqrt{\frac{2}{p^{+}}}\frac{1}{\left(\mathbf{k}_{\perp}-z\mathbf{p}_{\perp}\right)^{2}+\epsilon_{f}^{2}}\\
 & \times\begin{cases}
\left(z\delta_{\alpha+}\delta_{\beta+}-\left(1-z\right)\delta_{\alpha-}\delta_{\beta-}\right)\left(\mathbf{k}_{\perp}-z\mathbf{p}_{\perp}\right)\cdot\boldsymbol{\epsilon}_{\perp}^{1}+\frac{m}{\sqrt{2}}\delta_{\alpha+}\delta_{\beta-} & \lambda=1\\
\left(z\delta_{\alpha-}\delta_{\beta-}-\left(1-z\right)\delta_{\alpha+}\delta_{\beta+}\right)\left(\mathbf{k}_{\perp}-z\mathbf{p}_{\perp}\right)\cdot\boldsymbol{\epsilon}_{\perp}^{2}-\frac{m}{\sqrt{2}}\delta_{\alpha-}\delta_{\beta+} & \lambda=2
\end{cases},
\end{aligned}
\label{eq:gqqresultmomentum}
\end{equation}
and
\begin{equation}
\begin{aligned}\psi_{\alpha\beta}^{L}\left(p,k\right) & =\sqrt{\frac{2}{p^{+}}}\frac{Qz\left(1-z\right)}{\left(\mathbf{k}_{\perp}-z\mathbf{p}_{\perp}\right)^{2}+\epsilon_{f}^{2}}\delta_{\alpha\beta},\end{aligned}
\label{eq:gqqresultmomentumL}
\end{equation}
where we defined:
\begin{equation}
\epsilon_{f}^{2}\equiv m^{2}+\left(1-z\right)zQ^{2}.
\end{equation}
In the mixed Fourier-coordinate representation, the wave functions
become:
\begin{equation}
\begin{aligned}\phi_{\alpha\beta}^{T\lambda}\left(p,z,\mathbf{r}\right) & \equiv\int\mathrm{d}^{2}\mathbf{k}_{\perp}e^{i\mathbf{k}_{\perp}\cdot\mathbf{r}}\psi_{\alpha\beta}^{T\lambda}\left(p,k\right),\end{aligned}
\end{equation}
\begin{equation}
\begin{aligned} & \phi_{\alpha\beta}^{T\lambda}\left(p,z,\mathbf{r}\right)=2\pi\sqrt{\frac{2}{p^{+}}}e^{iz\mathbf{p}_{\perp}\cdot\mathbf{r}}\\
 & \times\begin{cases}
i\epsilon_{f}K_{1}\left(\epsilon_{f}r\right)\frac{\mathbf{r}\cdot\boldsymbol{\epsilon}_{\perp}^{1}}{r}\left(z\delta_{\alpha-}\delta_{\beta-}-\left(1-z\right)\delta_{\alpha+}\delta_{\beta+}\right)+\frac{m}{\sqrt{2}}K_{0}\left(\epsilon_{f}r\right)\delta_{\alpha+}\delta_{\beta-} & \lambda=1\\
i\epsilon_{f}K_{1}\left(\epsilon_{f}r\right)\frac{\mathbf{r}\cdot\boldsymbol{\epsilon}_{\perp}^{2}}{r}\left(z\delta_{\alpha+}\delta_{\beta+}-\left(1-z\right)\delta_{\alpha-}\delta_{\beta-}\right)-\frac{m}{\sqrt{2}}K_{0}\left(\epsilon_{f}r\right)\delta_{\alpha-}\delta_{\beta+} & \lambda=2
\end{cases},
\end{aligned}
\label{eq:gqqmixedfourier}
\end{equation}
and
\begin{equation}
\begin{aligned}\phi_{\alpha\beta}^{L}\left(p,z,\mathbf{r}\right) & \equiv\int\mathrm{d}^{2}\mathbf{k}_{\perp}e^{i\mathbf{k}_{\perp}\cdot\mathbf{r}}\psi_{\alpha\beta}^{L}\left(p,k\right),\\
 & =\sqrt{\frac{2}{p^{+}}}\int\frac{\mathrm{d}^{2}\mathbf{k}_{\perp}}{\left(2\pi\right)^{2}}e^{i\mathbf{k}_{\perp}\cdot\mathbf{r}}\frac{Qz\left(1-z\right)}{\left(\mathbf{k}_{\perp}-z\mathbf{p}_{\perp}\right)^{2}+\epsilon_{f}^{2}}\delta_{\alpha\beta},\\
 & =2\pi\sqrt{\frac{2}{p^{+}}}e^{iz\mathbf{p_{\perp}}\cdot\mathbf{r}}Qz\left(1-z\right)K_{0}\left(\epsilon_{f}r\right)\delta_{\alpha\beta},
\end{aligned}
\end{equation}
where we made use of the following identities:
\begin{equation}
\begin{aligned}\int\mathrm{d}^{2}\mathbf{k}_{\perp}\frac{e^{i\mathbf{k}_{\perp}\cdot\mathbf{r}}}{k_{\perp}^{2}+m^{2}} & =2\pi K_{0}\left(mr\right),\\
\int\mathrm{d}^{2}\mathbf{k}_{\perp}\frac{k_{\perp}^{i}e^{i\mathbf{k}_{\perp}\cdot\mathbf{r}}}{k_{\perp}^{2}+m^{2}} & =2\pi im\frac{r^{i}}{r}K_{1}\left(mr\right).
\end{aligned}
\label{eq:waveBessel}
\end{equation}

\subsubsection*{The wave function squared}

Neglecting the phase $e^{iz\mathbf{p}_{\perp}\cdot\mathbf{r}}$, the
square of the absolute value of the wave function squared, summed
over all outgoing quantum numbers, and averaged over the polarization
and color charge of the incoming gluon is:
\begin{equation}
\begin{aligned}\left|\phi_{T}^{g\rightarrow q\bar{q}}\left(p^{+},z,\mathbf{r},\mathbf{r}'\right)\right|^{2} & =\frac{1}{2}\frac{1}{N_{c}^{2}-1}\sum_{ab}\mathrm{Tr}\left(t^{a}t^{b}\right)\sum_{\lambda\alpha\beta}\phi_{\alpha\beta}^{T\lambda}\left(p^{+},z,\mathbf{r}\right)\phi_{\alpha\beta}^{T\lambda*}\left(p^{+},z,\mathbf{r}'\right),\\
 & =\frac{2\pi^{2}}{p^{+}}\Biggl(\epsilon_{f}^{2}K_{1}\left(\epsilon_{f}r\right)K_{1}\left(\epsilon_{f}r'\right)\frac{\mathbf{r}\cdot\mathbf{r}'}{rr'}\left(z^{2}+\left(1-z\right)^{2}\right)\\
 & +m^{2}K_{0}\left(\epsilon_{f}r'\right)K_{0}\left(\epsilon_{f}r\right)\Biggr),
\end{aligned}
\label{eq:wavegqqsquared}
\end{equation}
where we used the easy to prove identity:
\begin{equation}
\begin{aligned}\frac{\mathbf{r}\cdot\boldsymbol{\epsilon}_{\perp}^{1}}{r}\frac{\mathbf{r}'\cdot\boldsymbol{\epsilon}_{\perp}^{1*}}{r'}+\frac{\mathbf{r}\cdot\boldsymbol{\epsilon}_{\perp}^{2}}{r}\frac{\mathbf{r'}\cdot\boldsymbol{\epsilon}_{\perp}^{2*}}{r'} & =\frac{\mathbf{r}\cdot\mathbf{r}'}{rr'},\end{aligned}
\end{equation}
and where we added the color matrices $t^{a}$. In the limit $m\rightarrow0$
and $Q^{2}\rightarrow0$, and therefore $\epsilon_{f}\rightarrow0$,
we have that:
\begin{equation}
\begin{aligned}\lim_{\epsilon_{f}\rightarrow0}\epsilon_{f}^{2}K_{1}\left(\epsilon_{f}r\right)K_{1}\left(\epsilon_{f}r'\right) & =\frac{1}{rr'},\\
\lim_{\epsilon_{f}\rightarrow0}\epsilon_{f}^{2}K_{0}\left(\epsilon_{f}r\right)K_{0}\left(\epsilon_{f}r'\right) & =0,
\end{aligned}
\end{equation}
and as a result Eq. (\ref{eq:wavegqqsquared}) simplifies to:
\begin{equation}
\lim_{\epsilon_{f}\rightarrow0}\left|\phi_{T}^{g\rightarrow q\bar{q}}\left(p^{+},z,\mathbf{r},\mathbf{r}'\right)\right|^{2}=\frac{\left(2\pi\right)^{2}}{p^{+}}\frac{\mathbf{r}\cdot\mathbf{r}'}{r^{2}r'^{2}}\left(z^{2}+\left(1-z\right)^{2}\right).
\end{equation}
Likewise, for the longitudinal polarized gluon, again neglecting the
phase factor $e^{iz\mathbf{p_{\perp}}\cdot\mathbf{r}}$, we obtain:
\begin{equation}
\begin{aligned}\left|\phi_{L}^{g\rightarrow q\bar{q}}\left(p^{+},z,\mathbf{r},\mathbf{r}'\right)\right|^{2} & =\frac{1}{N_{c}^{2}-1}\sum_{ab}\mathrm{Tr}\left(t^{a}t^{b}\right)\sum_{\alpha\beta}\phi_{\alpha\beta}^{L}\left(p^{+},z,\mathbf{r}\right)\phi_{\alpha\beta}^{L*}\left(p^{+},z,\mathbf{r}'\right),\\
 & =\frac{2\pi^{2}}{p^{+}}4Q^{2}z^{2}\left(1-z\right)^{2}K_{0}\left(\epsilon_{f}r\right)K_{0}\left(\epsilon_{f}r'\right),
\end{aligned}
\end{equation}
which obviously vanishes in the limit $Q^{2}\to0$. 

The absolute value of the wave function squared in momentum space,
and in the limits $Q^{2}\rightarrow0$ and $m^{2}\to0$ (hence the
longitudinally polarized gluon disappears) yields:
\begin{equation}
\begin{aligned}\left|\psi_{T}^{g\rightarrow q\bar{q}}\left(p,k\right)\right|^{2} & =\frac{1}{2}\sum_{\lambda\alpha\beta}\psi_{\alpha\beta}^{\lambda\dagger}\left(p,k\right)\psi_{\alpha\beta}^{\lambda}\left(p,k\right)=\frac{1}{p^{+}}\frac{z^{2}+\left(1-z\right)^{2}}{k_{\perp}^{2}},\end{aligned}
\label{eq:gqqsquared}
\end{equation}
where we used that:
\begin{equation}
\left(\mathbf{k}_{\perp}\cdot\boldsymbol{\epsilon}_{\perp}^{1}\right)\left(\mathbf{k}_{\perp}\cdot\boldsymbol{\epsilon}_{\perp}^{1*}\right)=\left(\mathbf{k}_{\perp}\cdot\boldsymbol{\epsilon}_{\perp}^{2}\right)\left(\mathbf{k}_{\perp}\cdot\boldsymbol{\epsilon}_{\perp}^{2*}\right)=\frac{k_{\perp}^{2}}{2}.
\end{equation}
Note that we chose our definitions in such a way that the wavefunctions
for $g\to q\bar{q}$ and $\gamma\to q\bar{q}$ are the same, which
is why we had to add the trace over color matrices and the average
over gluon color explicitely when computing the wavefunction squared
in the $g\to q\bar{q}$ case. The wavefunction squared for $\gamma\to q\bar{q}$
is now simply obtained from Eq. (\ref{eq:wavegqqsquared}) by the
substitution:
\begin{equation}
\begin{aligned}\frac{1}{N_{c}^{2}-1}\sum_{ab}\mathrm{Tr}\left(t^{a}t^{b}\right)\to & \;\delta^{ij}\delta^{ji},\end{aligned}
\end{equation}
or equivalently:
\begin{equation}
\frac{1}{2}\to N_{c}.
\end{equation}

\section{The $q\rightarrow gq$ wave function \label{subsec:q->qg}}
\begin{center}
\begin{tikzpicture}[scale=2] 
\tikzset{photon/.style={semithick,decorate,decoration={snake}}, electron/.style={ postaction={decorate},decoration={markings,mark=at position .5 with {\arrow[]{latex}}}},	positron/.style={ postaction={decorate},decoration={markings,mark=at position .5 with {\arrow[]{latex reversed}}}},      	gluon/.style={decorate,decoration={coil,amplitude=4pt, segment length=5pt}}}

\draw[semithick,electron] (0.5,-1.5) node [left]{$p,\alpha$} --(1,-1.5);
\draw[semithick,gluon] (1,-1.5).. controls (1.2,-1.3) and (1.8,-1.1)  .. (2,-1.1) node [right]{$k,\lambda$};
\draw[semithick,electron] (1,-1.5).. controls (1.2,-1.7) and (1.8,-1.9)  .. (2,-1.9) node [right]{$p-k,\beta$};

\end{tikzpicture} 
\par\end{center}

The wave function of the dressed quark is given by:

\begin{equation}
\psi_{\alpha\beta}^{\lambda}\left(p,k\right)=\frac{1}{\sqrt{8\left(p-k\right)^{+}p^{+}k^{+}}}\frac{\bar{u}_{\beta}\left(p-k\right)\gamma_{\mu}\epsilon_{\lambda}^{\mu}\left(k\right)u_{\alpha}\left(p\right)}{\left(p-k\right)^{-}+k^{-}-p^{-}},
\end{equation}
where, using the notation $z=k^{+}/p^{+}$, the energy denominator
reads:
\begin{equation}
\begin{aligned}\frac{1}{\left(p-k\right)^{-}+k^{-}-p^{-}} & =\frac{2z\left(p^{+}-k^{+}\right)}{\left(\mathbf{k}_{\perp}-z\mathbf{p}_{\perp}\right)^{2}+z^{2}m^{2}}.\end{aligned}
\end{equation}

\subsubsection*{Spin up - spin up}

The numerator is given by:
\begin{equation}
\begin{aligned} & \bar{u}_{+}\left(p-k\right)\gamma_{\mu}\epsilon_{\lambda}^{\mu}\left(k\right)u_{+}\left(p\right)\\
 & =\frac{2p^{+}}{z\sqrt{p^{+}\left(p-k\right)^{+}}}\left(\left(1-z\right)\delta_{\lambda1}\left(\mathbf{k}_{\perp}-z\mathbf{p}_{\perp}\right)\cdot\boldsymbol{\epsilon}_{\perp}^{\left(1\right)}+\delta_{\lambda2}\left(\mathbf{k}_{\perp}-z\mathbf{p}_{\perp}\right)\cdot\boldsymbol{\epsilon}_{\perp}^{\left(2\right)}\right),
\end{aligned}
\end{equation}
and the wave function for the spin up - spin up case yields:

\begin{align}
\psi_{++}^{\lambda}\left(p,k\right) & =\sqrt{\frac{2}{k^{+}}}\frac{\left(\left(1-z\right)\delta_{\lambda1}+\delta_{\lambda2}\right)\left(\mathbf{k}_{\perp}-z\mathbf{p}_{\perp}\right)\cdot\boldsymbol{\epsilon}_{\perp}^{\lambda}}{\left(\mathbf{k}_{\perp}-z\mathbf{p}_{\perp}\right)^{2}+z^{2}m^{2}}.
\end{align}

\subsubsection*{Spin down - spin down}

The numerator gives:
\begin{equation}
\begin{aligned} & \bar{u}_{-}\left(q\right)\gamma_{\mu}\epsilon_{\lambda}^{\mu}\left(k\right)u_{-}\left(p\right)\\
 & =\frac{2p^{+}}{z\sqrt{p^{+}\left(p-k\right)^{+}}}\left(\delta_{\lambda1}\left(\mathbf{k}_{\perp}-z\mathbf{p}_{\perp}\right)\cdot\boldsymbol{\epsilon}_{\perp}^{\left(1\right)}+\left(1-z\right)\delta_{\lambda2}\left(\mathbf{k}_{\perp}-z\mathbf{p}_{\perp}\right)\cdot\boldsymbol{\epsilon}_{\perp}^{\left(2\right)}\right),
\end{aligned}
\end{equation}
and we obtain the following wave function:
\begin{align}
\psi_{--}^{\lambda}\left(p,k\right) & =\sqrt{\frac{2}{k^{+}}}\frac{\left(\delta_{\lambda1}+\left(1-z\right)\delta_{\lambda2}\right)\left(\mathbf{k}_{\perp}-z\mathbf{p}_{\perp}\right)\cdot\boldsymbol{\epsilon}_{\perp}^{\lambda}}{\left(\mathbf{k}_{\perp}-z\mathbf{p}_{\perp}\right)^{2}+z^{2}m^{2}}.
\end{align}

\subsubsection*{The spin-flip cases}

The wave function for the spin-flip cases is much simpler, as is apparent
from the expressions we find for the numerators:

\begin{equation}
\begin{aligned}\bar{u}_{+}\left(q\right)\gamma_{\mu}\epsilon_{\lambda}^{\mu}\left(k\right)u_{-}\left(p\right) & =\frac{\sqrt{2}k^{+}m\delta_{\lambda1}}{\sqrt{p^{+}\left(p-k\right)^{+}}},\\
\bar{u}_{-}\left(q\right)\gamma_{\mu}\epsilon_{\lambda}^{\mu}\left(k\right)u_{+}\left(p\right) & =-\frac{\sqrt{2}k^{+}m\delta_{\lambda2}}{\sqrt{p^{+}\left(p-k\right)^{+}}}.
\end{aligned}
\end{equation}
From this, we find the following results for the wave functions:
\begin{equation}
\begin{aligned}\psi_{+-}^{\lambda}\left(p,k\right) & =\frac{1}{\sqrt{k^{+}}}\frac{z^{2}m\delta_{\lambda1}}{\left(\mathbf{k}_{\perp}-z\mathbf{p}_{\perp}\right)^{2}+z^{2}m^{2}},\\
\psi_{-+}^{\lambda}\left(p,k\right) & =\frac{1}{\sqrt{k^{+}}}\frac{-z^{2}m\delta_{\lambda2}}{\left(\mathbf{k}_{\perp}-z\mathbf{p}_{\perp}\right)^{2}+z^{2}m^{2}}.
\end{aligned}
\end{equation}

\subsubsection*{Total result for the wave function}

Taking all these pieces together, we finally arrive at:
\begin{equation}
\begin{aligned}\psi_{\alpha\beta}^{\lambda}\left(p,k\right) & =\frac{1}{\sqrt{k^{+}}}\frac{1}{\left(\mathbf{k}_{\perp}-z\mathbf{p}_{\perp}\right)^{2}+z^{2}m^{2}}\\
 & \times\begin{cases}
\sqrt{2}\left(\mathbf{k}_{\perp}-z\mathbf{p}_{\perp}\right)\cdot\boldsymbol{\epsilon}_{\perp}^{1}\left[\delta_{\alpha-}\delta_{\beta-}+\left(1-z\right)\delta_{\alpha+}\delta_{\beta+}\right]+mz^{2}\delta_{\alpha+}\delta_{\beta-} & \lambda=1\\
\sqrt{2}\left(\mathbf{k}_{\perp}-z\mathbf{p}_{\perp}\right)\cdot\boldsymbol{\epsilon}_{\perp}^{2}\left[\delta_{\alpha+}\delta_{\beta+}+\left(1-z\right)\delta_{\alpha-}\delta_{\beta-}\right]-mz^{2}\delta_{\alpha-}\delta_{\beta+} & \lambda=2
\end{cases}.
\end{aligned}
\end{equation}
Averaging over the incoming spin, and summing over the outgoing spin
and polarization, we obtain in the massless limit:
\begin{equation}
\begin{aligned}\left|\psi_{T}^{q\rightarrow qg}\left(p,k\right)\right|^{2} & =\frac{1}{2}\sum_{\alpha\beta\lambda}\psi_{\alpha\beta}^{\lambda}\left(p,k\right)\psi_{\alpha\beta}^{\lambda*}\left(p,k\right)=\frac{1}{k^{+}}\frac{1+\left(1-z\right)^{2}}{k_{\perp}^{2}}.\end{aligned}
\end{equation}
In mixed Fourier-coordinate representation, again making use of the
identities in Eq. (\ref{eq:waveBessel}), one obtains the expression
for the wave function encountered in Refs. \protect\cite{Dominguez2011,Marquet2007}:
\begin{equation}
\begin{aligned}\phi_{\alpha\beta}^{\lambda}\left(p,z,\mathbf{r}\right) & \equiv\int\mathrm{d}^{2}\mathbf{k}_{\perp}e^{i\mathbf{k}_{\perp}\cdot\mathbf{r}}\psi_{\alpha\beta}^{\lambda}\left(p,k\right),\end{aligned}
\end{equation}
\begin{equation}
\begin{aligned} & \phi_{\alpha\beta}^{\lambda}\left(p,z,\mathbf{r}\right)=\frac{2\pi m}{\sqrt{k^{+}}}e^{iz\mathbf{p}_{\perp}\cdot\mathbf{r}}\\
 & \times\begin{cases}
iz\sqrt{2}K_{1}\left(mz\left|\mathbf{r}\right|\right)\frac{\mathbf{r}\cdot\boldsymbol{\epsilon}_{\perp}^{1}}{\left|\mathbf{r}\right|}\left[\delta_{\alpha-}\delta_{\beta-}+\left(1-z\right)\delta_{\alpha+}\delta_{\beta+}\right]+z^{2}K_{0}\left(mz\left|\mathbf{r}\right|\right)\delta_{\alpha+}\delta_{\beta-} & \lambda=1\\
iz\sqrt{2}K_{1}\left(mz\left|\mathbf{r}\right|\right)\frac{\mathbf{r}\cdot\boldsymbol{\epsilon}_{\perp}^{2}}{\left|\mathbf{r}\right|}\left[\delta_{\alpha+}\delta_{\beta+}+\left(1-z\right)\delta_{\alpha-}\delta_{\beta-}\right]-z^{2}K_{0}\left(mz\left|\mathbf{r}\right|\right)\delta_{\alpha-}\delta_{\beta+} & \lambda=2
\end{cases}.
\end{aligned}
\end{equation}

\section{Analytical calculation of the gluon TMDs in the MV model}

\subsection{Weizsäcker-Williams gluon TMDs $\mathcal{F}_{gg}^{\left(3\right)}$
and $\mathcal{H}_{gg}^{\left(3\right)}$}

Let us start with the calculation of the Weizsäcker-Williams gluon
TMD $\mathcal{F}_{gg}^{\left(3\right)}$, as well as its partner $\mathcal{H}_{gg}^{\left(3\right)}$
which corresponds to the linearly polarized gluons in the target.
From their definition in Eq. (\ref{eq:F123}), we see that the central
object is the quadrupole operator:
\begin{equation}
\left.\frac{\partial}{\partial x^{i}}\frac{\partial}{\partial y^{j}}\frac{1}{N_{c}}\mathrm{Tr}\Bigl\langle U\left(\mathbf{x}\right)U^{\dagger}\left(\mathbf{v}'\right)U\left(\mathbf{y}\right)U^{\dagger}\left(\mathbf{v}\right)\Bigr\rangle_{x}\right|_{\mathbf{x}=\mathbf{v},\,\mathbf{y}=\mathbf{v}'}.\label{eq:quadrupfirststep}
\end{equation}
This operator was already calculated in McLerran-Venugopalan model
in Ref. \protect\cite{Dominguez2011}:
\begin{equation}
\begin{aligned} & \frac{1}{N_{c}}\mathrm{Tr}\Bigl\langle U\left(\mathbf{x}\right)U^{\dagger}\left(\mathbf{v}'\right)U\left(\mathbf{y}\right)U^{\dagger}\left(\mathbf{v}\right)\Bigr\rangle_{x}\\
 & =e^{-\frac{C_{F}}{2}\left(\Gamma\left(\mathbf{x}-\mathbf{v}\right)+\Gamma\left(\mathbf{y}-\mathbf{v}'\right)\right)}e^{-\frac{N_{c}}{4}\mu^{2}F\left(\mathbf{x},\mathbf{y};\mathbf{v},\mathbf{v}'\right)+\frac{1}{2N_{c}}\mu^{2}F\left(\mathbf{x},\mathbf{v};\mathbf{y},\mathbf{v}'\right)}\\
 & \times\Biggl[\left(\frac{\sqrt{\Delta}+F\left(\mathbf{x},\mathbf{y};\mathbf{v},\mathbf{v}'\right)}{2\sqrt{\Delta}}-\frac{F\left(\mathbf{x},\mathbf{v};\mathbf{y},\mathbf{v}'\right)}{\sqrt{\Delta}}\right)e^{\frac{N_{c}}{4}\mu^{2}\sqrt{\Delta}}\\
 & +\left(\frac{\sqrt{\Delta}-F\left(\mathbf{x},\mathbf{y};\mathbf{v},\mathbf{v}'\right)}{2\sqrt{\Delta}}+\frac{F\left(\mathbf{x},\mathbf{v};\mathbf{y},\mathbf{v}'\right)}{\sqrt{\Delta}}\right)e^{-\frac{N_{c}}{4}\mu^{2}\sqrt{\Delta}}\Biggr],
\end{aligned}
\label{eq:S4finiteNc}
\end{equation}
where
\begin{equation}
\begin{aligned}F\left(\mathbf{x},\mathbf{y},\mathbf{v},\mathbf{w}\right) & \equiv L_{\mathbf{xv}}-L_{\mathbf{xw}}+L_{\mathbf{yw}}-L_{\mathbf{yv}},\end{aligned}
\label{eq:Fxyzw}
\end{equation}
and
\begin{equation}
\Delta\equiv F^{2}\left(\mathbf{x},\mathbf{y};\mathbf{\mathbf{v}},\mathbf{v}'\right)+\frac{4}{N_{c}^{2}}F\left(\mathbf{x},\mathbf{\mathbf{v}};\mathbf{y},\mathbf{v}'\right)F\left(\mathbf{x},\mathbf{v}';\mathbf{y},\mathbf{\mathbf{v}}\right).
\end{equation}
Eq. (\ref{eq:quadrupfirststep}) becomes after a lot of tedious algebra:
\begin{equation}
\begin{aligned} & \left.\frac{\partial}{\partial x^{i}}\frac{\partial}{\partial y^{j}}\frac{1}{N_{c}}\mathrm{Tr}\Bigl\langle U\left(\mathbf{x}\right)U^{\dagger}\left(\mathbf{v}'\right)U\left(\mathbf{y}\right)U^{\dagger}\left(\mathbf{v}\right)\Bigr\rangle_{x}\right|_{\mathbf{x}=\mathbf{v},\,\mathbf{y}=\mathbf{v}'}\\
 & =\frac{C_{F}}{N_{c}}\frac{1-e^{-\frac{N_{c}}{2}\Gamma\left(\mathbf{v}-\mathbf{v}'\right)}}{\Gamma\left(\mathbf{v}-\mathbf{v}'\right)}\frac{\partial}{\partial v^{i}}\frac{\partial}{\partial v'^{j}}\Gamma\left(\mathbf{v}-\mathbf{v}'\right),
\end{aligned}
\end{equation}
and therefore $\mathcal{F}_{gg}^{\left(3\right)}\left(x,q_{\perp}\right)$
can be written as:
\begin{equation}
\begin{aligned}\mathcal{F}_{gg}^{\left(3\right)}\left(x,q_{\perp}\right) & =-\frac{4}{g_{s}^{2}}C_{F}\delta_{ij}\int\frac{\mathrm{d}^{2}\mathbf{v}\mathrm{d}^{2}\mathbf{v}'}{\left(2\pi\right)^{3}}e^{-i\mathbf{q}_{\perp}\cdot\left(\mathbf{v}-\mathbf{v}'\right)}\frac{1-e^{-\frac{N_{c}}{2}\Gamma\left(\mathbf{v}-\mathbf{v}'\right)}}{\Gamma\left(\mathbf{v}-\mathbf{v}'\right)}\frac{\partial}{\partial v^{i}}\frac{\partial}{\partial v'^{j}}\Gamma\left(\mathbf{v}-\mathbf{v}'\right).\end{aligned}
\label{eq:F3Gamma}
\end{equation}
Using the identity:
\begin{equation}
\begin{aligned}\frac{\partial}{\partial v^{i}}\frac{\partial}{\partial v'^{j}}\Gamma\left(\mathbf{v}-\mathbf{v}'\right) & =-2g_{s}^{2}\mu_{A}\int\frac{\mathrm{d}^{2}\mathbf{k}_{\perp}}{\left(2\pi\right)^{2}}\frac{k_{\perp}^{i}k_{\perp}^{j}}{k_{\perp}^{4}}e^{i\mathbf{k}_{\perp}\cdot\left(\mathbf{v}-\mathbf{v}'\right)},\end{aligned}
\end{equation}
as well as the expression for the saturation scale in the MV model,
Eq. (\ref{eq:GammaQ}), we obtain:
\begin{equation}
\begin{aligned} & \left.\frac{\partial}{\partial x^{i}}\frac{\partial}{\partial y^{j}}\frac{1}{N_{c}}\mathrm{Tr}\Bigl\langle U\left(\mathbf{x}\right)U^{\dagger}\left(\mathbf{v}'\right)U\left(\mathbf{y}\right)U^{\dagger}\left(\mathbf{v}\right)\Bigr\rangle_{x}\right|_{\mathbf{x}=\mathbf{v},\,\mathbf{y}=\mathbf{v}'}\\
 & =g_{s}^{2}\mu_{A}\frac{2C_{F}}{\left(\mathbf{v}-\mathbf{v}'\right)^{2}Q_{sg}^{2}\left(\mathbf{v}-\mathbf{v}'\right)}\left(e^{-\frac{\left(\mathbf{v}-\mathbf{v}'\right)^{2}}{4}Q_{sg}^{2}\left(\mathbf{v}-\mathbf{v}'\right)}-1\right)\int\frac{\mathrm{d}^{2}\mathbf{k}_{\perp}}{\left(2\pi\right)^{2}}\frac{k_{\perp}^{i}k_{\perp}^{j}}{k_{\perp}^{4}}e^{i\mathbf{k}_{\perp}\cdot\left(\mathbf{v}-\mathbf{v}'\right)}.
\end{aligned}
\label{eq:doublederivquad}
\end{equation}
Combining Eqs. (\ref{eq:F123}), (\ref{eq:quadrupfirststep}), and
(\ref{eq:doublederivquad}), as well as using Eq. (\ref{eq:fourier2log}),
we obtain the following expression for the Weizsäcker-Williams distribution:
\begin{equation}
\begin{aligned}\mathcal{F}_{gg}^{\left(3\right)}\left(x,q_{\perp}\right) & =16\mu_{A}C_{F}N_{c}S_{\perp}\int\frac{\mathrm{d}^{2}\mathbf{r}}{\left(2\pi\right)^{3}}e^{-i\mathbf{q}_{\perp}\cdot\mathbf{r}}\frac{1-e^{-\frac{r^{2}}{4}Q_{sg}^{2}\left(r\right)}}{\mathbf{r}^{2}Q_{sg}^{2}\left(r\right)}\int\frac{\mathrm{d}^{2}\mathbf{k}_{\perp}}{\left(2\pi\right)^{2}}\frac{e^{i\mathbf{k}_{\perp}\cdot\mathbf{r}}}{k_{\perp}^{2}},\\
 & =\frac{2C_{F}S_{\perp}}{\alpha_{s}\pi^{2}}\int\frac{\mathrm{d}^{2}\mathbf{r}}{\left(2\pi\right)^{2}}e^{-i\mathbf{q}_{\perp}\cdot\mathbf{r}}\frac{1}{r^{2}}\left(1-e^{-\frac{r^{2}}{4}Q_{g}^{2}\left(r\right)}\right),
\end{aligned}
\label{eq:WWfiniteNc}
\end{equation}
in accordance with what we found in Eq. (\ref{eq:WWMV}) using a slightly
different approach.

The linearly polarized partner of $\mathcal{F}_{gg}^{\left(3\right)}$:
$\mathcal{H}_{gg}^{\left(3\right)}$ (see Eq. (\ref{eq:H123})), can
also be calculated, combining Eq. (\ref{eq:H123}) with Eqs. (\ref{eq:quadrupfirststep})
and (\ref{eq:doublederivquad}):
\begin{equation}
\begin{aligned} & \mathcal{H}_{gg}^{\left(3\right)}\left(x,q_{\perp}\right)\\
 & =\frac{4C_{F}}{g_{s}^{2}}\int\frac{\mathrm{d}^{2}\mathbf{v}\mathrm{d}^{2}\mathbf{v}'}{\left(2\pi\right)^{3}}e^{-i\mathbf{q}_{\perp}\cdot\left(\mathbf{v}-\mathbf{v}'\right)}\frac{e^{-\frac{N_{c}}{2}\Gamma\left(\mathbf{v}-\mathbf{v}'\right)}-1}{\Gamma\left(\mathbf{v}-\mathbf{v}'\right)}\left(\frac{2q_{\perp}^{i}q_{\perp}^{j}}{q_{\perp}^{2}}-\delta^{ij}\right)\frac{\partial}{\partial v^{i}}\frac{\partial}{\partial v'^{j}}\Gamma\left(\mathbf{v}-\mathbf{v}'\right),\\
 & =-2g_{s}^{2}\mu_{A}\frac{8C_{F}}{g_{s}^{2}}S_{\perp}N_{c}\int\frac{\mathrm{d}^{2}\mathbf{v}\mathrm{d}^{2}\mathbf{v}'}{\left(2\pi\right)^{3}}e^{-i\mathbf{q}_{\perp}\cdot\mathbf{r}}\\
 & \frac{e^{-\frac{r^{2}}{4}Q_{sg}^{2}\left(r^{2}\right)}-1}{r^{2}Q_{g}^{2}\left(r^{2}\right)}\left(\frac{2q_{\perp}^{i}q_{\perp}^{j}}{q_{\perp}^{2}}-\delta^{ij}\right)\int\frac{\mathrm{d}^{2}\mathbf{k}_{\perp}}{\left(2\pi\right)^{2}}\frac{k_{\perp}^{i}k_{\perp}^{j}}{k_{\perp}^{4}}e^{i\mathbf{k}_{\perp}\cdot\left(\mathbf{v}-\mathbf{v}'\right)},\\
 & =\frac{2N_{c}C_{F}}{\pi^{3}}\mu_{A}S_{\perp}\int\mathrm{d}r\frac{1}{rQ_{sg}^{2}\left(r\right)}\left(1-e^{-\frac{r^{2}}{4}Q_{sg}^{2}\left(r\right)}\right)\\
 & \times\int\mathrm{d}\phi e^{-iq_{\perp}\cdot r\cos\alpha}\int\frac{\mathrm{d}^{2}\mathbf{k}_{\perp}}{\left(2\pi\right)^{2}}\frac{1}{k_{\perp}^{4}}e^{i\mathbf{k}_{\perp}\cdot\mathbf{r}}\left(\frac{2\left(\mathbf{k}_{\perp}\cdot\mathbf{q}_{\perp}\right)^{2}}{q_{\perp}^{2}}-k_{\perp}^{2}\right).
\end{aligned}
\end{equation}
Using the intermediate result:
\begin{equation}
\begin{aligned}\int\frac{\mathrm{d}^{2}\mathbf{k}_{\perp}}{\left(2\pi\right)^{2}}\frac{1}{k_{\perp}^{4}}e^{i\mathbf{k}_{\perp}\cdot\mathbf{r}}\left(\frac{2\left(\mathbf{k}_{\perp}\cdot\mathbf{q}_{\perp}\right)^{2}}{q_{\perp}^{2}}-k_{\perp}^{2}\right) & =\int\frac{\mathrm{d}k_{\perp}\mathrm{d}\theta}{\left(2\pi\right)^{2}}\frac{1}{k_{\perp}}e^{i\mathbf{k}_{\perp}\cdot\mathbf{r}\cos\theta}\left(2\cos^{2}\left(\theta+\alpha\right)-1\right),\\
 & =-\frac{1}{2\pi}\cos\left(2\alpha\right)\int_{0}^{\infty}\frac{\mathrm{d}k_{\perp}}{k_{\perp}}J_{2}\left(k_{\perp}r\right),\\
 & =-\frac{\cos\left(2\alpha\right)}{4\pi},
\end{aligned}
\label{eq:dxidyjGamma}
\end{equation}
where $\alpha=\widehat{\mathbf{q}_{\perp}\mathbf{r}}$, as well as
employing the integral representation of the Bessel function of the
first kind:
\begin{equation}
\begin{aligned}\int_{0}^{2\pi}\mathrm{d}\phi\,e^{-iq_{\perp}r\cos\phi}\cos2\phi & =-2\pi J_{2}\left(q_{\perp}r\right),\end{aligned}
\label{eq:J2}
\end{equation}
we obtain:
\begin{equation}
\begin{aligned}\mathcal{H}_{gg}^{\left(3\right)}\left(x,q_{\perp}\right) & =\frac{C_{F}S_{\perp}}{\alpha_{s}\pi^{3}}\int\mathrm{d}r\frac{J_{2}\left(q_{\perp}r\right)}{r\ln\frac{1}{r^{2}\Lambda^{2}}}\left(1-e^{-\frac{r^{2}}{4}Q_{sg}^{2}\left(r\right)}\right).\end{aligned}
\label{eq:H3finiteNc}
\end{equation}

\subsection{Distributions built from dipoles:\textmd{\normalsize{} $\mathcal{F}_{gg}^{\left(1\right)}$,
$\mathcal{H}_{gg}^{\left(1\right)}$, $\mathcal{F}_{gg}^{\left(2\right)}$
and $\mathcal{H}_{gg}^{\left(2\right)}$}}

Just like the correlator of four Wilson lines in the previous section,
the correlator of the product of two dipoles is also known in the
MV model (see Ref. \protect\cite{Dominguez2009}):

\begin{equation}
\begin{aligned} & \frac{1}{N_{c}^{2}}\Bigl\langle\mathrm{Tr}\left(U\left(\mathbf{x}\right)U^{\dagger}\left(\mathbf{y}\right)\right)\mathrm{Tr}\left(U\left(\mathbf{v}'\right)U^{\dagger}\left(\mathbf{v}\right)\right)\Bigr\rangle_{x}\\
 & =e^{-\frac{C_{F}}{2}\left(\Gamma\left(\mathbf{x}-\mathbf{y}\right)+\Gamma\left(\mathbf{v}'-\mathbf{v}\right)\right)}e^{-\frac{N_{c}}{4}\mu^{2}F\left(\mathbf{x},\mathbf{v}';\mathbf{y},\mathbf{v}\right)+\frac{1}{2N_{c}}\mu^{2}F\left(\mathbf{x},\mathbf{y};\mathbf{v}',\mathbf{v}\right)}\\
 & \times\Biggl[\left(\frac{F\left(\mathbf{x},\mathbf{v}';\mathbf{y},\mathbf{v}\right)+\sqrt{D}}{2\sqrt{D}}-\frac{F\left(\mathbf{x},\mathbf{y};\mathbf{v}',\mathbf{v}\right)}{N_{c}^{2}\sqrt{D}}\right)e^{\frac{N_{c}}{4}\mu^{2}\sqrt{D}}\\
 & -\left(\frac{F\left(\mathbf{x},\mathbf{v}';\mathbf{y},\mathbf{v}\right)-\sqrt{D}}{2\sqrt{D}}-\frac{F\left(\mathbf{x},\mathbf{y};\mathbf{v}',\mathbf{v}\right)}{N_{c}^{2}\sqrt{D}}\right)e^{-\frac{N_{c}}{4}\mu^{2}\sqrt{D}}\Biggr],
\end{aligned}
\label{eq:DipDipfiniteNc}
\end{equation}
where
\begin{equation}
D\equiv F^{2}\left(\mathbf{x},\mathbf{v}';\mathbf{y},\mathbf{v}\right)+\frac{4}{N_{c}^{2}}F\left(\mathbf{x},\mathbf{\mathbf{y}};\mathbf{v}',\mathbf{v}\right)F\left(\mathbf{x},\mathbf{v};\mathbf{v}',\mathbf{y}\right),
\end{equation}
and where $F$ is defined as in Eq. (\ref{eq:Fxyzw}). We have that:
\begin{equation}
\begin{aligned} & \frac{1}{N_{c}^{2}}\left.\frac{\partial}{\partial x^{i}}\frac{\partial}{\partial y^{j}}\Bigl\langle\mathrm{Tr}\left(U\left(\mathbf{x}\right)U^{\dagger}\left(\mathbf{y}\right)\right)\mathrm{Tr}\left(U\left(\mathbf{v}'\right)U^{\dagger}\left(\mathbf{v}\right)\right)\Bigr\rangle_{x}\right|_{\mathbf{x}=\mathbf{v},\,\mathbf{y}=\mathbf{v}'}\\
 & =\frac{C_{F}}{8N_{c}^{3}}\frac{e^{-\frac{N_{c}}{2}\Gamma\left(\mathbf{v}-\mathbf{v}'\right)}}{\Gamma\left(\mathbf{v}-\mathbf{v}'\right)}\Biggl[16\left(1-e^{\frac{N_{c}}{2}\Gamma\left(\mathbf{v}-\mathbf{v}'\right)}\right)\frac{\partial}{\partial x^{i}}\frac{\partial}{\partial y^{j}}\Gamma\left(\mathbf{x}-\mathbf{y}\right)\\
 & +\Gamma\left(\mathbf{v}-\mathbf{v}'\right)\Biggl(N_{c}^{4}\frac{\partial}{\partial x^{i}}\Gamma\left(\mathbf{x}-\mathbf{v}'\right)\frac{\partial}{\partial y^{j}}\Gamma\left(\mathbf{v}-\mathbf{y}\right)-4N_{c}\left(N_{c}^{2}-2\right)\frac{\partial}{\partial x^{i}}\frac{\partial}{\partial y^{j}}\Gamma\left(\mathbf{x}-\mathbf{y}\right)\Biggr)\Biggr].
\end{aligned}
\label{eq:deriv2dipdip}
\end{equation}
The gluon TMD $\mathcal{F}_{gg}^{\left(1\right)}\left(x,q_{\perp}\right)$
(see Eq. (\ref{eq:F123})) then becomes:
\begin{equation}
\begin{aligned}\mathcal{F}_{gg}^{\left(1\right)}\left(x,q_{\perp}\right) & =\frac{S_{\perp}}{\alpha_{s}}\frac{C_{F}}{N_{c}^{2}}\frac{1}{16\pi^{2}}\int\frac{\mathrm{d}^{2}\mathbf{r}}{\left(2\pi\right)^{2}}e^{-i\mathbf{q}_{\perp}\cdot\mathbf{r}}\frac{e^{-\frac{N_{c}}{2}\Gamma\left(\mathbf{r}\right)}}{\Gamma\left(\mathbf{r}\right)}\\
 & \Biggl[-32\mu_{A}\alpha_{s}\left(1-e^{\frac{N_{c}}{2}\Gamma\left(\mathbf{r}\right)}\right)\frac{1}{\pi}\int\mathrm{d}^{2}\mathbf{k}_{\perp}\frac{e^{i\mathbf{k}_{\perp}\cdot\mathbf{r}}}{k_{\perp}^{2}}\\
 & +\Gamma\left(\mathbf{r}\right)\Biggl(4\alpha_{s}^{2}N_{c}^{4}\mu_{A}^{2}\frac{1}{\pi^{2}}\int\mathrm{d}^{2}\mathbf{k}_{\perp}\int\mathrm{d}^{2}\mathbf{l}{}_{\perp}\frac{\mathbf{k}_{\perp}\cdot\mathbf{l}_{\perp}}{l_{\perp}^{4}k_{\perp}^{4}}e^{i\left(\mathbf{k}_{\perp}+\mathbf{l}_{\perp}\right)\cdot\mathbf{\mathbf{r}}}\\
 & +8\alpha_{s}N_{c}\left(N_{c}^{2}-2\right)\mu_{A}\frac{1}{\pi}\int\mathrm{d}^{2}\mathbf{k}_{\perp}\frac{e^{i\mathbf{k}_{\perp}\cdot\mathbf{r}}}{k_{\perp}^{2}}\Biggr)\Biggr],
\end{aligned}
\label{eq:F1finiteNctssstap}
\end{equation}
where we used the following intermediate results for the derivatives
of $\Gamma\left(\mathbf{x}-\mathbf{y}\right)$:

\begin{equation}
\begin{aligned}\left.\frac{\partial}{\partial x^{i}}\frac{\partial}{\partial y^{i}}\Gamma\left(\mathbf{x}-\mathbf{y}\right)\right|_{\mathbf{x}=\mathbf{v},\mathbf{y}=\mathbf{v}'} & =-2\mu_{A}g_{s}^{2}\int\frac{\mathrm{d}^{2}\mathbf{k}_{\perp}}{\left(2\pi\right)^{2}}\frac{e^{i\mathbf{k}_{\perp}\cdot\left(\mathbf{v}-\mathbf{v}'\right)}}{k_{\perp}^{2}},\\
\left.\frac{\partial}{\partial x^{i}}\Gamma\left(\mathbf{x}-\mathbf{v}'\right)\frac{\partial}{\partial y^{i}}\Gamma\left(\mathbf{v}-\mathbf{y}\right)\right|_{\mathbf{x}=\mathbf{v},\mathbf{y}=\mathbf{v}'} & =\left(2\mu_{A}g_{s}^{2}\right)^{2}\int\frac{\mathrm{d}^{2}\mathbf{k}_{\perp}}{\left(2\pi\right)^{2}}\int\frac{\mathrm{d}^{2}\mathbf{l}{}_{\perp}}{\left(2\pi\right)^{2}}\frac{\mathbf{k}_{\perp}\cdot\mathbf{l}_{\perp}}{l_{\perp}^{4}k_{\perp}^{4}}e^{i\left(\mathbf{k}_{\perp}+\mathbf{l}_{\perp}\right)\mathbf{\left(\mathbf{v}-\mathbf{v}'\right)}}.
\end{aligned}
\label{eq:derivGamma}
\end{equation}
Finally, using Eq. (\ref{eq:fourier2log}), as well as:
\begin{equation}
\begin{aligned}\int\mathrm{d}^{2}\mathbf{k}_{\perp}\int\mathrm{d}^{2}\mathbf{l}{}_{\perp}\frac{\mathbf{k}_{\perp}\cdot\mathbf{l}_{\perp}}{l_{\perp}^{4}k_{\perp}^{4}}e^{i\left(\mathbf{k}_{\perp}+\mathbf{l}_{\perp}\right)\cdot\mathbf{\mathbf{r}}} & =\int\frac{\mathrm{d}k_{\perp}\mathrm{d}\theta}{k_{\perp}^{2}}\int\frac{\mathrm{d}l{}_{\perp}\mathrm{d}\phi}{l_{\perp}^{2}}\cos\left(\phi-\theta\right)e^{ik_{\perp}r\cos\left(\theta\right)}e^{il_{\perp}r\mathbf{\cos\left(\phi\right)}},\\
 & =-4\pi^{2}\int_{\Lambda}^{\infty}\frac{\mathrm{d}k_{\perp}}{k_{\perp}^{2}}\int_{\Lambda}^{\infty}\frac{\mathrm{d}l{}_{\perp}}{l_{\perp}^{2}}J_{1}\left(k_{\perp}r\right)J_{1}\left(l_{\perp}r\right),\\
 & =-\frac{1}{4}\pi^{2}r^{2}\left(1-2\gamma_{E}+\ln4+\ln\frac{1}{r^{2}\Lambda^{2}}\right)^{2},
\end{aligned}
\label{eq:helpint1}
\end{equation}
Eq. (\ref{eq:F1finiteNctssstap}) becomes:
\begin{equation}
\begin{aligned}\mathcal{F}_{gg}^{\left(1\right)}\left(x,q_{\perp}\right) & =\frac{S_{\perp}}{\alpha_{s}}\frac{C_{F}}{N_{c}^{2}}\frac{1}{32\pi^{3}}\int\mathrm{d}r\frac{J_{0}\left(q_{\perp}r\right)}{r}e^{-\frac{N_{c}}{2}\Gamma\left(r\right)}\Biggl[64\left(e^{\frac{N_{c}}{2}\Gamma\left(r\right)}-1\right)\\
 & -\alpha_{s}^{2}N_{c}^{4}\mu_{A}^{2}r^{4}\left(1-2\gamma_{E}+\ln4+\ln\frac{1}{r^{2}\Lambda^{2}}\right)^{2}+8\alpha_{s}N_{c}\left(N_{c}^{2}-2\right)\mu_{A}r^{2}\ln\frac{1}{r^{2}\Lambda^{2}}\Biggr)\Biggr].
\end{aligned}
\label{eq:F1finiteNc}
\end{equation}
As usual, the computation of $\mathcal{H}_{gg}^{\left(1\right)}\left(x,q_{\perp}\right)$
requires a bit more effort. Combining Eq. (\ref{eq:H123}) with Eq.
(\ref{eq:deriv2dipdip}), we find:
\begin{equation}
\begin{aligned} & \mathcal{H}_{gg}^{\left(1\right)}\left(x,q_{\perp}\right)=\left(\frac{2q_{\perp}^{i}q_{\perp}^{j}}{q_{\perp}^{2}}-\delta^{ij}\right)\frac{4N_{c}}{g_{s}^{2}}\int\frac{\mathrm{d}^{2}\mathbf{v}\mathrm{d}^{2}\mathbf{v}'}{\left(2\pi\right)^{3}}e^{-i\mathbf{q}_{\perp}\cdot\left(\mathbf{v}-\mathbf{v}'\right)}\\
 & \frac{C_{F}}{8N_{c}^{3}}\frac{e^{-\frac{N_{c}}{2}\Gamma\left(\mathbf{v}-\mathbf{v}'\right)}}{\Gamma\left(\mathbf{v}-\mathbf{v}'\right)}\Biggl[16\left(1-e^{\frac{N_{c}}{2}\Gamma\left(\mathbf{v}-\mathbf{v}'\right)}\right)\frac{\partial}{\partial x^{i}}\frac{\partial}{\partial y^{j}}\Gamma\left(\mathbf{x}-\mathbf{y}\right)\\
 & +\Gamma\left(\mathbf{v}-\mathbf{v}'\right)\Biggl(N_{c}^{4}\frac{\partial}{\partial x^{i}}\Gamma\left(\mathbf{x}-\mathbf{v}'\right)\frac{\partial}{\partial y^{j}}\Gamma\left(\mathbf{v}-\mathbf{y}\right)-4N_{c}\left(N_{c}^{2}-2\right)\frac{\partial}{\partial x^{i}}\frac{\partial}{\partial y^{j}}\Gamma\left(\mathbf{x}-\mathbf{y}\right)\Biggr)\Biggr].
\end{aligned}
\end{equation}
With the help of the following results:
\begin{equation}
\begin{aligned} & \left(\frac{2q_{\perp}^{i}q_{\perp}^{j}}{q_{\perp}^{2}}-\delta^{ij}\right)\frac{\partial}{\partial x^{i}}\frac{\partial}{\partial y^{j}}\Gamma\left(\mathbf{x}-\mathbf{y}\right)\\
 & =-2\mu_{A}g_{s}^{2}\int\frac{\mathrm{d}^{2}\mathbf{k}_{\perp}}{\left(2\pi\right)^{2}}\frac{1}{k_{\perp}^{2}}\left(\frac{2\left(\mathbf{q}_{\perp}\cdot\mathbf{k}_{\perp}\right)^{2}}{q_{\perp}^{2}k_{\perp}^{2}}-1\right)e^{i\mathbf{k}_{\perp}\left(\mathbf{v}-\mathbf{v}'\right)}\\
 & \times\left(\frac{2q_{\perp}^{i}q_{\perp}^{j}}{q_{\perp}^{2}}-\delta^{ij}\right)\frac{\partial}{\partial x^{i}}\Gamma\left(\mathbf{x}-\mathbf{v}'\right)\frac{\partial}{\partial y^{j}}\Gamma\left(\mathbf{v}-\mathbf{y}\right),\\
 & =\left(2\mu_{A}g_{s}^{2}\right)^{2}\int\frac{\mathrm{d}^{2}\mathbf{k}_{\perp}}{\left(2\pi\right)^{2}}\int\frac{\mathrm{d}^{2}\mathbf{l}_{\perp}}{\left(2\pi\right)^{2}}\frac{1}{k_{\perp}^{4}l_{\perp}^{4}}\left(\frac{2\left(\mathbf{q}_{\perp}\cdot\mathbf{k}_{\perp}\right)\left(\mathbf{q}_{\perp}\cdot\mathbf{l}_{\perp}\right)}{q_{\perp}^{2}}-\mathbf{k}_{\perp}\cdot\mathbf{l}_{\perp}\right)e^{i\left(\mathbf{k}_{\perp}+\mathbf{l}_{\perp}\right)\mathbf{\left(\mathbf{v}-\mathbf{v}'\right)}},
\end{aligned}
\label{eq:H12intermediates}
\end{equation}
which can be evaluated further with the help of Eq. (\ref{eq:dxidyjGamma}),
and with:
\begin{equation}
\begin{aligned} & \int\mathrm{d}^{2}\mathbf{k}_{\perp}\int\mathrm{d}^{2}\mathbf{l}{}_{\perp}\frac{1}{k_{\perp}^{4}l_{\perp}^{4}}\left(\frac{2\left(\mathbf{q}_{\perp}\cdot\mathbf{k}_{\perp}\right)\left(\mathbf{q}_{\perp}\cdot\mathbf{l}_{\perp}\right)}{q_{\perp}^{2}}-\mathbf{k}_{\perp}\cdot\mathbf{l}_{\perp}\right)e^{i\left(\mathbf{k}_{\perp}+\mathbf{l}_{\perp}\right)\cdot\mathbf{\mathbf{r}}}\\
 & =\int\frac{\mathrm{d}k_{\perp}\mathrm{d}\theta}{k_{\perp}^{2}}\int\frac{\mathrm{d}l{}_{\perp}\mathrm{d}\phi}{l_{\perp}^{2}}\left(2\cos\left(\theta+\alpha\right)\cos\left(\phi+\alpha\right)-\cos\left(\theta-\phi\right)\right)e^{ik_{\perp}r\cos\left(\theta\right)}e^{il_{\perp}r\mathbf{\cos\left(\phi\right)}},\\
 & =-4\pi^{2}\int_{\Lambda}^{\infty}\frac{\mathrm{d}k_{\perp}}{k_{\perp}^{2}}\int_{\Lambda}^{\infty}\frac{\mathrm{d}l{}_{\perp}}{l_{\perp}^{2}}J_{1}\left(k_{\perp}r\right)J_{1}\left(l_{\perp}r\right)\cos\left(2\alpha\right),\\
 & =-\frac{1}{4}\pi^{2}r^{2}\cos\left(2\alpha\right)\left(1-2\gamma_{E}+\ln4+\ln\frac{1}{r^{2}\Lambda^{2}}\right)^{2},
\end{aligned}
\label{eq:H12integral}
\end{equation}
we finally obtain:
\begin{equation}
\begin{aligned}\mathcal{H}_{gg}^{\left(1\right)}\left(x,q_{\perp}\right) & =\frac{S_{\perp}}{\alpha_{s}}\frac{C_{F}}{N_{c}^{2}}\frac{1}{32\pi^{3}}\int\mathrm{d}r\frac{J_{2}\left(q_{\perp}r\right)}{r}e^{-\frac{N_{c}}{2}\Gamma\left(r\right)}\Biggl[64\frac{1}{\ln\frac{1}{r^{2}\Lambda^{2}}}\left(e^{\frac{N_{c}}{2}\Gamma\left(r\right)}-1\right)\\
 & +\alpha_{s}^{2}N_{c}^{4}\mu_{A}^{2}r^{4}\left(1-2\gamma_{E}+\ln4+\ln\frac{1}{r^{2}\Lambda^{2}}\right)^{2}+8\alpha_{s}\mu_{A}N_{c}\left(N_{c}^{2}-2\right)r^{2}\Biggr].
\end{aligned}
\label{eq:H1finiteNc}
\end{equation}

We can follow the same procedure to calculate $\mathcal{F}_{gg}^{\left(2\right)}$
and $\mathcal{H}_{gg}^{\left(2\right)}$. Indeed, from Eq. (\ref{eq:DipDipfiniteNc})
we obtain:
\begin{equation}
\begin{aligned} & \frac{1}{N_{c}^{2}}\frac{\partial^{2}}{\partial x^{i}\partial y^{j}}\left.\mathrm{Re}\Bigl\langle\mathrm{Tr}\left(U\left(\mathbf{x}\right)U^{\dagger}\left(\mathbf{v}'\right)\right)\mathrm{Tr}\left(U\left(\mathbf{y}\right)U^{\dagger}\left(\mathbf{v}\right)\right)\Bigr\rangle_{x}\right|_{\mathbf{x}=\mathbf{v},\mathbf{y}=\mathbf{v}'}\\
 & =\frac{C_{F}}{8N_{c}^{3}}\frac{e^{-\frac{N_{c}}{2}\Gamma\left(\mathbf{v}-\mathbf{v}'\right)}}{\Gamma\left(\mathbf{v}-\mathbf{v}'\right)}\Biggl[16\left(e^{\frac{N_{c}}{2}\Gamma\left(\mathbf{v}-\mathbf{v}'\right)}-1\right)\frac{\partial^{2}}{\partial v^{i}\partial v'^{j}}\Gamma\left(\mathbf{v}-\mathbf{v}'\right)\\
 & +\Gamma\left(\mathbf{v}-\mathbf{v}'\right)\left(N_{c}^{4}\left.\frac{\partial}{\partial x^{i}}\Gamma\left(\mathbf{x}-\mathbf{v}'\right)\frac{\partial}{\partial y^{j}}\Gamma\left(\mathbf{v}-\mathbf{y}\right)\right|_{\mathbf{x}=\mathbf{v},\mathbf{y}=\mathbf{v}'}-8N_{c}\frac{\partial^{2}}{\partial v^{i}\partial v'^{j}}\Gamma\left(\mathbf{v}-\mathbf{v}'\right)\right)\Biggr].
\end{aligned}
\label{eq:dipdipderiv2}
\end{equation}
Plugging this into the definition of $\mathcal{F}_{gg}^{\left(2\right)}$,
Eq. (\ref{eq:F123}), and using the intermediate results Eqs. (\ref{eq:derivGamma}),
(\ref{eq:helpint1}) and (\ref{eq:fourier2log}), we end up with:
\begin{equation}
\begin{aligned}\mathcal{F}_{gg}^{\left(2\right)}\left(x,q_{\perp}\right) & =\frac{S_{\perp}}{\alpha_{s}}\frac{C_{F}}{N_{c}^{2}}\frac{1}{32\pi^{3}}\int\mathrm{d}r\frac{J_{0}\left(q_{\perp}r\right)}{r}e^{-\frac{N_{c}}{2}\Gamma\left(r\right)}\Biggl[64\left(e^{\frac{N_{c}}{2}\Gamma\left(r\right)}-1\right)\\
 & +N_{c}^{4}\mu_{A}^{2}\alpha_{s}^{2}r^{4}\left(1-2\gamma_{E}+\ln4+\ln\frac{1}{r^{2}\Lambda^{2}}\right)^{2}-16\alpha_{s}N_{c}\mu_{A}r^{2}\ln\frac{1}{r^{2}\Lambda^{2}}\Biggr].
\end{aligned}
\label{eq:F2finiteNc}
\end{equation}
Likewise, combining Eq. (\ref{eq:H123}) with Eqs. (\ref{eq:dipdipderiv2}),
(\ref{eq:H12intermediates}), (\ref{eq:H12integral}), (\ref{eq:dxidyjGamma})
and (\ref{eq:J2}), we obtain the following expression for $\mathcal{H}_{gg}^{\left(2\right)}$:
\begin{equation}
\begin{aligned} & \mathcal{H}_{gg}^{\left(2\right)}\left(x,q_{\perp}\right)=\frac{S_{\perp}}{\alpha_{s}}\frac{C_{F}}{N_{c}^{2}}\frac{1}{32\pi^{3}}\int\mathrm{d}r\frac{J_{2}\left(q_{\perp}r\right)}{r}e^{-\frac{N_{c}}{2}\Gamma\left(r\right)}\\
 & \Biggl[64\left(e^{\frac{N_{c}}{2}\Gamma\left(r\right)}-1\right)\frac{1}{\ln\frac{1}{r^{2}\Lambda^{2}}}-\alpha_{s}^{2}N_{c}^{4}\mu_{A}^{2}r^{4}\left(1-2\gamma_{E}+\ln4+\ln\frac{1}{r^{2}\Lambda^{2}}\right)^{2}-16\alpha_{s}N_{c}\mu_{A}r^{2}\Biggr].
\end{aligned}
\label{eq:H2finiteNc}
\end{equation}

\subsection{Relations between the gluon TMDs}

Let us first calculate the adjoint dipole gluon TMD $xG_{A}^{\left(2\right)}\left(x,q_{\perp}\right)$,
defined earlier in Eq. (\ref{eq:xG2A}). Taking the derivatives of
the correlator of two dipoles, Eq. (\ref{eq:DipDipfiniteNc}), we
obtain:
\begin{equation}
\begin{aligned} & \frac{\partial^{2}}{\partial x^{i}\partial y^{j}}\left.\Bigl\langle\mathrm{Tr}\left(U\left(\mathbf{x}\right)U^{\dagger}\left(\mathbf{y}\right)\right)\mathrm{Tr}\left(U\left(\mathbf{y}\right)U^{\dagger}\left(\mathbf{x}\right)\right)\Bigr\rangle_{x}\right|_{\mathbf{x}=\mathbf{v},\,\mathbf{y}=\mathbf{v}'}\\
 & =\frac{C_{F}}{2}N_{c}^{2}e^{-\frac{N_{c}}{2}\Gamma\left(\mathbf{v}-\mathbf{v}'\right)}\left(N_{c}\frac{\partial}{\partial x^{i}}\Gamma\left(\mathbf{x}-\mathbf{v}'\right)\frac{\partial}{\partial y^{i}}\Gamma\left(\mathbf{v}-\mathbf{y}\right)-2\frac{\partial}{\partial v^{i}}\frac{\partial}{\partial v'^{i}}\Gamma\left(\mathbf{v}-\mathbf{v}'\right)\right),\\
 & =2C_{F}N_{c}\frac{\partial}{\partial v^{i}}\frac{\partial}{\partial v'^{i}}e^{-\frac{N_{c}}{2}\Gamma\left(\mathbf{v}-\mathbf{v}'\right)},
\end{aligned}
\end{equation}
from which we find the following expression for $xG_{A}^{\left(2\right)}\left(x,q_{\perp}\right)$:
\begin{equation}
\begin{aligned}xG_{A}^{\left(2\right)}\left(x,q_{\perp}\right) & =4\frac{C_{F}}{g_{s}^{2}}\frac{1}{N_{c}^{2}-1}2C_{F}N_{c}\int\frac{\mathrm{d}^{2}\mathbf{v}\mathrm{d}^{2}\mathbf{v}'}{\left(2\pi\right)^{3}}e^{-i\mathbf{q}_{\perp}\left(\mathbf{v}-\mathbf{v}'\right)}\frac{\partial}{\partial v^{i}}\frac{\partial}{\partial v'^{i}}e^{-\frac{N_{c}}{2}\Gamma\left(\mathbf{v}-\mathbf{v}'\right)},\\
 & =4\frac{C_{F}}{g_{s}^{2}}\int\frac{\mathrm{d}^{2}\mathbf{v}\mathrm{d}^{2}\mathbf{v}'}{\left(2\pi\right)^{3}}q_{\perp}^{2}e^{-i\mathbf{q}_{\perp}\left(\mathbf{v}-\mathbf{v}'\right)}e^{-\frac{N_{c}}{2}\Gamma\left(\mathbf{v}-\mathbf{v}'\right)},\\
 & =\frac{S_{\perp}C_{F}}{2\alpha_{s}\pi^{2}}\int\frac{\mathrm{d}^{2}\mathbf{r}}{\left(2\pi\right)^{2}}q_{\perp}^{2}e^{-i\mathbf{q}_{\perp}\cdot\mathbf{r}}e^{-\frac{N_{c}}{2}\Gamma\left(\mathbf{r}\right)}.
\end{aligned}
\label{eq:GAcompact}
\end{equation}
From the other derivatives of the correlator of two dipoles, Eqs.
(\ref{eq:deriv2dipdip}) and (\ref{eq:dipdipderiv2}), together with
the definitions of $\mathcal{F}_{gg}^{\left(1\right)}$ and $\mathcal{F}_{gg}^{\left(2\right)}$,
Eq. (\ref{eq:F123}), we easily find:

\begin{equation}
\begin{aligned}\mathcal{F}_{gg}^{\left(1\right)}\left(x,q_{\perp}\right)-\mathcal{F}_{gg}^{\left(2\right)}\left(x,q_{\perp}\right) & =xG_{A}^{\left(2\right)}\left(x,q_{\perp}\right),\end{aligned}
\label{eq:F1-F2}
\end{equation}
and hence the rule Eq. (\ref{eq:F1-F2general}) still holds in the
MV model. Using Eq. (\ref{eq:PIGA}), it is easy to show that the
same is true for the difference between $\mathcal{H}_{gg}^{\left(1\right)}\left(x,q_{\perp}\right)$
and $\mathcal{H}_{gg}^{\left(2\right)}\left(x,q_{\perp}\right)$:
\begin{equation}
\begin{aligned}\mathcal{H}_{gg}^{\left(1\right)}\left(x,q_{\perp}\right)-\mathcal{H}_{gg}^{\left(2\right)}\left(x,q_{\perp}\right) & =xG_{A}^{\left(2\right)}\left(x,q_{\perp}\right).\end{aligned}
\label{eq:H1-H2}
\end{equation}
For the sum of $\mathcal{F}_{gg}^{\left(1\right)}\left(x,q_{\perp}\right)$
and $\mathcal{F}_{gg}^{\left(1\right)}\left(x,q_{\perp}\right)$,
we find:
\begin{equation}
\begin{aligned} & \mathcal{F}_{gg}^{\left(1\right)}\left(x,q_{\perp}\right)+\mathcal{F}_{gg}^{\left(2\right)}\left(x,q_{\perp}\right)\\
 & =-\frac{16}{g_{s}^{2}}\frac{C_{F}}{N_{c}^{2}}\int\frac{\mathrm{d}^{2}\mathbf{v}\mathrm{d}^{2}\mathbf{v}'}{\left(2\pi\right)^{3}}e^{-i\mathbf{q}_{\perp}\cdot\left(\mathbf{v}-\mathbf{v}'\right)}\frac{1-e^{-\frac{N_{c}}{2}\Gamma\left(\mathbf{v}-\mathbf{v}'\right)}}{\Gamma\left(\mathbf{v}-\mathbf{v}'\right)}\frac{\partial}{\partial v^{i}}\frac{\partial}{\partial v'^{i}}\Gamma\left(\mathbf{v}-\mathbf{v}'\right),\\
 & -\frac{2}{g_{s}^{2}}\frac{C_{F}}{N_{c}}\left(N_{c}^{2}-4\right)\int\frac{\mathrm{d}^{2}\mathbf{v}\mathrm{d}^{2}\mathbf{v}'}{\left(2\pi\right)^{3}}e^{-i\mathbf{q}_{\perp}\cdot\left(\mathbf{v}-\mathbf{v}'\right)}e^{-\frac{N_{c}}{2}\Gamma\left(\mathbf{v}-\mathbf{v}'\right)}\frac{\partial}{\partial v^{i}}\frac{\partial}{\partial v'^{i}}\Gamma\left(\mathbf{v}-\mathbf{v}'\right).
\end{aligned}
\end{equation}
In the first term of the above formula, we recognize the Weizsäcker-Williams
gluon TMD $\mathcal{F}_{gg}^{\left(3\right)}\left(x,q_{\perp}\right)$,
with the help of its expression in terms of $\Gamma\left(\mathbf{v}-\mathbf{v}'\right)$,
Eq. (\ref{eq:F3Gamma}). Furthermore, using the formula for the double
derivative of $\Gamma\left(\mathbf{v}-\mathbf{v}'\right)$, Eq. (\ref{eq:derivGamma})
in combination with Eq. (\ref{eq:fourier2log}), as well as the definition
of the saturation scale Eq. (\ref{eq:SaturationScale}), we can easily
show that
\begin{equation}
\begin{aligned}\mathcal{F}_{gg}^{\left(1\right)}\left(x,q_{\perp}\right)+\mathcal{F}_{gg}^{\left(2\right)}\left(x,q_{\perp}\right) & =\frac{4}{N_{c}^{2}}\mathcal{F}_{gg}^{\left(3\right)}\left(x,q_{\perp}\right)+\left(1-\frac{4}{N_{c}^{2}}\right)xG_{q\bar{q}}\left(x,q\right),\end{aligned}
\label{eq:F1+F2}
\end{equation}
where we defined the following auxiliary TMD:
\begin{equation}
\begin{aligned}xG_{q\bar{q}}\left(x,q\right) & \equiv\frac{S_{\perp}N_{c}}{2\pi^{2}\alpha_{s}}\int\frac{\mathrm{d}^{2}\mathbf{r}}{\left(2\pi\right)^{2}}Q_{s}^{2}\left(r^{2}\right)e^{-i\mathbf{q}_{\perp}\cdot\mathbf{r}}e^{-\frac{N_{c}}{2}\Gamma\left(\mathbf{r}\right)}.\end{aligned}
\label{eq:xGqq}
\end{equation}
Likewise, we have for $\mathcal{H}_{gg}^{\left(1\right)}\left(x,q_{\perp}\right)$
and $\mathcal{H}_{gg}^{\left(2\right)}\left(x,q_{\perp}\right)$:
\begin{equation}
\begin{aligned}\mathcal{H}_{gg}^{\left(1\right)}\left(x,q_{\perp}\right)+\mathcal{H}_{gg}^{\left(2\right)}\left(x,q_{\perp}\right) & =\frac{4}{N_{c}^{2}}\mathcal{H}_{gg}^{\left(3\right)}\left(x,q_{\perp}\right)+\left(1-\frac{4}{N_{c}^{2}}\right)xH_{q\bar{q}}\left(x,q\right),\end{aligned}
\label{eq:H1+H2}
\end{equation}
where $xH_{q\bar{q}}\left(x,q\right)$ is defined as follows:
\begin{equation}
xH_{q\bar{q}}\left(x,q\right)=\frac{N_{c}^{2}-1}{8\pi^{3}}S_{\perp}\mu_{A}\int\mathrm{d}r\,rJ_{2}\left(q_{\perp}r\right)e^{-\frac{N_{c}}{2}\Gamma\left(\mathbf{r}\right)}.\label{eq:xHqq}
\end{equation}

\subsection{Large-$N_{c}$ limit}

In the large-$N_{c}$ limit, the computation of $\mathcal{F}_{gg}^{\left(1\right)}\left(x,q_{\perp}\right)$,
$\mathcal{F}_{gg}^{\left(2\right)}\left(x,q_{\perp}\right)$, $\mathcal{H}_{gg}^{\left(1\right)}\left(x,q_{\perp}\right)$
and $\mathcal{H}_{gg}^{\left(2\right)}\left(x,q_{\perp}\right)$ is
considerably easier, since the correlator of two dipoles factorizes,
as we already saw in the context of the BK equation (see Eq. (\ref{eq:largeNcfactorization})).

In addition, the large-$N_{c}$ limit allows us to derive some simple
expressions for $\mathcal{F}_{gg}^{\left(1\right)}$, $\mathcal{F}_{gg}^{\left(2\right)}$,
$\mathcal{H}_{gg}^{\left(1\right)}$ and $\mathcal{H}_{gg}^{\left(2\right)}$,
which are useful to compare with the literature. To this end, we introduce
the dipole gluon distribution $xG^{\left(2\right)}\left(x,q_{\perp}\right)$,
this time in the fundamental representation:
\begin{equation}
\begin{aligned}xG^{\left(2\right)}\left(x,q_{\perp}\right) & \equiv\frac{4N_{c}}{g_{s}^{2}}\int\frac{\mathrm{d}^{2}\mathbf{v}\mathrm{d}^{2}\mathbf{v}'}{\left(2\pi\right)^{3}}e^{-i\mathbf{q}_{\perp}\left(\mathbf{v}-\mathbf{v}'\right)}\left.\frac{\partial^{2}}{\partial x^{i}\partial y^{i}}D\left(\mathbf{x}-\mathbf{y}\right)\right|_{\mathbf{x}=\mathbf{v},\,\mathbf{y}=\mathbf{v}'},\end{aligned}
\label{eq:dipoleTMD}
\end{equation}
which will take over the role of $xG_{A}^{\left(2\right)}\left(x,q_{\perp}\right)$
in the finite-$N_{c}$ calculations. Just like in the case of $xG_{A}^{\left(2\right)}\left(x,q_{\perp}\right)$,
Eq. (\ref{eq:GAcompact}), instead of computing the derivatives of
$\Gamma\left(\mathbf{v}-\mathbf{v}'\right)$ directly, it is more
elegant to perform a partial integration, which yields the compact
formula:
\begin{equation}
\begin{aligned}xG^{\left(2\right)}\left(x,q_{\perp}\right) & =\frac{S_{\perp}q_{\perp}^{2}N_{c}}{2\pi^{2}\alpha_{s}}F\left(x,q_{\perp}\right),\end{aligned}
\label{eq:G2F}
\end{equation}
where we introduced the following auxiliary Gaussian function:
\begin{equation}
\begin{aligned}F\left(x,q_{\perp}\right) & \equiv\int\frac{\mathrm{d}^{2}\mathbf{x}}{\left(2\pi\right)^{2}}e^{i\mathbf{q_{\perp}x}}D\left(\mathbf{x}\right).\end{aligned}
\label{eq:Gaussian}
\end{equation}
We can use this to rewrite $\mathcal{F}_{gg}^{\left(1\right)}$ and
$\mathcal{F}_{gg}^{\left(2\right)}$ as convolutions of $xG^{\left(2\right)}$
and $F$. Indeed, using the large-$N_{c}$ factorization of a correlator
of two dipoles, Eq. (\ref{eq:largeNcfactorization}), we write for
$\mathcal{F}_{gg}^{\left(1\right)}$:
\begin{equation}
\begin{aligned} & \mathcal{F}_{gg}^{\left(1\right)}\left(x,q_{\perp}\right)\\
 & =\frac{4N_{c}}{g_{s}^{2}}\int\frac{\mathrm{d}^{2}\mathbf{v}\mathrm{d}^{2}\mathbf{v}'}{\left(2\pi\right)^{3}}e^{-i\mathbf{q}_{\perp}\left(\mathbf{v}-\mathbf{v}'\right)}\left(\frac{\partial}{\partial v^{i}}\frac{\partial}{\partial v'^{j}}D\left(\mathbf{v}-\mathbf{v}'\right)\right)\int\mathrm{d}^{2}\mathbf{x}\,\delta^{\left(2\right)}\left(\mathbf{x}-\mathbf{v}'+\mathbf{v}\right)D\left(\mathbf{x}\right),\\
 & =\frac{4N_{c}}{g_{s}^{2}}\int\frac{\mathrm{d}^{2}\mathbf{v}\mathrm{d}^{2}\mathbf{v}'}{\left(2\pi\right)^{3}}e^{-i\mathbf{q}_{\perp}\left(\mathbf{v}-\mathbf{v}'\right)}\left(\frac{\partial}{\partial v^{i}}\frac{\partial}{\partial v'^{j}}D\left(\mathbf{v}-\mathbf{v}'\right)\right)\int\mathrm{d}^{2}\mathbf{x}\int\frac{\mathrm{d}^{2}\mathbf{k}_{\perp}}{\left(2\pi\right)^{2}}e^{i\mathbf{k}_{\perp}\left(\mathbf{x}-\mathbf{v}'+\mathbf{v}\right)}D\left(\mathbf{x}\right),\\
 & =\frac{4N_{c}}{g_{s}^{2}}\int\mathrm{d}^{2}\mathbf{k}_{\perp}\int\frac{\mathrm{d}^{2}\mathbf{v}\mathrm{d}^{2}\mathbf{v}'}{\left(2\pi\right)^{3}}e^{-i\left(\mathbf{q}_{\perp}-\mathbf{k}_{\perp}\right)\left(\mathbf{v}-\mathbf{v}'\right)}\left(\frac{\partial}{\partial v^{i}}\frac{\partial}{\partial v'^{j}}D\left(\mathbf{v}-\mathbf{v}'\right)\right)F\left(x,k_{\perp}\right),\\
 & =\int\mathrm{d}^{2}\mathbf{k}_{\perp}xG^{\left(2\right)}\left(x,q_{\perp}-k_{\perp}\right)F\left(x,k_{\perp}\right)=\int\mathrm{d}^{2}\mathbf{k}_{\perp}xG^{\left(2\right)}\left(x,k_{\perp}\right)F\left(x,q_{\perp}-k_{\perp}\right).
\end{aligned}
\label{eq:F1conv}
\end{equation}
In the same fashion, it is straightforward to obtain the following
relations between $\mathcal{F}_{gg}^{\left(1\right)}$, $\mathcal{H}_{gg}^{\left(1\right)}$,
$\mathcal{F}_{gg}^{\left(2\right)}$, $\mathcal{H}_{gg}^{\left(2\right)}$,
$xG^{\left(2\right)}$ and $F$:
\begin{equation}
\begin{aligned}\mathcal{H}_{gg}^{\left(1\right)}\left(x,q_{\perp}\right) & =2\int\mathrm{d}^{2}\mathbf{k}_{\perp}\frac{\left(\mathbf{q}_{\perp}\cdot\mathbf{k}_{\perp}\right)^{2}}{k_{\perp}^{2}q_{\perp}^{2}}xG^{\left(2\right)}\left(x,k_{\perp}\right)F\left(x,q_{\perp}-k_{\perp}\right)-\mathcal{F}_{gg}^{\left(1\right)}\left(x,q_{\perp}\right),\\
\mathcal{F}_{gg}^{\left(2\right)}\left(x,q_{\perp}\right) & =-\int\mathrm{d}^{2}\mathbf{k}_{\perp}\frac{\mathbf{q}_{\perp}\cdot\mathbf{k}_{\perp}}{k_{\perp}^{2}}xG^{\left(2\right)}\left(x,k_{\perp}\right)F\left(x,q_{\perp}-k_{\perp}\right)+\mathcal{F}_{gg}^{\left(1\right)}\left(x,q_{\perp}\right),
\end{aligned}
\label{eq:H1F2H2conv}
\end{equation}
\begin{equation}
\begin{aligned}\mathcal{F}_{gg}^{\left(1\right)}\left(x,q_{\perp}\right)-\mathcal{F}_{gg}^{\left(2\right)}\left(x,q_{\perp}\right) & =\mathcal{H}_{gg}^{\left(1\right)}\left(x,q_{\perp}\right)-\mathcal{H}_{gg}^{\left(2\right)}\left(x,q_{\perp}\right).\end{aligned}
\end{equation}

\section{\label{subsec:GBW}The gluon TMDs in the GBW model}

The GBW model differs from the MV model in that there is no logarithmic
$r$-dependence in the saturation scale:
\begin{equation}
\begin{aligned}Q_{s}^{2} & =\alpha_{s}C_{F}\mu_{A},\quad\mathrm{and}\quad Q_{sg}^{2}\equiv\frac{N_{c}}{C_{F}}Q_{s}^{2}=\alpha_{s}\mu_{A}N_{c},\end{aligned}
\end{equation}
which greatly simplifies calculations. 

Just like in the MV model, starting from the double derivative of
the correlator of two dipoles, Eqs. (\ref{eq:deriv2dipdip}) and (\ref{eq:dipdipderiv2}),
and making the identification: $\Gamma\left(r\right)=r^{2}Q_{sg}^{2}/2N_{c}$,
we obtain the following expressions for the gluon TMDs, only valid
in the GBW model:
\begin{equation}
\begin{aligned}\mathcal{F}_{gg}^{\left(1\right)}\left(x,q_{\perp}\right) & =\frac{1}{16\pi^{2}\alpha_{s}}\frac{C_{F}}{N_{c}^{2}}S_{\perp}\int\frac{\mathrm{d}^{2}\mathbf{r}}{\left(2\pi\right)^{2}}e^{-i\mathbf{q}_{\perp}\cdot\mathbf{r}}e^{-\frac{r^{2}Q_{sg}^{2}}{4}}\\
 & \Biggl[\frac{64}{r^{2}}\left(e^{\frac{r^{2}Q_{sg}^{2}}{4}}-1\right)-N_{c}^{2}r^{2}Q_{g}^{4}+8\left(N_{c}^{2}-2\right)Q_{sg}^{2}\Biggr],\\
\mathcal{H}_{gg}^{\left(1\right)}\left(x,q_{\perp}\right) & =\frac{S_{\perp}}{\alpha_{s}}\frac{1}{32\pi^{3}}C_{F}Q_{sg}^{4}\int\mathrm{d}r\,r^{3}J_{2}\left(q_{\perp}r\right)e^{-\frac{r^{2}Q_{sg}^{2}}{4}},\\
\mathcal{F}_{gg}^{\left(2\right)}\left(x,q_{\perp}\right) & =\frac{1}{16\pi^{2}\alpha_{s}}\frac{C_{F}}{N_{c}^{2}}S_{\perp}\int\frac{\mathrm{d}^{2}\mathbf{r}}{\left(2\pi\right)^{2}}e^{-i\mathbf{q}_{\perp}\cdot\mathbf{r}}e^{-\frac{r^{2}Q_{sg}^{2}}{4}}\\
 & \Biggl[\frac{64}{r^{2}}\left(e^{\frac{r^{2}Q_{sg}^{2}}{4}}-1\right)+N_{c}^{2}r^{2}Q_{sg}^{4}-16Q_{sg}^{2}\Biggr],\\
\mathcal{H}_{gg}^{\left(2\right)}\left(x,q_{\perp}\right) & =-\mathcal{H}_{gg}^{\left(1\right)}\left(x,q_{\perp}\right).
\end{aligned}
\label{eq:F1GBW}
\end{equation}
Moreover, from the formulas we found for $\mathcal{F}_{gg}^{\left(3\right)}\left(x,q_{\perp}\right)$
and $\mathcal{H}_{gg}^{\left(3\right)}\left(x,q_{\perp}\right)$ in
the MV model, Eqs. (\ref{eq:WWfiniteNc}) and (\ref{eq:H3finiteNc}),
we immediately obtain the corresponding expression in the GBW model
by simply neglecting the logarithm in the saturation scale.

In addition, it is easy to derive the following relations, useful
when comparing with the literature:
\begin{equation}
\begin{aligned}\mathcal{F}_{gg}^{\left(1\right)}\left(x,q_{\perp}\right)+\mathcal{F}_{gg}^{\left(2\right)}\left(x,q_{\perp}\right) & =\frac{4}{N_{c}^{2}}\mathcal{F}_{gg}^{\left(3\right)}\left(x,q_{\perp}\right)+\left(1-\frac{4}{N_{c}^{2}}\right)xG_{q\bar{q}}\left(x,q_{\perp}\right),\\
\mathcal{F}_{gg}^{\left(1\right)}\left(x,q_{\perp}\right)-\mathcal{F}_{gg}^{\left(2\right)}\left(x,q_{\perp}\right) & =xG_{A}^{\left(2\right)}\left(x,q_{\perp}\right),\\
\mathcal{H}_{gg}^{\left(1\right)}\left(x,q_{\perp}\right)+\mathcal{H}_{gg}^{\left(2\right)}\left(x,q_{\perp}\right) & =0,\\
\mathcal{H}_{gg}^{\left(1\right)}\left(x,q_{\perp}\right)-\mathcal{H}_{gg}^{\left(2\right)}\left(x,q_{\perp}\right) & =xG_{A}^{\left(2\right)}\left(x,q_{\perp}\right),
\end{aligned}
\label{eq:GBWrelations}
\end{equation}
where we introduced the TMD $xG_{q\bar{q}}\left(x,q_{\perp}\right)$,
defined earlier in Eq. (\ref{eq:xGqq}).

Using the above identities, the unpolarized part, Eq. (\ref{eq:TMDcrosssectionunp}),
of the $pA\to q\bar{q}X$ cross section becomes:
\begin{equation}
\begin{aligned} & \left.\frac{\mathrm{d}\sigma^{pA\rightarrow q\bar{q}X}}{\mathrm{d}\mathcal{P}.\mathcal{S}.}\right|_{GBW}=\frac{\alpha_{s}^{2}}{2C_{F}}\frac{1}{\hat{s}^{2}}x_{p}\mathcal{G}\left(x_{p},\mu^{2}\right)\frac{\hat{t}^{2}+\hat{u}^{2}}{4\hat{u}\hat{t}}\\
 & \times\Biggl\{\frac{\left(\hat{t}-\hat{u}\right)^{2}}{\hat{s}^{2}}xG_{A}\left(x,q_{\perp}\right)+\left(1-\frac{4}{N_{c}^{2}}\right)xG_{q\bar{q}}\left(x,q_{\perp}\right)+\frac{2}{N_{c}^{2}}\mathcal{F}_{gg}^{\left(3\right)}\left(x,q_{\perp}\right)\Biggr\},
\end{aligned}
\end{equation}
confirming the result found in Eq. (54) of Ref. \protect\cite{zhou}. Whether
we agree with the result for the polarization-dependent part, Eq.
(55) in the same reference, depends on the order of operations. Starting
from the GBW model, as we do in the present subsection, the polarized
part, Eq. (\ref{eq:resultcrosssectionpol}) of our CGC cross section
gives:
\begin{equation}
\begin{aligned}\left.\frac{\mathrm{d}\sigma^{pA\rightarrow q\bar{q}X}}{\mathrm{d}\mathcal{P}.\mathcal{S}.}\right|_{\phi,GBW} & =\frac{\alpha_{s}^{2}}{2C_{F}}\frac{1}{\hat{s}^{2}}x_{p}\mathcal{G}\left(x_{p},\mu^{2}\right)\frac{m^{2}}{\tilde{P}_{\perp}^{2}}\cos\left(2\phi\right)\\
 & \times\left\{ \frac{\left(\hat{t}-\hat{u}\right)^{2}}{\hat{s}^{2}}xG_{A}\left(x,q_{\perp}\right)-\frac{2}{N_{c}^{2}}\mathcal{H}_{gg}^{\left(3\right)}\left(x,q_{\perp}\right)\right\} ,
\end{aligned}
\label{eq:crosssectionpolGBW}
\end{equation}
in contradiction with the result quoted in \protect\cite{zhou}:
\begin{equation}
\begin{aligned}\left.\frac{\mathrm{d}\sigma^{pA\rightarrow q\bar{q}X}}{\mathrm{d}\mathcal{P}.\mathcal{S}.}\right|_{\phi} & =\frac{\alpha_{s}^{2}}{2C_{F}}\frac{1}{\hat{s}^{2}}x_{p}\mathcal{G}\left(x_{p},\mu^{2}\right)\frac{m^{2}}{\tilde{P}_{\perp}^{2}}\cos\left(2\phi\right)\\
 & \times\Biggl\{\frac{\left(\hat{t}-\hat{u}\right)^{2}}{\hat{s}^{2}}xG_{A}^{\left(2\right)}\left(x,q_{\perp}\right)+\left(1-\frac{4}{N_{c}^{2}}\right)xH_{q\bar{q}}\left(x,q_{\perp}\right)+\frac{2}{N_{c}^{2}}\mathcal{H}_{gg}^{\left(3\right)}\left(x,q_{\perp}\right)\Biggr\}.
\end{aligned}
\label{eq:resultcrosssectionpolGA}
\end{equation}
However, if we evaluate the cross section (\ref{eq:resultcrosssectionpol})
first in the MV model, and afterwards make the transition to the GBW
model by simply neglecting the logarithms $\ln1/r^{2}\Lambda^{2}$
in the expressions for the gluon TMDs, we recover precisely the above
expression.

\newpage{}

\thispagestyle{simple}
\addcontentsline{toc}{part}{}
\addcontentsline{toc}{section}{References}

\bibliographystyle{TAQEUstyle}
\bibliography{bibliografie}

\end{document}